\documentclass[12pt]{article}

\usepackage{latexsym}
\usepackage{amsmath}
\usepackage{mathrsfs}
\usepackage{multibox}
\usepackage{verbatim}

 \csname
@addtoreset\endcsname{equation}{section}

\def\prd{\pr \cdot}

\def\gz0{\gamma^{0}}

\def\nn{\nonumber}

\def\ket#1{|#1\rangle}

\def\a{\alpha}
\def\b{\beta}
\def\g{\gamma}

\def\d{\delta}
\def\D{\Delta}
\def\e{\epsilon}

\def\z{\zeta}
\def\h{\eta}
\def\th{\theta}
\def\Th{\Theta}

\def\l{\lambda}
\def\L{\Lambda}
\def\m{\mu}
\def\n{\nu}
\def\x{\xi}

\def\P{\Pi}

\def\r{\rho}

\def\s{\sigma}

\def\t{\tau}

\def\vf{\varphi}
\def\c{\chi}

\def\O{\Omega}

\def\cA{{\cal A}}
\def\cB{{\cal B}}
\def\cC{{\cal C}}

\def\cE{{\cal E}}
\def\cF{{\cal F}}

\def\cL{{\cal L}}

\def\cN{{\cal N}}
\def\cO{{\cal O}}

\def\cS{{\cal S}}

\def\cW{{\cal W}}

\def\cY{{\cal Y}}
\def\cZ{{\cal Z}}

\def\be{\begin{equation}}
\def\ee{\end{equation}}
\def\bea{\begin{eqnarray}}
\def\eea{\end{eqnarray}}
\def\ba{\begin{array}}
\def\ea{\end{array}}
\def\bec{\begin{center}}
\def\ec{\end{center}}
\def\ba{\begin{align}}
\def\ena{\end{align}}

\def\pe{\prime}
\def\12{\frac{1}{2}}
\def\fr{\frac}
\def\pr{\partial}
\def\prd{\partial \cdot}
\def\bra{\langle \,}
\def\ket{\, \rangle}
\def\comma{\,,\,}

\def\ra{\rightarrow}

\def\dsl{\not {\! \pr}}
\def\dsll{\not {\! \pr}}

\def\psisl{\not {\!\! \psi}}

\def\esl{\not {\! \epsilon}}

\def\ssl{\not {\! \cal S}}

\def\ssl{\not {\! \cal S}}

\def\psl{\not {\! p}}

\begin{document}

\begin{flushright}
AEI-2009-104
\end{flushright}

\vspace{30pt}

\begin{center}


{\Large\sc Metric-like Lagrangian Formulations for\\[5pt] Higher-Spin Fields of Mixed Symmetry}\\


\vspace{25pt}
{\sc A.~Campoleoni}

{\sl\small
Max-Planck-Institut f\"ur Gravitationsphysik\\
Albert-Einstein-Institut\\
Am M\"uhlenberg 1\\
DE-14476 Golm,\ GERMANY\\[5pt]
e-mail: {\small \it andrea.campoleoni@aei.mpg.de}}

\vspace{80pt} {\sc\large Abstract}\end{center}

We review the structure of local Lagrangians and field equations for free bosonic and fermionic gauge fields of mixed symmetry in flat space. These are first presented in a constrained setting extending the metric formulation of linearized gravity, and then the \mbox{($\g$-)}trace constraints on fields and gauge parameters are eliminated via the introduction of auxiliary fields. We also display the emergence of Weyl-like symmetries in particular classes of models in low space-time dimensions.

\newpage

{\linespread{0.6}\tableofcontents}

\newpage


\section{Introduction}\label{sec:intro}


When combined with General Relativity, the Standard Model of elementary particles provides a description of the fundamental forces of Nature that is in astonishing agreement with the experimental data available up to now. However, these theories rely upon two completely different frameworks, since General Relativity describes \emph{classical} gravitational interactions, while the Standard Model describes the other three fundamental forces at the \emph{quantum} level. On the other hand, some of the most spectacular advances in the history of physics originated from the search for a unified description of previously separated phenomena. Trying to overcome this distinction could thus provide a deep insight on both subjects and even on the very notions of space and time. Moreover, gravity couples to matter and, from a fundamental point of view, it is impossible to forego the need for a quantum description of the gravitational interaction, even if it is supposed to play a relevant role in regimes that lie far away from experimental possibilities. In fact, quantum-gravity effects are expected to occur at energies akin to the Plank scale
$G_N^{\,-1/2} \simeq \, 10^{\,19}$ GeV, to be compared, for instance, with the Fermi scale $G_F^{\,-1/2} \simeq \, 10^{\,2}$ GeV that is still under investigation in current accelerator experiments. At any rate, even without direct experimental inputs, one can try to delineate the key features of a consistent quantum theory of gravity that reduces to General Relativity at low energies.

Various approaches were developed over the years to deal with this problem, but up to now String Theory appears most promising. This may sound surprising looking at the origin of the subject, since string models were introduced at the end of the sixties to describe strong interactions \cite{veneziano}. Their most evident signature is indeed a spectrum with infinitely many massive excitations, of increasing mass and spin. These states are in some sense related to ``vibrational modes'' of the strings, and they immediately recall the trend observed starting from the fifties for hadronic resonances. In the seventies, however, deep inelastic scattering experiments robustly supported the QCD description of strong interactions, where hadronic resonances appear as composite objects. At the same time, Yoneya \cite{yoneda} and independently Scherk and Schwarz \cite{scherk_schwarz} pointed out that one can identify the single massless spin-$2$ particle contained in \emph{all} closed string spectra with the graviton. Furthermore, with a proper tuning of the string tension, the energy needed to excite massive string states becomes so big to justify the lack of their experimental observation. The ubiquitous presence of a spin-$2$ mode in closed string spectra and the UV softness displayed by string models, to be compared with the UV divergent behavior of Einstein gravity \cite{uv}, shortly led to consider String Theory as a paradigm for quantum gravity. This gave rise to an enormous effort aimed at clarifying the properties of this complicated framework\footnote{For a non-technical review of the present status of String Theory, that contains many references to books and technical reviews on the subject, see for instance \cite{review}. Some books on String Theory are also signaled in \cite{books}.}.

Despite the understanding thus attained of some particular aspects of String
Theory, it is worth stressing that a number of fundamental problems are still open. For
instance, from the space-time viewpoint String Theory nowadays is still formulated in a
\emph{background-dependent} manner, without any analogue of the powerful geometric
background independent presentation of Einstein gravity. Aside from a handful of
examples, all solvable string models describe infinitely many fields of increasing mass
and spin interacting in a Minkowski background. This view was made more precise starting
from the eighties, with the development of the second quantized reformulation of String
Theory, called String Field Theory\footnote{While at the free level the second quantized
formulation of all string models was completed in the eighties \cite{free_sft}, the
situation is slightly different at the interacting level. In fact, a Chern-Simons-like
action proposed by Witten \cite{open_sft} nicely describes interacting bosonic open
strings, while for closed strings a long term research activity culminated in a proposal
by Zwiebach \cite{closed_sft}, that presents some unusual features with respect to
ordinary Quantum Field Theory.}. This framework connects somehow the string proposal
to Quantum Field Theory on a flat background, despite the presence of rather
unconventional unbounded mass spectra and of higher-derivative interactions. Einstein
gravity is thus recovered via the self-interactions of massless spin-$2$ quanta in a
Minkowski background, while the pure-gravity UV divergences \cite{uv} are cured via the
exchange of infinite towers of massive modes with spin $s > 2$.

In conclusion, String Theory somehow restates the old idea that the framework of Quantum Field Theory could also account for gravitational phenomena, but at the price of considering also \emph{higher-spin excitations}. The new states that appear at the string scale weaken the gravitational potential and restore the high-energy predictivity that is lost in the non-renormalizable spin-$2$ model. This path qualitatively resembles that leading from the Fermi theory of weak interactions to the Standard Model. It is thus natural to ask whether the string picture is compelling in order to obtain a UV completion of Einstein gravity, or whether the key is rather to be found in the coupling with higher spins\footnote{Actually, the answer could be even simpler, since the recent improvement of the computational techniques for loop amplitudes provides some encouraging evidence to the effect that at least $\cN = 8$ supergravity \cite{sugra8} might be finite in $D = 4$. Recent results on this subject are reported in \cite{loop_sugra}, that also contains a wide list of references to the previous literature. The possible finiteness of $\cN = 8$ supergravity could drastically simplify the present understanding of the features requested by a quantum theory of gravity, but the string UV completion could continue to play an important role in view of its possible phenomenological applications, related to lower values of $\cN$.}. Moreover, a careful analysis of the properties of these complicated systems of interacting higher-spin fields could also unveil the path toward a potential background-independent framework. The quest for possible generalizations of the string setup and for a deeper understanding of its peculiar features is indeed one of the strongest motivations for the work reviewed here, as well as for much of the current literature on higher spins\footnote{For reviews on the current status of higher-spin field theories see \cite{reviewHS,reviews_vasiliev} and references therein.}.

The study of higher-spin field theories has nevertheless a long history: Majorana proposed Lorentz covariant wave equations for arbitrary spin particles already in 1932 \cite{majorana}, and some years later Dirac independently re-examined the problem \cite{dirac}. At the end of the thirties Wigner put the subject on a more solid basis, and clearly pointed out that elementary particles can be classified by the unitary irreducible representations of the isometry group of space-time \cite{wigner}. In the four-dimensional Minkowski case this is the Poincar\'e group $ISO(1,3)$, whose irreducible representations are labeled by mass squared and spin. These values are not restricted a priori, and the analysis of all corresponding linear covariant wave equations was completed at the end of the forties \cite{bargmann_wigner}. In the meantime the first Lagrangians for spin-$2$ and spin-($3/2$) particles were obtained by Fierz and Pauli \cite{fierz,fierz_pauli} and by Rarita and Schwinger \cite{rarita_schwinger}. However, even for the free theory, the search for a Lagrangian formulation for arbitrary higher-spin fields was not completed in this first stage. The problem was reconsidered only starting from the fifties, and it was then solved for spin $\frac{5}{2} \leq s \leq 4$ in \cite{moldauer,chang}.

Finally, in the mid seventies Singh and Hagen proposed Lagrangians for free \emph{massive} particles of arbitrary spin in \cite{singh_bose,singh_fermi}. In a four-dimensional Minkowski background the Lagrangian description of the free dynamics was then completed in 1978 by Fronsdal and Fang \cite{fronsdal,fang_fronsdal}, who considered the \emph{massless} limit of the Singh-Hagen construction and showed the expected emergence of a gauge symmetry. Shortly thereafter the Fang-Fronsdal results were actually recovered conjecturing the presence of an abelian gauge symmetry \cite{curt_gauge,dewit_fr} and analyzing the consequences of this request. The resulting massless free theory follows the lines of the metric formalism for gravity, and involves the smallest field content giving a covariant description. The basic fields are fully symmetric tensors $\vf_{\,\m_1 \ldots\, \m_s}$ and spinor-tensors $\psi^{\,\a}{}_{\m_1 \ldots\, \m_s}$, that generalize the vector potential $A_{\,\m}$, the linearized metric fluctuation $h_{\,\m\n}$ and the gravitino field $\psi^{\,\a}{}_{\m}$. Their gauge transformations read
\be \label{intro_gauge}
\d\, \vf_{\,\m_1 \ldots\, \m_s} \, = \, \pr_{\,(\,\m_1\,} \L_{\,\m_2 \ldots\, \m_s\,)} \, , \qquad\qquad
\d\, \psi^{\,\a}{}_{\m_1 \ldots\, \m_s} \, = \, \pr_{\,(\,\m_1\,} \e^{\,\a}{}_{\m_2 \ldots\, \m_s\,)} \, ,
\ee
where here and in the following a couple of parentheses denotes a complete symmetrization of the indices it encloses, with the minimum possible number of terms and with \emph{unit} overall normalization.
However, differently from the ``low-spin'' examples with $s \leq 2$, in the general case the fields are subjected to algebraic \emph{\mbox{($\g$-)}trace constraints} inherited from the massive free theory,
\be \label{intro_fieldconstr}
\vf_{\,\m_1 \ldots\, \m_{s-4}\,\l\r}{}^{\,\l\r} \, = \, 0 \, , \qquad\qquad\qquad \psisl{}_{\,\m_1 \ldots\, \m_{s-3}\,\l}{}^{\,\l} \, = \, 0 \, .
\ee
Similar constraints also appear for the gauge parameters,
\be \label{intro_gaugeconstr}
\L_{\,\m_1 \ldots\, \m_{s-3}\,\l}{}^{\,\l} \, = \, 0 \, , \qquad\quad\qquad\qquad \esl_{\,\m_1 \ldots \m_{s-2}} \, = \, 0 \, .
\ee
The role played by these constraints will be discussed in great detail in the following. To conclude this first overview of the subject, let us recall that a frame-like formulation of the free theory \cite{vasiliev_frame} was also developed by Vasiliev and others starting from the beginning of the eighties. While the Fang-Fronsdal setup extends to higher-spin fields the linearized Einstein-Hilbert presentation of gravity, the Vasiliev setup extends somehow its Cartan-Weyl form in the McDowell-Mansouri presentation \cite{mcdowell}.

Looking at this brief summary of the evolution of the \emph{free} theory, the large time span separating Majorana's paper and Fronsdal's work could appear surprising. Aside from technical issues and from the changing fortunes of field theory, this can be reasonably related to the difficulties that were soon encountered when trying to introduce interactions for higher-spin particles. But before commenting further on interactions, let us continue for a while to focus on the free theory, in order to better delineate the content of this review.

In fact, the Singh-Hagen \cite{singh_bose,singh_fermi} and the Fang-Fronsdal results \cite{fronsdal,fang_fronsdal} can be improved in two directions. On the one hand, one can look for alternative formulations of the free dynamics that are more likely to lead to an interacting theory. The already recalled frame-like formulation of \cite{vasiliev_frame} is the first example. One the other hand, one can follow the old idea of Kaluza and Klein \cite{kaluza_klein}, and consider a space-time with additional dimensions. This is possible if the extra dimensions are compact and have a characteristic scale so small to explain why they have not been detected so far in experiments. Let us start considering this latter possibility.

Models with extra dimensions were originally developed to unify Einstein gravity with electromagnetism, but this idea is crucial in String Theory. Indeed, string models can be quantized without breaking the Lorentz symmetry only in $D=26$ for bosonic strings and only in $D = 10$ for superstrings\footnote{For a clear explanation of this statement see for instance \cite{books}, where detailed references to the papers where the existence of critical dimensions was first noticed can be also found.}. If the dimension of space-time is greater than five the recalled results do not suffice to depict all representations of the isometry group of a flat background. Further steps are thus needed in order to covariantly describe the propagation of arbitrary free modes. For instance, in the massless case the irreducible representations of the Poincar\'e group $ISO(1,D-1)$ are built upon the irreducible representations of the little group $SO(D-2)$. These are no longer labeled by a single half-integer, the spin, but rather by Young tableaux with $N \leq \frac{D-2}{2}$ rows. The Fang-Fronsdal results only apply to single-row tableaux in arbitrary dimensions, corresponding to fully symmetric fields $\vf_{\,\m_1 \ldots\, \m_s}$ or $\psi^{\,\a}{}_{\m_1 \ldots\, \m_s}$. The other \emph{``mixed-symmetry''} cases, corresponding to arbitrary Young tableaux, are covariantly described by more general \mbox{(spinor-)}tensors $\vf_{\,\m_1 \ldots\, \m_{s_1},\, \n_1 \ldots\, \n_{s_2}, \,\ldots}$ and $\psi^{\,\alpha}{}_{\m_1 \ldots\, \m_{s_1},\, \n_1 \ldots\, \n_{s_2}, \,\ldots}$, that possess several families of fully \mbox{(anti)}symmetric space-time indices and that require a dedicated treatment.

The development of String Field Theory stimulated the study of mixed-symmetry fields in the mid eighties, even if some discussions date back to previous years \cite{curt_mixed}. In fact, all string spectra contain fields of mixed symmetry, that are associated to products of different types of bosonic or fermionic oscillators. Free String Field Theory thus provides a covariant description of arbitrary massive modes, at least in $D = 10$ or in $D = 26$. A number of authors then worked on how to extend this result also to the massless case. This should not only be considered as an ``academic problem'', since the dependence of couplings and masses on the same parameter $\a^{\,\pe}$ has long suggested the idea that String Theory could be a broken phase of a higher-spin \emph{gauge} theory. However, the breaking mechanism at work still lacks a precise formulation, and a proper analysis of this problem requires a full control of mixed-symmetry gauge fields.

After some partial results regarding particular classes of mixed-symmetry fields \cite{mixed1}, several covariant formulations for arbitrary massless modes were proposed. In \cite{mixed2,triplets} gauge-invariant actions for Bose fields were derived using BRST techniques. These covariant formulations are related to the $\alpha^{\,\prime} \to \infty$ limit of Free String Field Theory, and thus involve a number of auxiliary fields. Furthermore, they naturally describe reducible representations of the Poincar\'e group. Another approach was then developed for both Bose and Fermi fields via an extension of the light-cone representations of the Poincar\'e algebra \cite{siegel}.

Finally, Labastida proposed a set of covariant field equations for arbitrary massless Bose and Fermi fields of mixed symmetry exploiting an abelian gauge symmetry \cite{laba_bose,laba_fermi}. His formulation extends the Fang-Fronsdal setting of \cite{fronsdal,fang_fronsdal}: it involves multi-symmetric tensors $\vf_{\,\m_1 \ldots\, \m_{s_1},\, \n_1 \ldots\, \n_{s_2}, \,\ldots}$ and \mbox{(spinor-)}tensors $\psi^{\,\alpha}{}_{\m_1 \ldots\, \m_{s_1},\, \n_1 \ldots\, \n_{s_2}, \,\ldots}$, possessing several families of symmetrized space-time indices. It thus relies on the minimal field content required by a covariant description\footnote{Actually, without any symmetry relating the various index families these fields only describe reducible representations of the Lorentz group. At any rate, as we shall see more in detail in the following, performing standard Young projections they can be easily adapted to describe irreducible representations.}. On the other hand, the gauge transformations take a more complicated form in the mixed-symmetry case, since they involve one independent gauge parameter for each index family,
\begin{align}
& \d \, \vf_{\,\m_1 \ldots\, \m_{s_1},\, \n_1 \ldots\, \n_{s_2}, \,\ldots} =\, \pr_{\,(\,\m_1\,} \L^{(1)}{}_{\!\m_2 \ldots\, \m_s\,)\,,\,\n_1 \ldots\, \n_{s_2},\,\ldots} + \, \pr_{\,(\,\n_1\,|\,} \L^{(2)}{}_{\!\m_1 \ldots\, \m_s\,,\,|\,\n_2 \ldots\, \n_{s_2}\,)\,,\,\ldots} + \ldots \,, \nn \\[5pt]
& \d\, \psi^{\,\a}{}_{\,\m_1 \ldots\, \m_{s_1},\, \n_1 \ldots\, \n_{s_2}, \,\ldots} = \, \pr_{\,(\,\m_1\,} \e^{\,(1)\,\a}{}_{\!\m_2 \ldots\, \m_s\,)\,,\,\n_1 \ldots\, \n_{s_2},\,\ldots} + \,\ldots \, .
\end{align}
Qualitative features of the fully symmetric case still persist, and fields and gauge parameters satisfy some \mbox{($\g$-)}trace constraints. However, in this case it is possible to consider different \mbox{($\g$-)}traces, related to different index families, and only some of their linear combinations are forced to vanish.

The field equations of \cite{laba_bose,laba_fermi} are non-Lagrangian\footnote{In this review the term denotes a set of field equations that were originally obtained not starting from an action principle, but may or may not be eventually connected to a less conventional one.}, but in \cite{laba_lag} Labastida also built Lagrangians leading to equivalent equations of motion for Bose fields\footnote{In some low-dimensional cases the equivalence between the two setups actually involves some subtleties that were discussed in \cite{mixed_bose} and that will be analyzed in Section \ref{chap:weyl}.}. On the other hand, for Fermi fields the search for a Lagrangian formulation initiated long ago by Fierz and Pauli \cite{fierz_pauli} was completed only seventy years later in \cite{mixed_fermi}. While \cite{laba_lag} and \cite{mixed_fermi} contain only Lagrangians for massless fields of mixed symmetry, the corresponding massive theories can be obtained by a standard Kaluza-Klein reduction.

The structure of the Lagrangians for mixed-symmetry fields is rather involved, since they
contain a number of higher \mbox{($\g$-)}traces of the fields that grows proportionally
to the number of index families. This is a consequence of the more complicated form of
the \mbox{($\g$-)}trace constraints. The procedure to build the Lagrangians and
their structure will be discussed in detail in Section \ref{chap:constrained}, while
Section \ref{chap:weyl} will show how this involved form leads to the emergence of
Weyl-like symmetries in low enough space-time dimensions. This review is in fact mainly
based on the content of \cite{mixed_fermi} and of its companion paper \cite{mixed_bose},
where the Labastida formulation for Bose fields was reconsidered in order to embed it in
an \emph{unconstrained} framework.

In fact, the Fang-Fronsdal \cite{fronsdal,fang_fronsdal} and Labastida
\cite{laba_lag,mixed_fermi} descriptions of the free dynamics are compact and elegant,
but they present at least an undesirable feature. It is the unusual presence of off-shell
constraints, needed to ensure the propagation of the correct number of degrees of
freedom. These constraints are algebraic and thus in principle should not obstruct a
non-linear deformation of the theory, but in the tensionless limit $\alpha^{\,\prime} \to \infty$ string spectra do not quite result in sums of fields subject to the Labastida constraints. For instance, in this limit the fully symmetric modes of the first Regge trajectory of the open bosonic string are described by interesting systems of fields usually called ``bosonic triplets'' \cite{triplets,fs1,st}, that propagate reducible representations including, together with a given set of spin-$s$ modes, lower spin chains as well and do not rest on Fronsdal-like trace constraints. In a similar fashion, the fermionic counterparts of these triplets, defined in \cite{fs1,st}, play a role in orientifolds \cite{orientifolds} of the 0B theory \cite{0b}, although not in superstrings as a result of their GSO projections \cite{gso}. In general, it is thus natural to ask how one can eliminate the constraints and, as the string example suggests, whether their removal could be of help when attempting to switch on interactions.
The first concrete step in this direction was performed by Pashnev, Tsulaia, Buchbinder and collaborators \cite{brst} starting from the late nineties. These authors developed further the BRST techniques
adopted in \cite{mixed2,triplets}, in order to describe the propagation of a single
massless spin-$s$ mode. They also treated fermions and extended these results to constant
curvature backgrounds, but for a spin-$s$ field in their construction the constraints are
removed at the price of introducing \emph{$\cO(s)$ auxiliary fields}.

An alternative solution to the problem was proposed in 2002 by Francia and Sagnotti for fully symmetric fields \cite{nonlocal}. They eliminated the Fang-Fronsdal constraints without introducing any additional fields, via Lagrangians and field equations that are \emph{non local}. On the other hand, the non-localities can be eliminated by a partial gauge fixing involving the \mbox{($\g$-)}trace of the gauge parameters, that recovers the Fang-Fronsdal presentation. The non-local terms are thus harmless, and their introduction also gives an additional benefit: in fact, the resulting field equations and Lagrangians find a role for the higher-spin curvatures introduced by de Wit and Freedman \cite{dewit_fr,damour_deser}. These objects are gauge invariant under unconstrained gauge transformations, and sit at the top of a recursive chain of generalized connections that are characterized by simple gauge variations. Higher-spin connections and curvatures thus naturally extend the linearized Christoffel symbols and the Riemann tensor of General Relativity. Even if at the free level one can only display these algebraic similarities, they provide hints on a possible geometrical non-linear realization of higher-spin models. Further details and comments on this interesting alternative can be found in \cite{fs1,tesi_dario}. This formulation of the free theory for symmetric fields reached its final form in \cite{fms1}, with the identification of non-local Lagrangians leading to the correct current-current exchanges. Shortly after \cite{nonlocal} appeared, the Francia-Sagnotti results were instead extended to arbitrary mixed-symmetry massless Bose fields in \cite{nonlocal_mixed}, and more recently also to the case of massive fully symmetric Bose and Fermi fields \cite{dario_massive}. Similarly, unconstrained Lagrangians for both massless and massive mixed-symmetry fields were proposed in the context of the already mentioned BRST approach \cite{brst_mixed}.

The two presentations recalled here eliminate the constraints, but present again some unusual features: in the fully symmetric case, for instance, one has to face field equations containing either $\cO(s)$ auxiliary fields or $\cO(s)$ derivatives of the basic spin-$s$ field. A simpler solution was developed for fully symmetric fields by Francia and Sagnotti in \cite{fs1}, observing that the Fang-Fronsdal field equations
\begin{align}
& \cF_{\,\m_1 \ldots\, \m_s} \, \equiv \ \Box \, \vf_{\,\m_1 \ldots\, \m_s} \, - \, \pr_{\,(\,\m_1\,} \prd \vf_{\,\m_2 \ldots\, \m_s\,)} \, + \, \pr_{\,(\,\m_1\,}\pr_{\phantom{(}\!\m_2}\, \vf_{\,\m_3 \ldots\, \m_s\,)\,\l}{}^{\,\l} \, = \, 0 \, , \nn \\[5pt]
& \cS^{\,\a}{}_{\m_1 \ldots\, \m_s} \, \equiv \ i \, \left\{\, \dsl \ \psi^{\,\a}{}_{\m_1 \ldots\, \m_s} \, - \, \pr_{\,(\,\m_1}\! \psisl^{\,\, \a}{}_{\m_2 \ldots\, \m_s\,)} \,\right\} \, = \, 0 \, , \label{intro_eq}
\end{align}
vary under the \emph{unconstrained} gauge transformations \eqref{intro_gauge} as
\begin{align}
& \d \, \cF_{\, \m_1 \ldots\, \m_s} \, = \, 3\  \pr_{\,(\,\m_1\,}\pr_{\phantom{(}\!\m_2\,}\pr_{\phantom{(}\!\m_3}\, \L_{\,\m_4 \ldots\, \m_s\,)\,\l}{}^{\,\l} \, , \nn \\[5pt]
& \d \, \cS^{\,\a}{}_{\m_1 \ldots\, \m_s} \, = \, - \, 2\,i\ \pr_{\,(\,\m_1\,}\pr_{\phantom{(}\!\m_2}\! \esl^{\,\,\a}{}_{\m_3 \ldots\, \m_s\,)} \, . \label{intro_gaugevar}
\end{align}
For both Bose and Fermi fields, introducing a \emph{single} auxiliary field one can thus consider the equivalent field equations\footnote{For a spin-$3$ field an equation of motion of the form \eqref{intro_equnc} was actually proposed long ago by Schwinger in \cite{schwinger}.}
\begin{align}
& \cA_{\,\m_1 \ldots\, \m_s} \, \equiv \ \cF_{\,\m_1 \ldots\, \m_s} \, - \, 3\  \pr_{\,(\,\m_1\,}\pr_{\phantom{(}\!\m_2\,}\pr_{\phantom{(}\!\m_3}\, \a_{\,\m_4 \ldots\, \m_{s}\,)} \, = \, 0 \, , \nn \\[5pt]
& \cW^{\,\a}{}_{\m_1 \ldots\, \m_s} \, \equiv \ \cS^{\,\a}{}_{\m_1 \ldots\, \m_s} \, + \, 2\,i\ \pr_{\,(\,\m_1\,}\pr_{\phantom{(}\!\m_2}\, \x^{\,\,\a}{}_{\m_3 \ldots\, \m_s\,)} \, = \, 0 \, , \label{intro_equnc}
\end{align}
that display the unconstrained gauge symmetry
\be \label{intro_gaugeunc}
\left\{
\begin{array}{l}
\d \, \vf_{\,\m_1 \ldots\, \m_s} = \, \pr_{\,(\,\m_1\,} \L_{\,\m_2 \ldots\, \m_s\,)} \, , \\[5pt]
\d \, \a_{\,\m_1 \ldots\, \m_{s-3}} = \, \L_{\,\m_1 \ldots\, \m_{s-3}\,\l}{}^{\,\l} \, ,
\end{array}
\right.
\qquad\quad\qquad
\left\{
\begin{array}{l}
\d \, \psi^{\,\a}{}_{\m_1 \ldots\, \m_s} = \, \pr_{\,(\,\m_1\,} \e^{\,\a}{}_{\,\m_2 \ldots\, \m_s\,)} \, , \\[5pt]
\d \, \x^{\,\a}{}_{\m_1 \ldots\, \m_{s-2}} = \ \, \esl^{\,\,\a}{}_{\m_1 \ldots\, \m_{s-2}} \, .
\end{array}
\right.
\ee
Eqs.~\eqref{intro_equnc} reduce to the Fang-Fronsdal field equations after a partial gauge-fixing that does not spoil the original constrained gauge symmetry, since it involves the \mbox{($\g$-)trace} of the gauge parameter. As those in eq.~\eqref{intro_eq}, the equations of motion \eqref{intro_equnc} are non-Lagrangian, but in \cite{fs2} Lagrangians leading to equivalent field equations were also presented (for reviews see \cite{fs3,fms1}). For Bose fields they involve a further \emph{single} additional Lagrange multiplier transforming as the triple divergence of the gauge parameter,
\be
\d \, \b_{\,\m_1 \ldots\, \m_{s-4}} \, = \, \pr\cdot\pr\cdot\pr\cdot \L_{\,\m_1 \ldots\, \m_{s-4}} \, ,
\ee
while for Fermi fields they involve a \emph{single} additional Lagrange multiplier transforming as
\be
\d \, \l^{\,\a}{}_{\m_1 \ldots\, \m_{s-3}} \, = \, \pr\cdot\pr\cdot \e^{\,\a}{}_{\m_1 \ldots\, \m_{s-3}} \, .
\ee
In this context current-current exchanges were also considered in both Minkowski \cite{fms1} and constant curvature backgrounds \cite{fms2}.

In conclusion, in the fully symmetric case an unconstrained metric-like formulation can be obtained at the modest price of introducing two auxiliary fields, for both Bose and Fermi fields. With the same logic it is also possible to obtain a ``minimal'' unconstrained formulation for arbitrary fields of mixed symmetry, introducing at most one compensator for each Labastida constraint on the gauge parameters and at most one Lagrange multiplier for each Labastida constraint on the field. This program was completed in \cite{mixed_bose,mixed_fermi}, and the first part of Section \ref{chap:unconstrained} will be devoted to this issue. To conclude, let us stress that from an orthodox point of view the field equations \eqref{intro_equnc} also contain higher-derivative terms, and the same is true for the corresponding Lagrangian field equations. However, all higher derivative terms involve the compensators and as such are manifestly pure gauge contributions. They are thus not expected to create any problems, but for fully symmetric Bose fields a really ``orthodox solution'' to the problem, with only unconstrained fields, a fixed number of auxiliary fields and a two-derivative Lagrangian was proposed in \cite{buch_tripl}. It was obtained via an elegant off-shell truncation of the triplets of String Field Theory \cite{triplets}, and was followed by another proposal \cite{dario_massive} that is more akin to the spirit of the minimal unconstrained formulation. Although these solutions rely upon a wider field content, they represent an interesting alternative to the minimal setup. Such an alternative formulation was also presented for Bose and Fermi fields of mixed symmetry in \cite{mixed_bose,mixed_fermi}. It was obtained via a proper reformulation of the approach of \cite{dario_massive}, and will be discussed in the second part of Section \ref{chap:unconstrained}.

Aside from its simplicity, the minimal formulation of \cite{fs2,fs3,fms1,mixed_bose,mixed_fermi} has the advantage of leading easily to the non-local formulation via the elimination of the auxiliary fields in terms of the basic field. On the other hand, it was linked to the triplets systems emerging from the $\a^{\,\pe} \to \infty$ limit of String Field Theory in \cite{st,fs2}, and so also to the related BRST-like constructions. It thus somehow bridges the gap between the algebraic ``string inspired'' framework and the geometrical non-local setup.

We can now conclude this overview of the present status of the free theory recalling that
Lagrangians and field equations for mixed-symmetry fields were also recently obtained in
a frame-like approach \cite{mixed_frame} and in its unfolded reformulation
\cite{mixed_unfolded}, both in flat and in constant curvature spaces. In general,
higher-spin interactions present technical (and conceptual) difficulties related to the
presence of higher derivatives, and the term ``unfolding'' denotes a systematic
procedure aimed at eliminating them via the introduction of additional fields
\cite{vasiliev,vasiliev_final,reviews_vasiliev}. While this happens also in String Field
Theory, at present there is a crucial difference, since in the resulting unfolded
higher-spin systems the field equations are of first order. For fully symmetric fields
this line of development led to the Vasiliev equations, that describe non-linear
interactions in an \mbox{(Anti-)}de Sitter background
\cite{vasiliev,vasiliev_final,reviews_vasiliev}, and one natural goal is clearly to
understand how a similar result can be also attained for mixed-symmetry fields. On the
other hand, some steps toward the extension of the Labastida results to constant
curvature spaces were made in \cite{demedeiros}, while these efforts have been
accompanied during the years by a world-line first-quantized formulation of the dynamics
of higher-spin fields \cite{worldline}. Finally, mixed-symmetry fields have also been treated
in a light-cone formalism \cite{mixed_lightcone}. Although not covariant, this approach
can in fact efficiently deal with some interaction issues.

This excursus on higher spins was mainly driven by analogies with String Theory. They strongly motivated the study of mixed-symmetry representations of the higher-dimensional Lorentz groups, and also the analysis of the possible unconstrained realizations of the free theory. On the other hand, independent motivations for the development of new tools for higher spins come from the difficulties encountered when dealing with their interactions. Even if this review focuses on some aspects of the free theory, let us conclude this introduction by briefly recalling the present constraints on the features of a would be systematic development of the non-linear theory.

Amusingly, some rough intuitions of possible difficulties appeared about eighty years ago in the first, but almost ignored\footnote{A notable exception was given by Wigner, who referred to the Majorana paper in his time-honored 1939 paper \cite{wigner}. The Majorana approach was also recently reconsidered in \cite{majorana_upgrade}.}, paper on the subject \cite{majorana}. The Majorana equations indeed rely on an infinite chain of higher-spin fields, and the author noticed that switching on interactions may forbid the decoupling of a single spin-$s$ mode from the infinite chain! Clearer hints on the differences between interacting systems of higher-spin fields and their $s \leq 2$ counterparts came from the work of Fierz and Pauli \cite{fierz_pauli}. Actually, they looked for an action principle in order to overcome the problems displayed by the field equations proposed by Fierz \cite{fierz} for spin-$s$ fields,
\begin{align}
& \!\left(\,\Box \, + \, m^{\,2}\,\right) \vf_{\,\m_1 \ldots\, \m_{s}} \, = \, 0 \, , \nn \\[2pt]
& \pr\cdot \vf_{\,\m_1 \ldots\, \m_{s-1}} \, = \, 0 \, , \nn \\[2pt]
& \vf_{\,\m_1 \ldots\, \m_{s-2}\,\l}{}^{\,\l} \, = \, 0 \, ,
\end{align}
that are no longer compatible when one tries to couple them to an electromagnetic field via the minimal coupling $\pr_{\,\m} \to \, \pr_{\,\m} - \,e\,A_{\,\m}$. Similar complications also emerged in the work of Dirac \cite{dirac}.

As we have already seen in the previous pages, the activity on higher spins was then frozen for some years, probably also due to the lack of clearcut phenomenological motivations. The situation drastically changed with the discovery of higher-spin resonances, and the problem was then reconsidered in more modern terms. At any rate, rather than surpassing the old difficulties the new techniques signaled even more severe drawbacks. For instance, in \cite{velo_zwanziger} it was pointed out that the minimal coupling with an external electromagnetic field leads to noncausalities or to the propagation of spurious degrees of freedom even for spin-($3/2$) and spin-$2$ particles\footnote{Actually, more recently it was shown that, at least for spin-$2$ \cite{nappi} and spin-($3/2$) particles \cite{porrati_raman}, one can overcome these problems via a judicious choice of non-minimal interactions.}. Similar problems were identified in the minimal coupling of spin-$2$ particles with gravity  \cite{aragone_deser1}. Moreover, in the same decade Weinberg realized that, under reasonable assumptions essentially related to analogies with gravity, massless higher-spin particles cannot give rise to long-range interactions in an S-matrix scheme \cite{weinberg}. Coleman and Mandula then proved a famous theorem \cite{coleman_mandula} stating that, again under reasonable assumptions, a Quantum Field Theory leading to a non-trivial S-matrix can only admit a symmetry algebra given by the direct sum of the Poincar\'e algebra and of an internal algebra, whose generators commute with the Poincar\'e ones. This rules out conventional gauge theories with a \emph{finite} number of gauge fields of spin $s \geq 3$, that would be associated to forbidden symmetry generators.

More vigorous attempts followed the discovery of supergravity \cite{sugra}, that displayed a new mechanism enabling the coupling of gravity with spin-($3/2$) particles. On the other hand, it was soon realized that the supergravity ``miracle'' cannot be extended to higher-spin particles \cite{aragone_deser2,aragone_deser3}. Schematically, the main difference is that, after the minimal coupling with gravity, the gauge variation of the Rarita-Schwinger action is proportional to the Ricci tensor, and as such can be compensated by the variation of the gravitational action. On the other hand, starting from spin $5/2$ this is no longer true \cite{aragone_deser2}, and the whole Riemann tensor appears in the gauge variation of the higher-spin action. The variation of the gravitational action cannot cancel the Weyl tensor, while the loss of gauge invariance would lead at least to the propagation of spurious degrees of freedom. For spin $2$ the situation is slightly more subtle, but the conclusions remain the same \cite{aragone_deser3}. Moreover, up to now no consistent modifications of the couplings or of the transformation laws were found, so that a proper generalization of the known gauge symmetry could be recovered. This fact is at the origin of the common lore about the impossibility of consistently coupling higher-spin fields with gravity\footnote{It is also the origin of the upper limit of 32 supercharges considered in supergravity theories. Crossing the border of $\cN=8$ the graviton multiplet would indeed contain also higher-spin helicities.}, and all these results were then summarized in the extension of the Coleman-Mandula theorem due to Haag, Lopuszanski and Sohniuns \cite{haag}. These authors essentially confirmed the conclusions of \cite{coleman_mandula}, barring the inclusion of supersymmetry, that can be taken into account considering graded algebras rather than only ordinary ones. Other no-go arguments that confirm and extend the classical results recalled here were also presented more recently in \cite{nogo_new}.

In spite of these apparently discouraging no-go results, the problem was reconsidered at the beginning of the eighties by the G\"oteborg group, that identified surprising cubic vertices for self-interacting massless particles of any spin in a light-cone formalism \cite{bbb}. This result and the advent of String Field Theory gave a new mighty input to the field, and various cubic vertices were found in flat-space in a light-cone formalism \cite{vertices_light,mixed_lightcone}, in a covariant approach \cite{bbvd1,bbvd2,bbvd3} or resorting to BRST techniques \cite{vertices_brst}. On the other hand, the severe problems encountered at the quartic level led to the realization that consistent interactions for spin $s > 2$ particles should involve \emph{infinitely} many higher-spin fields at the same time \cite{bbvd2,nogo_spin3}. This observation perfectly agrees with the most evident signature of string spectra, and also bypasses the Coleman-Mandula theorem, since it contrasts one of its basic assumptions.

After these positive results, also supported and completed by more recent analyses
\cite{interactions_recent}, Fradkin and Vasiliev discovered a cubic vertex for
higher-spin fields in interaction with gravity \cite{fv}. They forewent the problems
identified by Aragone and Deser \cite{aragone_deser1,aragone_deser2} introducing a
cosmological constant $\L$, which allows the introduction of higher-derivative couplings
depending on inverse powers of $\L$. Notice that the presence of higher-derivative
interactions suffices to bypass the available no-go results, and indeed non-minimal cubic
vertices describing the interactions of massless higher-spin fields with gravity were
recently singled out even in a Minkowski background \cite{seeds}. A non-vanishing
cosmological constant $\L$ simply leads to them in a natural way, without the need of
introducing a new dimensionful coupling constant. At any rate, the Fradkin-Vasiliev
result is a milestone for the modern view of higher-spin theories. In fact, in the
following years Vasiliev built consistent non-linear field equations for an infinite set
of fully symmetric higher-spin fields interacting on an \mbox{(Anti-)}de Sitter background of
arbitrary dimension \cite{vasiliev,vasiliev_final}. To reach this goal he
essentially gauged an infinite-dimensional extension of the Lorentz algebra
\cite{HSalgebra} within an unfolded formulation of the dynamics (for a review of the
Vasiliev setup see, for instance, \cite{reviews_vasiliev}, while interesting related
contributions are due to Sezgin and Sundell \cite{ss}). The properties of the
resulting non-linear equations have been under active investigation in recent years and,
for instance, the first exact solutions were displayed by various groups
\cite{solutions}. On the other hand, up to now the Vasiliev equations still lack an
action principle, and they exploit a formalism that is rather remote from that usually
adopted to describe low-spin systems. Furthermore, a similar set of non-linear field
equations until now was not obtained for mixed-symmetry fields, for which only no-go
results are available in literature \cite{mixed_nogo} if one looks for non-abelian
deformations of the gauge algebra. Alternative approaches to the problem, as those
presented in this review, could thus be of help in overcoming these difficulties. At
the same time, it is quite natural to compare the frame-like path followed by Vasiliev
and collaborators in studying higher-spin interactions to a metric-like description.
While the first approach somehow extends the algebraic Yang-Mills setup, the second could
shed light on the geometry underlying these interactions, as the metric formulation does
in the spin-$2$ case.

In conclusion, one might say that higher-spin interactions have presently two vastly different realizations. In the first, provided by the Vasiliev setting, their gauge symmetry is unbroken, while in the second, provided by String Theory, it is somehow spontaneously broken. Unfortunately, as anticipated, we do not understand precisely how to bridge the gap between the two, or even how to attach a precise meaning to the breaking mechanism at work in String Theory. The higher-spin gauge symmetries should be restored in a proper high-energy regime of String Theory, but nowadays this remains a rather obscure corner of this framework. Some important steps forward came from the study of the high-energy limits of the string scattering amplitudes \cite{scattering}, but a detailed characterization of this regime is still missing\footnote{The AdS/CFT correspondence \cite{maldacena} could also provide a natural arena for investigating these relations \cite{sundborg}. A proper tensionless limit in the bulk should correspond to a weak limit on the boundary, and some investigations of this regime were pursued analyzing the higher-spin conserved currents of free super Yang-Mills models on the boundary \cite{anselmi,bouffe}.}. At any rate, String Theory and the Vasiliev model share some neat signatures, to wit spectra containing infinitely many higher-spin modes and higher-derivative interactions. Clearly, these features restate the old border between the $s \leq 2$ and the higher-spin physics, but these two examples at least show that systems of interacting higher-spin fields coupled to gravity exist. Still, the present understanding of their properties is far from being fully satisfactory, and its improvement provides a fascinating task for the future.

\subsubsection*{Structure of this review}

For the reader's benefit let us outline the structure of this review, summarizing the recent results that it elaborates upon.
\begin{itemize}
\item
In Section \ref{chap:constrained} the constrained formulation of the free dynamics for arbitrary mixed-symmetry fields will be discussed in detail. First of all, the identification of the non-Lagrangian Labastida equations will be reviewed with the aid of the compact notation proposed in \cite{mixed_bose,mixed_fermi}. Furthermore, it will be shown that they indeed single out the correct physical polarizations, thus filling a gap of their first presentation in \cite{laba_bose,laba_fermi}. Then, the Labastida Lagrangians for Bose fields \cite{laba_lag} will be constructed in an original way, exploiting the consequences of the Bianchi identities satisfied by the constrained fields \cite{mixed_bose}. The Lagrangians for mixed-symmetry Fermi fields of \cite{mixed_fermi} will be built with the same technique. Section \ref{chap:constrained} closes with some comments on the Lagrangian field equations, and on their reduction to the Labastida form. Finally, we shall also describe how to adapt the whole construction to irreducible fields propagating the degrees of freedom of an irreducible representation of the Lorentz group.
\item
In Section \ref{chap:unconstrained} the constrained Lagrangian free theory will be reformulated in various unconstrained settings, following \cite{mixed_bose,mixed_fermi} and introducing suitable sets of auxiliary fields. We shall begin by considering a minimal setup, introducing at most one compensator for each Labastida constraint on the gauge parameters and at most one Lagrange multiplier for each Labastida constraint on the field. Then, at the price of slightly enlarging the field content, we shall also present a formulation that is free of higher-derivative terms, but still contains a number of auxiliary fields only depending on the number of index families carried by the basic fields and not on their Lorentz labels.
\item
In Section \ref{chap:weyl} the Lagrangians for mixed-symmetry fields will be reconsidered, in order to show the emergence of Weyl-like symmetries in backgrounds with low-enough dimensions. Their presence was first recognized in \cite{mixed_bose,mixed_fermi}, even if a complete classification is still lacking when more than two index families are present. In the bosonic case we shall also identify here some Weyl-invariant models in $D > 4$ that were not discussed in \cite{mixed_bose}.
\item
Section \ref{chap:conclusions} closes this review with a brief summary and a
discussion of some possible developments of the work detailed here.
\end{itemize}

This work mainly reviews the original results obtained by the author in collaboration with Dario Francia, Jihad Mourad and Augusto Sagnotti. They were presented in \cite{mixed_bose,mixed_fermi} and they were also concisely reviewed in \cite{review_SIGRAV}.


\section{Constrained theory}\label{chap:constrained}


In the mid eighties Labastida proposed a set of \emph{covariant} field equations describing the propagation of the free \emph{massless} modes associated to generic representations of the Lorentz group in arbitrary space-time dimensions \cite{laba_bose,laba_fermi}. As anticipated in the Introduction, in order to achieve this result he considered multi-symmetric \emph{gauge} fields $\vf_{\,\m_1 \ldots\, \m_{s_1},\, \n_1 \ldots\, \n_{s_2}, \,\ldots}$ and $\psi^{\,\alpha}{}_{\m_1 \ldots\, \m_{s_1},\, \n_1 \ldots\, \n_{s_2}, \,\ldots}$, so that his ``metric-like'' formulation generalizes that obtained by Fronsdal and Fang for fully symmetric fields at the end of the seventies \cite{fronsdal,fang_fronsdal}. As in the fully symmetric case, in the Labastida approach fields and gauge parameters are subjected to algebraic \emph{\mbox{($\g$-)}trace constraints}.

In the Fang-Fronsdal works the constraints are inherited from the description of massive fields proposed by Singh and Hagen \cite{singh_bose,singh_fermi}. In fact, Fronsdal and Fang identified the massless Lagrangians considering the limit of the massive ones. On the other hand, Labastida derived his results conjecturing a set of gauge transformations and then looking for gauge invariant field equations, following the re-examination of the Fang-Fronsdal construction proposed by Curtright \cite{curt_gauge}. He selected his equations of motion without referring to an action principle, but in a successive paper he also derived Lagrangians leading to equivalent field equations for Bose fields \cite{laba_lag}. Nevertheless, even at the free level, for Fermi fields the construction of quadratic Lagrangians initiated long ago by Fierz and Pauli \cite{fierz_pauli} was left incomplete until recently. In fact, Lagrangians for Fermi fields of mixed symmetry were presented only in \cite{mixed_fermi}.

In this section we shall review the identification of the gauge transformations, the
constraints and the field equations forcing multi-symmetric fields to propagate free
massless modes \cite{laba_bose,laba_fermi}. Following a simple argument presented in
\cite{mixed_fermi}, we shall also verify that these conditions describe indeed the correct
physical polarizations. Furthermore, we shall present an original derivation of the
Labastida bosonic Lagrangians \cite{laba_lag,mixed_bose}, based on the Bianchi identities
satisfied by the gauge fields. This technique will also prove useful in the construction
of the fermionic Lagrangians \cite{mixed_fermi} that follows, and that represents one of
the main original results of the project that led to \cite{mixed_bose,mixed_fermi}. Then we shall make some comments on the related
Lagrangian field equations, and on their link with the non-Lagrangian ones proposed by
Labastida. Since in the following we are going to deal mainly with reducible fields, this
section closes with a description of how the theory adapts to irreducible fields
subjected to suitable Young projections.

While the results presented here refer only to massless fields, we would like to call to the reader's attention that a Stueckelberg-like description of the corresponding massive fields can be obtained via a standard Kaluza-Klein circle reduction \cite{reduction}.


\subsection{Non-Lagrangian field equations and constraints}\label{sec:labastida}


The non-Lagrangian field equations for Bose and Fermi fields of mixed-symmetry can be derived following similar logical steps. The choice of describing them separately is aimed at simplifying the presentation, but the reader is invited to notice the similarities between the two setups.


\subsubsection{Bose fields}\label{sec:laba-bose}


The well-known example of linearized gravity already exhibits the structure of the covariant field equations for arbitrary free higher-spin fields. In fact, the equation of motion for the linearized metric fluctuation $h_{\,\m\n}$ reads
\be
\label{ricci}
R_{\,\m\n} \, \equiv \ \Box \, h_{\,\m\n} \, - \, \pr_{\,\m}\, \prd h_{\,\n} \, - \, \pr_{\,\n}\, \prd h_{\,\m} \, + \, \pr_{\,\m\,} \pr_{\,\n} \, h_{\,\l}{}^{\,\l} \, = \, 0 \, .
\ee
It enforces the vanishing of the linearized Ricci tensor, that provides a gauge-invariant completion of $\Box\, h_{\,\m\n}$ under the linearized diffeomorphisms $\d h_{\,\m\n} = \pr_{\,\m}\,\x_{\,\n} + \, \pr_{\,\n}\,\x_{\,\m}$. The latter observation can be taken as a guiding principle in order to look for the equations of motion for arbitrary massless fields. For instance, the non-Lagrangian form of the Fronsdal equations \cite{fronsdal} for fully symmetric Bose fields reads
\be \label{fronsdal-b}
\cF_{\,\m_1 \ldots\, \m_s} \, \equiv \ \Box \, \vf_{\,\m_1 \ldots\, \m_s} \, - \, \pr_{\,(\,\m_1\,} \prd \vf_{\,\m_2 \ldots\, \m_s\,)} \, + \, \pr_{\,(\,\m_1\,}\pr_{\phantom{(}\!\m_2}\, \vf_{\,\m_3 \ldots\, \m_s\,)\,\l}{}^{\,\l} \, = \, 0 \, ,
\ee
where we would like to stress again that here and in the following a couple of parentheses denotes a complete symmetrization of the indices it encloses, with the minimum possible number of terms and with unit overall normalization. Eq.~\eqref{fronsdal-b} is invariant under the transformations
\be \label{gauge_fronsdal-b}
\d\, \vf_{\,\m_1 \ldots\, \m_s} \, = \, \pr_{\,(\,\m_1\,} \L_{\,\m_2 \ldots\, \m_s\,)}
\ee
with \emph{traceless} gauge parameters. As pointed out by Curtright \cite{curt_gauge}, one can actually identify the field equations \eqref{fronsdal-b} and the constraints
\be \label{constr_fronsdal-b}
\L_{\,\m_1 \ldots\, \m_{s-3}\,\l}{}^{\,\l} \, = \, 0
\ee
looking for a \emph{second-order} kinetic tensor that contains $\Box \, \vf_{\,\m_1 \ldots\, \m_s}$ and that is invariant under the gauge transformations \eqref{gauge_fronsdal-b}. In fact, they provide the natural generalization of the abelian linearized diffeomorphisms $\d h_{\,\m\n} = \pr_{\,(\,\m}\,\x_{\,\n\,)}$. The only possibility to achieve such goal is to constrain the gauge parameters according to \eqref{constr_fronsdal-b}, and the result is unique if one avoids to introduce traces of $\Box \, \vf_{\,\m_1 \ldots\, \m_s}$. This latter hypothesis is dictated by the quest for the simplest solution, and leads to eq.~\eqref{fronsdal-b}. On the other hand, it hampers the construction of Lagrangian field equations, so that it will be relaxed in Section \ref{sec:labalag-bose}. At any rate, the correctness of the resulting field equations can then be checked a posteriori, via a counting of the degrees of freedom propagated by their solutions. For the Fronsdal equations \eqref{fronsdal-b} this check was performed by a number of authors with different techniques \cite{curt_gauge,dewit_fr}.

For multi-symmetric Bose fields one can follow a similar path. In particular, when dealing with \emph{unprojected} tensors of rank $(s_1,\ldots,s_N)$, one can generalize eq.~\eqref{gauge_fronsdal-b} introducing an independent gauge parameter for each index family. These carry $(s_i-1)$ vector indices in the $i$-th index family, so that
\be \label{gauge_indices-b}
\d \, \vf_{\,\m_1 \ldots\, \m_{s_1},\, \n_1 \ldots\, \n_{s_2}, \,\ldots} = \, \pr_{\,(\,\m_1} \L^{(1)}{}_{\!\!\m_2 \ldots\, \m_{s_1})\,,\,\n_1 \ldots\, \n_{s_2},\,\ldots} + \, \pr_{\,(\,\n_1\,|\,} \L^{(2)}{}_{\!\!\m_1 \ldots\, \m_{s_1},\,|\,\n_2 \ldots\, \n_{s_2})\,,\,\ldots} + \,\ldots \, ,
\ee
where here and in the following a vertical bar signals that the symmetrization also encompasses the indices lying between the next bar and the closing parenthesis. Looking at eq.~\eqref{gauge_indices-b}, the need for an efficient and compact notation to deal with mixed-symmetry fields becomes evident, and in the following we shall resort to that introduced in \cite{mixed_bose,mixed_fermi}. It is based on the removal of all space-time indices, but \emph{``family'' indices} are needed in order to select the groups of space-time indices where the operators are acting upon. For instance, in the following a gradient which carries an index symmetrized with those in the $i$-th family will be denoted by
\be
\pr^{\,i} \, \vf \, \equiv \, \pr_{\,(\,\m^i_1|} \, \vf_{\,\ldots \,,\, | \, \m^i_2 \, \ldots \, \m^i_{s_i+1} ) \,,\, \ldots} \, ,
\ee
while a divergence contracting an index in the $i$-th family will be denoted by
\be
\pr_{\,i} \, \vf \, \equiv \, \pr^{\,\l} \, \vf_{\,\ldots \,,\, \l \, \m^i_1 \, \ldots \, \m^i_{s_i-1} \,,\, \ldots} \, .
\ee
The position of the family indices respects a useful convention of \cite{mixed_bose,mixed_fermi}, according to which \emph{lower} family indices are associated to operators removing Lorentz indices, while \emph{upper} family indices are associated to operators adding Lorentz indices, to be symmetrized with all other indices belonging to the group identified by the family label. Consistently with this rule, the gauge parameter lacking one space-time index in the $i$-th family will be denoted by $\L_{\,i}$. In conclusion, one can rewrite eq.~\eqref{gauge_indices-b} in the synthetic form
\be \label{gauge-b}
\d \, \vf \, = \, \pr^{\,i}\,\L_{\,i} \, .
\ee
This compact notation makes also manifest the existence of gauge for gauge transformations for mixed-symmetry fields. Differential forms already display this behavior, and being fully antisymmetric objects they are but a particular class of mixed-symmetry fields. At any rate, for arbitrary mixed-symmetry fields one can easily recognize that
\be
\d \, \L_{\,i} \, = \, \pr^{\,j}\, \L_{\,[\,ij\,]}
\ee
leaves $\vf$ invariant provided the $\L_{\,[\,ij\,]}$ are antisymmetric in their family indices. In general the chain would continue, since transformations of the $\L_{\,[\,ij\,]}$ leaving invariant the $\L_{\,i}$ could also exist. One can proceed in this fashion for a number of steps equal to the number of index families carried by $\vf$.

Having fixed the notation, we can now look for equations of motion that are invariant under the gauge transformations \eqref{gauge-b}. Eqs.~\eqref{ricci} and \eqref{fronsdal-b} suggest that it is necessary to consider also traces of the gauge fields in order to obtain them. In the mixed-symmetry case it is thus convenient to introduce the $T_{ij}$ operators acting as
\be
T_{ij} \, \vf \, \equiv \, \vf^{\phantom{\,\ldots \,,\,}\, \l}{}_{\hspace{-18pt} \ldots \,,\,\phantom{\l} \ \m^i_1 \, \ldots \, \m^i_{s_i-1} \,,\, \ldots \,,\, \l \,
\m^j_1 \ldots \, \m^j_{s_j-1} \,,\, \ldots} \, .
\ee
One can then verify that the Fronsdal-Labastida kinetic tensor \cite{laba_bose}
\be \label{laba-b}
\cF  \, \equiv \, \Box \, \vf \, - \, \pr^{\,i}\,\pr_{\,i} \, \vf \, + \, \12 \ \pr^{\,i}\pr^{\,j}\, T_{ij}\, \vf
\ee
is invariant under the transformations \eqref{gauge-b}, provided the gauge parameters satisfy the constraints
\be \label{constr_gauge-b}
T_{(\,ij}\, \L_{\,k\,)} \, = \, 0 \, .
\ee
All needed computational rules follow from the algebra of the operators $T_{ij}$, $\pr_{\,i}$ and $\pr^{\,i}$, that is presented in Appendix \ref{app:MIX}. Moreover, this is the unique \emph{two-derivative} gauge-invariant completion of $\Box\, \vf$ that does not contain its traces\footnote{As recalled in the Introduction, relaxing the upper bound on the number of derivatives and accepting the presence of non-local terms, one can actually obtain a gauge-invariant completion of $\Box\, \vf$ without any need for constraints \cite{nonlocal,fs1,fms1,nonlocal_mixed}.}. This led Labastida to propose in \cite{laba_bose} the field equation
\be \label{laba_eq-b}
\cF \, = \, 0
\ee
to describe the dynamics of massless mixed-symmetry Bose fields. In \cite{laba_lag} he also verified the correctness of eq.~\eqref{laba_eq-b} in a number of simple examples, performing the counting of the propagated degrees of freedom. The first proof of the Labastida conjecture was then provided much later in the third reference of \cite{nonlocal_mixed}, while we shall now show that eqs.~\eqref{gauge-b}, \eqref{constr_gauge-b} and \eqref{laba_eq-b} lead to the propagations of the correct polarizations for generic mixed-symmetry bosons reviewing the simple argument presented in \cite{mixed_fermi}.

In order to proceed, it is convenient to resort momentarily to an explicit oscillator realization of the index-free notation. This is attained folding a generic multi-symmetric gauge field $\vf_{\,\m_1 \ldots\, \m_{s_1},\,\n_1 \ldots\, \n_{s_2}\,,\, \ldots}$ with commuting vectors $u^{\,i\,\m}$ according to
\be \label{psiosc}
\vf \, \equiv \, \frac{1}{s_{1}\,! \,\ldots\, s_{N}\,!} \ u^{\,1\,\m_1} \ldots\, u^{\,1\,\m_{s_1}} \, u^{\,2\,\n_1} \ldots\, u^{\,2\,\n_{s_2}} \ldots \ \vf_{\,\m_1 \ldots\, \m_{s_1},\,\n_1 \ldots\, \n_{s_2}\,,\, \ldots} \, .
\ee
The form taken by the various operators introduced until now is described in detail in Appendix \ref{app:MIX} and, for instance, divergences and gradients can be cast in the form
\be
\pr_{\,i} \, \equiv \, \pr^{\,\m} \, \frac{\pr}{\pr \, u^{\,i\,\m}} \, , \qquad\qquad \pr^{\,i} \, \equiv \, \pr_{\,\m} \, u^{\,i\,\m} \, . \label{osc_notation}
\ee

In studying the plane-wave solutions of the Labastida equation \eqref{laba_eq-b}, that in momentum space takes the form
\be \label{momlab}
p^{\,2}\, \vf \, - \, (p \cdot u^i) \left(p \cdot \frac{\pr}{\pr\, u^i}\right) \vf \, +\, \frac{1}{2} \ (p \cdot u^i) \,(p \cdot u^j) \, T_{ij} \, \vf \, = \, 0 \, ,
\ee
it is then convenient to decompose $\vf$ according to
\be\label{vfdecomp}
\vf \, = \, (p \cdot u^i)\, \z_{\,i} \, + \, \widehat{\vf} \, ,
\ee
sorting out its portion $\widehat{\vf}$ that is independent of $(p \cdot u^i)$,
\be \label{transvb1}
\frac{\pr}{\pr\,(p\cdot u^i)} \, \widehat{\vf} \, = \, 0 \, .
\ee
Substituting in \eqref{momlab} then gives
\begin{align}
& p^{\,2}\, \widehat{\vf} \, - \, (p \cdot u^i) \left(p \cdot \frac{\pr}{\pr\, u^i}\right) \widehat{\vf} \, +\, \frac{1}{2} \ (p \cdot u^i) \,(p \cdot u^j) \, T_{ij} \, \widehat{\vf}  \nn \\
& +\,  \frac{1}{6} \ (p \cdot u^i)\, (p \cdot u^j)\, (p \cdot u^k)\, T_{(\,ij}\, \zeta_{\,k\,)} \, = \, 0 \, . \label{eqpsub}
\end{align}
The four terms appearing in this sum must vanish independently, since they carry different powers of $(p \cdot u^i)$. Therefore, eq.~\eqref{eqpsub} is equivalent to the four conditions
\begin{align}
& p^{\,2}\, \widehat{\vf} \, = \, 0 \ , \label{sys1pb} \\[1pt]
& \!\left(p \cdot \frac{\pr}{\pr\, u^i}\right) \widehat{\vf} \, = \, 0 \  , \label{sys2pb} \\[1pt]
& T_{ij} \, \widehat{\vf} \, = \, 0  \ , \label{sys3pb} \\[1pt]
& T_{(\,ij}\, \zeta_{\,k\,)} \, = \, 0 \ . \label{sys4pb}
\end{align}
Eq.~\eqref{sys4pb} now implies that one can gauge away the $\zeta_{\, i}$, since they are subject to the
Labastida constraints, and eq.~\eqref{sys1pb} then leads to the mass-shell condition $p^{\,2} = 0$. One can thus let $p=p^{\, +}$ and the conditions \eqref{transvb1} and \eqref{sys2pb} imply respectively
\be \label{notransvb}
\frac{\pr}{\pr\, u^{i\,-}}\, \widehat{\vf} \, = \, 0 \, , \qquad\qquad \frac{\pr}{\pr\, u^{i\,+}}\, \widehat{\vf} \, = \, 0 \, .
\ee
All components of $\widehat{\vf}$ along the two light-cone directions are thus absent on account of eqs.~\eqref{notransvb}, while eq.~\eqref{sys3pb} forces the remaining transverse components to be traceless. In conclusion, the Labastida equation \eqref{laba_eq-b} propagates the degrees of freedom of representations of the little group $O(D-2)$, as expected on account of its uniqueness. At this stage these are reducible representations, since we did not enforce any projection on $\vf$, but in Section \ref{sec:irreducible} we shall see how the theory can be easily adapted to describe irreducible representations.

Let us stress that in order to reach this result one does not have to impose any condition on the gauge field $\vf$, but eventually all its traces vanish \emph{on-shell}. On the other hand, Labastida verified the correctness of eq.~\eqref{laba_eq-b} in a number of examples adopting the \emph{off-shell} techniques introduced in \cite{siegel_count}. In order to match the correct number of degrees of freedom, he thus also identified a set of double trace constraints on the fields, generalizing the one identified by Fronsdal for a given spin-$s$ field,
\be \label{constr_fronsdal2-b}
\vf_{\,\m_1 \ldots\, \m_{s-4}\,\l\r}{}^{\,\l\r} \, = \, 0 \, .
\ee
The origin of the constraints can be better understood recalling that in the fully symmetric case one must impose the constraint \eqref{constr_fronsdal2-b} in order to obtain a gauge-invariant Lagrangian leading to a field equation equivalent to \eqref{fronsdal-b}. This was first pointed out in \cite{curt_gauge,dewit_fr}, and indeed the double trace of the field appears on the right-hand side of the Bianchi identity
\be \label{bianchi_fronsdal}
\prd \cF_{\,\m_1 \ldots\, \m_{s-1}} \, - \, \12 \ \pr_{\,(\,\m_1}\, \cF_{\,\m_2 \ldots\, \m_{s-1}\,)\,\l}{}^{\,\l} \, = \, - \, \frac{3}{2}\ \pr_{\,(\,\m_1\,}\pr_{\phantom{(}\!\m_2\,}\pr_{\phantom{(}\!\m_3}\, \vf_{\,\m_4 \ldots\, \m_{s-1}\,)\,\l\r}{}^{\,\l\r} \, ,
\ee
that generalizes the contracted Bianchi identity of linearized gravity,
\be \label{bianchi_grav}
\prd R_{\,\m} \, - \, \12 \ \pr_{\,\m}\, R_{\,\l}{}^{\,\l} \, = \, 0 \, .
\ee
As we shall see in Section \ref{sec:labalag-bose}, eq.~\eqref{bianchi_fronsdal} indeed plays a crucial role in the construction of the Lagrangian for a spin-$s$ field, as eq.~\eqref{bianchi_grav} does in the spin-$2$ case.
In a similar fashion the Labastida constraints on the field $\vf$ can be identified looking at the Bianchi identities
\be \label{bianchi-b1}
\pr_{\,i}\, \cF \, - \, \12 \ \pr^{\,j}\, T_{ij} \, \cF \, = \, - \, \frac{1}{12} \ \pr^{\,j}\pr^{\,k}\pr^{\,l} \, T_{(\,ij}\,T_{kl\,)} \, \vf \, ,
\ee
that provide a neat rationale for the Labastida Lagrangians of \cite{laba_lag}, as showed in \cite{mixed_bose}. In fact, in the present notation the constraints on the field introduced in \cite{laba_bose} read
\be \label{constr_field-b}
T_{(\,ij}\,T_{kl\,)} \, \vf \, = \, 0 \, ,
\ee
so that they cancel the right-hand side of eq.~\eqref{bianchi-b1}, precisely as the constraint \eqref{constr_fronsdal2-b} does in the fully symmetric setting.

Notice that, quite differently from what Fronsdal's case could naively suggest, \emph{not} all traces of the gauge parameters and \emph{not} all double traces of the field vanish. Only the linear combinations appearing in eqs.~\eqref{constr_gauge-b} and \eqref{constr_field-b} do, and this simple observation is at the origin of the main differences in the structure of the Lagrangians for fully symmetric and for mixed-symmetry fields that we are about to exhibit. Notice also that, as in Fronsdal's case, those in eq.~\eqref{constr_field-b} are the strongest gauge-invariant constraints available since
\be \label{gaugedouble}
T_{(\,ij}\,T_{kl\,)}\, \d \, \vf \, = \, \pr_{\,(\,i}\, T_{jk}\, \L_{\,l\,)} \, + \, \pr^{\,m} \left\{\, T_{(\,ij}\,T_{kl}\,\L_{\,m\,)} \, - \, T_{m\,(\,i}\,T_{jk}\,\L_{\,l\,)} \,\right\} \, .
\ee
The last terms can be cast in this form manifestly related to the constraints \eqref{constr_gauge-b} via the rewriting
\be
T_{(\,ij}\,T_{kl\,)} \, \L_{\,m} \, = \, T_{(\,ij}\,T_{kl}\,\L_{\,m\,)} \, - \, T_{m\,(\,i}\,T_{jk}\,\L_{\,l\,)} \, .
\ee

The Labastida constraints \eqref{constr_gauge-b} and \eqref{constr_field-b} are nicely ``covariant'' in the family index language, but they are \emph{not} independent. For instance, considering a further trace of $T_{(\,ij}\,T_{kl\,)}\, \vf$ one obtains
\be \label{51comb}
Y_{\{5,1\}}\, T_{mn}\, T_{(\,ij}\,T_{kl\,)}\, \vf \, = \, 0
\ee
without any need of enforcing the Labastida constraints. In this expression $Y_{\{5,1\}}$ denotes the Young projector onto the irreducible $\{5,1\}$ representation of the permutation
group acting on the family indices\footnote{The conventions
for the permutation group and related tools are spelled out at the end of Appendix A. Further details on these matters can be found, for instance, in \cite{group}.}. Eq.~\eqref{51comb} reflects a special property of products of identical tensors, that in the two-index $T_{ij}$ case can only build Young diagrams in family-index space with even numbers of boxes in each row. This fact will have important consequences in Section \ref{chap:unconstrained}, where we shall describe how one can remove the Labastida constraints adding a suitable set of auxiliary fields. Roughly speaking, one would like to associate an auxiliary field to each constraint, but their linear dependence gives rise to a number of subtleties.


\subsubsection{Fermi fields}\label{sec:laba-fermi}


As anticipated, when dealing with Fermi fields one can follow a similar path. To begin with, let us stress that here and in the following all spinor indices are hidden for simplicity. Then, notice that even for Fermi fields a well-know paradigmatic example already displays some key features of the arbitrary-spin construction. In fact, the Rarita-Schwinger equation for a massless field of spin $3/2$, that is usually presented in the form
\be \label{rarita_eq1}
i\ \g^{\,\m\n\r}\, \pr_{\,\n}\, \psi_{\,\r} \, = \, 0 \, ,
\ee
is manifestly invariant under the gauge transformation
\be
\d \, \psi_{\,\m} \, = \, \pr_{\,\m}\, \e \, ,
\ee
simply because two derivatives commute. Moreover, in general it can be combined with its $\g$-trace in order to obtain the equation
\be \label{rarita_eq2}
i\, \left\{\, \dsl \ \psi_{\,\m} \, - \, \pr_{\,\m} \psisl \,\right\} = \, 0 \, ,
\ee
that provides a gauge-invariant completion of $\!\pr\!\!\!/\, \psi_{\,\m}$. In complete analogy with the bosonic construction of the previous section, one can thus recover the field equations for arbitrary higher-spin fermions postulating a proper set of gauge transformations and looking for a gauge invariant completion of the ``naive'' generalization of the Dirac equation. In the fully symmetric setting the natural gauge transformation for a spin-($s+\12$) field is
\be \label{gauge_fang}
\d\, \psi_{\,\m_1 \ldots\, \m_s} \, = \, \pr_{\,(\,\m_1\,} \e_{\,\m_2 \ldots\, \m_s\,)} \, .
\ee
The corresponding Fang-Fronsdal field equation \cite{fang_fronsdal} is
\be \label{fang}
\cS_{\,\m_1 \ldots\, \m_s} \, \equiv \ i \, \left\{\, \dsl \ \psi_{\,\m_1 \ldots\, \m_s} \, - \, \pr_{\,(\,\m_1}\! \psisl{}_{\,\m_2 \ldots\, \m_s\,)} \,\right\} \, = \, 0 \, ,
\ee
that is invariant under \eqref{gauge_fang} provided the parameter satisfies
\be \label{constr_fang}
\esl_{\,\m_1 \ldots \m_{s-2}} \, = \, 0 \, .
\ee
Again, the result is unique up to possible combinations with its $\g$-traces, that are anyway needed to obtain Lagrangian field equations.

In the mixed-symmetry case it is natural to extend eq.~\eqref{gauge_fang} considering, for \emph{unprojected} spinor-tensors, an independent gauge parameter for each index family. It is clearly convenient to abide by the notation presented in the previous section, that permits to cast the gauge transformations in the form
\be \label{gauge-f}
\d \, \psi \, = \, \pr^{\,i}\, \e_{\,i} \, .
\ee
Notice that this expression has the same form as that considered for Bose fields, so that gauge for gauge transformations are generically present also in this setup. However, an ingredient is still missing in the family-index notation: one would also like to compute $\g$-traces, and this can be done introducing
\be
\g_{\, i}\, \psi \, \equiv \, \g^{\, \l} \, \psi_{\,\ldots \,,\, \l \, \m^i_1 \, \ldots \, \m^i_{s_i-1} ,\, \ldots} \ .
\ee
The Fang-Fronsdal-Labastida kinetic tensor \cite{laba_fermi}
\be \label{laba-f}
\cS \, \equiv \, i \, \left\{\, \dsl \, \psi \, - \, \pr^{\,i} \, \g_{\,i}\, \psi \,\right\}
\ee
thus gives a gauge-invariant completion of $\dsl\, \psi$, provided one enforces the constraints
\be \label{constr_gauge-f}
\g_{\,(\,i}\, \e_{\,j\,)} \, = \, 0 \, .
\ee
Again, this statement can be verified looking at the algebra of the operators introduced until now, that is presented in Appendix \ref{app:MIX}. The answer is unique if one considers only first-order differential operators\footnote{Even for Fermi fields, relaxing the upper bound on the number of derivatives one can actually obtain gauge-invariant completions of $\dsl\, \psi$ without any need for constraints \cite{nonlocal,fs1,fms1,nonlocal_mixed}.} and avoids to introduce $\g$-traces of $\dsl\, \psi$.

This led Labastida to propose in \cite{laba_fermi} the field equation
\be \label{laba_eq-f}
\cS \, = \, 0
\ee
in order to describe the dynamics of massless mixed-symmetry Fermi fields. Performing the counting of the propagated degrees of freedom, in \cite{laba_lag} he also verified the correctness of eq.~\eqref{laba_eq-f} in a number of simple examples. For generic mixed-symmetry fermions this issue was finally clarified in \cite{mixed_fermi}, following the lines already depicted for Bose fields in Section \ref{sec:laba-bose}.

Let us thus review how the argument adapts to Fermi fields. Even in this setup it is
convenient to resort to an oscillator realization of the index-free notation, introducing
a commuting vector $u^{i\,\m}$ for each index family. Eqs.~\eqref{osc_notation} then
enable to cast the momentum-space Labastida equation in the form
\be \label{plabaf}
\psl \, \psi \, - \, (p\cdot u^i)\, \g_{\,i} \, \psi \, = \, 0 \, .
\ee
As in the bosonic case, one can then decompose the gauge field $\psi$ according to
\be
\psi \, = \, (p\cdot u^i)\, \c_{\,i} \, + \, \widehat{\psi} \, ,
\ee
with
\be
\frac{\pr}{\pr\,(p\cdot u^i)} \, \widehat{\psi} \, = \, 0 \, .
\ee
Substituting in eq.~\eqref{plabaf} then gives
\be
\psl \, \widehat{\psi} \, - \, (p\cdot u^i)\, \g_{\,i} \, \widehat{\psi} \, - \, \12 \ (p\cdot u^i)\, (p\cdot u^j)\, \g_{\,(\,i}\,\c_{\,j\,)} \, = \, 0 \, ,
\ee
and one is led to the three independent conditions
\begin{align}
& \!\!\psl \, \widehat{\psi} \, = \, 0 \, , \label{pcondf1} \\[1pt]
& \g_{\,i}\, \widehat{\psi} \, = \, 0 \, , \label{pcondf2} \\[1pt]
& \g_{\,(\,i}\,\c_{\,j\,)} \, = \, 0 \, , \label{pcondf3}
\end{align}
since the terms in the sum carry different powers of $(p\cdot u^i)$. The condition \eqref{pcondf3} then ensures the possibility of gauging away altogether the $\c_{\,i}$, and one is left with the Dirac equation \eqref{pcondf1}. It leads again to the mass-shell condition $p^{\,2} = 0$, so that one can consider a momentum $p$ with a single non-vanishing light-cone component as, for instance, $p^{\,+}$. Therefore, the lack of dependence of $\widehat{\psi}$ on $(p \cdot u^i)$ translates into the condition
\be \label{no-}
\frac{\pr}{\pr \, u^{\, i\, -}}\ \, \widehat{\psi} \, = \, 0 \ ,
\ee
while the Dirac equation \eqref{pcondf1} turns into the standard projection
\be\label{dirac}
\g_{\, +} \, \widehat{\psi} \, = \, 0 \ ,
\ee
that halves the on-shell components of $\widehat{\psi}$. Finally,
eq.~\eqref{pcondf2} becomes
\be\label{eqgamma+}
\left(\, \g^{\,+} \ \frac{\pr}{\pr \, u^{\, i\, +}} \, + \, \g^{\,\underline{m}} \ \frac{\pr}{\pr \, u^{\, i\, \underline{m}}} \,\right) \, \widehat{\psi} \, = \, 0 \ ,
\ee
where we have used $\underline{m}$ to denote the transverse Lorentz index in order to
distinguish it from the family indices. Multiplying eq.~\eqref{eqgamma+} by
$\g_+$ and making use of eq.~\eqref{dirac} one is finally led to the conditions
\begin{align}
& \frac{\pr}{\pr \, u^{\, i\, +}} \ \widehat{\psi} \, = \, 0 \ ,\label{no+}\\
& \g^{\,\underline{m}} \ \frac{\pr}{\pr \, u^{\, i\, \underline{m}}} \ \widehat{\psi}\, = \, 0 \ . \label{nogamma}
\end{align}

All components of $\widehat{\psi}$ along the two light-cone directions thus are absent on
account of eqs.~\eqref{no-} and \eqref{no+}, while eq.~\eqref{nogamma} forces
the remaining transverse components to be $\g$-traceless. In conclusion, the Labastida
equation \eqref{laba_eq-f} propagates in general representations of the little group $O(D-2)$,
as expected on account of its uniqueness. As already stressed when dealing with the same issue
for bosons, at this stage these are reducible representations, since we
did not enforce any projection on $\psi$. On the other hand, it is possible
to describe irreducible representations following the path outlined in Section \ref{sec:irreducible}.

In order to reach this result one does not have to impose any additional condition on the gauge field $\psi$. However, in order to match the correct number of degrees of freedom via the techniques of \cite{siegel_count} Labastida also identified a set of triple $\g$-trace constraints on the fields. These generalize the one identified by Fang and Fronsdal for a given spin-($s+\frac{1}{2}$) field,
\be \label{constr_fronsdal2-f}
\psisl_{\,\m_1 \ldots\, \m_{s-3}\,\l}{}^{\,\l} = \, 0 \, ,
\ee
that, as in the bosonic case, is needed in order to obtain a gauge-invariant Lagrangian leading to a field equation equivalent to \eqref{fang}. In fact, the triple $\g$-trace of the field appears on the right-hand side of the Bianchi identity
\be \label{bianchi_fang}
\pr\,\cdot\, \cS_{\,\m_1 \ldots\, \m_{s-1}} \, - \, \12 \dsl \ssl_{\,\m_1 \ldots\, \m_{s-1}} - \, \12 \ \pr_{\,(\,\m_1}\, \cS_{\,\m_2 \ldots\, \m_{s-1}\,)\,\l}{}^{\,\l} = \, i \ \pr_{\,(\,\m_1\,}\pr_{\phantom{(}\!\m_2} \psisl{}_{\,\m_3 \ldots\, \m_{s-1}\,)\,\l}{}^{\,\l} \, .
\ee
In a similar fashion the Labastida constraints on mixed-symmetry fields can be identified looking at the Bianchi identities
\be \label{bianchi-f1}
\pr_{\,i}\, \cS \, - \, \12 \dsl \, \g_{\,i}\, \cS \, - \, \12 \ \pr^{\,j}\,T_{ij}\, \cS \, - \, \frac{1}{6} \ \pr^{\,j}\,\g_{\,ij}\,\cS \, = \, \frac{i}{6} \ \pr^{\,j}\pr^{\,k}\, T_{(\,ij}\,\g_{\,k\,)}\, \psi \, ,
\ee
that in Section \ref{sec:labalag-fermi} will play a crucial role in the construction of the Lagrangians of \cite{mixed_fermi}. Indeed, in the present notation the constraints on the fields of \cite{laba_fermi} read
\be \label{constr_field-f}
T_{(\,ij}\,\g_{\,k\,)} \, \psi \, = \, 0 \, ,
\ee
and thus cancel the right-hand side of eq.~\eqref{bianchi-f1}, precisely as the constraint \eqref{constr_fronsdal2-f} does in the fully symmetric setting.
The new symbol appearing in eq.~\eqref{bianchi-f1} simply denotes an antisymmetric combination of two ``family'' $\g$-matrices,
\be
\g_{\,ij} \, = \, \12 \, \left(\, \g_{\,i}\,\g_{\,j} \, - \, \g_{\,j}\,\g_{\,i} \,\right) \, .
\ee
The idea underlying this choice is to extend to operators carrying family indices the common convention of working with an antisymmetric basis for $\g$-matrices.

Notice that, again in sharp contrast with the fully symmetric setting, \emph{not} all $\g$-traces of the gauge parameters and \emph{not} all triple $\g$-traces of the fields vanish. Only the linear combinations appearing in eqs.~\eqref{constr_gauge-f} and \eqref{constr_field-f} do, and even for fermions this simple observation is at the origin of the main differences in the structure of the Lagrangians for fully symmetric and for mixed-symmetry fields. Notice also that, as in the Fang-Fronsdal case, those in eq.~\eqref{constr_field-f} are the strongest gauge-invariant constraints available since
\begin{align}
T_{(\,ij}\,\g_{\,k\,)}\, \d \, \psi \, & = \ \dsl\ T_{(\,ij}\, \e_{\,k\,)} \, + \, \pr_{\,(\,i}\, \g_{\,j}\, \e_{\,k\,)} \nn \\
& + \, \pr^{\,l} \left\{\, T_{(\,ij}\,\g_{\,k}\,\e_{\,l\,)} \, - \, T_{l\,(\,i}\,\g_{j}\,\e_{\,k\,)} \, - \, \g_{\,l}\, T_{(\,ij}\, \e_{\,k\,)} \,\right\} \, .
\end{align}
The last terms can be cast in this form manifestly related to the constraints \eqref{constr_gauge-f} via the rewriting
\be
T_{(\,ij}\,\g_{\,k\,)} \, \e_{\,l} \, = \, T_{(\,ij}\,\g_{\,k}\,\e_{\,l\,)} \, - \, T_{l\,(\,i}\,\g_{j}\,\e_{\,k\,)} \, - \, \g_{\,l}\, T_{(\,ij}\, \e_{\,k\,)} \, .
\ee

As in the bosonic case, the Labastida constraints \eqref{constr_gauge-f} and \eqref{constr_field-f} are nicely ``covariant'' in the family index language, but they are \emph{not} independent. In the fermionic case the simplest example of this fact is slightly more involved, but it can be again identified resorting to the permutation group acting on family indices. In this respect, the Labastida constraints \eqref{constr_field-f} carry a $\{3\}$ representation, since they are fully symmetric in the family indices. On the other hand, a trace $T_{ij}$ carries a $\{2\}$ representation while an antisymmetrized $\g$-trace $\g_{\,ij}$ carries a $\{1,1\}$ representation. Thus, the combination $T_{lm}\,T_{(\,ij}\,\g_{\,k\,)}$ admits the decomposition
\be
\{3\} \, \otimes \, \{2\} \, = \, \{5\} \, \oplus \, \{4,1\} \, \oplus \, \{3,2\} \, ,
\ee
while the combination $\g_{\,lm}\,T_{(\,ij}\,\g_{\,k\,)}$ admits the decomposition
\be
\{3\} \, \otimes \, \{1,1\} \, = \, \{4,1\} \, \oplus \, \{3,1,1\} \, .
\ee
It is thus natural to expect the existence of an identically vanishing combination of $\{4,1\}$-projected traces and antisymmetrized $\g$-traces of $T_{(\,ij}\,\g_{\,k\,)}\,\psi$. This is indeed the case, since
\begin{align}
& \left(\, \g_{\,mi}\, T_{(\,jk}\,\g_{\,l\,)} + \, \g_{\,mj}\, T_{(\,ik}\,\g_{\,l\,)} \,\right) \psi \, - \, 15 \ Y_{\{4,1\}} \, T_{ij}\, T_{(\,kl}\,\g_{\,m\,)}\,\psi \nn \\[2pt]
& + \, \left[\,(\,i,j\,)\,\leftrightarrow\,(\,k,l\,)\,\right] \, = \, 0 \label{41comb}
\end{align}
without any need of imposing the constraints. The first term is automatically $\{4,1\}$-projected on account of the interchange between the couples of indices $(\,i,j\,)$ and $(\,k,l\,)$ indicated in the last line. Even in this setting, the lack of linear dependence for the constraints will give rise to some subtleties when we shall proceed to an unconstrained description of the dynamics in Section \ref{chap:unconstrained}.


\subsection{Lagrangians}\label{sec:labalag}


Moving toward a Lagrangian formulation, Bose and Fermi fields begin to follow different paths. As we shall see, the request for gauge invariance and the Bianchi identities satisfied by the fields provide a clearcut rationale for the construction of the Lagrangians for both types of fields. However, the structure of the results is rather different, and for Fermi fields a number of subtleties related to the presence of $\g$-matrices emerge.


\subsubsection{Bose fields}\label{sec:labalag-bose}


Even when dealing with the Lagrangians, the example of linearized gravity already displays the logic that will drive the treatment of the arbitrary case. The equation of motion \eqref{ricci} is indeed non-Lagrangian, but in general it is equivalent to the one following from the linearized Einstein-Hilbert Lagrangian
\be \label{lag_grav}
\cL \, = \, \12 \ h^{\,\m\n} \left(\, R_{\,\m\n} \, - \, \12 \ \h_{\,\m\n}\, R_{\,\l}{}^{\,\l} \,\right) \, ,
\ee
that contains the linearized Ricci tensor appearing in eq.~\eqref{ricci} and its only available trace. Furthermore, the relative coefficient entering eq.~\eqref{lag_grav} is fixed uniquely by the requirement of gauge invariance. In fact, up to total derivatives, the gauge variation of \eqref{lag_grav} under the linearized diffeomorphisms $\d\, h_{\,\m\n} = \, \pr_{\,(\,\m}\,\x_{\,\n\,)}$ is
\be \label{var_grav}
\d\, \cL \, = \, - \ \x^{\,\m} \left(\, \prd R_{\,\m} \, - \, \12 \ \pr_{\,\m}\, R_{\,\l}{}^{\,\l} \,\right) \, ,
\ee
and vanishes on account of the Bianchi identity \eqref{bianchi_grav}.

The Lagrangians for fully symmetric bosons take a very similar form, since the Fronsdal constraints \eqref{constr_fronsdal2-b} on the fields induce similar constraints on the kinetic tensors,
\be \label{constrF_symm}
\cF_{\,\m_1 \ldots\, \m_{s-4}\,\l\r}{}^{\,\l\r} \, = \, 0 \, .
\ee
Thus, the Lagrangians can only contain $\cF_{\,\m_1 \ldots\, \m_s}$ and its trace. The structure of the Bianchi identities \eqref{bianchi_fronsdal} eventually fixes the relative coefficient to be the same entering the Einstein-Hilbert Lagrangian. The final result,
\be
\cL \, = \, \12 \ \vf^{\,\m_1 \ldots\, \m_s} \left(\, \cF_{\,\m_1 \ldots\, \m_s} \, - \, \12 \ \h_{\,(\,\m_1\m_2}\, \cF_{\,\m_3 \ldots\, \m_{s}\,)\,\l}{}^{\,\l} \,\right) \, ,
\ee
is indeed gauge invariant since
\begin{align}
\d \, \cL \, & = \, - \, \frac{s}{2} \ \L^{\,\m_1 \ldots\, \m_{s-1}} \left(\, \prd \cF_{\,\m_1 \ldots\, \m_{s-1}} \, - \, \12 \ \pr_{\,(\,\m_1}\, \cF_{\,\m_2 \ldots\, \m_{s-1}\,)\,\l}{}^{\,\l} \,\right) \nn \\
& + \, \frac{3}{4}\,\binom{s}{3} \, \L^{\,\m_1 \ldots\, \m_{s-3}\,\l}{}_{\,\l} \ \prd \cF_{\,\m_1 \ldots\, \m_{s-3}\,\l}{}^{\,\l} \, .  \label{var_fronsdal}
\end{align}
The first line builds the Bianchi identity \eqref{bianchi_fronsdal}, whose right-hand side vanishes in the constrained theory, while the second one is proportional to the Fronsdal constraint \eqref{constr_fronsdal-b} on the gauge parameter. Notice that, even ignoring the condition \eqref{constrF_symm}, the cancelation of the first two terms in the gauge variation of an ansatz of the form
\begin{align}
\cL \, & = \, \12 \ \vf^{\,\m_1 \ldots\, \m_s} \left(\, \cF_{\m_1 \ldots\, \m_s} + \, a_1 \, \h_{\,(\,\m_1\m_2}\, \cF_{\m_3 \ldots\, \m_{s}\,)\,\l}{}^{\l} \right. \nn \\
& \left. + \ a_2 \ \h_{\,(\,\m_1\m_2} \h_{\,\m_3\m_4}\, \cF_{\m_5 \ldots\, \m_s\,)\,\l\r}{}^{\l\r} + \, \ldots \ \right)
\end{align}
would require a Bianchi identity free of the ``classical anomaly'' appearing in the right-hand side of eq.~\eqref{bianchi_fronsdal}. This leads to the Fronsdal constraint and eventually to eq.~\eqref{constrF_symm}.

In the mixed-symmetry case the structure of the Lagrangians drastically changes, since \emph{not} all double traces of the kinetic tensor $\cF$ are forced to vanish. Hence, the Lagrangians can and indeed do contain additional contributions with multiple traces of $\cF$. On the other hand, the correspondence between constraints on the fields and on the kinetic tensors is still true. In fact, the Labastida constraints on the fields \eqref{constr_field-b} induce the conditions
\be \label{constrF}
T_{(\,ij}\,T_{kl\,)}\, \cF \, = \, 0 \, ,
\ee
so that only particular linear combinations of traces of $\cF$ can enter the Lagrangians. Before identifying the relevant ones, let us briefly comment on eq.~\eqref{constrF}. The commutation rules of Appendix \ref{app:MIX} lead to
\begin{align}
T_{(\,ij}\,T_{kl\,)}\, \cF \, & = \, 3 \, \Box \, T_{(\,ij} \, T_{kl\,)} \, \vf \, - \, \pr^{\,m} \pr_{\,m} \, T_{(\,ij} \, T_{kl\,)} \, \vf \, + \,
\pr^{\,m} \pr_{\,(\,i}\, T_{jk} \, T_{l\,)\,m} \,  \vf \nn \\
& + \, \12 \ \pr^{\,m}\pr^{\,n}\, T_{mn} \, T_{(\,ij} \, T_{kl\,)} \, \vf \, . \label{compute_constrF}
\end{align}
All terms barring the last one in the first line are manifestly annihilated by the Labastida constraints \eqref{constr_field-b}. However, the remaining one can be also related to the constraints on $\vf$ noticing that for a couple of traces a symmetrization over three indices induces a symmetrization over the whole four indices. This is again a consequence of the presence of two identical $T_{ij}$ tensors.

In order to select the non-vanishing contributions that enter the Lagrangians we can now rely upon the \emph{permutation group acting on family indices}. One can indeed classify the multiple traces of $\cF$ of the form
\be
T_{i_1j_1} \ldots\, T_{i_pj_p}\, \cF
\ee
according to its irreducible representations. The final result is that only a particular class of representations does not vanish on account of eq.~\eqref{constrF}. To reach this conclusion, let us begin by noticing that a product of identical $T_{ij}$ tensors admits only components labeled by Young tableaux with an even number of boxes in each row. This follows from an already mentioned theorem discussed, for instance, in \cite{plethysm}. Then, those associated to a tableau with at least four boxes in the first row are built upon a sum of contributions containing at least a symmetrization over four indices. These terms can be divided into three sets, according to the different displacements of the four symmetrized indices over the $T_{ij}$ tensors. They can indeed contain a combination of the form $T_{(\,ij}\,T_{kl\,)}$, or of the form  $T_{(\,ij}\,T_{k\,|\,m}\,T_{|\,l\,)\,n}$ or, finally, of the form $T_{m\,(\,i}\,T_{j\,|\,n}\,T_{p\,|\,k}\,T_{l\,)\,q}$. Since they act on $\cF$, the first two terms are manifestly annihilated by the conditions \eqref{constrF}, either directly or via the enlargement of the symmetrizations guaranteed by the presence of identical tensors. The remaining one, with the symmetrized indices spread over four traces, can be related to the previous ones noticing that
\begin{align}
T_{m\,(\,i}\,T_{j\,|\,n}\,T_{p\,|\,k}\,T_{l\,)\,q} & = \, T_{(\,ij}\,T_{kl}\,T_{mn}\,T_{pq\,)} - \,
T_{(\,ij}\,T_{k\,|\,(\,m\,|}\,T_{|\,l\,)\,|\,n}\,T_{pq\,)} \nn \\
& - \, T_{(\,ij}\,T_{kl\,)}\,T_{(\,mn}\,T_{pq\,)} \, .
\end{align}
Just to clarify the notation, we would like to recall that in the previous expressions the vertical bars are used to stress that the symmetrization also applies to the indices after the next bar and so on, until a parenthesis signals the end of the group.

In conclusion, only \emph{two-column projected} multiple traces of $\cF$ can enter the Lagrangians, and it is convenient to denote them concisely as
\be
\cF^{\,[\,p\,]}{}_{\,i_1j_1,\,\ldots\,,\,i_pj_p} \, \equiv \, Y_{\{2^p\}}\, T_{i_1j_1} \ldots\, T_{i_pj_p} \, \cF \, .
\ee
Here $Y_{\{2^p\}}$ denotes the Young projector onto the $\{2,\ldots,2\}$ irreducible representation of the permutation group acting on the family indices. Hence, $\cF^{\,[\,1\,]}{}_{\,ij}$ simply denotes a trace $T_{ij}$ while, for instance,
\be
\cF^{\,[\,2\,]}{}_{\,i_1j_1,\,i_2j_2} \, = \, \frac{1}{3} \, \left\{\, 2\, T_{i_1j_1}\,T_{i_2j_2} \, - \, T_{i_1(\,i_2}T_{j_2)\,j_1} \,\right\} \cF \, .
\ee
An ansatz for the bosonic Lagrangians should contain all these terms, contracted with suitable products of invariant tensors. It is thus natural to introduce the new operators
\begin{align}
& \h^{\,ii} \, \vf \, = \, \h_{\,(\,\m^i_1\m^i_2\,|}\, \vf_{\,\ldots\,,\,|\,\m^i_3 \ldots\, \m^i_{s_i+2}\,)\,,\,\ldots} \,, \nn \\
& \h^{\,ij} \, \vf \, = \, \12 \, \sum_{n\,=\,1}^{s_i+1}  \, \h_{\,\m^i_n \, (\,\m^j_1\,|}\, \vf_{\,\ldots \,,\, \ldots \, \m^i_{n-1} \, \m^i_{n+1}\, \ldots  \,,\, \ldots
\,,\, |\,\m^j_2 \ldots\, \m^j_{s_j+1}) \,,\, \ldots} \, , \qquad i\, \neq j \, , \label{hij}
\end{align}
where the somehow unconventional factor $1/2$ entering the second line will prove convenient in the presentation of a number of results.

We now have all ingredients to propose an ansatz displaying the structure of the Lagrangians for arbitrary Bose fields of mixed symmetry \cite{laba_lag,mixed_bose}. It reads
\be \label{ansatz-b}
\cL \, = \, \12 \ \bra\, \vf \,\comma\, \sum_{p\,=\,0}^N \, k_{\,p} \ \h^{i_1j_1} \ldots\, \h^{i_pj_p}\, \cF^{\,[\,p\,]}{}_{\,i_1j_1,\,\ldots\,,\,i_pj_p} \,\ket \, ,
\ee
where we have introduced a convenient scalar product that is described in detail in Appendix \ref{app:MIX}. The upper limit of the sum is the number of index families, simply because it is not possible to antisymmetrize over more than $N$ family indices, as a $\{2^{\,N+1}\}$ Young projection would require. Barring some unimportant differences in the notation, the ansatz \eqref{ansatz-b} is equivalent to that proposed by Labastida in \cite{laba_lag}. He did not explicitly consider the two-column projections but, as we have seen, for any given product of traces only a single two-column component exists. Thus, one can also present \eqref{ansatz-b} in a redundant fashion, without displaying the projections explicitly. At any rate, Labastida fixed the coefficients $k_{\,p}$ looking for a self-adjoint Einstein-like tensor, while in the following we shall directly look for a gauge invariant action. A self-adjoint extension of $\cF$ is clearly automatically gauge invariant, but the approach we are reviewing here has the advantage of leading to an unconstrained extension of the theory in a clearcut way. Moreover, it also definitely shows that the Bianchi identities provide the conceptual frame for the Lagrangians.

As we learned from gravity and from the Fronsdal examples, the gauge invariance of the Lagrangians is guaranteed by the Bianchi identities, that in the \emph{constrained} mixed-symmetry case read
\be \label{bianchi-b2}
\pr_{\,i}\, \cF \, - \, \12 \ \pr^{\,j}\, T_{ij} \, \cF \, = \, 0 \, ,
\ee
since the ``classical anomaly'' appearing on the right-hand side of eq.~\eqref{bianchi-b1} cancels imposing the Labastida constraints \eqref{constr_field-b}. Clearly, eq.~\eqref{bianchi-b2} cannot directly deal with the higher traces contained in the ansatz \eqref{ansatz-b}, but we shall see in a while that suitable combinations of multiple traces of the Bianchi identities provide all the needed information. In fact, the gauge variation of eq.~\eqref{ansatz-b} reads
\begin{align}
\d \, \cL \, = \, & - \, \sum_{p\,=\,0}^N \, \frac{1}{2^{\,p+1}} \, \bra\, T_{i_1j_1} \ldots\, T_{i_pj_p}\, \L_{\,k} \,\comma\, k_{\,p}\ \pr_{\,k}\, \cF^{\,[\,p\,]}{}_{\,i_1j_1,\,\ldots\,,\,i_pj_p}  \nn \\
& + \, (\,p+1\,)\, k_{\,p+1}\ \pr^{\,l}\, \cF^{\,[\,p+1\,]}{}_{\,i_1j_1,\,\ldots\,,\,i_pj_p,\,kl} \ket \, , \label{lag_var-b}
\end{align}
where the factors $1/2$ come from the definition \eqref{hij} of the $\h^{ij}$ tensors.
In analogy with the Fronsdal case, part of this gauge variation is proportional to the constraints \eqref{constr_gauge-b} on the gauge parameters. Again, one can identify these contributions with the aid of the permutation group acting on the family indices. The divergence terms in eq.~\eqref{lag_var-b} indeed admit $\{3,2^{\,p-1}\}$ and $\{2^{\,p},1\}$ Young projections, but the first one would produce terms annihilated by the constraints \eqref{constr_gauge-b} in the left entries of the scalar products. This can be realized resorting to  arguments similar to those used to discard from the ansatz \eqref{ansatz-b} the tableaux with more than two columns. In conclusion, one can cast the gauge variation \eqref{lag_var-b} in the form
\begin{align}
\d \, \cL \, = \, & - \, \sum_{p\,=\,0}^N \, \frac{1}{2^{\,p+1}} \, \bra\, Y_{\{2^p,1\}}\, T_{i_1j_1} \ldots\, T_{i_pj_p}\, \L_{\,k} \,\comma\, k_{\,p}\ Y_{\{2^p,1\}}\, \pr_{\,k}\, \cF^{\,[\,p\,]}{}_{\,i_1j_1,\,\ldots\,,\,i_pj_p} \nn \\
& + \, (\,p+1\,)\, k_{\,p+1}\ \pr^{\,l}\, \cF^{\,[\,p+1\,]}{}_{\,i_1j_1,\,\ldots\,,\,i_pj_p,\,kl} \ket \, , \label{lag_var-b2}
\end{align}
since the gradient terms are already $\{2^{\,p},1\}$-projected in their free family indices. Notice that for $p=0$ the terms
\be
k_{\,0}\ \pr_{\,k}\, \cF \, + \, k_{\,1} \ \pr^{\,l}\, T_{kl}\, \cF
\ee
appear in the right entry of the scalar product. Fixing conventionally $k_{\,0}=1$, the Bianchi identities \eqref{bianchi-b2} annihilate this sum provided $k_{\,1}=-\12$. The other coefficients can be similarly fixed resorting to suitable combinations of traces of the Bianchi identities.

The multiple traces of eq.~\eqref{bianchi-b2} read
\begin{align}
& \pr_{\,k}\, T_{i_1j_1} \ldots\, T_{i_pj_p} \, \cF \, - \, \12 \, \sum_{n\,=\,1}^p \, \pr_{\,(\,i_n}\,T_{j_n\,)\,k}\, \prod_{r\,\neq\,n}^p \, T_{i_rj_r}\, \cF \nn \\
& - \, \12 \, \pr^{\,l}\, T_{i_1j_1} \ldots\, T_{i_pj_p}\, T_{kl}\, \cF \, = \, 0 \, , \label{trbianchinp-b}
\end{align}
and in order to compare them with eq.~\eqref{lag_var-b2} one has to compute their $\{2^{\,p},1\}$ projections. Due to the symmetries of these expressions, the full Young projector can be chosen to coincide with that associated to the single standard Young tableau
\be \label{youngid1b}
\textrm{
\begin{picture}(30,70)(0,-15)
\multiframe(0,35)(15,0){1}(15,15){{\footnotesize $i_1$}}
\multiframe(15.5,35)(15,0){1}(15,15){{\footnotesize $j_1$}}
\multiframe(0,15.5)(15,0){1}(15,19){\vspace{7pt}$\vdots$}
\multiframe(15.5,15.5)(15,0){1}(15,19){\vspace{7pt}$\vdots$}
\multiframe(0,0)(15,0){1}(15,14.8){{\footnotesize $i_p$}}
\multiframe(15.5,0)(15,0){1}(15,14.8){{\footnotesize $j_p$}}
\multiframe(0,-15.3)(15,0){1}(15,15){{\footnotesize $k$}}
\end{picture}
}
\ee
which corresponds to the choice of standard labeling $i_n = 2\,n - 1$, $j_n = 2\,n$ for $1 \leq n \leq p$ and
$k = 2\,p + 1$. Thus, thanks to a standard property of Young tableaux, when acting with $Y_{\{2^p,1\}}$ all symmetrizations in $(\,i_n,j_n,k\,)$ vanish. This means that the terms in the sum can be manipulated according to
\be \label{manipulation}
Y_{\{2^p,1\}}\, \pr_{\,(\,i_n}\,T_{j_n\,)\,k}\, \prod_{r\,\neq\,n}^p \, T_{i_rj_r}\, \cF \, = \, - \ Y_{\{2^p,1\}}\, \pr_{\,k}\, T_{i_1j_1} \ldots\, T_{i_pj_p} \, \cF \, ,
\ee
while the products of traces can be everywhere replaced with their two-column projections on account of eq.~\eqref{constrF}. The $\{2^{\,p},1\}$ Young projection of eq.~\eqref{trbianchinp-b} thus reads
\be
(\,p+2\,) \ Y_{\{2^p,1\}} \, \pr_{\,k} \, \cF^{\,[\,p\,]}{}_{\,i_1
j_1,\,\ldots\,,\,i_p j_p}\, - \, \pr^{\,l} \,
\cF^{\,[\,p+1\,]}{}_{\,i_1
j_1,\,\ldots\,,\,i_p j_p\,,\,k\,l} \, = \, 0 \, . \label{trbianchi-b}
\ee
Notice that this result was obtained using only elementary properties of the involved two-column projections, without any need of knowing their detailed form. At any rate, the identities \eqref{trbianchi-b} contain the same terms entering eq.~\eqref{lag_var-b2}, and one can use them to eliminate the gauge variation of the ansatz \eqref{ansatz-b} provided the coefficients satisfy
\be
k_{\,p+1} \, = \, - \ \frac{k_{\,p}}{(\,p+1\,)(\,p+2\,)} \, .
\ee
Since $k_{\,0}=1$, this recursion relation is solved by
\be
k_{\,p} \, = \, \frac{(-1)^{\,p}}{p\,!\,(\,p+1\,)\,!} \, , \label{coeflab}
\ee
and indeed the coefficients in eq.~\eqref{coeflab} are those first identified by Labastida in \cite{laba_lag}, barring a slight change of conventions and a typo in their relative signs.

In order to identify the coefficients \eqref{coeflab} we only kept track of a particular class of traces of the Bianchi identities \eqref{bianchi-b2}. Notice, however, that in the constrained theory this is the only non-trivial one, since the combination
\be
\pr_{\,k}\, \cF^{\,[\,p\,]}{}_{\,i_1j_1,\,\ldots\,,\,i_pj_p} \, - \, \12 \, \sum_{n\,=\,1}^p \, \pr_{\,(\,i_n}\, \cF^{\,[\,p\,]}{}_{j_n\,)\,k\,,\,\ldots\,,\,i_{r\neq n}j_{r\neq n},\,\ldots}
\ee
appearing in eq.~\eqref{trbianchinp-b} is actually $\{2^p,1\}$-projected when eq.~\eqref{constrF} holds. The same is true also for the gradient terms. A simpler manifestation of this fact can be recognized in the fully symmetric case where, for instance, the trace of the Bianchi identity \eqref{bianchi_fronsdal} simply reduce to the gradient of the double trace of $\cF$, that vanishes on account of eq.~\eqref{constrF_symm}.

Before closing this section, let us also review Labastida's derivation of the coefficients \eqref{coeflab}, exploiting the self-adjointness of the Einstein-like tensors that will be useful to identify the field equations following from the Lagrangians \eqref{ansatz-b}. Starting from
\be
\cE \, = \, \sum_{p\,=\,0}^N \, k_{\,p}\ \h^{i_1j_1} \!\ldots\, \h^{i_pj_p}\, T_{i_1j_1} \ldots\, T_{i_pj_p} \, \cF
\ee
and resorting to the expression for the multiple traces of $\cF$ that is displayed in eq.~\eqref{fgentrace} one finds
\begin{align}
& \bra\, \vf \,\comma\, \cE\,(\,\phi\,)\,\ket \, = \, \bra\, \cE\,(\,\vf\,) \,\comma\, \phi \,\ket \nn \\
& - \sum_{p\,=\,0}^N \, \frac{1}{2^{\,p}} \, \bra\, Y_{\{2^p\}}\, T_{i_1j_1} \ldots\, T_{i_pj_p}\, \vf \,\comma  \left[\, (\,p+1\,)\,k_{\,p} \, + \, \frac{k_{\,p-1}}{p} \,\right] Y_{\{2^p\}} \sum_{n\,=\,1}^p \, \pr_{\,i_n}\pr_{\,j_n} \prod_{r\,\neq\,n}^p\, T_{i_rj_r}\, \phi \nn \\
& - \, \12\, \bigg[\, k_{\,p}\, + \, (\,p+1\,)\,(\,p+2\,)\,k_{\,p+1} \,\bigg]\, \pr^{\,k}\pr^{\,l}\, Y_{\{2^{p+1}\}}\, T_{i_1j_1} \ldots\, T_{i_pj_p}\,T_{kl}\, \phi \,\ket \, .
\label{selfadjbose}
\end{align}
This rewriting holds up to constraints and total derivatives. It follows from the combination of some of the contributions entering eq.~\eqref{fgentrace}, that is admitted when the multiple traces of $\vf$ or $\phi$ are substituted with their two-column projections. The relevant manipulations are similar to those adopted in the treatment of multiple traces of the Bianchi identities. In conclusion, if the $k_{\,p}$ are those in eq.~\eqref{coeflab}, the $\cE$ tensors are self-adjoint since eq.~\eqref{selfadjbose} up to total derivatives reduces to
\be
\bra\, \vf \,\comma\, \cE\,(\,\phi\,)\,\ket \, = \, \bra\, \cE\,(\,\vf\,) \,\comma\, \phi \,\ket \, .
\ee

To recapitulate, the Lagrangian for an arbitrary Bose field of mixed-symmetry is
\be \label{lagconstr_bose}
\cL \, = \, \12 \ \bra\, \vf \,\comma\, \sum_{p\,=\,0}^N \, \frac{(-1)^{\,p}}{p\,!\,(\,p+1\,)\,!} \ \h^{i_1j_1} \ldots\, \h^{i_pj_p}\, \cF^{\,[\,p\,]}{}_{\,i_1j_1,\,\ldots\,,\,i_pj_p} \,\ket \, .
\ee
Its most evident signature is the presence of higher traces of $\cF$. Let us stress that this property is definitely related to the presence of more than one family of symmetrized space-time indices or, equivalently, to the number of rows of the Young tableau describing the representation of the Lorentz group under scrutiny. Therefore, it is independent of the need for constraints, that is dictated by the number of columns in the Young tableau.

A clarifying example is provided by irreducible mixed-symmetry bosons with at most two columns. In the present symmetric convention, these fields carry an arbitrary number of families comprising at most two space-time indices, while all symmetrizations over any group of three indices vanish. They are the simplest class of mixed-symmetry Bose fields and they are actually \emph{unconstrained}. Their Lagrangians can be easily obtained generalizing the expression
\be \label{antigrav}
\cL \, = \, - \, \12 \ \pr^{\,\m}\, h^{\,\n}{}_{\,[\,\m}\, \pr_{\phantom{[}\!\n}\, h_{\,\r\,]}{}^{\,\r} \, ,
\ee
where the couple of square brackets denotes a complete antisymmetrization of the indices it encloses, with the minimum possible number of terms and with unit overall normalization. The Lagrangian \eqref{antigrav} coincides with the linearized Einstein-Hilbert one \eqref{lag_grav}, up to total derivatives. Furthermore, it is manifestly gauge-invariant under linearized diffeomorphisms, simply because two derivatives commute. In a similar fashion, when written directly in terms of the field, the Lagrangian for an arbitrary two-column boson\footnote{Notice that the indices coming after the semicolon are actually fully antisymmetrized between each others. As warned in Appendix \ref{app:MIX}, we shall often abide on this convention also in the following.} $\vf_{\,\m^1_1\m^1_2\,,\,\ldots\,,\,\m^p_1\m^p_2\,;\,\n_1\,\ldots\,\n_q}$ must be of the form
\begin{align}
\cL \, \sim & \ \pr^{\,\m_1} \, \vf^{\,\m_2}{}_{\,[\,\m_1\,|\,, \,\ldots\, ,}{}^{\,\m_{p+1}}{}_{\,|\,\m_p\,;\,\m_{p+1} \ldots\, \m_{p+q}\,|} \nn \\
& \, \times \, \pr_{\,|\,\m_{p+q+1}}\, \vf_{\,\m_{p+q+2}\,|}{}^{\,\m_{p+2}}{}_{\,, \,\ldots\, ,\,|\,\m_{2p+q+1}\,]}{}^{\,\m_{2p+1}; \,\m_{2p+2} \ldots\, \m_{2p+q+1}} \, . \label{antilagb}
\end{align}
This expression is again manifestly gauge invariant due to the antisymmetrizations it contains and, up to total derivatives, must be proportional to the Lagrangian \eqref{lagconstr_bose}, that is uniquely fixed by the request for gauge invariance. Furthermore, in eq.~\eqref{antilagb} the presence of higher traces of the field is manifest.


\subsubsection{Fermi fields}\label{sec:labalag-fermi}


In the fully symmetric case Bose and Fermi fields behave in a very similar way. Actually, the Fang-Fronsdal constraint \eqref{constr_fronsdal2-f} on the field induces a similar condition on the kinetic spinor-tensor,
\be \label{constrS_symm}
\ssl_{\,\m_1 \ldots\, \m_{s-3}\,\l}{}^{\,\l} \, = \, 0 \, ,
\ee
and the Lagrangians thus take the form
\be \label{lag_fang}
\cL \, = \, \12 \ \bar{\psi}^{\,\m_1 \ldots\, \m_s} \left(\, \cS_{\,\m_1 \ldots\, \m_s} \, - \, \12 \ \g_{\,(\,\m_1}\! \ssl_{\,\m_2 \ldots\, \m_s\,)} \, - \, \12 \ \h_{\,(\,\m_1\m_2}\, \cS_{\,\m_3 \ldots\, \m_s\,)\,\l}{}^{\,\l} \,\right) \, + \, \textrm{h.c.} \, .
\ee
The gauge variation of eq.~\eqref{lag_fang} does not entail particularly subtleties and reads
\begin{align}
& \d \, \cL \, = \, - \, \frac{s}{2} \ \bar{\e}^{\,\m_1 \ldots\, \m_{s-1}} \left(\, \pr\cdot \cS_{\,\m_1 \ldots\, \m_{s-1}} \, - \, \12 \dsl \ssl_{\,\m_1 \ldots\, \m_{s-1}} - \, \12 \ \pr_{\,(\,\m_1}\, \cS_{\,\m_2 \ldots\, \m_{s-1}\,)\,\l}{}^{\,\l} \,\right) \nn \\
& + \, \12 \, \binom{s}{2} \bar{\esl}^{\ \m_1 \ldots\, \m_{s-2}} \ \pr\,\cdot\! \ssl_{\, \m_1 \ldots\, \m_{s-2}} \, + \, \frac{3}{4} \, \binom{s}{3} \, \bar{\e}^{\,\m_1 \ldots \m_{s-3}\,\l}{}_{\,\l} \ \pr\cdot \cS_{\,\m_1 \ldots\, \m_{s-3}\,\l}{}^{\,\l} \, + \, \textrm{h.c.} \, .
\end{align}
The first group of terms cancels on account of the constrained Bianchi identity coming from eq.~\eqref{bianchi_fang}, while the others cancel on account of the constraint \eqref{constr_fronsdal2-f} on the gauge parameter.

In the mixed-symmetry case, the structure of the Lagrangians is roughly modified as in the bosonic case, since \emph{not} all triple $\g$-traces of the kinetic tensor $\cS$ are forced to vanish. Hence, the Lagrangians can and indeed do contain additional contributions with multiple $\g$-traces of $\cS$. Still in analogy with the bosonic case, the Labastida constraints on the fields \eqref{constr_field-f} induce similar constraints
\be \label{constrS}
T_{(\,ij}\,\g_{\,k\,)}\, \cS \, = \, 0
\ee
on the kinetic tensors. As a consequence, only particular linear combinations of $\g$-traces of $\cS$ can enter the Lagrangians. Notice that eq.~\eqref{constrS} can be easily checked via a direct computation:
\be \label{comp_constrS}
T_{(\,ij}\,\g_{\,k\,)}\, \cS \, = \, - \, i \, \Big(\, 2 \dsl \ T_{(\,ij} \, \g_{\,k\,)} \, \psi \, + \, \pr^{\,l} \, T_{(\,ij}\,T_{kl\,)} \, \psi \, + \, \pr^{\,l} \, T_{(\,ij} \, \g_{\,k\,)\,l} \, \psi \,\Big) \, .
\ee
The first two terms in this expression are directly related to the constraints \eqref{constr_field-f}, while the last one can be linked to them expanding the antisymmetric $\g$-trace according to
\be
T_{(\,ij} \, \g_{\,k\,)\,l}\, \psi \, = \, - \ \g_{\,l}\, T_{(\,ij}\,\g_{\,k\,)} \, \psi \, - \, T_{(\,ij}\,T_{kl\,)} \, \psi \, .
\ee

In order to select the non-vanishing combinations entering the Lagrangians one can again classify all candidate terms according to the irreducible representations of the permutation group acting on the family indices. Moreover, as anticipated in Section~\ref{sec:laba-fermi}, it is convenient to work in an antisymmetric basis for $\g$-matrices, introducing the fully antisymmetrized $\g$-traces
\be
\g_{\,k_1 \ldots\, k_q} \, = \, \frac{1}{q\,!}\ \g_{\,[\,k_1}\g_{\phantom{[}k_2} \ldots\, \g_{\,k_q\,]} \, .
\ee
The final outcome of this analysis is that, for any given number of
traces and antisymmetric $\g$-traces of the kinetic tensor $\cS$,
all Young projections in family indices with more than two
columns vanish in the constrained theory on account of
eq.~\eqref{constrS}. This condition identifies the wide class of
terms that can enter fermionic Lagrangians which, as a result, are
far more involved than their bosonic counterparts. For $N$ families
the Lagrangians are in fact bound to rest on ${\cal O}(N^2)$
distinct Young projections, with a first column of length $l_1\leq
N$ and a second column of length $l_2 \leq l_1$, that in general is
actually shorter, since antisymmetric $\g$-traces $\g_{\,k_1
\ldots\, k_q}$ of $\cS$ can accompany the ordinary $T_{ij}$ traces
that already enter the bosonic Lagrangians. Before using this information to present an ansatz for the Lagrangians for mixed-symmetry Fermi fields, let us sketch a proof of this statement. This goal can be reached repeating the analysis performed in the bosonic case. First, one can recognize that all Young projections associated to tableaux with at least three columns involve a sum of terms containing at least a symmetrization over three family indices. One can then study how the symmetrized indices can be spread over the products of traces and antisymmetrized $\g$-traces, to eventually recognize that all different possibilities can be related to the Labastida-like constraints \eqref{constrS}. This detailed analysis is performed in \cite{mixed_fermi}, but here we would like to refer to a simpler argument, that was also introduced in the same paper. Indeed, one can notice that
\be \label{expr1}
T_{i_1j_1} \ldots\, T_{i_{p-1}j_{p-1}} T_{(\,ab}\, \g_{\,c\,)\,k_1 \ldots\, k_{q-1}}
\ee
admits all Young projections allowed by the very general expression
\be \label{expr2}
T_{i_1j_1} \ldots\, T_{i_{p-1}j_{p-1}} T_{ab}\, \g_{\,c\,k_1 \ldots\, k_{q-1}} \, ,
\ee
aside from the two-column one. Furthermore, acting on each
irreducible component of \eqref{expr1} the permutation group can
generate the entire corresponding irreducible subspace, and in
particular the irreducible component of \eqref{expr2} that it
contains. This suffices to show that all Young projections
of the generic expression \eqref{expr2} with more than two columns are
actually related to terms containing an expression annihilated by the constraints. In fact, the previous
observation shows that they can be expressed as linear combinations
of \eqref{expr1} with similar quantities obtained permuting indices, while
\eqref{expr1} is related to the constraints via the identity
\be \label{expr3}
T_{(\,ab}\, \g_{\,c\,)\,k_1 \ldots\, k_{q-1}} = \, (-1)^{\,q+1} \, \g_{\,k_1 \ldots\, k_{q-1}}\, \g_{\,(\,a}\, T_{\,bc\,)} \, + \, (-1)^{\,q} \, \g_{\,[\,k_1 \ldots\, k_{q-2}}\, T_{\,k_{q-1}\,]\,(\,a}\, T_{\,bc\,)} \, .
\ee
The first term in eq.~\eqref{expr3} indeed contains manifestly symmetrized triple $\g$-traces, while the second can be related to them extending the symmetrizations as pertains to a
product of identical $T$ tensors.

In conclusion, only \emph{two-column projected} multiple ($\g$-)traces of $\cS$ can enter the constrained Lagrangians, and it is convenient to denote them as
\be \label{spq}
(\,\g^{\,[\,q\,]}\,\cS^{\,[\,p\,]}\,)_{\ i_1 j_1,\,\ldots\,,\,i_p
j_p \, ;\, k_1 \ldots\, k_q} \, \equiv \ Y_{\{\,2^p,\,1^q\}} \ T_{i_1
j_1} \ldots \, T_{i_p j_p} \, \g_{\,k_1 \ldots\, k_q} \, \cS \, .
\ee
Here $Y_{\{2^p,1^q\}}$ denotes the Young projector onto the $\{2,\ldots,2,1,\ldots,1\}$ irreducible representation of the permutation group acting on the family indices. Furthermore, the reader should appreciate that in eq.~\eqref{spq} colons separate again groups of symmetric indices, while a semicolon precedes the left-over group of antisymmetric ones. Hence, barring the substitution $\cF \to \cS$, for $q=0$ these terms reduce to those introduced in the previous section, while for $p=0$ they simply reduce to fully antisymmetrized $\g$-traces,
\be
(\,\g^{\,[\,q\,]}\,\cS^{\,[\,0\,]}\,)_{\,k_1 \ldots\, k_q} \, = \ \g_{\, k_1 \ldots\, k_q} \, \cS \, .
\ee
All other terms contain both traces and antisymmetrized $\g$-traces, as for instance
\be
(\,\g^{\,[\,1\,]}\,\cS^{\,[\,1\,]}\,)_{\,ij\,;\,k} \, = \, \frac{1}{3} \, \left\{\, 2\, T_{ij}\,\g_{\,k} \, - \, T_{k\,(\,i}\,\g_{\,j\,)}\, \,\right\} \cS \, .
\ee
An ansatz for the fermionic Lagrangians should contain all these terms, contracted with suitable products of invariant tensors. It is of course natural to consider both the $\h^{ij}$ operators introduced in eq.~\eqref{hij} and the operators
\be
\g^{\, i} \, \psi \, \equiv \, \g_{\,(\,\m^i_1|} \, \psi_{\,\ldots \,,\, |\, \m^i_2 \, \ldots \, \m^i_{s_i+1} ) \,,\, \ldots} \, ,
\ee
together with their antisymmetric combinations
\be
\g^{\,k_1 \ldots\, k_q} \, = \, \frac{1}{q\,!}\ \g^{\,[\,k_1}\g^{\phantom{[}k_2} \ldots\, \g^{\,k_q\,]} \, .
\ee

We now have all ingredients to propose an ansatz for the structure of the Lagrangians for Fermi fields of mixed symmetry \cite{mixed_fermi}. It reads
\be
\cL \, = \, \frac{1}{2} \ \bra\, \bar{\psi} \,\comma\!\! \sum_{p\,,\,q\,=\,0}^{N} k_{\,p\,,\,q} \ \h^{i_1 j_1} \!\ldots \h^{i_p j_p} \, \g^{\,k_1 \ldots\, k_q} \, (\,\g^{\,[\,q\,]}\,\cS^{\,[\,p\,]}\,)_{\, i_1 j_1\,,\,\ldots\,,\,i_p j_p \, ;\, k_1 \ldots\, k_q} \ket \, + \, \textrm{h.c.}\, ,
\label{lagferconstr}
\ee
where we are resorting to the scalar product that was already introduced for bosons, and that is described in detail in Appendix \ref{app:MIX}. Again, the upper limit of the sum is the number of index families, simply because it is not possible to antisymmetrize over more than $N$ family indices, as would be required by a Young tableaux with $N+1$ rows. We shall see shortly that the $k_{\,p\,,\,q}$ coefficients
in eq.~\eqref{lagferconstr} are uniquely determined by the condition
that $\cL$ be gauge invariant, up to the convenient choice $k_{\,0\,,\,0} \, = \, 1$.

As in the bosonic case and in the fully symmetric example, the gauge invariance of the Lagrangians is guaranteed by the Bianchi identities, that for \emph{constrained} mixed-symmetry fermions read
\be \label{bianchi_cf}
\pr_{\,i}\,\cS \, - \, \12 \dsl \, \g_{\,i}\, \cS \, - \, \12 \ \pr^{\,j}\,T_{ij}\,\cS \, - \, \frac{1}{6} \ \pr^{\,j}\,\g_{\,ij}\,\cS \, = \, 0 \, .
\ee
In fact, the ``classical anomaly'' appearing in the right-hand side of eq.~\eqref{bianchi-f1} cancels after imposing the Labastida constraints \eqref{constr_field-f} on the field. Again, eq.~\eqref{bianchi_cf} cannot directly deal with the higher traces contained in the ansatz \eqref{lagferconstr}, but suitable combinations of multiple $\g$-traces of the Bianchi identities can, barring some new subtleties that we shall dwell upon shortly. In order to proceed, one can begin by noticing that the identities
collected in Appendix \ref{app:identities} make it possible to recast the
gauge variation of eq.~\eqref{lagferconstr}, up to total
derivatives, in the form
\begin{align}
\d \, \cL \, & = \, - \sum_{p\,,\,q\,=\,0}^{N} \, \frac{1}{2^{\,p+1}} \ \bra \, T_{i_1j_1} \ldots T_{i_pj_p} \, \bar{\e}_{\,l} \, \g_{\,k_1 \ldots\, k_q} \,\comma\, k_{\,p\,,\,q} \ \pr_{\,l} \, (\,\g^{\,[\,q\,]}\,\cS^{\,[\,p\,]}\,)_{\, i_1 j_1,\,\ldots\,,\,i_p j_p \, ;\, k_1 \ldots\, k_q} \nn \\
& + \,(\,q+1\,)\,k_{\,p\,,\,q+1} \dsl \, (\,\g^{\,[\,q+1\,]}\,\cS^{\,[\,p\,]}\,)_{\, i_1 j_1,\,\ldots\,,\,i_p j_p \, ;\, k_1 \ldots\, k_q \, l} \nn \\[5pt]
& + \,(\,p+1\,)\,k_{\,p+1,\,q} \ \pr^{\,m}\, (\,\g^{\,[\,q\,]}\,\cS^{\,[\,p+1\,]}\,)_{\, i_1 j_1,\,\ldots\,,\,i_p j_p\,,\,lm \, ;\, k_1 \ldots\, k_q} \nn \\[5pt]
& + \,(\,q+1\,)\,(\,q+2\,)\,k_{\,p\,,\,q+2} \ \pr^{\,m}\, (\,\g^{\,[\,q+2\,]}\,\cS^{\,[\,p\,]}\,)_{\, i_1 j_1,\,\ldots\,,\,i_p j_p \, ;\, k_1 \ldots\, k_q\,lm} \, \ket \, + \, \textrm{h.c.} \, . \label{gaugefgen}
\end{align}
All $(\,\g^{\,[\,m\,]}\,\cS^{\,[\,n\,]}\,)$ terms that are
right entries of the scalar products are two-column projected by
assumption. This statement applies to all free indices carried by
the last three terms, while the first is slightly
more complicated, since due to the $l$ index carried by the divergence
it also allows projections with three boxes
in the first row. However, the left entry of the scalar product
makes this type of projections proportional to the constraints on the
gauge parameters \eqref{constr_gauge-f}. As usual, this can be realized resorting to arguments similar to those used to discard from the ansatz \eqref{lagferconstr} the tableaux with more than two columns. Thus, up to the Labastida constraints one can effectively restrict the attention to two-column projections of
the gauge variation \eqref{gaugefgen}.

On the other hand, in sharp contrast with the bosonic case, here two types of two-column projections are allowed for the terms of the type
\be \label{tpgammaq}
T_{i_1j_1} \ldots\, T_{i_pj_p} \, \bar{\e}_{\,l} \, \g_{\,k_1 \ldots\, k_q} \ee
that are left entries of the scalar products. One can obtain for them a two-column projection acting on $\bar{\e}_{\,l}$ with an already two-column projected combination of traces and antisymmetric $\g$-traces, but the possible outcomes are
\be \label{deceps}
\{\,2^{\,p},\,1^q\,\} \, \otimes \, \{1\} \, = \,  \{\,2^{\,p},\,1^{q+1}\} \, \oplus \, \{\,2^{\,p+1},\,1^{q-1}\} \, \oplus \, \{3,\,2^{\,p-1},\,1^q\} \, ,
\ee
where the last term is related to the constraints on the gauge parameters. Thus, after using the constraints \eqref{constr_gauge-f}, the gauge variation \eqref{gaugefgen} can be cast in the form
\begin{align}
& \d \, \cL \, = \, - \sum_{p\,,\,q\,=\,0}^{N} \, \frac{1}{2^{\,p+1}} \ \bra \, Y_{\{2^p,1^{q+1}\}}\, T_{i_1j_1} \ldots T_{i_pj_p} \, \bar{\e}_{\,l} \, \g_{\,k_1 \ldots\, k_q} \,\comma \nn \\
& k_{\,p\,,\,q} \ \pr_{\,l} \, (\,\g^{\,[\,q\,]}\,\cS^{\,[\,p\,]}\,)_{\, i_1 j_1,\,\ldots\,,\,i_p j_p \, ;\, k_1 \ldots\, k_q} + \,(\,q+1\,)\,k_{\,p\,,\,q+1}\! \dsl \, (\,\g^{\,[\,q+1\,]}\,\cS^{\,[\,p\,]}\,)_{\, i_1 j_1,\,\ldots\,,\,i_p j_p \, ;\, k_1 \ldots\, k_q \, l} \nn \\[5pt]
& + \,(\,p+1\,)\,k_{\,p+1,\,q} \ \pr^{\,m}\, (\,\g^{\,[\,q\,]}\,\cS^{\,[\,p+1\,]}\,)_{\, i_1 j_1,\,\ldots\,,\,i_p j_p\,,\,lm \, ;\, k_1 \ldots\, k_q} \nn \\[5pt]
& + \,(\,q+1\,)\,(\,q+2\,)\,k_{\,p\,,\,q+2} \ \pr^{\,m}\, (\,\g^{\,[\,q+2\,]}\,\cS^{\,[\,p\,]}\,)_{\, i_1 j_1,\,\ldots\,,\,i_p j_p \, ;\, k_1 \ldots\, k_q\,lm} \, \ket \nn \\[5pt]
& - \sum_{p\,,\,q\,=\,0}^{N} \, \frac{1}{2^{\,p+1}} \ \bra \, Y_{\{2^{p+1},1^{q-1}\}}\, T_{i_1j_1} \ldots T_{i_pj_p} \, \bar{\e}_{\,l} \, \g_{\,k_1 \ldots\, k_q} \,\comma \nn \\
& \phantom{+}\, k_{\,p\,,\,q} \ \pr_{\,l} \, (\,\g^{\,[\,q\,]}\,\cS^{\,[\,p\,]}\,)_{\, i_1 j_1,\,\ldots\,,\,i_p j_p \, ;\, k_1 \ldots\, k_q} \nn \\[5pt]
& + \, (\,p+1\,)\, k_{\,p+1\,,\,q} \ \pr^{\,m}\, (\,\g^{\,[\,q\,]}\,\cS^{\,[\,p+1\,]}\,)_{\, i_1 j_1,\,\ldots\,,\,i_p j_p\,,\,lm \, ;\, k_1 \ldots\, k_q} \, \ket \, + \, \textrm{h.c.}\, . \label{gaugefgen2}
\end{align}
Notice that the first term in eq.~\eqref{gaugefgen} gives contributions to both groups of projected terms above, while the second term gives a single contribution. The analysis of the gradient terms is slightly more subtle, but the end result is that one of them, the third term in eq.~\eqref{gaugefgen}, contributes to both. Indeed,
\begin{align}
& \{2^{\,p},\,1^{q+1}\} \, \otimes \, \{1\} \, = \,  \{\,2^{\,p},\,1^{q+2}\} \, \oplus \, \{\,2^{\,p+1},\,1^{q}\} \, \oplus \, \{3,\,2^{\,p-1},\,1^{q+1}\} \, , \nn \\
& \{2^{\,p+1},\,1^{q-1}\} \, \otimes \, \{1\} \, = \,  \{\,2^{\,p+1},\,1^{q}\} \, \oplus \, \{\,2^{\,p+2},\,1^{q-2}\} \, \oplus \, \{3,\,2^{\,p},\,1^{q-1}\} \, ,
\end{align}
so that both types of projections are compatible with the $\{2^{p+1},1^q\}$ projection carried by $(\,\g^{\,[\,q\,]}\,\cS^{\,[\,p+1\,]}\,)$. This is no longer true for the fourth term in eq.~\eqref{gaugefgen}.

At this stage fermionic fields bring about
a novel type of complication. Indeed, the $\g$-traces of the Bianchi identities do \emph{not} suffice to
directly set to zero the two groups of terms of eq.~\eqref{gaugefgen2}, but
these, on the other hand, are not fully independent and can be
partly combined using the constraints \eqref{constr_gauge-f} on the
gauge parameters. In fact the projections $\{\,2^{\,p},\,1^{q+1}\}$ and $\{\,2^{\,p+1},\,1^{q-1}\}$ appearing in eq.~\eqref{deceps} can actually arise from \emph{pairs} of neighboring contributions of the form \eqref{tpgammaq}, with values of $p$ differing by one unit and with values of $q$ differing by two units. The key step to proceed is to relate these two types
of terms, that are only apparently different, making use of the
constraints \eqref{constr_gauge-f} on the gauge parameters and of proper relabelings. To this
end, it proves particularly convenient to manipulate the
combinations $\bar{\e}_{\,l}\, \g_{\,k_1 \ldots\, k_q}$ according to
\be\label{q-terms}
\bar{\e}_{\,l} \, \g_{\,k_1 \ldots\, k_q} \, = \, \frac{1}{q+1} \ \bar{\e}_{\,[\,l} \, \g_{\,k_1 \ldots\, k_q\,]} \, + \, \frac{1}{q+1}\, \sum_{n\,=\,1}^{q} \, (-1)^{(n+1)} \ \bar{\e}_{\,(\,l} \, \g_{\,k_n\,)\,k_1 \ldots\, k_{r\neq n} \ldots\, k_q} \, .
\ee
The terms entering the sum can then be rewritten more conveniently as
\be \label{littlewood}
\bar{\e}_{\,(\,l} \, \g_{\,k_1\,) \, k_2 \ldots\, k_q} \, = \, \bar{\e}_{\,(\,l} \, \g_{\,k_1\,)} \, \g_{\, k_2 \ldots\, k_q} \, - \, T_{[\,k_2\,|\,(\,l} \, \bar{\e}_{\,k_1\,)} \, \g_{\,|\,k_3 \ldots\, k_q\,]} \, ,
\ee
where the first contribution on the right-hand side clearly vanishes
in the constrained setting, on account of eq.~\eqref{constr_gauge-f}.
Enforcing the constraints thus leads to
\be
\sum_{n\,=\,1}^{q} \, (-1)^{(n+1)} \ \bar{\e}_{\,(\,l} \, \g_{\,k_n\,)\,k_1 \ldots\, k_{r\neq n} \ldots\, k_q} \, \to \ T_{l\,[\,k_1} \, \bar{\e}_{\,k_2} \, \g_{\,k_3 \ldots\, k_q\,]} \, ,
\ee
since the sum with oscillating signs induces a complete antisymmetry in the $k_r$ indices, that as such forbids the simultaneous presence of two of them on the trace $T$. As a result, in the constrained theory eq.~\eqref{q-terms} implies
\be \label{relstruct}
T_{i_1j_1} \ldots T_{i_pj_p} \, \bar{\e}_{\,l} \, \g_{\,k_1 \ldots\, k_q} = \, \frac{1}{q+1} \ T_{i_1j_1} \ldots\, T_{i_pj_p} \left(\, \bar{\e}_{\,[\,l} \, \g_{\,k_1 \ldots\, k_q\,]} \, + \, T_{\,l\,[\,k_1} \bar{\e}_{\,k_2} \, \g_{\,k_{3} \ldots\, k_q\,]} \,\right) ,
\ee
where the first term is annihilated by the $\{2^{p+1},1^{q-1}\}$
projection, while the second is similarly annihilated by the
$\{2^{p},1^{q+1}\}$ projection. In conclusion one is led to
\begin{align}
& Y_{\{\,2^{p+1},\,1^{q-1}\}}\, T_{i_1j_1}
\ldots T_{i_pj_p} \, \bar{\e}_{\,l} \, \g_{\,k_1 \ldots\, k_q} \, =
\, \frac{1}{q+1} \ Y_{\{\,2^{p+1},\,1^{q-1}\}} \, T_{i_1j_1} \ldots
T_{i_pj_p} \, T_{\,l\,[\,k_1} \bar{\e}_{\,k_2} \, \g_{\,k_{3}
\ldots\, k_q\,]} \, , \nn \\[5pt]
& Y_{\{\,2^p,\,1^{q+1}\}}\, T_{i_1j_1} \ldots T_{i_pj_p} \, \bar{\e}_{\,l} \, \g_{\,k_1 \ldots\, k_q} \, = \, \frac{1}{q+1} \ Y_{\{\,2^p,\,1^{q+1}\}} \, T_{i_1j_1} \ldots T_{i_pj_p} \, \bar{\e}_{\,[\,l} \, \g_{\,k_1 \ldots\, k_q\,]} \, .\label{projfgauge}
\end{align}

While the second of these relations is rather trivial, the first is quite
interesting, since it connects a relatively complicated projection
of $p$ traces and a single $q$-fold antisymmetric $\g$-trace to a
simpler one involving ${p+1}$ traces and a $(q-2)$-fold
antisymmetric $\g$-trace with the maximum possible number of
antisymmetrizations. Let us stress that one can also proceed in the opposite direction, decreasing the value of $p$. The relevant identity,
that can be proved with similar techniques is then
\begin{align}
& Y_{\{\,2^p,1^{q+1}\}} \, T_{i_1j_1} \ldots T_{i_pj_p} \,
\bar{\e}_{\,l}
\, \g_{\,k_1 \ldots\, k_q} \nn \\
& = \, \frac{(-1)^{\,q+1}}{p\,(\,q+1\,)} \, Y_{\{\,2^p,1^{q+1}\}} \sum_{n\,=\,1}^{p} \,
\left(\,\prod_{r\,\ne\,n}^p \, T_{i_rj_r}\, \right) \, \bar{\e}_{\,(\,i_n} \, \g_{\,j_n\,)\,k_1 \,
\ldots\, k_q \, l} \, ,
\end{align}
but it is simpler to work with terms whose projections
involve the maximum number of antisymmetrizations compatible with
their tensorial character.

One can now manipulate the second group of terms in eq.~\eqref{gaugefgen2} using the key relation appearing in the first line of eq.~\eqref{projfgauge}. As a result, after a proper relabeling eq.~\eqref{gaugefgen2} can be finally turned into
\begin{align}
\d \, \cL \, & = \, - \sum_{p\,,\,q\,=\,0}^{N} \, \frac{1}{2^{\,p+1}} \ \bra \, Y_{\{\,2^p,\,1^{q+1}\}}\, T_{i_1j_1} \ldots T_{i_pj_p} \, \bar{\e}_{\,l} \, \g_{\,k_1 \ldots\, k_q} \,\comma \nn \\
& \phantom{+} \, k_{\,p\,,\,q} \ \pr_{\,l} \, (\,\g^{\,[\,q\,]}\,\cS^{\,[\,p\,]}\,)_{\, i_1 j_1,\,\ldots\,,\,i_p j_p \, ;\, k_1 \ldots\, k_q} \nn \\[2pt]
& + \, \frac{(\,q+1\,)\,(\,q+2\,)}{p\,(\,q+3\,)} \, k_{\,p-1,\,q+2}  \ \sum_{n\,=\,1}^{p}\, \pr_{\,(\,i_n\,|}\, (\,\g^{\,[\,q+2\,]}\,\cS^{\,[\,p-1\,]}\,)_{\,\ldots\,,\,i_{r\,\ne\,n} j_{r\,\ne\,n} \,,\,\ldots\, ;\,|\,j_n\,)\,l\, k_1 \ldots\, k_q} \nn \\[2pt]
& + \,(\,q+1\,)\,k_{\,p\,,\,q+1} \dsl \, (\,\g^{\,[\,q+1\,]}\,\cS^{\,[\,p\,]}\,)_{\, i_1 j_1,\,\ldots\,,\,i_p j_p \, ;\, k_1 \ldots\, k_q \, l} \nn \\[5pt]
& + \,(\,p+1\,)\,k_{\,p+1,\,q} \ \pr^{\,m}\, (\,\g^{\,[\,q\,]}\,\cS^{\,[\,p+1\,]}\,)_{\, i_1 j_1,\,\ldots\,,\,i_p j_p\,,\,lm \, ;\, k_1 \ldots\, k_q} \nn \\[3pt]
& + \, \frac{(\,q+1\,)\,(\,q+2\,)}{q+3} \, k_{\,p\,,\,q+2} \, \sum_{n\,=\,1}^{p} \, \pr^{\,m}\, (\,\g^{\,[\,q+2\,]}\,\cS^{\,[\,p\,]}\,)_{\,\ldots\,,\,i_{r\,\ne\,n} j_{r\,\ne\,n},\,\ldots\,,\,m\,(\,i_n ;\, j_n\,)\, l \, k_1 \ldots\, k_q} \nn \\[2pt]
& + \,(\,q+1\,)\,(\,q+2\,)\,k_{\,p\,,\,q+2} \ \pr^{\,m}\, (\,\g^{\,[\,q+2\,]}\,\cS^{\,[\,p\,]}\,)_{\, i_1 j_1,\,\ldots\,,\,i_p j_p \, ;\, k_1 \ldots\, k_q\,lm} \ket \, + \, \textrm{h.c.}\, ,
\end{align}
where we enforced the symmetry under interchanges of the $(\,i_n,j_n\,)$ pairs brought about by the products of identical $T$ tensors present on the right-hand sides of the scalar products. Furthermore, the contributions in the last two terms are proportional, since
\be
(\,\g^{\,[\,q+2\,]}\,\cS^{\,[\,p\,]}\,)_{\, i_1 j_1,\,\ldots\,,\,i_p j_p \, ;\, k_1 \ldots\, k_q\,lm} = (\,\g^{\,[\,q+2\,]}\,\cS^{\,[\,p\,]}\,)_{\, i_1 j_1,\,\ldots\,,\,i_{p-1} j_{p-1} \,,\,m\,(\,i_p\, ;\,j_p\,)\, l \, k_1 \ldots\, k_q} \, ,
\ee
because a symmetrization in $(\,i_p,j_p,m\,)$ vanishes on account of the two-column projection. Therefore, on account of the constraints \eqref{constr_gauge-f}, eq.~\eqref{gaugefgen} finally reduces to
\begin{align}
\d \, \cL \, &  = \, - \sum_{p\,,\,q\,=\,0}^{N} \, \frac{1}{2^{\,p+1}} \ \bra \, Y_{\{\,2^p,\,1^{q+1}\}}\, T_{i_1j_1} \ldots T_{i_pj_p} \, \bar{\e}_{\,l} \, \g_{\,k_1 \ldots\, k_q} \,\comma \nn \\
& \phantom{+} \, k_{\,p\,,\,q} \ \pr_{\,l} \, (\,\g^{\,[\,q\,]}\,\cS^{\,[\,p\,]}\,)_{\, i_1 j_1,\,\ldots\,,\,i_p j_p \, ;\, k_1 \ldots\, k_q} \nn \\[2pt]
& + \, \frac{(\,q+1\,)\,(\,q+2\,)}{p\,(\,q+3\,)} \, k_{\,p-1,\,q+2} \ \sum_{n\,=\,1}^{p}\, \pr_{\,(\,i_n\,|}\, (\,\g^{\,[\,q+2\,]}\,\cS^{\,[\,p-1\,]}\,)_{\,\ldots\,,\,i_{r\,\ne\,n} j_{r\,\ne\,n}\,,\,\ldots \, ;\,|\,j_n\,)\,l \, k_1 \ldots\, k_q} \nn \\[2pt]
& + \, \frac{(\,q+1\,)\,(\,q+2\,)\,(\,q+p+3\,)}{q+3} \, k_{\,p\,,\,q+2}  \, \pr^{\,m} \, (\,\g^{\,[\,q+2\,]}\cS^{\,[\,p\,]}\,)_{\, i_1 j_1,\,\ldots\,,\,i_p j_p \, ;\, k_1 \ldots\, k_q\,lm} \nn \\[2pt]
& + \,(\,q+1\,)\,k_{\,p\,,\,q+1} \dsl \, (\,\g^{\,[\,q+1\,]}\,\cS^{\,[\,p\,]}\,)_{\, i_1 j_1,\,\ldots\,,\,i_p j_p \, ;\, k_1 \ldots\, k_q \, l} \nn \\[5pt]
& + \,(\,p+1\,)\,k_{\,p+1,\,q} \ \pr^{\,m}\, (\,\g^{\,[\,q\,]}\,\cS^{\,[\,p+1\,]}\,)_{\, i_1 j_1,\,\ldots\,,\,i_p j_p\,,\,lm \, ;\, k_1 \ldots\, k_q} \ket + \textrm{h.c.}\, . \label{finalgaugef}
\end{align}

In the fermionic case, the $\g$-traces of the Bianchi identities
take the rather involved form presented in eq.~\eqref{gtrace_bianchi}, but it is possible to extract from them the far simpler
chain of $\{\,2^{\,p},1^{q+1}\}$-projected identities
\begin{align}
& (\,p+q+2\,) \ Y_{\{\,2^p,\,1^{q+1}\}} \, \pr_{\,l} \, (\,\g^{\,[\,q\,]}\,\cS^{\,[\,p\,]}\,)_{\, i_1 j_1,\,\ldots\,,\,i_p j_p \, ;\, k_1 \ldots\, k_q} \nn \\
+ \ & \frac{1}{q+3} \ Y_{\{\,2^p,\,1^{q+1}\}} \, \sum_{n\,=\,1}^{p}\, \pr_{\,(\,i_n\,|}\, (\,\g^{\,[\,q+2\,]}\,\cS^{\,[\,p-1\,]}\,)_{\,\ldots\,,\,i_{r\,\ne\,n} j_{r\,\ne\,n} \,,\,\ldots\, ;\,|\,j_n\,)\,l \, k_1 \ldots\, k_q}  \nn \\[2pt]
+ \ & (-1)^{\,q+1} \dsl \, (\,\g^{\,[\,q+1\,]}\,\cS^{\,[\,p\,]}\,)_{\, i_1 j_1,\,\ldots\,,\,i_p j_p \, ;\, k_1 \ldots\, k_q \, l}  \, - \, \pr^{\,m}\, (\,\g^{\,[\,q\,]}\,\cS^{\,[\,p+1\,]}\,)_{\, i_1 j_1,\,\ldots\,,\,i_p j_p\,,\,lm \, ;\, k_1 \ldots\, k_q} \nn \\[5pt]
- \ & \frac{1}{q+3} \ \pr^{\,m}\,
(\,\g^{\,[\,q+2\,]}\,\cS^{\,[\,p\,]}\,)_{\, i_1 j_1,\,\ldots\,,\,i_p
j_p \, ;\, k_1 \ldots\, k_q\,lm} \, = \, 0\, . \label{finalbianchif}
\end{align}
This can be done resorting to techniques similar to those applied to manipulate eq.~\eqref{trbianchinp-b} in the bosonic case, and more details are presented in Appendix \ref{app:proofs}.
Comparing eqs.~\eqref{finalgaugef} and \eqref{finalbianchif} one can
now recognize that, in order to obtain a gauge invariant Lagrangian, the
coefficients must satisfy the two recursion relations
\bea
k_{\,p+1,\,q} & = & - \, \frac{1}{(\,p+1\,)\,(\,p+q+2\,)} \, k_{\,p\,,\,q} \, , \nonumber \\
k_{\,p\,,\,q+1} & = & \frac{(-1)^{\,q+1}}{(\,q+1\,)\,(\,p+q+2\,)} \, k_{\,p\,,\,q} \, ,
\eea
whose solution is
\be \label{solgenf}
k_{\,p\,,\,q} \, =  \,
\frac{(-1)^{\,p\,+\,\frac{q\,(q+1)}{2}}}{p\,!\,q\,!\,(\,p+q+1\,)\,!} \
\ee
if $k_{\,0\,,\,0}=1$. Notice that this expression includes as special cases the coefficients first obtained by Labastida for Bose fields: in fact, letting $q=0$ in eq.~\eqref{solgenf} recovers
the result presented in eq.~\eqref{coeflab}.

The coefficients given in eq.~\eqref{solgenf} determine completely the constrained Lagrangians \eqref{lagferconstr} for mixed-symmetry Fermi fields. Together with the unconstrained Lagrangians that we shall come across in Section \ref{chap:unconstrained}, they are the main results of \cite{mixed_bose,mixed_fermi}.

Interestingly, eq.~\eqref{solgenf} can be also obtained demanding that the
kinetic operator be self-adjoint, along the lines of what was done
by Labastida for Bose fields in \cite{laba_lag}. The idea is
to build a Rarita-Schwinger-like tensor such that
\be \label{selfadjoint}
\bra \bar{\psi} \comma \cE \ket \, + \, \textrm{h.c.} \, = \, 2 \, \bra \bar{\psi} \comma \cE
\ket \, ,
\ee
up to constraints and total derivatives, starting from a general
ansatz of the type
\be \label{einsteingen}
\cE \, = \, \sum_{p\,,\,q\,=\,0}^N \, k_{\,p\,,\,q} \ \h^{i_1j_1} \ldots\, \h^{i_pj_p} \,
\g^{\,k_1 \ldots\, k_q} \ T_{i_1j_1} \ldots\, T_{i_pj_p} \, \g_{\,k_1 \ldots\, k_q} \,
\cS \, .
\ee
Whereas this sum apparently differs from the one contained in
eq.~\eqref{lagferconstr}, in the constrained case they actually
coincide, since each term gives rise to a \emph{unique} two-column
projection. Hence, making use of the result for the general
$\g$-trace of the $\cS$ tensor given in eq.~\eqref{sgentrace}, up to
total derivatives and constraints one is led to the condition
\begin{align}
& \bra \bar{\psi} \comma \cE \ket \, + \, \textrm{h.c.} \, = \, 2 \,
\bra \bar{\psi} \comma \cE \ket \, + \, \sum_{p\,,\,q\,=\,0}^N \,
\frac{i}{2^{\,p+1}} \, \bra\, Y_{\{2^p,1^q\}} \, T_{i_1j_1} \ldots\,
T_{i_pj_p} \, \bar{\psi} \, \g_{\,k_1 \ldots\, k_q} \comma \nn \\
& -
\left[\, (\,p+q+1\,)\, k_{\,p\,,\,q} \, - \,
\frac{(-1)^{\,q}}{q} \, k_{\,p\,,\,q-1} \,\right] \, Y_{\{2^p,1^q\}}
\,\g_{\,[\,k_1 \ldots\, k_{q-1}\,} \pr_{\,k_q\,]} \prod_{r\,=\,1}^p
\, T_{i_rj_r} \, \psi \nn \\
& +  \left[\, k_{\,p\,,\,q} \, - \,
(-1)^{\,q} \, \frac{q+1}{p}
\, k_{\,p-1\,,\,q+1}\,\right] \, Y_{\{2^p,1^q\}} \, \sum_{n\,=\,1}^p
\, \g_{\,k_1 \ldots\, k_q \, (\,i_n} \, \pr_{\,j_n\,)}
\prod_{r\,\ne\,n}^p \, T_{i_rj_r} \, \psi \nn \\
& +  \bigg[\, k_{\,p\,,\,q} \, + \,
(-1)^{\,q} \, (\,q+1\,)\,(\,p+q+2\,) \, k_{\,p\,,\,q+1} \,\bigg] \,
\pr^{\,l} \, Y_{\{2^p,1^{q+1}\}} \, \g_{k_1 \ldots\, k_q\, l}
\prod_{r\,=\,1}^p \, T_{i_rj_r} \, \psi \nn \\
& + \left[\,
k_{\,p\,,\,q} \, + \, (-1)^{\,q} \, \frac{p+1}{q}
\, k_{\,p+1\,,\,q-1} \,\right] \pr^{\,l} \, Y_{\{2^{p+1},1^{q-1}\}}
\,  \g_{\,[\,k_1 \ldots\, k_{q-1}\,} T_{k_q\,]\,l} \prod_{r\,=\,1}^p
\, T_{i_rj_r} \, \psi \ket \, ,
\label{selfadjferm}
\end{align}
so that the remainder vanishes precisely if the coefficients
$k_{\,p\,,\,q}$ are those of eq.~\eqref{solgenf}. To reiterate, the
coefficients for the general fermionic Lagrangians can be derived in
two distinct ways: enforcing the gauge symmetry via the Bianchi identities as in \cite{mixed_fermi}, or alternatively requiring that the kinetic operator be self adjoint, as was done only for Bose fields in \cite{laba_lag}. In Section \ref{sec:minimal-fermi} we shall see that the first method appears more natural in the unconstrained case. Moreover, as in the bosonic case, it also definitely shows that the Bianchi identities provide the frame of the far more involved Lagrangians \eqref{lagferconstr}.

In order to summarize the result of this rather long discussion, let us briefly comment on the structure of the Lagrangian for an arbitrary Fermi field of mixed-symmetry, that reads
\be
\cL \, = \, \12 \, \bra\, \bar{\psi} \,\comma \!\sum_{p\,,\,q\,=\,0}^{N} \, \frac{(-1)^{\,p\,+\,\frac{q\,(q+1)}{2}}}{p\,!\,q\,!\,(\,p+q+1\,)\,!} \ \h^{\,p} \, \g^{\,q}\, (\,\g^{\,[\,q\,]}\,\cS^{\,[\,p\,]}\,) \,\ket \, + \, \textrm{h.c.}\, ,
\label{lagconstr_fermi}
\ee
where for brevity we have introduced the shorthand
\be \label{shorthand}
\h^{\,p} \, \g^{\,q}\, (\,\g^{\,[\,q\,]}\,\cS^{\,[\,p\,]}\,) \, \equiv \, \h^{i_1 j_1} \ldots \, \h^{i_p j_p} \, \g^{\,k_1 \ldots\, k_q} \ (\,\g^{\,[\,q\,]}\,\cS^{\,[\,p\,]}\,)_{\, i_1 j_1\,,\,\ldots\,,\,i_p j_p \, ;\, k_1 \ldots\, k_q} \, .
\ee
As in the bosonic case, the most evident signature of the Lagrangian \eqref{lagconstr_fermi} is the presence of higher $\g$-traces of $\cS$. Again, this property is definitely related to the presence of more than one family of symmetrized space-time indices or, equivalently, to the number of rows of the Young tableau describing the representation of the Lorentz group under scrutiny. On the contrary, the need for constraints is dictated by the number of columns in the Young tableau.

A clarifying example is provided by fully antisymmetric, ``single-column'' fermions. They are the simplest class of mixed-symmetry Fermi fields and they are actually \emph{unconstrained}. Their Lagrangians can be easily obtained generalizing the usual presentation of the Rarita-Schwinger Lagrangian, that takes the form
\be \label{antirarita}
\cL \, = \, \frac{i}{2} \ \bar{\psi}_{\,\m}\, \g^{\,\m\n\r}\, \pr_{\,\n}\, \psi_{\,\r} \, + \, \textrm{h.c.} \ .
\ee
Up to total derivatives, it actually coincides with the Lagrangian that one can extract from eq.~\eqref{lag_fang} for a spin-($3/2$) fermion. Eq.~\eqref{antirarita} is manifestly left invariant by the gauge transformation $\d\, \psi_{\,\m} \, = \, \pr_{\,\m}\, \e$, simply because two derivatives commute. In a similar fashion, when written directly in terms of the field, the Lagrangian for an arbitrary single-column fermion\footnote{Notice the slight abuse of notation: here the indices carried by the field are fully antisymmetrized, rather than fully symmetrized as, for instance, in eq.~\eqref{lag_fang}.} $\psi_{\,\n_1 \ldots\, \n_q}$ can be cast in the form
\be \label{antilagf}
\cL \, = \, i \ \frac{(-1)}{2\,q\,!}^{\frac{q\,(q-1)}{2}} \ \bar{\psi}_{\,\m_1 \ldots\, \m_q}\, \g^{\,\m_1 \ldots\, \m_q  \l \, \n_1 \ldots\, \n_q} \, \pr_{\,\l}\, \psi_{\,\n_1 \ldots\, \n_q}  + \, \textrm{h.c.} \ .
\ee
This expression is again manifestly gauge invariant due to the presence of the fully antisymmetrized $\g$-matrix and, up to total derivatives, must be proportional to the Lagrangian \eqref{lagconstr_fermi}, that is uniquely fixed by the request for gauge invariance. Furthermore, in eq.~\eqref{antilagf} the presence of higher $\g$-traces of the field is manifest.


\subsection{Lagrangian field equations}\label{sec:lag-eq}


In the previous sections, together with a set of gauge transformations and a set of constraints to be enforced on fields and gauge parameters, we first presented gauge-invariant differential operators and then gauge-invariant quadratic Lagrangians built upon them. We also saw that the null eigenfunctions of these differential operators propagate free massless modes. The next step is to ask whether the quadratic Lagrangians \eqref{lagconstr_bose} and \eqref{lagconstr_fermi} give rise to field equations equivalent to the non-Lagrangian ones of eqs.~\eqref{laba_eq-b} and \eqref{laba_eq-f}, that set to zero these differential operators.

First of all one has to vary the Lagrangians in order to extract the corresponding field equations. In eq.~\eqref{selfadjbose} we showed that the Einstein-like tensor
\be \label{Ebose}
\cE_{\,\textrm{Bose}} \, = \, \frac{(-1)^{\,p}}{p\,!\,(\,p+1\,)\,!} \ \h^{i_1j_1} \ldots\, \h^{i_pj_p}\, \cF^{\,[\,p\,]}{}_{\,i_1j_1,\,\ldots\,,\,i_pj_p}
\ee
entering eq.~\eqref{lagconstr_bose} is self-adjoint, as first pointed out by Labastida \cite{laba_lag}. As a consequence, up to total derivatives, the $\d\,\vf$ variation of the Lagrangians \eqref{lagconstr_bose} can be cast in the form
\be \label{varself_bose}
\d \, \cL_{\,\textrm{Bose}} \, = \, \bra\, \d \, \vf \,\comma\, \cE_{\,\textrm{Bose}} \,\ket \, .
\ee
Furthermore, in eq.~\eqref{selfadjferm} we proved that similar considerations apply to the Rarita-Schwinger-like spinor-tensor
\be \label{Efermi}
\cE_{\,\textrm{Fermi}} \, = \, \frac{(-1)^{\,p\,+\,\frac{q\,(q+1)}{2}}}{p\,!\,q\,!\,(\,p+q+1\,)\,!} \ \h^{i_1 j_1} \!\ldots \h^{i_p j_p} \, \g^{\,k_1 \ldots\, k_q} \, (\,\g^{\,[\,q\,]}\,\cS^{\,[\,p\,]}\,)_{\, i_1 j_1\,,\,\ldots\,,\,i_p j_p \, ;\, k_1 \ldots\, k_q} \, .
\ee
Again up to total derivatives, the $\d\, \bar{\psi}$ variation of the Lagrangians \eqref{lagconstr_fermi} can thus be cast in the form
\be \label{varself_fermi}
\d \, \cL_{\,\textrm{Fermi}} \, = \, \bra\, \d \, \bar{\psi} \,\comma\, \cE_{\,\textrm{Fermi}} \,\ket \, ,
\ee
where the $\d\, \bar{\psi}$ dependence of $\d \, \cL$ is now manifest.

At this point one would be tempted to declare that the Lagrangian field equations are given by $\cE_{\,\textrm{Bose}} \, = \, 0$ and $\cE_{\,\textrm{Fermi}} \, = \, 0$. However, a bit of caution is needed, since we are dealing with \emph{constrained} variations. In the Fang-Fronsdal case the free variation actually coincides with the constrained one, since the double traces of the Einstein-like tensors and the triple $\g$-traces of the Rarita-Schwinger-like tensors vanish. This is almost obvious recalling eqs.~\eqref{constrF_symm} and \eqref{constrS_symm}. On the other hand, as first\footnote{Actually, in \cite{laba_lag} Labastida briefly commented on this issue, but he did not seem to fully realize the subtleties that we are about to describe.} noticed in \cite{mixed_bose,mixed_fermi}, in the mixed-symmetry case some subtleties arise, since
\be \label{traceE}
T_{(\,ij}\,T_{kl\,)} \, \cE_{\,\textrm{Bose}} \, \neq \, 0 \, , \qquad\qquad T_{(\,ij}\,\g_{\,k\,)} \, \cE_{\,\textrm{Fermi}} \, \neq \, 0 \, ,
\ee
even if
\be \label{constrFS}
T_{(\,ij}\,T_{kl\,)} \, \cF \, = \, 0 \, , \qquad\qquad\qquad T_{(\,ij}\,\g_{\,k\,)} \, \cS \, = \, 0 \, .
\ee
At first sight this result could look strange. Nevertheless, in the mixed-symmetry case a new class of operators enters the \mbox{($\g$-)}traces of expressions containing a bunch of invariant tensors. In fact, as shown in Appendix \ref{app:MIX}, the operators
\be \label{flip1}
S^{\,i}{}_{j} \, \vf \, \equiv \, \vf_{\ldots \,,\, (\, \m^i_1 \ldots\, \m^i_{s_i} | \,,\, \ldots \,,\, |\, \m^i_{s_i+1} \,) \, \m^j_1
\ldots\, \m^j_{s_j-1} \,,\, \ldots}  \, .
\ee
are needed to close the algebra of the operators $T_{ij}$ and $\h^{ij}$ or $\g_{\,i}$ and $\g^{\,i}$. Looking at eq.~\eqref{flip1} one can see that, for $i \, \neq j$, the net effect of the $S^{\,i}{}_j$ operators is to displace indices from one family to another. Their ubiquitous presence in the \mbox{($\g$-)}traces of the $\cE$ tensors can thus be manifested looking at very simple examples, such as
\be
\h^{\,\l\r} \, \h_{\,\l\,(\,\m}\, \vf_{\,\n\,)\,,\,\r} \, = \, \vf_{\,(\,\m\,,\,\n\,)} \, .
\ee
These new operators play a crucial role in leading to eqs.~\eqref{traceE}. In fact, they permit to spread the symmetrized indices also over the $S^{\,i}{}_j$, rather than only over products of \mbox{($\g$-)}traces. This avoids the reconstruction of $T_{(\,ij}\,T_{kl\,)}\, \cF$ or $T_{(\,ij}\,\g_{\,k\,)}\, \cS$ in some terms, that as such are \emph{not} annihilated by eqs.~\eqref{constrFS}. To be more concrete, let us dwell further upon these subtleties in a simplified setting. For instance, Bose fields with only two index families already display this behavior, but their Einstein-like tensors take the relatively simple form
\be
\cE \, = \, \cF \, - \, \12 \, \h^{ij}\, T_{ij} \, \cF \, + \, \frac{1}{36}\, \h^{ij}\,\h^{kl} \left(\, 2\, T_{ij}\,T_{kl} \, - \, T_{i\,(\,k}\,T_{l\,)\,j} \,\right)\, \cF \, .
\ee
Enforcing the Labastida constraints on the field, the symmetrized double traces of $\cE$ read
\be
T_{(\,ij}\,T_{kl\,)} \, \cE \, = \, - \, \frac{1}{36} \ S^{\,m}{}_{(\,i}\, S^{\,n}{}_{j\,|}\, \left(\, 2\, T_{|\,kl\,)}\,T_{mn} \, - \, T_{m\,|\,k}\,T_{l\,)\,n} \,\right) \, \cF \, .
\ee
The combination of double traces of $\cF$ entering this expression is actually $\{2,2\}$-projected in its family indices, so that it is orthogonal to the $\{4\}$-projected constraints \eqref{constrF}. This explains why the result does not vanish even in the constrained theory. As a consequence, in order to obtain a field equation satisfying the same constraints as the gauge field a projection is needed. However, the projected field equations would take a very involved form, so that it is more convenient to reach them via the addition of some independent contributions to the naive equations, coming from suitable Lagrange multiplier terms to be added to the Lagrangians. We shall postpone this discussion to Section \ref{chap:unconstrained}, where these terms will naturally emerge in the unconstrained framework that we shall present there.

Independently of the precise form taken by the Lagrangian field equations, some very general comments regarding their equivalence to the non-Lagrangian Labastida ones can be already made. Even in this respect, the paradigmatic example of linearized gravity provides important hints. In fact, in all space-time dimensions $D \neq 2$ the Lagrangian field equation
\be \label{lageq_grav}
R_{\,\m\n} \, - \, \12 \ \h_{\,\m\n}\, R_{\,\l}{}^{\,\l} \, = \, 0
\ee
can be reduced to the non-Lagrangian one \eqref{ricci} combining it with its trace. On the other hand, in two dimensions a couple of pathologies appear: the Einstein-Hilbert action actually vanishes, even if the linearized Ricci tensor $R_{\,\m\n}$ does not, while the formal expression \eqref{lageq_grav} is traceless. The first statement becomes manifest writing the Lagrangian directly in terms of the fields as in eq.~\eqref{antigrav}, where an antisymmetrization over three indices is present. Therefore, in two dimensions we should compare a non-Lagrangian setup where the field is bounded to satisfy the non-trivial equation of motion
\be
R_{\,\m\n} \, = \, 0 \, ,
\ee
and a Lagrangian setup, where no equations of motion are present. On the other hand, in $D=2$ the little group is degenerate and no local degrees of freedom are present. A three-component symmetric tensor $h_{\,\m\n}$ can thus be forced to propagate no degrees of freedom in two different ways, either imposing a gauge-invariant field equation, or stating directly that all its components are pure gauge. Indeed, in $D = 2$ an arbitrary shift of the field is a symmetry of the action! More precisely, the usual gauge transformations in $D=2$ are accompanied by Weyl transformations, that in their linearized form read
\be
\d \, h_{\,\m\n} \, = \, \h_{\,\m\n}\, \O \, .
\ee
When combined, these two sets of gauge transformations suffice to rebuild an arbitrary shift of the field. Even without resorting to eq.~\eqref{antigrav}, the presence of a pathological behavior can be recognized looking at eq.~\eqref{lageq_grav} and considering $R_{\,\m\n}$ as an independent tensor. Since the combination in eq.~\eqref{lageq_grav} is traceless in $D=2$, it is evident that the Lagrangian field equation is at least weaker than the non-Lagrangian one, because it displays an additional symmetry. In $D=2$ the traceless part of the linearized Ricci tensor $R_{\,\m\n}$ then disappears altogether, and this brings together the two observations we made.

In conclusion, in some degenerate cases the Lagrangian field equations can not be directly equivalent to the non-Lagrangian ones, but at the same time new Weyl-like symmetries emerge. The rich structure of two-dimensional gravity naturally leads to wonder whether it is the only example presenting this pathological behavior. Actually, another example should be rather familiar, since the Rarita-Schwinger Lagrangian \eqref{antirarita} manifestly vanishes in $D = 2$, even if $\cS_{\,\m}$ does not. However, these two-dimensional examples are the only pathological cases in the fully symmetric setting, for both Bose and Fermi fields.

Moving to the mixed-symmetry setup, the far more involved form of the Lagrangians makes it plausible that in particular space-time dimensions the tensors \eqref{Ebose} and \eqref{Efermi} could coincide by chance with some particular $o\,(D)$ components of the tensors $\cF$ or $\cS$. This is indeed the only way to obtain a new symmetry with respect to the non-Lagrangian field equations. The resulting $o\,(D)$ component could then also vanish when written in terms of the fields. A detailed discussion of these issues leading to the identification of a rich set of pathological cases will be presented in Section \ref{chap:weyl}. On the other hand, we can already identify some mixed-symmetry fields behaving exactly like spin-$2$ or spin-($3/2$) fields in two space-time dimensions. Indeed, the actions \eqref{antilagb} for $\{2^{\,p},1^q\}$-projected Bose fields manifestly vanish in $D \leq 2\,p+q$, while these tensors are available for $D \geq p+q$. In a similar fashion, the actions \eqref{antilagf} for fully antisymmetric Fermi fields with $q$ vector indices manifestly vanish in \mbox{$q \leq D \leq 2\,q$} even if the corresponding $\cS$ spinor-tensors do not vanish for $D \geq q+1$. In all these pathological cases the fields do not propagate any degrees of freedom\footnote{The gauge transformations, the field equations and, possibly, the constraints introduced until now force $gl(D)$ tensors and spinor-tensors to propagate the degrees of freedom of the \mbox{$o\,(D-2)$} representation associated to the same Young tableau identifying them. On the other hand, a general result of representation theory (see, for instance, the first reference in \cite{group}, \textsection $10$-$6$) implies that, for $O(n)$ groups, if the total number of boxes in the first two columns of a tableau exceeds $n$, the corresponding traceless tensor vanishes. A similar result also applies to spinor-tensor representations.}, and this is compatible with the vanishing of their actions. Since any shift of the fields leaves invariant the actions, the usual gauge transformations must be accompanied by additional Weyl-like symmetries as in the case of two-dimensional gravity. Indeed, let us stress that the formal expressions \eqref{Ebose} and \eqref{Efermi} for the Einstein-like tensors can vanish when they are written in terms of the fields only if they coincide with a vanishing $o\,(D)$ component of the non-trivial kinetic tensors.


\subsection{Irreducible fields}\label{sec:irreducible}


The gauge fields considered in the previous sections were in general only symmetric under the
interchange of pairs of vector indices belonging to the same set, but did not possess any symmetry relating different sets. In other words, they were \emph{reducible} $gl(D)$ tensors or
spinor-tensors. Fields of this type naturally emerge in String Theory, where they are
associated with products of bosonic oscillators, together with more general reducible
$gl(D)$ tensors, where some sets of indices are actually antisymmetrized rather than
symmetrized. In \cite{mixed_bose,mixed_fermi} we also described how the previous results can be rewritten in an antisymmetric framework, while in the following we shall briefly concentrate on
Young projected $gl(D)$ spinor-tensors, still with manifestly symmetric index sets.
They propagate the degrees of freedom of \emph{irreducible} representations of the Lorentz
group, up to possible (Majorana-)Weyl projections in the fermionic case,
and in this respect they are more along the lines of the usual low-spin examples.
Since Young projections only affect in a non-trivial fashion the vector
indices carried by the gauge fields, the changes
occurring in this case were essentially the same for both Bose and Fermi fields.

The condition of irreducibility states that the symmetrization of a given line of the
corresponding Young tableaux with any additional index belonging to one of the lower
lines gives a vanishing result. This condition can be neatly formulated using the
$S^{\,i}{}_j$ operators introduced in Appendix \ref{app:MIX} as
\be \label{condirr}
S^{\,i}{}_j \, \vf \, = \, 0 \, , \qquad S^{\,i}{}_j \, \psi \, = \, 0 \, , \qquad \ \, i < j \, , \quad 2 \leq j \leq N \, ,
\ee
where $N$ is the number of index families. The Fronsdal-Labastida kinetic operator \eqref{laba-b} and the
Fang-Fronsdal-Labastida kinetic operator \eqref{laba-f} \emph{commute with the $S^{\,i}{}_j$ operators}. In a similar fashion, the operators building the Einstein-like tensors \eqref{Ebose} and the Rarita-Schwinger-like tensors \eqref{Efermi} also commute with the $S^{\,i}{}_j$ operators, on account of the relations
\be
[\,S^{\,m}{}_n \comma \h^{i_1j_1} \ldots\, \h^{i_pj_p}\, \g^{\,k_1 \ldots\, k_q}\, T_{i_1j_1} \ldots\, T_{i_pj_p}\, \g_{\,k_1 \ldots\, k_q} \,] \, = \, 0 \, .
\ee
As a consequence, in the constrained theory Lagrangians and field equations
take exactly the same form for reducible and irreducible gauge fields. On the other hand, for
irreducible fields not all traces of $\vf$ or $\cF$ and not all $\g$-traces of $\psi$ or $\cS$ remain independent, and therefore
it is possible to reduce the number of those appearing in the field equations and in the Lagrangians, although the redundant description remains correct.

Moreover, in general irreducible fields involve fewer gauge parameters than their reducible counterparts, and how this number is reduced was already stated in the second reference of \cite{mixed1}: the relevant irreducible gauge parameters can be associated to all admissible Young diagrams obtained stripping one box from the diagram corresponding to the gauge field. The difference could be sizeable: if an irreducible gauge
field is characterized by a Young diagram containing a number of identical rows, only a single gauge parameter is associated to all of them.

It is instructive to recover these results in an algebraic fashion, that will also prove very useful in Section \ref{chap:weyl}. They follow in fact rather directly from the
structure of the $S^{\,i}{}_j$ operators that implement the
irreducibility condition. For instance, let us begin by considering how to select independent gauge parameters in the bosonic case, since no relevant changes occur in the fermionic case. First of all, eq.~\eqref{condirr} must be preserved under gauge transformations, and this implies that
\be
\partial^{\,k} \left(\, S^{\,i}{}_j \, \L_{\,k} + \d^{\,i}{}_k \, \L_{\,j} \,\right) = \, 0 \, , \qquad  i <j \, ,
\ee
which forces the terms between parentheses to define a gauge-for-gauge transformation:
\be
S^{\,i}{}_j \, \L_{\,k} \, + \,  \d^{\,i}{}_k \, \L_{\,j} \, = \,
\partial^{\,l} \, \L^{i}{}_{\,[\,kl\,]\,j} \, ,  \qquad  i \,<\,j \, .
\label{gaugeparcondirr}
\ee
Up to the last irrelevant term, one thus obtains a set of
constraints that are conveniently analyzed starting from the highest
available value of $k$. In this case, in fact, $i$ is necessarily
less than $k$, being constrained to be less than $j$, so that
eq.~\eqref{gaugeparcondirr} reduces to the more familiar condition
that the last gauge parameter in the chain, $\L_k$, be irreducible,
\be
S^{\,i}{}_j \, \L_{\,k} \,=\, 0     \, .
\label{lastcondirr}
\ee
The remaining contributions can then be used to determine the other
parameters corresponding to lower values of $k$ in terms of this
solution and of additional independent ones that emerge, one for
each step, from the homogeneous parts of eqs.~\eqref{gaugeparcondirr}.
Let us stress that the independent parameters solve further
irreducibility conditions, and therefore can only exist if the
corresponding Young diagrams are admissible. In conclusion, one thus
obtains an irreducible gauge parameter for each admissible Young
diagram built stripping one box from the original diagram for the
gauge field, precisely as anticipated.
Furthermore, the independent parameters lead by construction to
irreducible gauge transformations where no explicit Young projectors
are needed. The general solution of the conditions \eqref{gaugeparcondirr} can be found in \cite{mixed_fermi}, but here we can see how this works referring to a simple example. For instance, for an irreducible $\{4,2\}$ gauge field the conditions \eqref{gaugeparcondirr} read
\begin{align}
& S^{\,1}{}_2\, \L_{\,1} \, + \, \L_{\,2} \, = \, 0 \, ,  \nonumber\\
& S^{\,1}{}_2\, \L_{\,2} \, = \, 0 \, . \label{irredgauge42}
\end{align}
To begin with, $\L_1$ could have three irreducible components, a
$\{3,2\}$, a $\{4,1\}$ and a $\{5\}$, while $\L_2$ could have a
$\{4,1\}$ and a $\{5\}$. Now the second of eqs.~\eqref{irredgauge42}
eliminates directly the $\{5\}$ component of $\L_2$, and then the
first eliminates the $\{5\}$ component of $\L_1$. The two $\{4,1\}$
components are then connected by the first equation, while the
$\{3,2\}$ component is a zero mode of the first equation, and as
such is not constrained. In space-time notation the conclusion is
that, when adapted to an irreducible $\{4,2\}$ field, the original
reducible gauge transformation of the form \eqref{gauge_indices-b} can be recast as
\begin{align}
\d\, \vf_{\m_1\m_2\m_3\m_4\,,\, \n_1\n_2} \, & = \,
\pr_{\,(\,\m_1\,}\L^{(1)\{3,2\}}{}_{\m_2\m_3\m_4\,)\,,\, \n_1\n_2}\, \nn \\
&+ \, \left(\, \pr_{\,(\,\m_1}\L^{(1)\{4,1\}}{}_{\m_2\m_3\m_4\,)\,,\, \n_1\n_2}\,-\, \pr_{\,(\,\n_1|\,}\L^{(1)\{4,1\}}{}_
{(\, \m_1\m_2\m_3\,,\, \m_4)\,|\,\n_2\,)} \,\right)\, .
\end{align}
Let us stress again that no explicit Young projector is needed here,
since both combinations are already $\{4,2\}$ projected, as one can see enlarging the cycle that contains the $\m$ indices to comprise also a $\n$ index.

In full analogy, one can relate to each other the irreducible components of the \mbox{($\g$-)}traces of $\vf$ and $\psi$, or of $\cF$ and $\cS$, simply computing the \mbox{($\g$-)}traces of eq.~\eqref{condirr} and moving to the left the $S^{\,i}{}_j$ operators. For instance, for Bose fields the relevant conditions are
\be \label{idirrF}
S^{\,m}{}_n\, T_{i_1j_1} \ldots\, T_{i_pj_p}\, \cF \, + \, \sum_{i\,=\,1}^p \, \d^{\,m}{}_{(\,i_n}\, T_{j_n\,)\,n}\, \prod_{r\,\neq\,n}^p\, T_{i_rj_r}\, \cF \, = \, 0 \, ,
\ee
that can be solved resorting to the arguments presented for eq.~\eqref{gaugeparcondirr}. These identities should not be regarded as a mere curiosity, since they will play a crucial role in simplifying some expressions that we shall come across in Section \ref{chap:weyl}.


\section{Unconstrained theory}\label{chap:unconstrained}


In Section \ref{chap:constrained} we have presented second-order field equations and Lagrangians for massless Bose fields of mixed symmetry and first-order field equations and Lagrangians for massless Fermi fields of mixed symmetry. The field content was the minimal one leading to a covariant description, since the representations of the Lorentz group associated with Young tableaux $\{s_1,s_2,\ldots\}$ were described via single Young-projected multi-symmetric tensors $\vf_{\,\m_1 \ldots\, \m_{s_1},\, \n_1 \ldots\, \n_{s_2}, \,\ldots}$ or spinor-tensors $\psi^{\,\alpha}{}_{\m_1 \ldots\, \m_{s_1},\, \n_1 \ldots\, \n_{s_2}, \,\ldots}$. As repeatedly stressed, however, in the formulation presented in Section \ref{chap:constrained} fields and gauge parameters must satisfy some algebraic \mbox{($\g$-)}trace \emph{constraints}. In this section we shall describe how one can recover a more conventional description, removing the constraints at the price of introducing a number of \emph{auxiliary fields}. Aside from bringing these systems closer to their ``low-spin'' counterparts, this operation could well remove some complications in the treatment of these models at the interacting level. With this aim in mind, we shall present a way to forego the need for constraints preserving as much as possible the simplicity of the construction developed in the previous sections. Thus, we shall try to contain the enhancement of the field content, introducing at most one auxiliary field for each constraint. To do that, we shall follow the main lines of the strategy adopted by Francia and Sagnotti to build their local unconstrained formulation for fully symmetric fields \cite{fs1,fs2,fs3,fms1}. This ``minimal'' procedure will lead to higher-derivative terms involving some of the new auxiliary fields, but this does not create any problem since the extra derivatives appear in pure-gauge contributions. At any rate, we shall also show how to remove them enlarging a bit the field content, via additional auxiliary fields whose number only depends on the number of index families, and not on the total number of space-time indices. These unconstrained formulations will be obtained re-examining in a critical fashion the consequences of the principle of gauge invariance underlying the constrained theory.

Let us close these cursory remarks by recalling that the introduction of these auxiliary fields has another important benefit. As anticipated in the Introduction, eliminating them in terms of the basic fields $\vf$ or $\psi$ one indeed ends up with the geometrical non-local formulation of the dynamics \cite{nonlocal,nonlocal_mixed,fs1,tesi_dario}. While we shall not deal with this point, we would like to stress that the formulation we are going to present is in this sense the minimal local one allowing a contact with the \emph{higher-spin geometry}.


\subsection{Minimal Lagrangian formulation}\label{sec:minimal}


In the Lagrangian theory of Section \ref{chap:constrained}, the Labastida constraints can be eliminated in a ``minimal'' way in two steps:
\begin{itemize}
\item the constraints \eqref{constr_gauge-b} or \eqref{constr_gauge-f} on the gauge parameters can be eliminated via the introduction of at most one compensator field for each constraint;
\item the constraints \eqref{constr_field-b} or \eqref{constr_field-f} on the gauge fields can be eliminated via the introduction of at most one Lagrange multiplier for each constraint.
\end{itemize}
This program can be pursued following similar steps for both Bose and Fermi fields. Nevertheless, the two types of fields will be treated separately as in Section \ref{chap:constrained}, also because Fermi fields involve again some additional subtleties related to the presence of $\g$-matrices. At any rate, the reader is invited to notice the similarities between the two setups.


\subsubsection{Bose fields}\label{sec:minimal-bose}


In Section \ref{chap:constrained} we showed that the Labastida constraints
\be \label{constrL}
T_{(\,ij}\,\L_{\,k\,)} \, = \, 0
\ee
lead to second-order gauge-invariant field equations. Indeed, computing the gauge variation of the Fronsdal-Labastida tensor \eqref{laba-b} one obtains
\be \label{gaugeF}
\d \, \cF \, = \, \frac{1}{6} \ \pr^{\,i}\pr^{\,j}\pr^{\,k}\, T_{(\,ij}\, \L_{\,k\,)} \, .
\ee
We also showed that the Labastida constraints
\be \label{constrvf}
T_{(\,ij}\,T_{kl\,)} \, \vf \, = \, 0
\ee
lead to gauge-invariant Lagrangians, since
\be \label{bianchiF}
\pr_{\,i}\, \cF \, - \, \12 \ \pr^{\,j}\, T_{ij} \, \cF \, = \, - \, \frac{1}{12} \ \pr^{\,j}\pr^{\,k}\pr^{\,l} \, T_{(\,ij}\,T_{kl\,)} \, \vf \, ,
\ee
and Bianchi identities free of the ``classical anomaly'' appearing on the right hand-side of eq.~\eqref{bianchiF} guarantee the gauge invariance of the Lagrangians \eqref{lagconstr_bose}.

Looking at eq.~\eqref{gaugeF}, it seems that the simplest solution to overcome the constraints \eqref{constrL} would be to introduce for each of them one auxiliary field transforming as
\be \label{aijk}
\d \, \a_{\,ijk} \, = \, \frac{1}{3} \ T_{(\,ij}\, \L_{\,k\,)} \, .
\ee
This was indeed the strategy followed in \cite{fs1,fs2,fs3,fms1} in order to obtain an unconstrained local formulation for fully symmetric Bose fields. Consistently with the notation adopted for the gauge parameters $\L_{\,i}$, the auxiliary fields $\a_{\,ijk}$ would carry one less space-time index in the $i$-th, $j$-th and $k$-th families with respect to the gauge field $\vf$. Introducing these auxiliary fields, unconstrained gauge-invariant field equations could be obtained defining the new kinetic tensors
\be \label{Atemp}
\cA \, \equiv \, \cF \, - \, \frac{1}{2} \ \pr^{\,i}\pr^{\,j}\pr^{\,k} \a_{\,ijk} \, ,
\ee
and forcing them to vanish\footnote{As already recalled in the Introduction, for a spin-$3$ field an equation of motion of this form was first proposed by Schwinger in \cite{schwinger}, and independently rediscovered by Francia and Sagnotti in \cite{nonlocal}. This result was then extended to arbitrary symmetric fields in \cite{fs1}. The other papers already recalled \cite{fs2,fs3,fms1} present a Lagrangian extension of the non-Lagrangian field equations of the form \eqref{Atemp} in the fully symmetric case.}. Furthermore, $\cA$ would satisfy the Bianchi identities
\be \label{bianchiA}
\pr_{\,i}\, \cA \, - \, \12 \ \pr^{\,j}\, T_{ij} \, \cA \, = \, - \, \frac{1}{4} \ \pr^{\,j}\pr^{\,k}\pr^{\,l} \, \cC_{\,ijkl} \, ,
\ee
where the tensors
\be \label{Ctemp}
\cC_{\,ijkl} \, = \, \frac{1}{3} \, \left\{ \, T_{(\,ij}\,T_{kl\,)}
\, \vf \, - \, 3 \, \pr_{\,(\,i}\, \a_{\,jkl\,)} \, - \, \frac{3}{2} \ \pr^{\,m}
\left(\, T_{(\,ij} \, \a_{\,kl\,)\,m} \, - \, T_{m\,(\,i} \, \a_{\,jkl\,)} \,\right) \,\right\}
\ee
provide a gauge invariant completion of the Labastida constraints \eqref{constrvf}.

On the other hand, as anticipated in Section \ref{chap:constrained}, the Labastida constraints \eqref{constrL} or \eqref{constrvf} are \emph{not} independent. As a consequence, gauge-invariant combinations of the $\a_{\,ijk}$ exist. This means that, if regarded as independent, they would \emph{not} be pure-gauge fields as the single compensator introduced for a given spin-$s$ field in \cite{fs1,fs2,fs3,fms1}. This is however a crucial requirement, since it enables one to eliminate the auxiliary fields by a partial gauge fixing, while still maintaining the constrained gauge symmetry needed to force the gauge fields to propagate the proper massless modes. Before showing how it is possible to overcome the problem, let us support these arguments by displaying explicitly a gauge-invariant combination of the $\a_{\,ijk}$. In order to do that, it is convenient to first exploit the linear dependence of the Labastida constraints on the fields. In Section \ref{sec:laba-bose} we have already seen an example of this fact, and in the present setup the result shown in eq.~\eqref{51comb} implies that the $\vf$ part of a $\{5,1\}$-projected combination of traces of the $\cC_{\,ijkl}$ tensors vanishes. On the contrary, there are no group-theoretical arguments forcing its $\a_{\,ijk}$ portion to vanish as well. Moreover, if it does not vanish, it must define a gauge-invariant combination of these fields, since the $\cC_{\,ijkl}$ are gauge invariant. A direct computation finally shows that the divergence terms in \mbox{$Y_{\{5,1\}} \, T_{mn}\, \cC_{\,ijkl}$} actually vanish, while the gradient ones do not. In conclusion,
\be \label{51unc}
Y_{\{5,1\}} \, T_{mn}\, \cC_{\,ijkl} \, = \, - \, \frac{3}{2}\ Y_{\{5,1\}}\, \pr^{\,p}\, T_{mn} \left(\, T_{(\,ij} \, \a_{\,kl\,)\,p} \, - \, T_{p\,(\,i} \, \a_{\,jkl\,)} \,\right) \, ,
\ee
where the $\{5,1\}$ projection is applied only on the free indices and not on the dumb index $p$. The combination of double traces of the $\a_{\,ijk}$ fields appearing in eq.~\eqref{51unc} is available in general from two-families onward and it is gauge invariant, as the reader can verify using its explicit expression
\begin{align}
& Y_{\{5,1\}}\, T_{mn}\, \cC_{\,ijkl} \, \sim \ \pr^{\,p} \Big\{ \,
\left(
\, T_{n(\,i} \, T_{jk} \,
\a_{\,lm\,)\,p} + \, T_{m(\,i} \, T_{jk} \, \a_{\,ln\,)p} \, \right) \nn \\
& - \,2 \left( \, T_{p\,(\,i\,|} \, T_{n\,|j} \, \a_{\,klm\,)} + \, T_{p\,(\,i\,|} \, T_{m\,|j} \, \a_{\,kln\,)} \, \right) + \left(\, T_{pn} \, T_{(\,ij} \, \a_{\,klm\,)} + \, T_{pm} \,
T_{(\,ij} \, \a_{\,kln\,)} \,\right)\ \nn \\
& - \,2 \left( \, T_{(\,ij} \, T_{kl} \, \a_{\,m\,)\,np}  + \, T_{(\,ij} \, T_{kl} \, \a_{\,n\,)\,mp} \, \right) + \left( \, T_{p\,(\,i\,} T_{jk} \, \a_{\,lm\,)\,n} + \,
T_{p\,(\,i\,} T_{jk} \, \a_{\,ln\,)\,m} \, \right) \!\Big\} .
\end{align}
Let us close this digression by noticing that the linear dependence of the Labastida
constraints does not manifest itself in simple mixed-symmetry models with small numbers of space-time indices. In fact, with fewer traces of the $\a_{\,ijk}$ it is not possible to obtain gauge-invariant combinations.

As we have argued before, the presence of gauge-invariant combinations of the $\a_{\,ijk}$ fields definitely shows that one cannot consider them independent. The most economical solution to the problem consists in regarding them as ``composite'' objects, introducing a set of fundamental compensators that transform as
\be \label{Phi}
\d \, \Phi_{\,i} \, = \, \L_{\,i} \, .
\ee
On the other hand, the new compensators enter field equations and Lagrangians only through their symmetrized traces, so that they can be gauged away without spoiling the Labastida constrained gauge symmetry. For this reason, for brevity in the following we shall often abide by the notation
\be \label{alpha_phi}
\a_{\,ijk} \, \equiv \, \frac{1}{3} \ T_{(\,ij}\, \Phi_{\,k\,)} \, ,
\ee
explicitly displaying the $\Phi$ dependence of the ``composite compensators'' $\a_{\,ijk}$ only where it is particularly relevant.

Furthermore, the $\Phi_{\,i}$ compensators naturally emerge when trying to enlarge the set of gauge transformations via a Stueckelberg-like realization of the missing unconstrained ones. In fact, this can be done shifting the gauge field as
\be \label{stueckelberg}
\vf \, \to \, \vf \, - \, \pr^{\,i}\,\Phi_{\,i} \, .
\ee
Notice that the gradients present in eq.~\eqref{stueckelberg} imply that the compensators cannot have the same dimension as the gauge field. This fact is at the origin of the (harmless) \emph{higher-derivative} term that appears in the new kinetic tensor of the unconstrained theory. Either directly or via the shift \eqref{stueckelberg}, one can indeed cast it in the form
\be \label{A}
\cA \, \equiv \, \cF \, - \, \frac{1}{6} \ \pr^{\,i}\pr^{\,j}\pr^{\,k}\, T_{(\,ij}\,\Phi_{\,k\,)} \, ,
\ee
and define the equation of motion
\be \label{A_eq}
\cA \, = \, 0 \, ,
\ee
that is invariant under the \emph{unconstrained} gauge transformations
\begin{align}
& \d \, \vf \, = \, \pr^{\,i}\, \L_{\,i} \, , \nn \\
& \d \, \Phi_{\,i} \, = \, \L_{\,i} \, . \label{gaugeuncb}
\end{align}
We shall comment more on the issue of higher derivatives in Section \ref{sec:ordinary-bose}, where we shall show how to obtain a two-derivative unconstrained Lagrangian formulation for Bose fields, joining the $\Phi_{\,i}$ compensators with other auxiliary fields of the same dimension as the gauge field $\vf$. At any rate, when working with the fundamental compensators $\Phi_{\,i}$, the $\cC_{\,ijkl}$ tensors appearing on the right-hand sides of the Bianchi identities \eqref{bianchiA} take the simpler form
\be \label{C}
\cC_{\,ijkl} \, = \, \frac{1}{3}\ T_{(\,ij}\,T_{kl\,)} \left(\, \vf \, - \, \pr^{\,m}\, \Phi_{\,m} \,\right) \, ,
\ee
as should be clear looking at the Stueckelberg method to introduce the $\Phi_{\,i}$. Furthermore, eq.~\eqref{alpha_phi} was introduced to eliminate all gauge invariant combinations of the $\a_{\,ijk}$ fields, so that it implies
\be \label{51C}
Y_{\{5,1\}} \, T_{mn}\, \cC_{\,ijkl}\,(\,\vf\,,\Phi_{\,i}\,) \, = \, 0 \, .
\ee

We have thus seen how to eliminate the constraints \eqref{constrL} on the gauge parameters and we can now move on and analyze the role of the constraints \eqref{constrvf} on the fields. Since they are related to the Lagrangians, whose complexity increases with the number of index families, let us return for a while to the fully symmetric case in order to better delineate the strategy that we shall follow. Moreover, we shall begin by keeping the constraint on the gauge parameter in order to focus only on the elimination of the double trace constraint on $\vf$. Using the Bianchi identity \eqref{bianchi_fronsdal} in eq.~\eqref{var_fronsdal}, in this framework the gauge variation of the Fronsdal Lagrangian can be cast in the form
\be
\d \, \cL \, = \, - \, 3\ \binom{s}{4} \ \prd\prd\prd \L^{\,\m_1 \ldots\, \m_{s-4}} \, \vf_{\,\m_1 \ldots\, \m_{s-4}\,\l\r}{}^{\,\l\r} \, .
\ee
Instead of imposing the constraint \eqref{constr_fronsdal2-b} on the gauge field, one can thus cancel it introducing a Lagrange multiplier with $(s-4)$ space-time indices that transforms as
\be \label{beta_fronsdal}
\d \, \b_{\,\m_1 \ldots\, \m_{s-4}} \, = \, \pr\cdot\pr\cdot\pr\cdot \L_{\,\m_1 \ldots\, \m_{s-4}} \, .
\ee
In fact, one can couple it to the double trace of the gauge field and add the new term to the Fronsdal Lagrangian:
\begin{align}
\cL \, & = \, \12 \ \vf^{\,\m_1 \ldots\, \m_s} \left(\, \cF_{\,\m_1 \ldots\, \m_s} \, - \, \12 \ \h_{\,(\,\m_1\m_2}\, \cF_{\,\m_3 \ldots\, \m_{s}\,)\,\l}{}^{\,\l} \,\right) \nn \\
&  + \, 3 \, \binom{s}{4}\, \b^{\,\m_1 \ldots\, \m_{s-4}} \, \vf_{\,\m_1 \ldots\, \m_{s-4}\,\l\r}{}^{\,\l\r} \, .
\end{align}
The resulting Lagrangian contains unconstrained fields, but is still gauge invariant only under constrained gauge transformations. One can also relax this requirement introducing suitable compensator terms that lead to the Lagrangian\footnote{Actually the Lagrangians of \cite{fs2,fs3,fms1} were presented in the far more effective index-free convention which inspired the one that we are using to describe mixed-symmetry fields. On the other hand, the space-time indices are here displayed in order to keep this cursory example immediately accessible.} of \cite{fs2,fs3,fms1}. Thus, in the fully symmetric setting a couple of auxiliary fields suffices to eliminate the constraints for any value of the spin $s$.

In the mixed-symmetry case we shall now attack the problem following the same strategy. To begin with, let us continue for a while to discard the terms proportional to the constraints on the gauge parameters \eqref{constrL}. As we have seen in eq.~\eqref{lag_var-b2}, the gauge variation of the Labastida Lagrangian \eqref{lagconstr_bose} thus reads
\begin{align}
\d \, \cL \, = \, & \sum_{p\,=\,0}^N \, \frac{(\,p+1\,)\, k_{\,p+1}}{2^{\,p+1}} \, \bra\, Y_{\{2^p,1\}}\, T_{i_1j_1} \ldots\, T_{i_pj_p}\, \L_{\,k} \,\comma \nn \\
& (\,p+2\,)\, Y_{\{2^p,1\}}\, \pr_{\,k}\, \cF^{\,[\,p\,]}{}_{\,i_1j_1,\,\ldots\,,\,i_pj_p} \, - \,  \pr^{\,l}\, \cF^{\,[\,p+1\,]}{}_{\,i_1j_1,\,\ldots\,,\,i_pj_p,\,kl} \ket \, , \label{varmult-b1}
\end{align}
where the $k_{\,p}$ are the coefficients of eq.~\eqref{coeflab}. In order to continue one has to identify the unconstrained analogues of the identities \eqref{trbianchi-b} that we used at this stage in the constrained setting. As a starting point one can consider again the multiple traces of the Bianchi identities of eq.~\eqref{trbianchinp-b}, but some of the manipulations that led from them to eq.~\eqref{trbianchi-b} were based on the constraints on $\cF$ of eq.~\eqref{constrF}. These are clearly not available here, since they were induced by the constraints on $\vf$ that we are eliminating. Some care is thus in order, but it is still true that the divergence terms of eq.~\eqref{trbianchinp-b} can be combined when computing a two-column projection. Thus, the $\{2^{\,p},1\}$ projection of a product of $p$ traces of the Bianchi identities reads
\begin{align}
& (\,p+2\,)\, Y_{\{2^p,1\}} \, \pr_{\,k}\, T_{i_1j_1} \ldots\, T_{i_pj_p}\, \cF \, - \, Y_{\{2^p,1\}} \, \pr^{\,l}\, T_{i_1j_1} \ldots\, T_{i_pj_p}\, T_{kl}\, \cF \nn \\
& = \, - \, \frac{1}{6} \ Y_{\{2^p,1\}} \, T_{i_1j_1} \ldots\, T_{i_pj_p}\, \pr^{\,l}\pr^{\,m}\pr^{\,n} \, T_{(\,kl}\,T_{mn\,)} \, \vf \, .
\end{align}
To compare it with the gauge variation \eqref{varmult-b1}, one must now reconstruct the two-column projections of the traces of $\cF$. This is the point where in Section \ref{sec:labalag-bose} we resorted for brevity to the constraints \eqref{constrF}.  On the other hand, one can notice that
\be
Y_{\{2^p,1\}}\, \pr_{\,k}\, T_{i_1j_1} \ldots\, T_{i_pj_p} \, \cF \, = \, Y_{\{2^p,1\}}\, \pr_{\,k}\, \cF^{\,[\,p\,]}{}_{\,i_1j_1,\,\ldots\,,\,i_pj_p} \, ,
\ee
since all other terms in the expansion
\be \label{expansion}
T_{i_1j_1} \ldots\, T_{i_pj_p} \, \cF \, = \, Y_{\{2^p\}}\, T_{i_1j_1} \ldots\, T_{i_pj_p} \, \cF \, + \, Y_{\{4,2^{p-1}\}}\, T_{i_1j_1} \ldots\, T_{i_pj_p} \, \cF \, + \, \ldots
\ee
would be annihilated by the $\{2^{\,p},1\}$ projection. Thus, the result that was obtained via the constraints \eqref{constrF} still holds. In a similar fashion, the presence of identical tensors promotes the $\{2^{\,p},1\}$ projection in the gradient term to a $\{2^{\,p+1}\}$ projection comprising also the lower index $l$. We can thus conclude that
\begin{align}
& (\,p+2\,) \ Y_{\{2^p,1\}} \, \pr_{\,k} \, \cF^{\,[\,p\,]}{}_{\,i_1
j_1,\,\ldots\,,\,i_p j_p}\, - \, \pr^{\,l} \,
\cF^{\,[\,p+1\,]}{}_{\,i_1
j_1,\,\ldots\,,\,i_p j_p\,,\,k\,l} \nn \\
& = \, - \, \frac{1}{6} \ Y_{\{2^p,1\}} \, T_{i_1j_1} \ldots\, T_{i_pj_p}\, \pr^{\,l}\pr^{\,m}\pr^{\,n} \, T_{(\,kl}\,T_{mn\,)} \, \vf \, ,
\end{align}
where for brevity we avoided to normal order the terms on the right-hand side. Notice that simple considerations fix again the form of the result, even without knowing explicitly the form taken by the projections involved.
Using these identities the gauge variation \eqref{varmult-b1} can be eventually cast in the form
\begin{align}
\d \, \cL \, = \, & \sum_{p\,=\,0}^N \, \frac{(\,p+1\,)\, k_{\,p+1}}{48} \, \times \nn \\
& \times\, \bra\, \pr_{\,(\,k\,}\pr_{\,l\,}\pr_{\,m\,|}\, \h^{i_1j_1} \!\ldots\, \h^{i_pj_p}\, Y_{\{2^p,1\}}\, T_{i_1j_1} \ldots\, T_{i_pj_p}\, \L_{\,|\,n\,)} \,\comma\, T_{(\,kl}\,T_{mn\,)} \, \vf \,\ket \, . \label{var-b2}
\end{align}

In analogy with the fully symmetric case, it is thus possible to deal with \emph{unconstrained fields} introducing the Lagrangians
\begin{align}
\cL \, & = \, \12 \ \bra\, \vf \,\comma\, \sum_{p\,=\,0}^N \, \frac{(-1)^{\,p}}{p\,!\,(\,p+1\,)\,!} \ \h^{i_1j_1} \!\ldots\, \h^{i_pj_p}\, \cF^{\,[\,p\,]}{}_{\,i_1j_1,\,\ldots\,,\,i_pj_p} \,\ket \nn \\
& + \, \frac{1}{24} \, \bra\, \b_{\,ijkl} \,\comma\, T_{(\,ij}\,T_{kl\,)}\, \vf \,\ket \, , \label{lagmult_bose}
\end{align}
that are invariant under \emph{constrained gauge transformations} provided the Lagrange multipliers $\b_{\,ijkl}$ transform as
\be \label{beta}
\d \, \b_{\,ijkl} = \frac{1}{2} \, \sum_{p\,=\,0}^N \, \frac{(-1)^{\,p}}{p\,!\,(\,p+2\,)\,!} \ \pr_{\,(\,i\,}\pr_{\,j\,}\pr_{\,k\,|}\, \h^{i_1j_1} \!\ldots\, \h^{i_pj_p}\, Y_{\{2^p,1\}}\, T_{i_1j_1} \ldots\, T_{i_pj_p} \, \L_{\,|\,l\,)} \, .
\ee
Indeed, let us recall that in eq.~\eqref{gaugedouble} we also saw that $T_{(\,ij}\,T_{kl\,)}\,\vf$ is left invariant by constrained gauge transformations. Notice that the new term added to the Lagrangian \eqref{lagconstr_bose} enforces on-shell the Labastida constraints \eqref{constrvf} via the equations of motion of the Lagrange multipliers. Moreover, the result is consistent with that in eq.~\eqref{beta_fronsdal}, since in the single-family case the $\{2^{\,p},1\}$ projections are clearly not available while the $p=0$ term in the sum reproduces eq.~\eqref{beta_fronsdal}.

The linear dependence of the Labastida constraints leads again to some subtleties: here it induces a symmetry of the Lagrangian \eqref{lagmult_bose} under shifts of the Lagrange multipliers of the form
\be \label{gaugebgen}
\d \, \b_{\,ijkl} \, = \, \h^{m_1n_1} \!\ldots\, \h^{m_kn_k} \, Y_{\,\textrm{odd}} \, L_{\,ijkl\,,\,m_1n_1,\,\ldots\,,\,m_kn_k} \, .
\ee
These transformations are characterized by $Y_{\,\textrm{odd}}$, that must be a Young projector associated to a tableau having an odd number of boxes in some rows. To be more concrete, let us consider the simplest example
\be \label{gaugeb}
\d \, \b_{\,ijkl} \, = \, \h^{mn}\, Y_{\{5,1\}}\, L_{\,ijkl\,,\,mn}
\ee
that is already visible for two-family fields, and from which all other transformations can be extracted as particular cases. The variation of the Lagrangian \eqref{lagmult_bose} under the shifts \eqref{gaugeb} is
\be
\d \, \cL \, = \, \frac{1}{48} \, \bra\, L_{\,ijkl\,,\,mn} \,\comma\, Y_{\{5,1\}}\, T_{mn}\, T_{(\,ij}\,T_{kl\,)}\, \vf \,\ket \, ,
\ee
and vanishes on account of the repeatedly stressed properties of products of identical $T_{ij}$ tensors (see for instance eq.~\eqref{51comb}). A similar cancelation clearly accompanies a generic transformation of the form \eqref{gaugebgen}. As a consequence, some combinations of traces of the $\b_{\,ijkl}$ Lagrange multipliers are left undetermined by the field equations, but they can be anyhow gauged away exploiting the new symmetry. Moreover, this observation also implies that part of the gauge transformations \eqref{beta} is actually ineffective in the Lagrangian \eqref{lagmult_bose}. One could indeed normal-order eq.~\eqref{beta} commuting the $\h^{ij}$ operators through the divergences, and a generic term of the sum would still start with a bunch of $\h^{ij}$ operators. When considering the gauge variation of the Lagrangian, these act as trace operators on $T_{(\,ij}\,T_{kl\,)}\, \vf$. As we have seen, their Young projections with odd numbers of boxes in some rows vanish. On the other hand, projections of this kind are admitted by the terms appearing in eq.~\eqref{beta}. Indeed, in general their lower family indices carry either components that can couple to the resulting products of traces or components that annihilate them. This can be recognized looking at the manifest symmetries of these terms, and selecting the compatible structures according to
\be
\{2^{\,p},1\} \, \otimes \, \{3\} \, = \, \{5,2^{\,p-1},1\} \, \oplus \, \{4,2^{\,p}\} \, \oplus \, \ldots \, .
\ee
This observation will play a role in the definition of a proper Stueckelberg-like shift for the Lagrangian theory, and more details and examples can be found in \cite{mixed_bose}.

To build the Lagrangians \eqref{lagmult_bose}, we revisited the gauge variation of the constrained Lagrangians \eqref{lagconstr_bose}. As a consequence, the Einstein-like tensors still only contain the projected traces of $\cF$ that are not \emph{fully} proportional to symmetrized double traces of $\vf$. On the other hand, if one relaxes the constraints some terms proportional to the double traces of $\vf$ pop up in eq.~\eqref{lagconstr_bose}. These are at the origin of the ``classical anomalies'' that in Section \ref{sec:labalag-bose} we canceled imposing the constraints. This observation suggests an alternative approach to the elimination of the constraints that leads to a different presentation of the Lagrangians for unconstrained fields. In fact, one can directly extract the terms proportional to $T_{(\,ij}\,T_{kl\,)}\,\vf$ in the Lagrangians \eqref{lagconstr_bose}. This can be done introducing the quantities
\be \label{newF}
\widehat{\cF}^{\,[\,p\,]}{}_{\,i_1j_1, \,\ldots\, ,\,i_pj_p} = \, \cF^{\,[\,p\,]}{}_{\,i_1j_1, \,\ldots\, ,\,i_pj_p} \, - \, \12 \ \pr^{\,k}\pr^{\,l}\, Y_{\{4,2^{p-1}\}}\, T_{i_1j_1} \!\ldots T_{i_pj_p} T_{kl} \, \vf \, ,
\ee
that coincide with the two-column projected traces of $\cF$ up to the terms proportional to the symmetrized double traces of $\vf$ that the latter contain (see for instance eq.~\eqref{fgentrace}). One can eventually use them to obtain the gauge-invariant Lagrangians
\begin{align}
\cL \, & = \, \12 \ \bra\, \vf \,\comma\, \sum_{p\,=\,0}^N \, \frac{(-1)^{\,p}}{p\,!\,(\,p+1\,)\,!} \ \h^{i_1j_1} \!\ldots\, \h^{i_pj_p}\, \widehat{\cF}^{\,[\,p\,]}{}_{\,i_1j_1,\,\ldots\,,\,i_pj_p} \,\ket \nn \\
& + \, \frac{1}{24} \, \bra\, \cB_{\,ijkl} \,\comma\, T_{(\,ij}\,T_{kl\,)}\, \vf \,\ket \, , \label{lagfield_bose}
\end{align}
where the $\cB_{\,ijkl}$ are \emph{gauge invariant} Lagrange multipliers enforcing on-shell the double trace constraints. These Lagrangians are not written in terms of the multiple traces of $\cF$ but their Einstein-like tensors are manifestly self-adjoint, since they contain \emph{only} the non-trivial terms that survive in eq.~\eqref{selfadjbose} when one imposes the constraints. Since the bare $\vf$ terms in eq.~\eqref{newF} can be cast in a form manifestly proportional to $T_{(\,ij}\,T_{kl\,)}\, \vf$, the new Lagrangians are clearly related to those presented in eq.~\eqref{lagmult_bose} via a field redefinition of the Lagrange multipliers. In this respect they do not introduce any real novelty, but this different viewpoint can be of some interest nonetheless. For instance, it manifestly shows that the complicated gauge transformation of eq.~\eqref{beta} are not an intrinsic feature, since they can be eliminated via a judicious choice of the terms quadratic in $\vf$. Finally, notice that the new tensors entering the Lagrangians satisfy the ``anomaly-free'' Bianchi identities
\be
(\,p+2\,) \ Y_{\{2^p,1\}} \, \pr_{\,k} \, \widehat{\cF}^{\,[\,p\,]}{}_{\,i_1
j_1,\,\ldots\,,\,i_p j_p}\, - \, \pr^{\,l} \,
\widehat{\cF}^{\,[\,p+1\,]}{}_{\,i_1
j_1,\,\ldots\,,\,i_p j_p\,,\,k\,l} \, = \, 0 \, ,
\ee
as the gauge invariance of eq.~\eqref{lagfield_bose} suggests.

To summarize, in eq.~\eqref{A_eq} we have presented non-Lagrangian field equations that are left invariant by the unconstrained gauge transformations \eqref{gaugeuncb}, while in eqs. \eqref{lagmult_bose} and \eqref{lagfield_bose} we have presented Lagrangians for unconstrained gauge fields that are still invariant only under constrained gauge transformations. We can now combine these two improvements and build a fully unconstrained Lagrangian theory. As a matter of principle, this can be done simply refining the Stueckelberg-like shift of eq.~\eqref{stueckelberg}. For instance, performing the shift
\begin{align}
& \vf \, \to \, \vf \, - \, \pr^{\,i}\,\Phi_{\,i} \, , \nn \\
& \b_{\,ijkl} \, \to \, \b_{\,ijkl} \, - \, \frac{1}{2} \, \sum_{p\,=\,0}^N \, \frac{k_{\,p}}{p+2} \ \pr_{\,(\,i\,}\pr_{\,j\,}\pr_{\,k\,|}\, \h^{i_1j_1} \!\ldots\, \h^{i_pj_p}\, Y_{\{2^p,1\}}\, T_{i_1j_1} \ldots\, T_{i_pj_p} \Phi_{\,|\,l\,)} \label{stueck_beta}
\end{align}
in eq.~\eqref{lagmult_bose}, one ends up with a fully unconstrained Lagrangian, where the compensators $\Phi_{\,i}$ only appear through their symmetrized traces. The shift of the Lagrange multipliers is crucial in order to fulfill this important condition, and it maps the $\b_{\,ijkl}$ into their gauge invariant completions. Since the Stueckelberg-like shift of $\vf$ has the same structure as a gauge transformation, the $\Phi_{\,i}$ portion of the $\b_{\,ijkl}$ shifts cancels the terms that in the gauge variation would be annihilated via the Bianchi identities. In fact, in the present context they would not be expressible in terms of $T_{(\,ij}\,\Phi_{\,k\,)}$. This approach makes it clear that a fully unconstrained Lagrangian exists, and will prove useful when dealing with its field equations. On the other hand, it does not provide a clear hint on the structure of the compensator terms, and further computations are needed. Since the steps involved are exactly the same, rather than dealing with the Stueckelberg-like shifts \eqref{stueck_beta} we shall now fix the precise form of the compensator terms reconsidering the gauge variation of the Lagrangian \eqref{lagmult_bose}. We shall thus compensate the terms proportional to $T_{(\,ij}\,\L_{k\,)}$ adding new contributions to the Lagrangian, rather than taking advantage of the constraints \eqref{constrL}.

To fix the structure of the fully unconstrained Lagrangians one can thus start from
\begin{align}
\cL_0 \, & = \, \12 \ \bra\, \vf \,\comma\, \sum_{p\,=\,0}^N \, \frac{(-1)^{\,p}}{p\,!\,(\,p+1\,)\,!} \ \h^{i_1j_1} \!\ldots\, \h^{i_pj_p}\, \cA^{\,[\,p\,]}{}_{\,i_1j_1;\,\ldots\,;\,i_pj_p} \,\ket \nn \\
& + \, \frac{1}{8} \, \bra\, \b_{\,ijkl} \,\comma\, \cC_{\,ijkl} \,\ket \, , \label{trial_bose}
\end{align}
that correspond to eq.~\eqref{lagmult_bose} up to the substitution of $\cF$ with $\cA$ and of $T_{(\,ij}\,T_{kl\,)}\vf$ with $\cC_{\,ijkl}$. Since the tensors $\cA$ and $\cC_{\,ijkl}$ of eqs.~\eqref{A} and \eqref{C} are gauge invariant under unconstrained gauge transformations, the gauge variation of the trial Lagrangian \eqref{trial_bose} can be cast in the form
\be
\d \, \cL \, = \, - \, \sum_{p\,=\,0}^N \, \frac{k_{\,p}}{2^{\,p+1}} \, \bra\, Y_{\{3,2^{\,p-1}\}}\, T_{i_1j_1} \ldots\, T_{i_pj_p}\, \L_{\,k} \,\comma\, Y_{\{3,2^{\,p-1}\}}\, \pr_{\,k}\, \cA^{\,[\,p\,]}{}_{\,i_1j_1,\,\ldots\,,\,i_pj_p} \,\ket \, . \label{varunc-b}
\ee
These are indeed the terms that were discarded in eq.~\eqref{varmult-b1}.
As we have already stressed in Section \ref{sec:labalag-bose}, the left entries of the scalar products are proportional to $T_{(\,ij}\,\L_{k\,)}$ on account of the $\{3,2^{\,p-1}\}$ projection that they carry. However, we now have to refine this argument, since we must identify the precise relation to compensate them adding new terms to the Lagrangians. The needed techniques are similar to those adopted to manipulate the two-column projections of the traces of the Bianchi identities. On the other hand, the chain of steps leading to the result is more involved, and in the following we shall only quote the final result. The interested reader can find more details in Appendix \ref{app:proofs}. At any rate, these manipulations recast the gauge variation \eqref{varunc-b} in the form
\be
\d \, \cL \, = \, - \, \sum_{p\,=\,0}^N \, \frac{p\,k_{\,p}}{2^{\,p+1}(\,p+2\,)} \, \bra\, T_{i_1j_1} \ldots\, T_{i_pj_p} \,
\L_{\,k} \,\comma\, \pr_{\,(\,k} \, \cA^{[\,p\,]}
{}_{\,i_1j_1\,),\,i_2j_2\,,\, \ldots\, ,\,i_pj_p} \,\ket \, , \label{varunc-b2}
\ee
where now only symmetrized traces of the gauge parameters appear. This suffices to identify the structure of the fully unconstrained Lagrangians
\begin{align}
\cL \, & = \, \12 \, \bra\, \vf \,\comma\, \sum_{p\,=\,0}^{N} \,
\frac{(-1)^{\,p}}{p\,!\,(\,p+1\,)\,!} \ \h^{i_1 j_1} \ldots \, \h^{i_{\,p} j_{\,p}} \,
\cA^{\,[\,p\,]}{}_{\,i_1 j_1,\,\ldots\,,\,i_{p} j_{p}} \,\ket \nn \\
& - \frac{1}{4} \, \bra\, \a_{\,ijk} \,\comma\, \sum_{p\,=\,0}^{N-1}
\,
\frac{(-1)^{\,p}}{p\,!\,(\,p+3\,)\,!} \ \h^{i_1 j_1} \ldots \, \h^{i_{p} j_{p}} \, \pr_{\,(\,i}\,
\cA^{\,[\,p+1\,]}{}_{jk\,),\,i_1 j_1,\,\ldots\,,\,i_{p} j_{p}}
\,\ket \nn \\
 &+ \, \frac{1}{8} \, \bra\, \b_{\,ijkl} \,\comma\, \cC_{\,ijkl}
 \,\ket\, . \label{lagunc_bose}
\end{align}
Notice that these Lagrangians possess the same symmetry under shifts of the Lagrange multipliers as their counterparts in eq.~\eqref{lagmult_bose}. Actually, in eq.~\eqref{51C} we have seen that, when working in terms of the compensators $\Phi_{\,i}$, the $\cC_{\,ijkl}$ tensors satisfy the same properties as the combinations $T_{(\,ij}\,T_{kl\,)}\vf$.

In the previous pages we have extended the constrained Labastida framework proposing the Lagrangians \eqref{lagmult_bose} and \eqref{lagfield_bose} for unconstrained fields still subjected to constrained gauge transformations, and the Lagrangians \eqref{lagunc_bose} for fully unconstrained fields. We can now close this section presenting their field equations. Those coming from the Lagrangians \eqref{lagfield_bose} can be obtained almost by inspection, since we stressed that their Einstein-like tensors are self-adjoint as in the constrained theory. On the other hand, the subtleties discussed in Section \ref{sec:lag-eq} do not arise in this framework since the fields are \emph{unconstrained}. The equations of motion following from the Lagrangian \eqref{lagfield_bose} thus read
\begin{align}
E_{\,\vf} \, & : \ \sum_{p\,=\,0}^{N} \,
\frac{(-1)^{\,p}}{p\,!\,(\,p+1\,)\,!} \ \h^{i_1 j_1}  \ldots \, \h^{i_{\,p} j_{\,p}} \,
\widehat{\cF}^{\,[\,p\,]}{}_{\,i_1 j_1,\,\ldots\,,\,i_{\,p} j_{\,p}}  \, + \, \12 \ \h^{ij}\,\h^{kl}\, \cB_{\,ijkl} \, = \, 0 \, , \label{Evf} \\[5pt]
E_{\,\cB} \, & : \ T_{(\,ij}\,T_{kl\,)}\,\vf \, = \, 0 \, , \label{Eb}
\end{align}
where the projections that we discussed in Section \ref{sec:lag-eq} can be recovered in $E_{\,\vf}$ via the elimination of the $\cB_{\,ijkl}$ tensors. This is the first example where the removal of the constraints simplifies, technically at least, the discussion of mixed-symmetry fields.

The equations of motion following from the Lagrangians \eqref{lagmult_bose} take a similar form, that reads
\begin{align}
E_{\,\vf} \, & : \ \sum_{p\,=\,0}^{N} \,
\frac{(-1)^{\,p}}{p\,!\,(\,p+1\,)\,!} \ \h^{i_1 j_1}  \ldots \, \h^{i_{\,p} j_{\,p}} \,
\cF^{\,[\,p\,]}{}_{\,i_1 j_1,\,\ldots\,,\,i_{\,p} j_{\,p}}  \, + \, \12 \ \h^{ij}\,\h^{kl}\, \cB_{\,ijkl} \nn \\
& - \, \frac{1}{4} \, \sum_{p\,=\,0}^{N} \,
\frac{(-1)^{\,p}}{p\,!\,(\,p+1\,)\,!} \ \h^{i_1 j_1}  \ldots \, \h^{i_{\,p} j_{\,p}} \, \pr^{\,k}\pr^{\,l} \, Y_{\{4,2^{p-1}\}} \, T_{kl} \,
\vf^{\,[\,p\,]}{}_{i_1 j_1,\,\ldots\,,\,i_{\,p} j_{\,p}} \, = \, 0 \, , \label{Evf1} \\[5pt]
E_{\,\b} \, & : \ T_{(\,ij}\,T_{kl\,)}\,\vf \, = \, 0 \, , \label{Eb1}
\end{align}
where the $\cB_{\,ijkl}$ are now the combinations
\be
\cB_{\,ijkl} \, = \, \b_{\,ijkl} \, - \, \frac{1}{2} \, \sum_{p\,=\,0}^{N-1} \, \frac{(-1)^{\,p}}{p\,!\,(\,p+3\,)\,!} \ \h^{i_1j_1} \ldots\,
\h^{j_{p}j_{p}} \,
\pr_{\,(\,i\,}\pr_{\,j}\,\vf^{\,[\,p+1\,]}{}_{kl\,),\,i_1j_1,\,\ldots\,,\,i_{p}j_{p}}
\, , \label{B}
\ee
that are invariant under constrained gauge transformations.
Notice that in these expressions we have extended the notation adopted for $\cF$ in order to denote the two-column projected traces of $\vf$. Moreover, when combined with eq.~\eqref{Eb1}, eq.~\eqref{Evf1} can be cast in the form \eqref{Evf}, so that in both cases we can effectively refer to these equations of motion.

Finally, the field equations following from the Lagrangians \eqref{lagunc_bose} can be also obtained from eqs.~\eqref{Evf} and \eqref{Eb}. In fact, these Lagrangians result from the action of the Stueckelberg-like shifts \eqref{stueck_beta} in eq.~\eqref{lagmult_bose}. One can thus identify the corresponding field equations for $\vf$ and the $\b_{\,ijkl}$ as
\begin{alignat}{2}
& E_{\,\vf} & & : \ \sum_{p\,=\,0}^{N} \,
\frac{(-1)^{\,p}}{p\,!\,(\,p+1\,)\,!} \ \h^{i_1 j_1}  \ldots \, \h^{i_{\,p} j_{\,p}} \,
\cA^{\,[\,p\,]}{}_{\,i_1 j_1;\,\ldots\,;\,i_{\,p} j_{\,p}}  \, + \, \12 \ \h^{ij}\,\h^{kl}\,
\cB_{\,ijkl}\, = \, 0
 ,\label{eqgenbvf} \\
& E_{\,\b} & & : \ \cC_{ijkl} \, = \, 0
\label{eqbgen}
\, ,
\end{alignat}
where, barring some subtleties that we shall dwell upon in a while, the $\cB_{\,ijkl}$ tensors are obtained shifting the corresponding tensors in eq.~\eqref{B} according to eq.~\eqref{stueck_beta}. Furthermore, the equations of motion for the compensators $\Phi_i$ must provide the conservation conditions for external unconstrained currents that are missing in the theory with constrained gauge invariance. In the general mixed symmetry case they thus read
\be
\pr_{\,i}\, E_{\,\vf} \, +  \sum_{p\,=\,0}^N \,
\frac{2}{p\,!\,(\,p+2\,)\,!}
\ \h^{m_1n_1} \ldots\, \h^{m_pn_p} \, Y_{\{2^p,1\}}\, T_{m_1n_1} \ldots\, T_{m_pn_p} \,
\pr^{\,j}\pr^{\,k}\pr^{\,l} \, (E_\b)_{\,ijkl}\,=\, 0\, .
\ee
The equations of motion \eqref{eqgenbvf} and \eqref{eqbgen} can then be reduced to eqs.~\eqref{Evf} and \eqref{Eb} undoing the Stueckelberg-like shift or, more precisely, performing a partial gauge fixing that does not spoil the Labastida constrained gauge symmetry. As a consequence, when in Section \ref{sec:weyl-bose} we shall discuss the reduction of these three sets of field equations to the Labastida form $\cF = 0$ we shall only deal with eqs.~\eqref{Evf} and \eqref{Eb}. These indeed suffice to capture the relevant features of all equations of motion following from the various approaches.

A last comment is in order, since the linear dependence of the Labastida constraints plays a role at this stage. In fact, the Stueckelberg-like shifts \eqref{stueck_beta} map the $\b_{\,ijkl}$ into their gauge invariant completions, but we have seen that the gauge transformations \eqref{beta} can be redefined eliminating all terms that do not affect the Lagrangians. Only the shifts built upon these refined transformations will lead to $\cB_{\,ijkl}$ tensors expressible in terms of the combinations $\a_{\,ijk}\,(\,\Phi\,)$, while the ``naive'' form presented in eq.~\eqref{stueck_beta} does not. A more detailed discussion supported by explicit two-family examples can be found in \cite{mixed_bose}.


\subsubsection{Fermi fields}\label{sec:minimal-fermi}


In Section \ref{chap:constrained} we showed that the Labastida constraints
\be \label{constre}
\g_{\,(\,i}\,\e_{\,j\,)} \, = \, 0
\ee
lead to first order gauge invariant field equations. Indeed, computing the gauge variation of the Fang-Fronsdal-Labastida tensor \eqref{laba-f} one obtains
\be
\d \, \cS \, = \, - \, \frac{i}{2} \ \pr^{\,i}\pr^{\,j}\, \g_{\,(\,i}\,\e_{\,j\,)} \, .
\ee
We also showed that the Labastida constraints
\be \label{constrpsi}
T_{(\,ij}\,\g_{\,k\,)} \, \psi \, = \, 0
\ee
lead to gauge-invariant Lagrangians, since
\be \label{bianchiS}
\pr_{\,i}\, \cS \, - \, \12 \dsl \, \g_{\,i}\, \cS \, - \, \12 \ \pr^{\,j}\,T_{ij}\, \cS \, - \, \frac{1}{6} \ \pr^{\,j}\,\g_{\,ij}\,\cS \, = \, \frac{i}{6} \ \pr^{\,j}\pr^{\,k}\, T_{(\,ij}\,\g_{\,k\,)}\, \psi \, ,
\ee
and Bianchi identities free of the ``classical anomaly'' appearing on the right hand-side of eq.~\eqref{bianchiS} guarantee the gauge invariance of the Lagrangians \eqref{lagconstr_fermi}.

The simplest solution to eliminate the constraints \eqref{constre} on the gauge parameters would be again to introduce for each of them one auxiliary field transforming as
\be
\d \, \x_{\,ij} \, = \, \12 \ \g_{\,(\,i}\,\e_{\,j\,)} \, .
\ee
However, even in this setup the linear dependence of the Labastida constraints \eqref{constre} would imply the existence of gauge invariant combinations of the $\x_{\,ij}$ fields. At any rate, in analogy with the bosonic case, this problem can be overcame resorting to a set of fundamental compensators transforming as
\be \label{Psi}
\d \, \Psi_{\,i} \, = \, \e_{\,i} \, .
\ee
The constrained gauge symmetry then guarantees that the $\Psi_{\,i}$ enter field equations and Lagrangians only through their symmetrized $\g$-traces. For this reason, in the following we shall often abide by the notation
\be \label{xi}
\x_{\,ij} \, \equiv \, \12 \ \g_{\,(\,i}\,\e_{\,j\,)} \, .
\ee
Before going on along the lines followed for bosons, let us support further these arguments by displaying a gauge invariant combination of the $\x_{\,ij}$. A convenient starting point is the combination \eqref{41comb}, that displays a vanishing linear combination of $\g$-traces of the constraints \eqref{constrpsi} on the fields $\psi$. Considering the gauge invariant completions of these constraints,
\begin{align}
\cZ_{\,ijk} & = \, \frac{i}{3} \left\{ \, T_{(\,ij} \psisl_{k\,)}  - \, 2 \, \pr_{\,(\,i} \, \xi_{\,jk\,)}\, -  \dsl \, \g_{\,(\,i} \, \xi_{\,jk\,)} \right. \nn \\
& \left. - \, \pr^{\,l} \left(\, 2 \, T_{(\,ij} \, \xi_{\,k\,)\,l}  - \, T_{l\,(\,i} \, \xi_{\,jk\,)}  - \, \g_{\,l\,(\,i} \, \xi_{\,jk\,)} \,\right) \right\} \, , \label{Ztemp}
\end{align}%
one can built the same combination of $\g$-traces acting on the $\cZ_{\,ijk}$ tensors. Yet, there are no group theoretical arguments forcing its $\x_{\,ij}$ part to vanish, and indeed
\begin{align}
& \left(\, \g_{\,m\,(\,i}\, \cZ_{\,j\,)\,kl} \, + \, \g_{\,m\,(\,k}\, \cZ_{\,l\,)\,ij} \,\right) \, - \, 15 \ Y_{\{4,1\}} \, \left(\, T_{ij}\, \cZ_{\,klm} \, + \, T_{kl}\, \cZ_{\,ijm} \,\right) \nn \\
& = \, \frac{i}{3} \ \pr^{\,n} \left\{\, 8 \, T_{(\,ij}\,T_{kl\,)}\, \x_{\,mn} \, - 4 \, T_{mn}\, T_{(\,ij}\, \x_{\,kl\,)} \, + \, 4\, T_{m\,(\,i\,|}\, T_{n\,|\,j}\, \x_{\,kl\,)} \, - \, 2\, T_{(\,m\,|\,(\,i}\,T_{jk}\, \x_{\,l\,)\,|\,n\,)} \,\right\} \nn \\
& + \, \frac{i}{3} \ \pr^{\,n} \left\{\, 2\, T_{(\,i\,|\,(\,m}\, \g_{\,n\,)\,|\,j}\, \x_{\,kl\,)} \, - \, 2 \, \g_{\,(\,m\,|\,(\,i}\, T_{jk}\, \x_{\,l\,)\,|\,n\,)} \,\right\} \, . \label{gauge_inv}
\end{align}
The terms on the right-hand side of this expression are available from two families onward and define a gauge invariant combination of the $\x_{\,ij}$ fields. In analogy with the bosonic case, linear combinations involving a smaller number of $\g$-traces cannot lead to gauge invariant combinations of the $\x_{\,ij}$, so that this problem can be avoided in simple mixed-symmetry models as in the fully symmetric case. Moreover, when the $\cZ_{\,ijk}$ are expressed in terms of the independent compensators $\Psi_{\,i}$, they clearly satisfy the same set of linear relations as the combinations $T_{(\,ij}\,\g_{\,k\,)}\,\psi$: for instance
\be \label{41unc}
\left(\, \g_{\,m\,(\,i}\, \cZ_{\,j\,)\,kl} \, + \, \g_{\,m\,(\,k}\, \cZ_{\,l\,)\,ij} \,\right) \, - \, 15 \ Y_{\{4,1\}} \, \left(\, T_{ij}\, \cZ_{\,klm} \, + \, T_{kl}\, \cZ_{\,ijm} \,\right) \, = \, 0 \, .
\ee

Now that we have identified the needed auxiliary fields, we can introduce the new kinetic tensors
\be \label{W}
\cW \, = \, \cS \, + \, \frac{i}{2} \ \pr^{\,i}\pr^{\,j}\, \g_{\,(\,i}\,\Psi_{\,j\,)} \, ,
\ee
that are invariant under the \emph{unconstrained} gauge transformations
\begin{align}
& \d \, \psi \, = \, \pr^{\,i}\, \e_{\,i} \, , \nn \\
& \d \, \Psi_{\,i} \, = \, \e_{\,i} \, . \label{gaugevarunc_fermi}
\end{align}
One can then obtain gauge-invariant field equations imposing
\be \label{W_eq}
\cW \, = \, 0 \, .
\ee
This ``minimal'' recipe to recover an unconstrained gauge symmetry leads again to a higher-derivative kinetic tensor. Indeed, $\cW$ contains two gradients acting on the $\Psi_{\,i}$, but the corresponding term is harmless since it can be gauged away without spoiling the constrained Labastida gauge symmetry. At any rate, ordinary single derivative field equations will be presented in Section \ref{sec:ordinary-fermi} at the price of increasing a bit the number of auxiliary fields.

Notice that the new kinetic tensors satisfy the Bianchi identities
\be \label{bianchiW}
\pr_{\,i}\, \cW \, - \, \12 \dsl \, \g_{\,i}\, \cW \, - \, \12 \ \pr^{\,j}\,T_{ij}\, \cW \, - \, \frac{1}{6} \ \pr^{\,j}\,\g_{\,ij}\,\cW \, = \, \frac{i}{2} \ \pr^{\,j}\pr^{\,k}\, \cZ_{\,ijk} \, ,
\ee
where the tensors
\be \label{Z}
\cZ_{\,ijk} \, = \, \frac{1}{3} \ T_{(\,ij}\,\g_{\,k\,)} \left(\, \psi \, - \, \pr^{\,l}\, \Psi_{\,l} \,\right)
\ee
coincide with those previously introduced in eq.~\eqref{Ztemp}, as can be seen expressing the $\x_{\,ij}$ in terms of the $\Psi_{\,i}$. Moreover, as suggested by the form taken by the $\cZ_{\,ijk}$, it is also possible to introduce the $\Psi_{\,i}$ compensators performing a Stueckelberg-like shift
\be \label{stueck_fermi}
\psi \, \to \, \psi \, - \, \pr^{\,i}\, \Psi_{\,i} \, .
\ee

Having seen how to eliminate the constraints \eqref{constre} on the gauge parameters, we can move on and analyze the constraints \eqref{constrpsi} on the field. The strategy will be that already delineated in the bosonic case. Relaxing the constraints \eqref{constrpsi}, the gauge variation of the Lagrangian \eqref{lagconstr_fermi} no longer vanishes, but using the Bianchi identities \eqref{bianchiS} one can control the remaining terms proportional to $T_{(\,ij}\,\g_{\,k\,)}\,\psi$. Rather than imposing the Labastida constraints \eqref{constrpsi}, these can then be canceled via the gauge variation of some extra terms containing a proper set of Lagrange multipliers. To identify them, let us continue to discard the contributions that are proportional to $\g_{\,(\,i}\,\e_{\,j\,)}$ as in Section \ref{sec:labalag-fermi}. In this fashion we can focus only on the portion of the gauge variation that is really linked to the Labastida constraints on the fields $\psi$. Under this hypothesis it is again possible to combine the two kinds of contributions appearing in eq.~\eqref{gaugefgen2}, as we did in the constrained theory. Thus, we can start to reconsider the gauge variation of the Lagrangian \eqref{lagconstr_fermi} from eq.~\eqref{finalgaugef}, that we write again here for clarity, denoting by $k_{\,p\,,\,q}$ the coefficients of eq.~\eqref{solgenf}:
\begin{align}
\d \, \cL &= \, - \sum_{\,p\,,\,q\,=\,0}^{N} \, \frac{1}{2^{\,p+1}} \, \frac{k_{\,p\,,\,q}}{(\,q+3\,)\,(\,p+q+2\,)} \ \bra \, T_{i_1j_1} \ldots T_{i_pj_p} \, \bar{\e}_{\,l} \, \g_{\,k_1 \ldots\, k_q} \,\comma \nn \\[2pt]
& \phantom{+} \, (\,q+3\,)\,(\,p+q+2\,) \ Y_{\{2^p,1^{q+1}\}} \, \pr_{\,l} \, (\,\g^{\,[\,q\,]}\,\cS^{\,[\,p\,]}\,)_{\, i_1 j_1,\,\ldots\,,\,i_p j_p \, ;\, k_1 \ldots\, k_q} \nn \\[2pt]
& + \, Y_{\{2^p,1^{q+1}\}} \, \sum_{n\,=\,1}^{p}\, \pr_{\,(\,i_n\,|}\, (\,\g^{\,[\,q+2\,]}\,\cS^{\,[\,p-1\,]}\,)_{\,\ldots\,,\,i_{r\,\ne\,n} j_{r\,\ne\,n}\,,\,\ldots \, ;\,|\,j_n\,)\,l \, k_1 \ldots\, k_q} \nn \\[2pt]
& + \, (-1)^{\,q+1}\, (\,q+3\,) \dsl \, (\,\g^{\,[\,q+1\,]}\,\cS^{\,[\,p\,]}\,)_{\, i_1 j_1,\,\ldots\,,\,i_p j_p \, ;\, k_1 \ldots\, k_q\, l} \nn \\[6pt]
& - \, (\,q+3\,)\ \pr^{\,m} \, (\,\g^{\,[\,q\,]}\,\cS^{\,[\,p+1\,]}\,)_{\, i_1 j_1,\,\ldots\,,\,i_p j_p \,,\, lm \, ;\, k_1 \ldots\, k_q} \nn \\[6pt]
& - \, \pr^{\,m} \, (\,\g^{\,[\,q+2\,]}\,\cS^{\,[\,p\,]}\,)_{\, i_1 j_1,\,\ldots\,,\,i_p j_p \, ;\, k_1 \ldots\, k_q\, lm} \, \ket \, + \, \textrm{h.c.} \ . \label{unc_finalvar}
\end{align}
In order to proceed it is then necessary to identify the unconstrained analogue of eq.~\eqref{finalbianchif}. In the fermionic case this step is more subtle than its bosonic counterpart. Indeed, a generic $\g$-trace $T_{i_1j_1} \ldots\, T_{i_pj_p}\, \g_{\,k_1 \ldots\, k_q}$ of the Bianchi identities \eqref{bianchiS} admits a couple of two-column projections since
\be
\{2^{\,p},1^q\} \, \otimes \, \{1\} \, = \, \{2^{\,p},1^{q+1}\} \, \oplus \, \{2^{\,p+1},1^{q-1}\} \, \oplus \, \{3,2^{\,p-1},1^q\} \, .
\ee
While in the constrained case the constraints on $\cS$ of eq.~\eqref{constrS} allow to simply consider the first of them to reach eq.~\eqref{finalbianchif}, in the unconstrained case one has to combine different consequences of the Bianchi identities in order to reproduce the terms entering eq.~\eqref{unc_finalvar}. This rather involved procedure is sketched in Appendix \ref{app:proofs}, but the final outcome is
\begin{align}
& (\,p+q+2\,) \ Y_{\{\,2^p,\,1^{q+1}\}} \, \pr_{\,l} \, (\,\g^{\,[\,q\,]}\,\cS^{\,[\,p\,]}\,)_{\, i_1 j_1,\,\ldots\,,\,i_p j_p \, ;\, k_1 \ldots\, k_q} \nn \\
& + \, \frac{1}{q+3} \ Y_{\{\,2^p,\,1^{q+1}\}} \, \sum_{n\,=\,1}^{p}\, \pr_{\,(\,i_n\,|}\, (\,\g^{\,[\,q+2\,]}\,\cS^{\,[\,p-1\,]}\,)_{\,\ldots\,,\,i_{r\,\ne\,n} j_{r\,\ne\,n} \,,\,\ldots\, ;\,|\,j_n\,)\,l \, k_1 \ldots\, k_q}  \nn \\[2pt]
& + \, (-1)^{\,q+1}\! \dsl \, (\,\g^{\,[\,q+1\,]}\,\cS^{\,[\,p\,]}\,)_{\, i_1 j_1,\,\ldots\,,\,i_p j_p \, ;\, k_1 \ldots\, k_q \, l}  \, - \, \pr^{\,m}\, (\,\g^{\,[\,q\,]}\,\cS^{\,[\,p+1\,]}\,)_{\, i_1 j_1,\,\ldots\,,\,i_p j_p\,,\,lm \, ;\, k_1 \ldots\, k_q} \nn \\[5pt]
& - \, \frac{1}{q+3} \ \pr^{\,m}\,
(\,\g^{\,[\,q+2\,]}\,\cS^{\,[\,p\,]}\,)_{\, i_1 j_1,\,\ldots\,,\,i_p
j_p \, ;\, k_1 \ldots\, k_q\,lm} \nn \\
& = \, \frac{i}{p+q+3}\ Y_{\{2^p,1^{q+1}\}} \bigg\{\, (\,p+2\,)\ T_{i_1j_1} \!\ldots T_{i_pj_p} \, \g_{\,k_1 \ldots\, k_q} \pr^{\,m}\pr^{\,n}\, T_{(\,lm}\,\g_{\,n\,)}\, \psi \nn \\
& - \, \frac{1}{q+3} \, \sum_{n\,=\,1}^p \, \prod_{r\,\ne\,n}^p \, T_{i_rj_r} \, \g_{\,k_1
\ldots\, k_q\,l\,(\,i_n\,|}\, \pr^{\,m}\pr^{\,n} \left(\, T_{\,|\,j_n\,)\,(\,m}\,\g_{\,n\,)} \, + \, T_{\,mn}\,\g_{\,|\,j_n\,)} \,\right)\, \psi \,\bigg\}\ . \label{trbianchiuncf}
\end{align}
The complicated structure of the constraint terms on the right-hand side actually comes from the combination of different $\g$-traces of the Bianchi identities. At any rate, substituting eq.~\eqref{trbianchiuncf} in eq.~\eqref{unc_finalvar} one can identify the gauge variation of the set of Lagrange multipliers to be added to the Lagrangian \eqref{lagconstr_fermi}. Removing the Labastida constraints on the fields thus leads to consider the Lagrangians
\begin{align}
\cL \, & = \, \12 \, \bra\, \bar{\psi} \,\comma \sum_{p\,,\,q\,=\,0}^{N} \, \frac{(-1)^{\,p\,+\,\frac{q\,(q+1)}{2}}}{p\,!\,q\,!\,(\,p+q+1\,)\,!} \ \h^{\,p} \, \g^{\,q}\, (\,\g^{\,[\,q\,]}\,\cS^{\,[\,p\,]}\,) \,\ket \nn \\
& + \, \frac{i}{12} \, \bra\, \bar{\l}_{\,ijk}\comma T_{(\,ij}\,\g_{\,k\,)}\, \psi \,\ket \, + \, \textrm{h.c.}\ , \label{lagmult_fermi}
\end{align}
where we resorted to the shorthand notation of eq.~\eqref{shorthand}. These Lagrangians are gauge invariant under constrained gauge transformations provided the $\l_{\,ijk}$ transform as
\begin{align}
& \d \, \l_{\,ijk} \, = \, \sum_{p\,,\,q\,=\,0}^N \, \frac{1}{p+q+2}\ \h^{m_1n_1} \ldots\, \h^{m_pn_p} \, \g^{\,l_1 \ldots\, l_q} \,\times \nn \\
& \times\, \pr_{\,(\,i\,}\pr_{\,j\,|} \, \bigg\{\, \frac{(\,p+2\,)\ k_{\,p\,,\,q}}{(\,p+q+3\,)} \ Y_{\{2^p,1^{q+1}\}} \, T_{m_1n_1} \ldots T_{m_pn_p} \, \g_{\,l_1 \ldots\, l_q} \, \e_{\,|\,k\,)} \nn \\[3pt]
& + \, \frac{(-1)^{\,q+1}\,(\,p+1\,)\ k_{\,p+1\,,\,q-2}}{q\,(\,q+1\,)\,(\,p+q+1\,)} \ Y_{\{2^{p+1},1^{q-1}\}} \, T_{m_1n_1} \ldots T_{m_pn_p} \,T_{|\,k\,)\,[\,l_1} \g_{\,l_2 \ldots\, l_{q-1}\,} \e_{\,l_q\,]} \,\bigg\} \, . \label{lagrangegen}
\end{align}

In analogy with the bosonic case, the linear dependence of the Labastida constraints displayed in eq.~\eqref{41comb} implies the existence of a symmetry of the Lagrangians \eqref{lagmult_fermi} under shifts of the Lagrange multipliers of the type
\be
\delta \l_{\,ijk} \, = \, \eta^{\,lm}\, M_{\,ijk,\,lm} \, +\, \g^{\,lm}\, N_{\,ijk;\,lm} \, . \label{lambdalsym}
\ee
The symmetry holds provided the $M_{\,ijk,\,lm}$ and $N_{\,ijk;\,lm}$ parameters are $\{4,1\}$-projected
in their family indices and are related according to
\be
N_{\,ij\,(\,k\,;\,l\,)\,m} \, + \, N_{\,kl\,(\,i\,;\,j\,)\,m} \, = \, - \, \12 \left(\, M_{\,ijm,\,kl} + \, M_{\,klm,\,ij} \,\right) \, . \label{MNrel}
\ee
In fact, the shift \eqref{lambdalsym} gives rise to
\begin{align}
\delta \, \cL \, & = \, \frac{i}{24} \ \bra \bar{M}_{\,ijk,\,mn} \comma T_{mn}\,
T_{(\, ij}\, \g_{\, k\,)} \, \psi \ket \nn \\
& - \, \frac{i}{12} \ \bra \bar{N}_{\,ijk;\,mn} \comma \left(\,
\g_{\,mn\,(\,i}\,T_{jk\,)} \, - \,
T_{(\,ij}\,T_{k\,)\,[\,m}\,\g_{\,n\,]} \,\right) \,
\psi  \ket \, + \, \textrm{h.c.} \, . \label{compmn}
\end{align}
The first scalar product can give
contributions in the $\{5\}$, $\{4,1\}$ and $\{3,2\}$ representations of the permutation group acting on the family indices. On the other
hand, the first term in the second scalar product can give contributions
in the $\{3,1,1\}$ and $\{2,1,1,1\}$, and finally the last can give
contributions in the $\{4,1\}$ and $\{3,1,1\}$. Therefore, if the
$M$ and $N$ parameters are $\{4,1\}$ projected and are related to
one another as in eq.~\eqref{MNrel}, an interesting symmetry of the
Lagrangians \eqref{lagmult_fermi} emerges.
It manifests itself rather clearly when one tries to reduce the Lagrangian field equations to
the Labastida form \eqref{laba_eq-f}, since all contributions that can be shifted away by
\eqref{lambdalsym} are left undetermined. Furthermore, as in the
bosonic case, an important lesson to be drawn from this symmetry is
that only part of the gauge transformation \eqref{lagrangegen} of the
Lagrange multipliers $\l_{\,ijk}$ is effective in the variation of the Lagrangian. As we shall see in a while, this is important when trying to extend at the Lagrangian level the Stueckelberg-like shift \eqref{stueck_fermi}.

For Bose fields we also saw an alternative presentation of the Lagrangians for unconstrained fields. One can follow the same approach even for Fermi fields. In particular, looking at eq.~\eqref{sgentrace} one can recognize that the combinations
\begin{align}
\widehat{\cS}^{\,[\,p\,,\,q\,]}{}_{\,i_1j_1, \,\ldots\, ,\,i_pj_p;\,k_1 \ldots\, k_q}\, & = \, Y_{\{2^p,1^q\}}\, T_{i_1j_1} \!\ldots T_{i_pj_p} \g_{\,k_1 \ldots\, k_q} \, \cS \nn \\
& + \, i \ \pr^{\,l}\, Y_{\{3,2^{p-1},1^q\}}\, T_{i_1j_1} \!\ldots T_{i_pj_p} \g_{\,k_1 \ldots\, k_q l}\, \psi \, , \label{Smod}
\end{align}
allow to define the Lagrangians
\begin{align}
\cL \, & = \, \12 \, \bra\, \bar{\psi} \,\comma \sum_{p\,,\,q\,=\,0}^{N} \, \frac{(-1)^{\,p\,+\,\frac{q\,(q+1)}{2}}}{p\,!\,q\,!\,(\,p+q+1\,)\,!} \ \h^{\,p} \, \g^{\,q}\, (\,\g^{\,[\,q\,]}\,\widehat{\cS}^{\,[\,p\,]}\,) \,\ket \nn \\
& + \, \frac{i}{12} \, \bra\, \bar{\cY}_{\,ijk}\comma T_{(\,ij}\,\g_{\,k\,)}\, \psi \,\ket \, + \, \textrm{h.c.}\ , \label{lagfield_fermi}
\end{align}
that are gauge invariant under constrained transformations even with gauge invariant Lagrange multipliers $\cY_{\,ijk}$. Notice that the portion of the new Lagrangians containing both $\bar{\psi}$ and $\psi$ is no more written in terms of the projected traces of $\cS$, but rather directly in terms of the fields. As in the bosonic case, these Lagrangians are related to those in eq.~\eqref{lagmult_fermi} via a field redefinition of the Lagrange multipliers, but this rewriting will prove useful when dealing with the field equations. Furthermore, the new tensors entering the Lagrangians satisfy ``anomaly-free'' Bianchi identities with the same structure as those in eq.~\eqref{finalbianchif}.

To summarize, in eq.~\eqref{W_eq} we have presented non-Lagrangian field equations that are invariant under the full unconstrained gauge transformations \eqref{gaugevarunc_fermi}, while in eqs.~\eqref{lagmult_fermi} and \eqref{lagfield_fermi} we have presented Lagrangians for unconstrained gauge fields that are still only invariant under constrained gauge transformations. As we did for Bose fields, we can now combine these two results and build a fully unconstrained Lagrangian theory. Again, in principle this can be done simply refining the Stueckelberg-like shift of eq.~\eqref{stueck_fermi}. This can be done defining also a proper shift of the Lagrange multipliers according to
\begin{align}
& \psi \, \rightarrow \, \psi \, - \, \pr^{\,i}\, \Psi_{\,i} \, , \nn \\
& \z_{\,ijk} \, \rightarrow \, \l_{\,ijk} \, - \, \Delta_{\,ijk}\,(\,\Psi\,) \, . \label{stueck_lambda}
\end{align}
The $\D_{\,ijk}(\Psi)$ that appear in eq.~\eqref{stueck_lambda} can be formally obtained
replacing $\e_{\,i}$ with $\Psi_{\,i}$ in the gauge variation of the
Lagrange multipliers $\l_{\,ijk}$, so that the $\z_{\,ijk}$ are
mapped into gauge-invariant combinations. Performing these shifts in eq.~\eqref{lagmult_fermi}, one ends up with a fully unconstrained Lagrangian, where the compensators $\Psi_{\,i}$ only appear through their symmetrized $\g$-traces. As for Bose fields, the shift of the Lagrange multipliers is crucial in order to satisfy this important condition, since it cancels the terms that would not be expressible in terms of $\g_{\,(\,i}\,\Psi_{\,j\,)}$. Again, this approach makes it clear that a fully unconstrained Lagrangian exists but, at least at this level, it does not provide a clear hint on the structure of the compensator terms. More work is needed to display these terms, and to this end we shall now follow the path outlined in Section \ref{sec:minimal-bose} for Bose fields. Rather than dealing directly with the shifts \eqref{stueck_lambda}, we shall in fact reconsider the gauge variation of a suitable modification of the Lagrangian \eqref{lagmult_fermi}, compensating the terms proportional to $\g_{\,(\,i}\,\e_{\,k\,)}$ via the addition of new compensator terms.

A convenient starting point is thus provided by the trial Lagrangian obtained substituting $\cS$ with $\cW$ and $T_{(\,ij}\,\g_{\,k\,)}\,\psi$ with $\cZ_{\,ijk}$ in eq.~\eqref{lagmult_fermi}:
\begin{align}
\cL_0 \, & = \, \12 \, \bra\, \bar{\psi} \,\comma \sum_{p\,,\,q\,=\,0}^{N} \, \frac{(-1)^{\,p\,+\,\frac{q\,(q+1)}{2}}}{p\,!\,q\,!\,(\,p+q+1\,)\,!} \ \h^{\,p} \, \g^{\,q}\, (\,\g^{\,[\,q\,]}\,\cW^{\,[\,p\,]}\,) \,\ket \nn \\
& + \, \frac{1}{4} \, \bra\, \bar{\l}_{\,ijk}\comma \cZ_{\,ijk} \,\ket \, + \, \textrm{h.c.}\ . \label{trial_fermi}
\end{align}
The gauge variation of the first line of eq.~\eqref{trial_fermi} takes exactly the form of eq.~\eqref{gaugefgen}, barring the substitution of $\cS$ with $\cW$. On the other hand, since we are relaxing the constraints \eqref{constre} on the gauge parameters, we now have to keep the terms that in Section \ref{sec:labalag-fermi} we discarded in going from eq.~\eqref{gaugefgen} to eq.~\eqref{finalgaugef}. These can eventually be canceled by the gauge variation of some compensator terms to be added to $\cL_0$. However, differently from the bosonic case, here the compensator terms originate from two types of remainders: in fact, in the constrained theory we first eliminated the $\{3,2^{\,p-1},1^q\}$-projected terms in eq.~\eqref{gaugefgen} and then we used again these constraints in order to combine the two classes of two-column projections appearing in eq.~\eqref{gaugefgen2}. In the unconstrained theory the first type of remainders can be compensated exploiting the precise dependence on the constraints of the $\{3,2^{\,p-1},1^q\}$ components of the $T_{i_1j_1} \ldots T_{i_pj_p} \, \bar{\e}_{\,l} \, \g_{\,k_1 \ldots\, k_q}$ combinations appearing on the left entries of the scalar products in eq.~\eqref{gaugefgen}. The precise relation is rather involved and it is derived in Appendix \ref{app:proofs}, but the end result is
\begin{align}
& \bra Y_{\{3,2^{p-1},1^q\}} \, T_{i_1j_1} \ldots T_{i_pj_p} \, \bar{\e}_{\,l} \, \g_{\,k_1 \ldots\, k_q} \comma \pr_{\,l} \, (\,\g^{\,[\,q\,]}\,\cW^{\,[\,p\,]}\,)_{\, i_1 j_1,\,\ldots\,,\,i_p j_p \, ;\, k_1 \ldots\, k_q} \ket \nn \\[5pt]
& = \, \bra \, (\,p+1\,) \ T_{i_2j_2} \ldots T_{i_pj_p}\, T_{(\,i_1j_1} \, \bar{\e}_{\,l\,)} \, \g_{\,k_1 \ldots\, k_q} \, - \, T_{i_2j_2} \ldots T_{i_pj_p} \, T_{\,[\,k_1\,|\,(\,l} \, \bar{\e}_{\,i_1} \, \g_{\,j_1\,)}\, \g_{\,|\,k_2 \ldots\, k_q\,]} \nn \comma \\
& \frac{p}{3\, (\,p+1\,)(\,p+q+2\,)} \ \pr_{\,(\,l\,|} \,
(\,\g^{\,[\,q\,]}\,\cW^{\,[\,p\,]}\,)_{\,|\, i_1
j_1\,),\,\ldots\,,\,i_p j_p \, ;\, k_1 \ldots\, k_q} \ket\, . \label{idconstrf}
\end{align}
Hence, the three-column projected terms appearing in the gauge variation of the trial Lagrangians \eqref{trial_fermi} can be compensated adding to them
\begin{align}
& \cL_1 \, = \, - \, \frac{1}{4} \ \bra\, \bar{\x}_{\,ij} \ ,  \sum_{p\,,\,q\,=\,0}^N \, \h^{i_1j_1} \ldots \h^{i_pj_p}\, \g^{\,k_1 \ldots\, k_q} \, \times \nn \\
& \times \bigg\{\, \frac{(\,p+1\,)\,k_{\,p+1\,,\,q-1}}{p+q+2} \ \pr_{\,(\,k_1\,|} (\,\g^{\,[\,q-1\,]}\,\cW^{\,[\,p+1\,]}\,)_{\,|\, ij\,),\,i_1j_1\,,\,\ldots\,,\,i_pj_p\,;\, k_2 \ldots\, k_q} \nn \\[3pt]
& + \, \frac{2\,p\,(\,p-1\,)\,q\, k_{\,p\,,\,q+1}}{(\,p+1\,)\,(\,p+q+3\,)} \ \pr_{\,(\,i_1\,|}
(\g^{\,[\,q+1\,]}\,\cW^{\,[\,p\,]})_{\,|\, ij\,),\,i_2j
_2\,,\,\ldots\,,\,i_{p} j_{p}\,;\,j_1\, k_1 \ldots\, k_q} \,\bigg\}\,\ket \, + \, \textrm{h.c.} \ , \label{comp1}
\end{align}

In the constrained theory we combined the remaining two-column projected terms in eq.~\eqref{gaugefgen2} using the first identity in eq.~\eqref{projfgauge}, whose unconstrained analogue is
\begin{align}
& Y_{\{2^{p+1},1^{q-1}\}} \, T_{i_1j_1} \ldots T_{i_pj_p} \, \bar{\e}_{\,l} \, \g_{\,k_1 \ldots\, k_q} \, = \, \frac{1}{q+1} \ Y_{\{2^{p+1},1^{q-1}\}} \, T_{i_1j_1} \ldots T_{i_pj_p} \, T_{\,l\,[\,k_1} \bar{\e}_{\,k_2\,} \g_{\,k_{3} \ldots\, k_q\,]} \nn \\
& - \, \frac{2}{q+1} \ Y_{\{2^{p+1},1^{q-1}\}} \, \, T_{i_1j_1} \ldots T_{i_pj_p} \, \d \, \bar{\x}_{\,l\,[\,k_1\,} \g_{\,k_{2} \ldots\, k_q\,]}\ . \label{projfgaugeunc}
\end{align}
The gauge variation of $(\,\cL_0+\cL_1\,)$ thus vanishes up to the terms proportional to $\d\, \x_{\,ij}$ coming from eq.~\eqref{projfgaugeunc}, that can be canceled by the gauge transformation of
\begin{align}
& \cL_2 = \, - \, \frac{1}{4} \ \bra\, \bar{\x}_{\,ij} \ ,  \sum_{p\,,\,q\,=\,0}^N \, \h^{i_1j_1} \ldots \h^{i_pj_p}\, \g^{\,k_1 \ldots\, k_q} \, \times \nn \\
& \times \, \frac{2\,(\,q+1\,)}{q+2}\, \bigg\{\, k_{\,p\,,\,q+1} \ Y_{\{2^{p+1},1^q\}} \, \pr_{\,(\,i\,|} (\,\g^{\,[\,q+1\,]}\,\cW^{\,[\,p\,]}\,)_{\,i_1j_1\,,\,\ldots\,,\,i_pj_p\,;\,|\,j\,)\, k_1 \ldots\, k_q} \nn \\[3pt]
& - \, (\,p+1\,)\,k_{\,p+1\,,\,q+1} \ \pr^{\,m}\,
(\g^{\,[\,q+1\,]}\,\cW^{\,[\,p+1\,]})_{\,i_1j
_1\,,\,\ldots\,,\,i_{p} j_{p}\,,\,ij\,;\,m\, k_1 \ldots\, k_q} \,\bigg\} \ket \, + \, \textrm{h.c.} \ . \label{comp2}
\end{align}

As a consequence, our fully unconstrained Lagrangians for mixed-symmetry Fermi fields take the form
\begin{align}
\cL \, & = \, \12 \, \bra \bar{\psi} \comma \!\sum_{p\,,\,q\,=\,0}^{N} \, \frac{(-1)^{\,p\,+\,\frac{q\,(q+1)}{2}}}{p\,!\,q\,!\,(\,p+q+1\,)\,!} \ \h^{\,p} \, \g^{\,q}\, (\,\g^{\,[\,q\,]}\,\cW^{\,[\,p\,]}\,) \ket \, + \, \frac{1}{4} \, \bra \bar{\x}_{\,ij} \comma \Xi_{\,ij} \ket \nn \\
& + \, \frac{1}{4} \, \bra \bar{\l}_{\,ijk}\comma \cZ_{\,ijk} \ket \, + \, \textrm{h.c.}\ , \label{lagunc_fermi}
\end{align}
where
\be
\frac{1}{4} \, \bra \bar{\x}_{\,ij} \comma \Xi_{\,ij} \ket \, = \, \cL_1 \, + \, \cL_2 \, ,
\ee
and where the gauge transformations of the Lagrange multipliers are
given in eq.~\eqref{lagrangegen}. Notice that the fully unconstrained Lagrangians also display a symmetry under shifts of the Lagrange multipliers of the form \eqref{lambdalsym}. This essentially follows from eq.~\eqref{41unc}, that holds when one works in terms of the independent compensators $\Psi_{\,i}$.

Following the lines already depicted in the bosonic case, in the previous pages we have extended the constrained framework proposing the Lagrangians \eqref{lagmult_fermi} and \eqref{lagfield_fermi} for unconstrained fields that still allow only constrained gauge transformations, and the Lagrangians \eqref{lagunc_fermi} for fully unconstrained fields. We can now close this section by presenting their field equations. As for Bose fields, those coming from the Lagrangians \eqref{lagfield_fermi} can be obtained almost by inspection, since these Lagrangians satisfy the identity \eqref{varself_fermi} as in the constrained theory. Furthermore, in this framework the fields are \emph{unconstrained}, so that the subtleties discussed in Section \ref{sec:lag-eq} are absent. The equations of motion thus read
\begin{align}
& E_{\, \bar{\psi}}\, : \ \sum_{p\,,\,q\,=\,0}^N \, \frac{(-1)^{\,p\,+\,\frac{q\,(q+1)}{2}}}{p\,!\,q\,!\,(\,p+q+1\,)\,!} \ \h^{\,p} \, \g^{\,q}\, (\,\g^{\,[\,q\,]}\,\widehat{\cS}^{\,[\,p\,]}\,)\,  - \, \12 \ \h^{ij}\,\g^{\,k}\, \cY_{\,ijk} \, = \, 0 \, , \label{Epsi}\\[5pt]
& E_{\, \bar{\cY}}\, : \ T_{(\,ij}\,\g_{\,k\,)}\, \cS \, = \, 0 \, , \label{El}
\end{align}
where the projections of Section \ref{sec:lag-eq} can be recovered via the eliminations of the $\cY_{\,ijk}$ tensors from eq.~\eqref{Epsi}. This is another example of the technical simplifications following from the elimination of the constraints \eqref{constrpsi}. Clearly, the equation of motion coming from the Lagrangians \eqref{lagmult_fermi} can be cast in the same form if one cancels from $E_{\,\bar{\psi}}$ the terms proportional to $E_{\,\bar{\l}}$. The only difference is that now the $\cY_{\,ijk}$ are expressed in terms of the basic fields via the combinations
\begin{align}
\cY_{\,ijk} \equiv \, i\ \z_{\,ijk} \,& + \,i \sum_{p\,,\,q\,=\,0}^N \, \frac{(-1)^{\,p\,+\,\frac{q\,(q+1)}{2}\,+\,1}}{p\,!\,q\,!\,(\,p+q+3\,)\,!}\ \h^{m_1n_1} \ldots\, \h^{m_pn_p}\, \g^{\,l_1 \ldots\, l_q} \, \times \nn \\
& \times \, \pr_{\,(\,i\,|}\, (\,\g^{\,[\,q\,]}\,\psi^{\,[\,p+1\,]}\,){}_{\,|\,jk\,)\,,\,m_1n_1,\,\ldots \,,\,m_pn_p\,;\,l_1 \ldots\, l_q}\, , \label{Y}
\end{align}
that are invariant under constrained gauge transformations.
Notice that in these expressions we have extended to $\psi$ the notation adopted in order to denote the two-column projected traces of $\cS$.

Finally, even the field equations coming from the Lagrangians \eqref{lagunc_fermi} can be obtained from eqs.~\eqref{Epsi} and \eqref{El}. In fact, these Lagrangians result from the action of the Stueckelberg-like shifts \eqref{stueck_lambda} in eq.~\eqref{lagmult_fermi}. One can thus identify the corresponding field equations for $\bar{\psi}$ and the $\bar{\l}_{\,ijk}$ shifting eqs.~\eqref{Epsi} and \eqref{El} into
\begin{align}
& E_{\, \bar{\psi}}\, : \ \sum_{p\,,\,q\,=\,0}^N \, \frac{(-1)^{\,p\,+\,\frac{q\,(q+1)}{2}}}{p\,!\,q\,!\,(\,p+q+1\,)\,!} \ \h^{\,p} \, \g^{\,q}\, (\,\g^{\,[\,q\,]}\,\cW^{\,[\,p\,]}\,)\,  - \, \12 \ \h^{ij}\,\g^{\,k}\, \cY_{\,ijk} \, = \, 0 \, , \label{eqgenpsi} \\[5pt]
& E_{\, \bar{\l}}\, : \ \cZ_{\,ijk} \, = \, 0 \, , \label{eqlgen}
\end{align}
where, barring some subtleties that we shall dwell upon in a while, the $\cY_{\,ijk}$ tensors are obtained shifting the corresponding tensors in eq.~\eqref{Y} according to eq.~\eqref{stueck_lambda}. Furthermore, the equations of motion for the compensators $\bar{\Psi}_i$ must provide the conservation conditions for external unconstrained currents that are missing in the theory with constrained gauge invariance. In the general mixed symmetry case they thus read
\begin{align}
& E_{\,\bar{\Psi}} \, : \ \pr_{\,l} \, E_{\,\bar{\psi}} \ - \, \sum_{\,p\,,\,q\,=\,0}^{N} \, \frac{3\, (\,q+1\,)\,(\,q+2\,)\, (-1)^{\,p\,+\,\frac{q\,(q+1)}{2}}}{p\,!\,(\,q+3\,)\,!\,(\,p+q+3\,)\,!}\ \h^{i_1j_1} \!\ldots \h^{i_pj_p}\, \g^{\,k_1 \ldots\, k_q}\, Y_{\{2^p,1^{q+1}\}}\, \times \nn \\[2pt]
& \phantom{E_{\,\bar{\Psi}} \, : \ \pr_{\,l} \, E_{\,\bar{\psi}} \ } \times \, \Big\{ \, (\,p+2\,)\,(\,q+3\,)\ T_{i_1j_1} \!\ldots T_{i_pj_p} \, \g_{\,k_1 \ldots\, k_q} \pr^{\,m}\pr^{\,n}\, (E_{\, \bar{\l}})_{\,lmn} \nn \\
& \phantom{E_{\,\bar{\Psi}} \, : \ \pr_{\,l} \, E_{\,\bar{\psi}} \ } - \, \sum_{n\,=\,1}^p \, \prod_{r\,\ne\,n}^p \, T_{i_rj_r} \, \g_{\,k_1
\ldots\, k_q\,l\,(\,i_n\,|}\, \pr^{\,m}\pr^{\,n}\, (E_{\, \bar{\l}})_{\,|\,j_n\,)\,mn} \, \Big\} \, = \, 0 \, .
\end{align}
The equations of motion \eqref{eqgenpsi} and \eqref{eqlgen} can then be reduced to eqs.~\eqref{Epsi} and \eqref{El} performing a partial gauge fixing that does not spoil the Labastida constrained gauge symmetry. As a consequence, when we shall discuss the reduction of these three sets of field equations to the Labastida form $\cS = 0$ we shall only deal with eqs.~\eqref{Epsi} and \eqref{El}. These equations indeed suffice to capture the relevant features of all those coming from the various approaches.

A last comment is in order, since the linear dependence of the Labastida constraints plays a role at this stage. In fact, the Stueckelberg-like shifts \eqref{stueck_lambda} map the $\l_{\,ijk}$ into their gauge invariant completions, but we have seen that the gauge transformations \eqref{lagrangegen} can be redefined eliminating all terms that do not affect the Lagrangians. Only the shifts built upon these refined gauge transformations will lead to $\cY_{\,ijk}$ tensors expressible in terms of the combinations $\x_{\,ij}\,(\,\Psi\,)$, while the ``naive'' form presented in eq.~\eqref{stueck_lambda} does not. A more detailed discussion supported by explicit two-family examples can be found in \cite{mixed_fermi}.


\subsubsection{Irreducible fields}\label{sec:minimal-irreducible}


As in most of Section \ref{chap:constrained}, the fields considered in the previous sections were in general only symmetric under the interchange of pairs of vector indices belonging to the same set, but did not possess any symmetry relating different sets. They were thus \emph{reducible} $gl(D)$ tensors or spinor-tensors. In Section \ref{sec:irreducible} we have already seen the kinds of modifications that emerge when dealing with \emph{irreducible} fields, that are subjected to suitable Young projections. The relations \eqref{gaugeparcondirr} involving the gauge parameters are still valid in the unconstrained framework, since we did not exploit the constraints \eqref{constre} to derive them. In a similar fashion, the Einstein-like tensors appearing in the Lagrangians still commute with the $S^{\,i}{}_j$ operators as in the constrained setup. Furthermore, the \mbox{($\g$-)}traces of the unconstrained kinetic tensors $\cA$ or $\cW$ still satisfy the same relations as the traces of the constrained kinetic tensors $\cF$ of $\cS$. For instance, in the bosonic case the identity \eqref{idirrF} simply becomes
\be \label{idirrA}
S^{\,m}{}_n\, T_{i_1j_1} \ldots\, T_{i_pj_p}\, \cA \, + \, \sum_{i\,=\,1}^p \, \d^{\,m}{}_{(\,i_n}\, T_{j_n\,)\,n}\, \prod_{r\,\neq\,n}^p\, T_{i_rj_r}\, \cA \, = \, 0 \, .
\ee

On the other hand, in general the irreducibility conditions \eqref{condirr} also restrict the extra field content introduced in the previous sections. For instance, eq.~\eqref{gaugeparcondirr} induces an identical set of conditions on the $\Phi_{\,i}$ compensators,
\be
S^{\,m}{}_n\, \Phi_{\,i} \, + \, \d^{\,m}{}_i\, \Phi_{\,n} \, = \, 0 \, .
\ee
As a consequence, the ``composite'' compensators $\a_{\,ijk}$ satisfy the relations
\be
S^{\,m}{}_n\, \a_{\,ijk} \, + \, \d^{\,m}{}_{(\,i}\, \a_{\,jk\,)\,n} \, = \, 0 \, ,
\ee
while computing the symmetrized double traces of \eqref{condirr} and moving to the left the $S^{\,m}{}_n$ operator one can obtain the relations satisfied by the $\cC_{\,ijkl}$ tensors. These induce identical conditions on the Lagrange multipliers, so that
\be
S^{\,m}{}_n\, \b_{\,ijkl} \, + \, \d^{\,m}{}_{(\,i}\, \b_{\,jkl\,)\,n} \, = \, 0 \, .
\ee

Similar conditions restrict the field content of the unconstrained fermionic models, so that the $\Psi_{\,i}$ are related according to
\be
S^{\,m}{}_n\, \Psi_{\,i} \, + \, \d^{\,m}{}_i\, \Psi_{\,n} \, = \, 0 \, ,
\ee
while their symmetrized $\g$-traces satisfy
\be
S^{\,m}{}_n\, \x_{\,ij} \, + \, \d^{\,m}{}_{(\,i}\, \x_{\,j\,)\,n} \, = \, 0 \, .
\ee
Finally, in the fermionic case the Lagrange multipliers are related according to
\be
S^{\,m}{}_n\, \l_{\,ijk} \, + \, \d^{\,m}{}_{(\,i}\, \l_{\,jk\,)\,n} \, = \, 0 \, .
\ee
For further details we refer the reader to \cite{mixed_fermi}, where we also sketched a general procedure for solving conditions with this structure.


\subsection{Low-derivative Lagrangian formulation}\label{sec:ordinary}


In Section \ref{sec:minimal} we have seen how the constraints can be eliminated introducing a ``minimal'' number of auxiliary fields \cite{fs2,fs3,fms1,mixed_bose,mixed_fermi}, that only depends on the number of index families and not on the total number of space-time indices carried by the gauge fields as in the BRST-like constructions of \cite{brst,brst_mixed}. However, while the constraints on the gauge fields were eliminated adding Lagrange multiplier terms to the Lagrangians, the unconstrained gauge symmetry was recovered adding higher-derivative terms involving extra compensator fields. As repeatedly stressed in the previous pages, these higher-derivative contributions do not represent a real problem since they can be gauged away without spoiling the Labastida constrained gauge symmetry, but they still represent an unconventional ingredient that is absent in lower-spin examples. On the contrary, the BRST-like constructions contain a number of auxiliary fields that grows with the ``spin'', but associate only the conventional number of derivatives even to the auxiliary fields. In this section we would like to show that the presence of one or more of these peculiarities is not an essential feature of higher-spin dynamics. In fact, in the following we shall explain how to combine the good properties of both unconstrained setups, presenting a more conventional description of the free dynamics for massless mixed-symmetry fields. It contains again only unconstrained $gl(D)$ \mbox{(spinor-)}tensors, but they have now second order field equations and Lagrangians in the case of Bose fields and first order field equations and Lagrangians in the case of Fermi fields. Furthermore, the field content is a bit larger with respect to the ``minimal'' construction, but the number of auxiliary fields still depends only on the number of index families. This means, for instance, that in the fully symmetric framework four auxiliary fields suffice to obtain an ordinary Lagrangian formulation for arbitrary higher-spin bosons and fermions.

Before moving to the details of the construction, let us recall that for fully symmetric Bose and Fermi fields an unconstrained formulation of the free dynamics without higher-derivative terms and with a fixed number of extra fields was first attained in \cite{buch_tripl}. It is an interesting off-shell
variant of the on-shell truncation of the ``triplets''
\cite{triplets} of String Field Theory \cite{free_sft} obtained in
\cite{fs1,st} (see also \cite{sorokvas} for some recent
developments in the ``frame-like'' formalism). For fully symmetric Bose fields it was then followed by the construction of \cite{dario_massive}, that is really tailored to the compensator constructions of \cite{fs2,fms1}. This also stimulated the development of the formulation that was presented in \cite{mixed_bose,mixed_fermi}, for both Bose and Fermi fields of mixed-symmetry, and that we are going to review.


\subsubsection{Bose fields}\label{sec:ordinary-bose}


The idea underlying the construction of the new unconstrained formulation is the replacement of the $\Phi_{\,i}$ compensators of eq.~\eqref{Phi} with other fields whose dimensions
are at least as high as that of the gauge potentials $\vf$. Let us
stress that such fields can not be pure gauge for this reason but, as first proposed in \cite{dario_massive}, they can be forced to vanish on shell adding a suitable set of Lagrange multipliers to the Lagrangians.

Whether one relaxes the request for pure-gauge compensators, it is simple to construct gauge invariant kinetic tensors similar to $\cA$ but free of higher derivatives, and indeed several options are available. However, it is less straightforward to make sure that the resulting Lagrangians do not propagate additional degrees of freedom, and in \cite{mixed_bose} the potential problems were discussed in some examples. On the other hand, a systematic way to forego the constraints without introducing higher-derivative terms can be attained refining the Stueckelberg way to enhance the gauge symmetry \cite{mixed_bose}. This procedure permits to follow closely the steps outlined in the previous sections, and in this fashion it also simplifies the check of its reduction to the constrained formulation. In Section \ref{sec:minimal-bose} we have indeed seen that the kinetic tensors $\cA$ of eq.~\eqref{A} can be obtained via the Stueckelberg-like shift
\be \label{shift_naive}
\vf \, \to \, \vf \, - \, \pr^{\,i}\, \Phi_{\,i} \, , \qquad \textrm{with} \qquad \d\, \Phi_{\,i}\, = \, \L_{\,i} \, .
\ee
As we have already pointed out, this operation makes it manifest that the $\Phi_{\,i}$ compensators have higher (mass) dimensions than the gauge field, and this leads directly to the higher-derivative term in eq.~\eqref{A}. Furthermore, the shift \eqref{shift_naive} maps $\vf$ into a gauge invariant combination, so that arbitrary functions of $\vf$ become gauge invariant under its action. However, this condition looks somehow unnatural, since our goal should be only to enhance the constrained gauge symmetry that is already present. Indeed, only the symmetrized traces of the $\Phi_{\,i}$ compensators play a role in the minimal unconstrained construction. Thus, even if the solution provided by the Stueckelberg-like shift \eqref{shift_naive} is technically very simple, it would be more natural to map $\vf$ into a combination that is gauge invariant only under the transformations that are forbidden in the constrained theory.

Actually, these two comments are related, and a proper redefinition of the shift \eqref{shift_naive} naturally outlines a safe way of introducing a set of compensators with the same dimensions as the gauge field. Indeed, the gauge parameters can be decomposed according to
\be
\L_{\, i} \, = \, \L^{(t)}{}_{i} \, + \, \h^{\, jk} \, \L^{(p)}{}_{i\,,\,jk} \, ,
\ee
where the $\L^{\,(t)}{}_i$ satisfy the conditions
\be
T_{(\,ij}\, \L^{\,(t)}{}_{k\,)} \, \equiv \, 0
\ee
while the $\L^{(p)}{}_{i\,,\,jk}$ carry the full amount of gauge symmetry that one would like to add, and are such that
\be \label{def_lambdap}
T_{(\,ij\,|}\, \h^{\,lm}\, \L^{\,(p)}{}_{|\,k\,)\,,\,lm} \, \equiv \, T_{(\,ij}\, \L_{\,k\,)} \, .
\ee
Hence, the Labastida constrained gauge transformations can be cast in the form
\be \label{labatransf_bose}
\d\, \vf \, = \, \pr^{\,i}\,\L^{\,(t)}{}_i \, ,
\ee
while the gauge transformations of the form
\be \label{unctransf_bose}
\d\, \vf \, = \, \h^{\,ij}\,\pr^{\,k}\, \L^{\,(p)}{}_{k\,,\,ij}
\ee
are those we would like to allow via the introduction of a proper set of auxiliary fields.

If one accepts compensators that are not pure gauge this goal can be achieved with a set of fields transforming as
\be \label{gauge_theta}
\d \, \theta_{\,ij} \, = \, \pr^{\,k}\, \L^{\,(p)}{}_{k\,,\,ij} \, .
\ee
They manifestly have the same dimension as the gauge potential, and one can use them to define the Stueckelberg-like shift
\be \label{shiftb_ok}
\vf \, \to \, \vf \, - \, \h^{\,ij}\, \th_{\,ij} \, .
\ee
The combination on the right-hand side of this expression transforms exactly like the gauge potential under \eqref{labatransf_bose}, while it is gauge invariant under \eqref{unctransf_bose}. Performing the shift \eqref{shiftb_ok} in the Fronsdal-Labastida tensor \eqref{laba-b} one thus obtains the new kinetic tensor
\be \label{kinlower}
\cA_{\,\th} \, = \, \cF \, - \, \12 \, \left[\, (\,D - 2\,)\,
\pr^{\,i} \pr^{\, j} \, + \, \pr^{\, k} \pr^{\, (\,i\,} S^{\, j\,)}{}_k \,\right] \, \th_{\, ij}
\, - \, \h^{\, ij} \, \cF\left(\,\theta_{\,ij}\,\right) \, ,
\ee
that is gauge invariant under the unconstrained gauge transformations
\begin{align}
& \d \, \vf \, = \, \pr^{\, i} \, \L_{\, i} \, ,\\
& \d \, \th_{\, ij}  \, = \, \pr^{\, k} \, \L^{(p)}{}_{ijk} \, ,
\end{align}
and where $\cF\,(\,\th_{\,ij}\,)$ is the Fronsdal-Labastida tensor for the $\th_{\,ij}$.
Moreover, $\cA_{\,\th}$ satisfies the Bianchi identities
\be
\pr_{\, i} \, \cA_{\,\th} \, - \, \12 \ \pr^{\, j} \, T_{\, i j} \, \cA_{\,\th} \, = \,
- \, \frac{1}{12} \ \pr^{\,j} \pr^{\,k} \pr^{\,l}
\, T_{(\,ij}\,T_{kl\,)}
\left(\,
\vf - \h^{\, mn} \, \th_{\,mn} \,\right)\, ,
\ee
that have the same form as those emerging in the minimal formulation. As a consequence, a gauge-invariant Lagrangian for fully unconstrained fields obtains performing the shift \eqref{shiftb_ok} in eq.~\eqref{lagmult_bose}, or in eq.~\eqref{lagfield_bose}.

This step does not conclude our analysis: one cannot gauge away the $\th_{\,ij}$ exploiting the new gauge symmetry because their transformations \eqref{gauge_theta} are not algebraic in the gauge parameters. However, a possible way out of this problem was proposed for the fully symmetric case in \cite{dario_massive}. In that framework, a spin-$s$ boson is described in an unconstrained fashion adding a compensator transforming as
\be
\d\, \th_{\,\m_1 \ldots\, \m_{s-2}} \, = \, \pr_{\,(\,\m_1\,} \L_{\,\m_2 \ldots\, \m_{s-2}\,)\,\l}{}^{\,\l} \, .
\ee
In analogy with the $\th_{\,ij}$ of the present setup, $\th_{\,\m_1 \ldots\, \m_{s-2}}$ is not pure gauge but it can be forced to vanish on shell via the addition of a Lagrange multiplier term of the form
\be \label{multsymm}
\c^{\,\m_1 \ldots \m_{s-2}}\, \left(\, \th_{\,\m_1 \ldots\, \m_{s-2}} \, - \, \pr_{\,(\,\m_1\,} \a_{\,\m_2 \ldots\, \m_{s-2}\,)} \,\right)
\ee
The $\a_{\,\m_1 \ldots\, \m_{s-3}}$ field entering eq.~\eqref{multsymm} is nothing but the pure-gauge compensator of the minimal construction, that transforms as
\be
\d \, \a_{\,\m_1 \ldots\, \m_{s-3}} \, = \, \L_{\,\m_2 \ldots\, \m_s\,\l}{}^{\,\l}
\ee
and can be gauged away without spoiling the Labastida gauge symmetry. This finally permits to effectively gauge away the  $\th_{\,\m_1 \ldots\, \m_{s-2}}$ field via the equation of motion of the gauge invariant Lagrange multiplier $\c_{\,\m_1 \ldots \m_{s-2}}$.

The same strategy leads to consider in the present framework the Lagrangians
\begin{align}
\cL \, & = \, \12 \, \bra\, \vf  \, - \, \h^{\, mn} \, \th_{\, mn} \,\comma\,
\sum_{p\,=\,0}^{N} \, \frac{(-1)^{\,p}}{p\,!\,(\,p+1\,)\,!}
 \ \h^{i_1j_1} \!\ldots\, \h^{i_pj_p}\, \cA_{\,\th}^{\,[\,p\,]}{}_{\,i_1j_1, \,\ldots\, ,\,i_pj_p} \,\ket \nn \\
& + \, \fr{1}{8} \, \bra\, \b_{\, i j k l} \comma T_{(\,ij}\,T_{kl\,)}
 \left(\,\vf \, - \, \h^{\, mn} \, \th_{\, mn}\,\right) \ket
\, + \, \bra\, \c_{\, ij} \comma \th_{\, ij} - \, \pr^{\, k} \, \P_{\, i j k}^{\, l m n} \, \a_{\, l m n} (\Phi) \ket\, , \label{lowerlagr}
\end{align}
that can be obtained shifting the $\vf$ fields in eq.~\eqref{lagmult_bose} according to eq.~\eqref{shiftb_ok} and adding a new set of gauge invariant Lagrange multipliers $\c_{\,ij}$. They play the same role as the $\c_{\,\m_1 \ldots \m_{s-2}}$ multiplier introduced in the previous example and relate each $\th_{\,ij}$ compensator to the combination of the pure-gauge compensators $\Psi_{\,i}$ displaying the same gauge transformation. As a consequence, this Lagrangian is invariant under the gauge transformations that we
already defined, and that we collect again here for the sake of clarity:
\begin{alignat}{2}
& \d \, \vf & & = \ \pr^{\, i} \, \L_{\, i} \, , \nn \\
& \d \, \th_{\, ij} & & = \ \pr^{\, k} \, \L^{(p)}{}_{k\,,\,ij} \, , \nn \\
& \d \, \a_{\, ijk} & & = \ \frac{1}{3} \ T_{(\,ij} \, \L_{\,k)} \, , \nn \\
& \d \, \b_{\, ijkl} & & = \ \frac{1}{2} \, \sum_{p\,=\,0}^{N} \, \frac{(-1)^{\,p}}{p\,!\,(\,p+2\,)\,!} \
\pr_{\,(\,i\,}\pr_{\,j\,}\pr_{\,k\,|} \, \h^{i_1j_1} \!\ldots\, \h^{i_pj_p} \, Y_{\{2^p,1\}}
\, T_{i_1j_1} \ldots\, T_{i_pj_p} \, \L^{(t)}{}_{|\,l\,)} \, , \nn \\
& \d \, \c_{\,ij} & & = \ 0 \, .
\end{alignat}
Notice that the $\P_{\, i j k}^{\, l m n}$ appearing in eq.~\eqref{lowerlagr} is the projector
defining the solution of eq.~\eqref{def_lambdap} for $\L^{(p)}{}_{i j
k}$ in the form
\be \label{sollambdap}
\L^{(p)}{}_{i j k }\, = \, \P_{\, i j k}^{\, l m n} \, T_{\, (\,l m } \, \L_{\, n\,)} \, .
\ee
Computing the projector $\P_{\, i j k}^{\, l m n}$ represents the
main technical difficulty of this construction, and indeed we cannot
display an explicit closed form for it in the general
case, although it is rather straightforward, if lengthy, to compute
it explicitly in specific cases of interest. Thus, for instance, in
the fully symmetric case the explicit relation
between $\L^{\, (p)}$ and the full gauge parameter $\L$ is
\be \label{explambdp}
\L^{(p)} \, = \, \sum_{k = 0}^{[\fr{s - 3}{2}]} \, \fr{(-1)^{\, k} \, k\, !}{\prod_{i =
0}^{k}
\, \sum_{j = 0}^{i} \, \left[D \, + \, 2 \, (s \, - 2 \, j \, - \, 3)\right]} \ \h^{\, k} \, \L^{\, [k+1]} \,
,
\ee
where $\L^{\, [k+1]}$ denotes the $(k+1)$-th trace of $\L$ and
$\h^{\, k}$ is a product of $k$ Minkowski metric tensors written
with unit overall normalization and with the minimal number of terms
needed to be totally symmetric.

Finally, the reduction of the equations of motion to the
Fronsdal-Labastida form $\cF = 0$ would result from steps similar to
those needed for the minimal higher-derivative Lagrangians. Indeed, the fields $\th_{\, ij}$ and
$\Phi_{\, i}$ can be gauged away simultaneously, while the Lagrange multipliers
$\b_{\,ijkl}$ and $\c_{\, ij}$ can be expressed in terms of the physical field $\vf$ by the
equations for $\vf$ and $\th_{\,ij}$, respectively. Clearly this last statement holds up to the already discussed symmetries of the Lagrangians \eqref{lowerlagr} under shifts of the Lagrange multipliers $\b_{\,ijkl}$. Furthermore, this construction can be adapted to describe irreducible fields following the same lines sketched for the minimal case in Section \ref{sec:minimal-irreducible}.


\subsubsection{Fermi fields}\label{sec:ordinary-fermi}


The extension of the previous results to mixed-symmetry fermions proceeds exactly
along the same lines: to begin with, in the gauge variation \eqref{gauge-f} of the field
$\psi$,
\be
\d \, \psi \, = \, \pr^{\,i} \, \e_{\,i} \, ,
\ee
one should isolate the contribution to the gauge parameters that is
to vanish in the constrained theory. To this end, let us consider
the decomposition
\be \label{gammadec}
\e_{\,i} \, = \, \e^{\,(t)}{}_{\,i} \, + \, \g^{\, k} \, \e^{\,(g)}{}_{\, i\, ; \, k} \, ,
\ee
where
\begin{align}
& \g_{\,(\, i}\,\e^{\,(t)}{}_{\,j\,)} \, \equiv \, 0 \, , \nn \\
& \g_{\,(\,i}\, \g^{\, k} \, \e^{\,(g)}{}_{\, j\,)\, ; \, k}  \, \equiv \,
\g_{\,(\,i}\,\e_{\,j\,)} \, .
\end{align}
In this fashion, the parameters $\e^{\,(g)}{}_{\, i\, ; \, k}$ embody
precisely the gauge freedom that is absent in the Labastida
formulation, and that we would like to allow here without
introducing new fields whose dimensions differ from that of $\psi$.
Guided by the solution of the problem for Bose fields presented in the previous
section, we are thus led to introduce new compensator fields $\s_{\, k}$ such that
\be
\d \, \s_{\, k} \, = \, \pr^{\, l} \, \e^{\,(g)}{}_{\, l\, ; \, k} \, .
\ee
Notice that we have to give up again pure gauge compensators in order to do that, but this allows the redefinition
\be \label{fermishift}
\psi \, \ra \, \psi \, - \, \g^{\, k} \, \s_{\, k}
\ee
of the field $\psi$, that has the virtue of identifying a combination of fields
possessing the same gauge transformation as the \emph{constrained} Labastida field under
variations involving \emph{unconstrained} gauge parameters. As we already observed,
this is to be contrasted with the Stueckelberg shift \eqref{stueck_fermi}. It is then possible
to verify that, after the substitution \eqref{fermishift}, the Labastida tensor
\eqref{laba-f} takes the form
\be \label{lowerkin}
\cW_{\, \s} \, = \, \cS \, + \, i \, \left[\,(\, D \, - \, 2 \, ) \, \pr^{\, j} \, + \,
2 \, \pr^{\, i} \, S^{\, j}{}_{\, i}\,\right]\, \s_{\, j} \, + \, \g^{\, k} \, \cS\left(\,\s_{\,k}\,\right) \, ,
\ee
where
\be
\cS\left(\,\s_{\,k}\,\right) \, = \, i \, (\, \dsll \, \s_{\, k} \, - \, \pr^{\, j} \, \g_{\, j} \, \s_{\, k} \, )
\ee
is the Fang-Fronsdal-Labastida tensor for the $\s_{\, k}$, and where, let us reiterate,
no higher derivatives appear, precisely because the compensators
$\s_{\, k}$ have the same physical dimension as $\psi$.

The Bianchi identities for $\cW_{\, \s}$,
\begin{align}
& \pr_{\,i}\, \cW_{\, \s} \, - \, \12 \dsl \, \g_{\, i} \, \cW_{\, \s} \, - \,
\12 \ \pr^{\,j} \, T_{ij} \, \cW_{\, \s} \, - \, \frac{1}{6} \ \pr^{\,j} \,
\g_{\,ij} \, \cW_{\, \s} \nn \\
& = \, \frac{i}{6} \ \pr^{\,j} \pr^{\,k} \, T_{(\,ij}\, \gamma_{\, k\,
)}\, ( \, \psi \, - \, \g^{\, l} \, \s_{\, l} \,) \, , \label{bianchi_Ws}
\end{align}
take a particularly nice form, as should be clear recalling eq.~\eqref{fermishift}.

One can thus follow a relatively straightforward route to
a Lagrangian based on the tensor \eqref{lowerkin},
starting from the Lagrangian defined by
eq.~\eqref{lagmult_fermi}, where the constraints on the symmetrized triple $\gamma$-trace of $\psi$ are enforced by
Lagrange multipliers. Since \eqref{lagmult_fermi} is gauge invariant under
\emph{constrained} gauge transformations, by construction it will become gauge invariant
under unconstrained ones if one performs everywhere the substitution \eqref{fermishift},
that leads to
\begin{align}
\cL \, &= \, \frac{1}{2} \ \bra\, \bar{\psi} \,+ \, \bar{\s}_{\, k}\, \g^{\, k}
 \comma\! \sum_{p\,,\,q\,=\,0}^{N} \, \frac{(-1)^{\,\frac{q\,(q+1)}{2}}}{p\,!\,q\,!\,(\,p+q+1\,)\,!} \ \h^{\,p} \, \g^{\,q}\, (\,\g^{\,[\,q\,]}\,
\cW_{\, \s}^{\,[\,p\,]}\,) \,\ket \nn \\
 & + \, \frac{i}{12} \, \bra \bar{\l}_{\,ijk}\comma T_{(\,ij}\, \gamma_{\, k\,
)}\, ( \, \psi \, - \, \g^{\, l} \, \s_{\, l} \,) \ket + \, \textrm{h.c.}\,  .
\label{3lagferconstr}
\end{align}
Finally, in order to guarantee that the new compensators $\s_{\, k}$ be effectively pure
gauge, eq.~\eqref{3lagferconstr} should be supplemented with further
constraints, meant to enforce a linear relation between the $\s_{\, k}$ and the
compensators $\xi_{\,ij}\,(\Psi)$ defined in \eqref{xi}, so that the complete
Lagrangian eventually takes the form
\begin{align}
\cL \, & = \, \frac{1}{2} \ \bra \bar{\psi} \,+ \, \bar{\s}_{\, k}\, \g^{\, k}
\, \comma\! \sum_{p\,,\,q\,=\,0}^{N} \, \frac{(-1)^{\,\frac{q\,(q+1)}{2}}}{p\,!\,q\,!\,(\,p+q+1\,)\,!} \ \h^{\,p} \, \g^{\,q}\, (\,\g^{\,[\,q\,]}\,
\cW_{\, \s}^{\,[\,p\,]}\,) \,\ket \nn \\
& + \, \frac{i}{12} \, \bra \bar{\l}_{\,ijk}\comma T_{(\,ij}\, \gamma_{\, k\,
)}\, ( \, \psi \, - \, \g^{\, l} \, \s_{\, l} \,) \ket \,
+ \, \bra \bar{\chi}_{\, k} \comma \s_{\, k} \, - \, \pr^{\, l} \, \P^{\, i j }_{\, k l } \ \xi_{\,ij}\,(\Psi) \ket \, + \,
\textrm{h.c.}\, . \label{basiclagfer}
\end{align}
$\P^{\, i j }_{\, k l }$ is the projector allowing to express the solution of
eq.~\eqref{gammadec} as
\be \label{fermiproj}
\e^{\,(g)}{}_{\, k\, ; \, l} \, = \,  \P^{\, i j }_{\, k l} \, \g_{\,(\,i}\,\e_{\,j\,)} \, ,
\ee
while the transformations that leave
eq.~\eqref{basiclagfer} invariant are
\begin{align}
&\d \, \psi \, = \, \pr^{\, i} \, \left(\,\e^{\, (t)}{}_{\, i} \, + \, \g^{\, k} \, \e^{\, (g)}{}_{\, i\, ; \, k}\,\right) \, , \nn \\
&\d \, \s_{\, k} \, = \, \pr^{\, l} \, \e^{\,(g)}{}_{\, l\, ; \, k} \, , \nn \\
&\d \, \xi_{\, ij} \, = \, \g_{\,(\,i}\,\e_{\,j\,)} \, , \nn \\
&\d \, \chi_{\, k} \, = \, 0 \, ,
\end{align}
with $\d \, \l_{\,ijk}$ given in \eqref{lagrangegen} after replacing $\e$ with $\e^{\,
(t)}$.

As in the bosonic case, constructing explicitly the projector $\P^{\, i j }_{\, k l}$ represents the
main technical difficulty of this construction, and we are again unable to
display an explicit closed form for it in the general
case. On the other hand, it is still rather straightforward, if lengthy, to compute
it explicitly in specific cases of interest. For instance, in
the fully symmetric case the explicit relation
between $\e^{\, (g)}$ and the full gauge parameter $\e$ is
\be \label{gtrproj}
\e^{\, (g)} \, = \,  \sum_{n=1}^{[\fr{s}{2}]}\, \rho_{n}(D, \, s)\,
\h^{\, n-1}\, \left\{\esl^{\ [n-1]}\, + \, \fr{1}{2 \, n} \, \g \, \e^{\, [n]} \right\}\, ,
\ee
with
\be
\r_{n}(D, \, s)\, = \, (- \, 1)^{\, n + 1} \, \prod_{k = 1}^n \, \fr{1}{D \, + \, 2 \, (s \, - \, k \, - \, 1)} \, .
\ee
Here $s$ is the tensorial rank of $\psi$ and $\e^{\, [n]}$ denotes the $n$-th
trace of $\e$, while $\h^{\, n}$ denotes a combination of products of $n$ Minkowski
metric tensors defined with unit overall normalization and with the minimal number of
terms needed in order to obtain a totally symmetric expression.

Let us close the present section by noticing that this derivation of the Lagrangian \eqref{basiclagfer} makes it also manifest that the equations of motion reduce generically to $\cS=0$, since the fields $\s_{\, i}$ and
$\Psi_{\, i}$ can be gauged away simultaneously, while the Lagrange multipliers
$\l_{\,ijk}$ and $\c_{\, k}$ can be expressed in terms of the physical field $\psi$ by the
equations for $\bar{\psi}$ and $\bar{\s}_{\, k}$, respectively. Clearly this last statement holds up to the already discussed symmetries of the Lagrangians \eqref{basiclagfer} under shifts of the Lagrange multipliers $\l_{\,ijk}$. Again, this formulation can be also adapted to irreducible fields following steps similar to those performed in the minimal case in Section \ref{sec:minimal-irreducible}.


\section{Weyl-like symmetries}\label{chap:weyl}


Inspired by the example of two-dimensional gravity, in Section \ref{sec:lag-eq} we identified a class of mixed-symmetry Bose and Fermi fields with vanishing actions. This is a manifestation of a more general phenomenon, since in low space-time dimensions field equations and Lagrangians can display additional accidental symmetries under Weyl-like shifts of the fields. The classification of these ``pathological'' cases is very interesting, since in most of them the Lagrangian field equations display extra symmetries with respect to the non-Lagrangian ones. Thus, the two sets of field equations are \emph{not} directly equivalent, and in the Lagrangian setup the correct description of the free dynamics involves in a crucial way the extra Weyl-like transformations. The rich structure displayed by two-dimensional gravity also stimulates an effort aimed at classifying the ``pathological'' examples. Even if at the free level we can only display these algebraic similarities, the analogy could well signal some peculiar features of possible non-linear completions of these models.

Following \cite{mixed_bose,mixed_fermi}, in Section \ref{sec:weyl} we shall thus derive the conditions that identify the mixed-symmetry models with additional Weyl-like symmetries, for both Bose and Fermi fields. In Section \ref{sec:classification} we shall then display some classes of solutions of these conditions in the general $N$-family case. The results of this analysis show that the two-column Bose fields and the single-column Fermi fields discussed in Section \ref{sec:lag-eq} are simple examples of a rather rich set of Weyl-invariant mixed-symmetry models. Let us also anticipate that some of these models have non-trivial actions, so that the existence of Weyl-like symmetries and the vanishing of the action do not necessarily go in hands in the mixed-symmetry case as in the fully symmetric setup. Indeed, in Section \ref{sec:classification} we shall exhibit some simple counterexamples to this effect.


\subsection[Reduction of the Lagrangian field equations to the Labastida form]{Reduction of the Lagrangian field equations to the\\ Labastida form}\label{sec:weyl}


In the fully symmetric framework, the minimal local unconstrained Lagrangian field equations of Francia and Sagnotti were shown to be generally equivalent to the Fang-Fronsdal ones in \cite{fs2}. This result was achieved computing successive traces of the equations of motion and using them to express the Lagrange multipliers in terms of the gauge fields. This information then enables one to set zero also the $\cA$ or $\cW$ tensors and their \mbox{($\g$-)}traces that are not directly proportional to the constraints. The other \mbox{($\g$-)}traces are in fact directly annihilated by the field equations of the Lagrange multipliers. The elimination of the compensators via a partial gauge fixing eventually recovers the Fang-Fronsdal form of eqs.~\eqref{fronsdal-b} or \eqref{fang}. As anticipated in Section \ref{sec:lag-eq}, this procedure reduces the Lagrangian field equations to the non-Lagrangian ones in all space-time dimensions and for all spins, with two exceptions: the spin-$2$ and the spin-($3/2$) cases in two space-time dimensions.

As we repeatedly pointed out, the elimination of the compensators via a partial gauge fixing does not create any problem even in the mixed-symmetry framework, but the procedure just delineated would become rather complicated for mixed-symmetry fields. Indeed, eq.~\eqref{traceE} implies that the kinetic tensors and the gauge-invariant completions of the multipliers cannot be simply decoupled as in the fully symmetric setup, while the rather involved structure of the quadratic portion of the Lagrangians complicates their simultaneous treatment. On the other hand, the existence of the symmetries \eqref{gaugeb} and \eqref{lambdalsym} under shifts of the Lagrange multipliers already shows that for generic mixed-symmetry fields some quantities are left undetermined by the Lagrangian field equations. At the same time it suggests that the classification of the symmetries of the equations of motions could be a more efficient way to treat the problem, and in the following we shall proceed along this direction. The logic behind this approach will be exactly the same for both Bose and Fermi fields, but in the fermionic framework the proofs of some statements in Section \ref{sec:weyl-fermi} are still missing due to the harder technical steps that they involve.


\subsubsection{Bose fields}\label{sec:weyl-bose}


At the end of Section \ref{sec:minimal-bose} we have seen that all field equations following from the unconstrained Lagrangians that we presented in eqs.~\eqref{lagmult_bose}, \eqref{lagfield_bose} and \eqref{lagunc_bose} can be cast in the form
\begin{align}
E_{\,\vf} \, & : \ \sum_{p\,=\,0}^{N} \,
\frac{(-1)^{\,p}}{p\,!\,(\,p+1\,)\,!} \ \h^{i_1 j_1}  \ldots \, \h^{i_{\,p} j_{\,p}} \,
\cF^{\,[\,p\,]}{}_{\,i_1 j_1,\,\ldots\,,\,i_{\,p} j_{\,p}}  \, + \, \12 \ \h^{ij}\,\h^{kl}\, \cB_{\,ijkl} \, = \, 0 \, , \label{Efield_bose} \\[5pt]
E_{\,\cB} \, & : \ T_{(\,ij}\,T_{kl\,)}\,\vf \, = \, 0 \quad \Longrightarrow \quad T_{(\,ij}\,T_{kl\,)}\,\cF \, = \, 0 \, , \label{Emult_bose}
\end{align}
possibly after a partial gauge fixing that eliminates the compensators $\Phi_{\,i}$. Moreover, these conditions can also conveniently describe the field equations following from the constrained Lagrangians \eqref{lagconstr_bose}. In that case the second condition holds by definition even off-shell, while a field equation for $\vf$ alone is obtained eliminating the $\cB_{\,ijkl}$ tensors from eq.~\eqref{Efield_bose}. Let us stress that, in this respect, the equivalence between the Lagrangian field equations and the non-Lagrangian Labastida ones
\be \label{nl_bose}
\cF \, = \, 0
\ee
must be verified even at the constrained level, as first pointed out in \cite{laba_lag}. Rather than adding complications, the unconstrained formulation simplifies the treatment of this problem providing the more manageable form of the field equations displayed in eq.~\eqref{Efield_bose}, that allowed the detailed discussion of \cite{mixed_bose}.

To better describe the framework, let us focus for a while on the fully symmetric case. There the field equations \eqref{Efield_bose} and \eqref{Emult_bose} take the simpler form
\begin{align}
E_{\,\vf} \, & : \ \cF_{\,\m_1 \ldots\, \m_s} \, - \, \12 \ \h_{\,(\,\m_1\m_2}\, \cF_{\,\m_3 \ldots\, \m_{s}\,)\,\l}{}^{\,\l} \, + \, \h_{\,(\,\m_1\m_2}\, \h_{\,\m_3\m_4}\, \cB_{\,\m_5 \ldots\, \m_s\,)} \, = \, 0 \, , \label{Esymm1}\\[5pt]
E_{\,\cB} \, & : \ \vf_{\,\m_1 \ldots\, \m_{s-4}\,\l\r}{}^{\,\l\r} \, = \, 0 \quad \Longrightarrow \quad \cF_{\,\m_1 \ldots\, \m_{s-4}\,\l\r}{}^{\,\l\r} \, = \, 0 \, ,
\end{align}
and they imply $\cF_{\,\m_1 \ldots\, \m_s}\, = \, 0$ and $\cB_{\,\m_1 \ldots\, \m_{s-4}}\, = \, 0$ unless $s = 2$ and $D = 2$. Even if the field equations take a universal form in any space-time dimension, an explicit $D$ dependence indeed emerges when computing their traces. One can better understand its role recalling that in the unique pathological case the combination of the kinetic tensor and its trace appearing in eq.~\eqref{Esymm1} becomes traceless. In the mixed-symmetry case a similar behavior is displayed by more complicated models so that, in general, eq.~\eqref{Efield_bose} \emph{does not} reduce directly to the conditions
\be
\cF \, = \, 0 \, , \qquad\qquad \, \cB_{\,ijkl} \, = \, 0 \, .
\ee
The existence of non-trivial solutions of the homogeneous equation \eqref{Efield_bose} calls for the existence of additional symmetries, and let us stress again that we already know an example of this kind. In fact, the symmetry of the Lagrangians under the shift \eqref{gaugeb} of the Lagrange multipliers induces a symmetry of the field equations under shifts
\be \label{shiftB}
\d \, \cB_{\,ijkl} \, = \, \h^{mn}\, Y_{\{5,1\}}\, L_{\,ijkl\,,\,mn}
\ee
even when the $\cB_{\,ijkl}$ are composite objects. For all mixed-symmetry fields carrying a large enough number of space-time indices the field equations thus do not set to zero all trace components of the $\cB_{\,ijkl}$ tensors. On the other hand, those left undetermined do not really enter the linear combinations that are present in the equations of motion or, equivalently, they can be gauged away exploiting the symmetry \eqref{shiftB}. As a result, the leftover components are immaterial insofar as the reduction to the Labastida form \eqref{nl_bose} is concerned. More explicit examples of this statement can be found in \cite{mixed_bose}, where the field equations of some simple models were reduced to the Labastida form dealing with all their traces.

We can now classify the more interesting models where a symmetry involving also the $\cF$ portion of eq.~\eqref{Efield_bose} is present. In these cases some $o\,(D)$ components of the kinetic tensors are not forced to vanish on-shell by eq.~\eqref{Efield_bose}, as the trace of the linearized Ricci tensor is left undetermined by eq.~\eqref{Esymm1} in $D=2$. Therefore, Lagrangian and non-Lagrangian field equations are \emph{not} directly equivalent. In order to proceed toward the identification of the pathological cases, let us remark that in these preliminary considerations we are treating $\cF$ as if it were a generic tensor. This is however \emph{not} true, since it is the specific combination of differential operators acting on the gauge field $\vf$ defined in eq.~\eqref{laba-b}. This fact has a couple of notable implications: first of all in some cases the non-Lagrangian field equations could also display additional symmetries, and in highly pathological models the $\cF$ tensors even vanish identically\footnote{A very simple example, even if highly degenerate, is given by a spin-$1$ field in one dimension, for which the Fronsdal tensor vanishes identically.}. In these examples, rather than discovering a lack of direct equivalence of the different field equations, one can thus only identify a set of symmetries that are common to both setups. Moreover, analyzing the symmetries of the equations of motion we are really interested only in those that can be realized in terms of the fields. This means that we can perform our analysis dealing directly with the expression \eqref{Efield_bose}, but we must consider only the shift symmetries of $\cF$ that can be realized shifting $\vf$. Other possible alternatives would be associated to tensors that cannot be written in the form \eqref{laba-b}, and that are thus of no interest. This information can be codified requesting that the $\d\,\cF$ transformations also preserve the Bianchi identities, that directly follow from eq.~\eqref{laba-b}. When enforcing the field equations \eqref{Emult_bose} these take the form
\be \label{bianchiredb}
\mathscr{B}_i\, : \ \pr_{\,i}\, \cF \, - \, \12 \ \pr^{\,j}\, T_{ij} \, \cF \, = \, 0 \, .
\ee
Furthermore, we must clearly consider only the transformations that preserve also the set of field equations \eqref{Emult_bose}, or equivalently that preserve
\be \label{constrredb}
T_{(\,ij}\,T_{kl\,)}\,\cF \, = \, 0 \, .
\ee

Thus, in the following we shall first classify the conditions that allow non-trivial shifts of $\cF$ preserving the Bianchi identities \eqref{bianchiredb} and the constraints \eqref{constrredb}, and then we shall show that all these transformations also leave invariant the field equation \eqref{Efield_bose}. We shall finally close the circle, showing explicitly that all these $\cF$ shifts can be induced by Weyl-like transformations of the gauge field $\vf$. These thus enable one to complete the reduction of the Lagrangian field equations to the Labastida form via a partial gauge fixing, and thus complete the proof of the equivalence of the two setups in this roundabout way. The conditions that we are going to identify will be then discussed in Section \ref{sec:class-bose}, but we can anticipate that all pathological models that we are going to present ``live'' in backgrounds with dimensions $D < \frac{N-2}{2}$, where $N$ is the number of index families. This means that they do not really propagate the degrees of freedom of a mixed-symmetry representation of the Lorentz group, and actually they do not propagate any local degrees of freedom. At any rate, this is true also for two-dimensional gravity, so that the appearance of Weyl-invariant models in arbitrary space-time dimensions could be of some interest in the future.

Looking at eq.~\eqref{Efield_bose} one can then recognize that, if $\cF$ does not vanish identically, in principle it can admit shift symmetries of the form\footnote{Since we are not considering the cases where $\cF$ is identically zero, we are thus going to identify only the conditions that select the models for which the Lagrangian field equations are not directly equivalent to the non-Lagrangian ones. An exhaustive classification of the symmetries of the actions should also take care of the highly degenerate cases where $\cF$ itself vanishes.}
\be \label{shift1_bose}
\d \, \cF \, = \, \h^{ij}\, \O_{\,ij} \, ,
\ee
and we now want to classify those that also preserve the Bianchi identities. In general under shifts of the form \eqref{shift1_bose} these vary according to\footnote{In this section we shall often introduce a slight abuse of notation, identifying with $\mathscr{B}_i$ and with ``Bianchi identities'' only the left-hand side of eq.~\eqref{bianchiredb}.}
\be \label{bianchi1_bose}
\d \, \mathscr{B}_i \, = \, - \, \12 \ \pr^{\,j} \left[\, (\,D-2\,)\,\O_{\,ij} \, + \, S^{\,k}{}_{(\,i}\,\O_{\,j\,)\,k} \,\right] \, + \, \h^{\,jk}\, \mathscr{B}_i \left(\,\O_{\,jk}\,\right) \, ,
\ee
where the $\mathscr{B}_i\,(\,\O_{\,jk}\,)$ obtain acting with the differential operator $\pr_{\,i} - \12 \, \pr^{\,j}\,T_{ij}$ on the $\O_{\,jk}$ parameters. Thus, if the conditions
\be \label{cond1_bose}
(\,D-2\,)\,\O_{\,ij} \, + \, S^{\,k}{}_{(\,i}\,\O_{\,j\,)\,k} \, = \, 0
\ee
are satisfied by non-trivial parameters $\O_{ij}$ that are themselves subject to the Bianchi identities, the corresponding shift \eqref{shift1_bose} preserves indeed eq.~\eqref{bianchiredb}. Notice that the $S^{\,i}{}_{j}$ operators already encountered in Section \ref{sec:lag-eq} and defined in detail in Appendix \ref{app:MIX} play a crucial role at this stage. Indeed, the conditions \eqref{cond1_bose} are nothing but eigenvalue problems for the $S^{\,i}{}_j$ operators, parametrized by the space-time dimension $D$. This outcome restates in a more precise algebraic language that, while field equations and Lagrangians take a universal form that does not depend on the space-time dimension and on the detailed Lorentz structure of the fields under scrutiny, their traces do not. They depend explicitly on $D$, while we shall see that the behavior of the $S^{\,i}{}_j$ operators is directly related to the Lorentz structure of the tensors on which they act. The eigenvalue problems \eqref{cond1_bose} will be discussed in Section \ref{sec:class-bose}, but let us stress that in general these conditions are only \emph{sufficient} to identify a symmetry of the Bianchi identities. In fact, the contributions that appear in eq.~\eqref{bianchi1_bose} could in principle balance each other. However, one can proceed to look for alternative solutions in an iterative manner, analyzing directly more complicated shifts of $\cF$ that involve more than one naked $\h^{ij}$ tensor.

The most general shift transformation that one can consider is of the form
\be \label{shiftn_bose}
\d \, \cF \, = \, \h^{i_1j_1} \!\ldots\, \h^{i_nj_n}\, \O^{\,(n)}{}_{\,i_1j_1,\, \ldots \,,\,i_nj_n} \, ,
\ee
and varies the Bianchi identities according to
\begin{align}
\d \, \mathscr{B}_k \, = \, & - \, \frac{n}{2}\ \h^{i_1j_1}\ldots\,
\h^{i_{n-1}j_{n-1}}\, \pr^{\,l}\Big[\,(\,D-2\,)\, \O^{\,(n)}{}_{kl\,,\,i_1j_1,\, \ldots \, ,\,
i_{n-1}j_{n-1}} \nn  \\
&+\,\sum_{r\,=\,1}^{n-1}\,\O^{\,(n)}{}_{i_r\,(\,k\,,\,
l\,)\,j_r,\,\ldots\,,\, i_{s\neq r}j_{s \neq r},\, \ldots}\,+\,
S^{\,m}{}_{(\,k}\, \O^{\,(n)}{}_{l\,)\,m\,,\,i_1j_1,\, \ldots \, ,\,
i_{n-1}j_{n-1}} \,\Big] \nn \\
&+\, \h^{i_1j_1}\ldots\,\h^{i_nj_n}\, \mathscr{B}_k\left(\,
\O^{\,(n)}{}_{i_1j_1,\, \ldots \, ,\, i_nj_n} \,\right) \, . \label{bianchin_bose}
\end{align}
Following this recursive approach, in order to identify the models that can display additional Weyl-like symmetries one is thus led to consider all conditions of the form
\begin{align}
& (\,D-2\,)\, \O^{\,(n)}{}_{kl\,,\,i_1j_1,\, \ldots \, ,\,
i_{n-1}j_{n-1}} +\,\sum_{r\,=\,1}^{n-1}\,\O^{\,(n)}{}_{i_r\,(\,k\,,\,
l\,)\,j_r,\,\ldots\,,\, i_{s\neq r}j_{s \neq r},\, \ldots} \nn \\
& +\, S^{\,m}{}_{(\,k}\, \O^{\,(n)}{}_{l\,)\,m\,,\,i_1j_1,\, \ldots \, ,\,
i_{n-1}j_{n-1}} \, = \, 0 \, . \label{condn_bose}
\end{align}
Their solutions depend again on the space-time dimension and on the Lorentz structure of the involved tensors. Notice also that these conditions are all inequivalent, since the one identified by a given value of $n$ cannot be obtained via the redefinition
\be
\O^{\,(n-1)}{}_{\,i_1j_1,\, \ldots \,,\,i_{n-1}j_{n-1}} \to \, \h^{i_nj_n}\, \O^{\,(n)}{}_{\,i_1j_1,\, \ldots \,,\,i_nj_n} \, ,
\ee
or extracting in a similar way $\h^{ij}$ tensors from the parameters characterized by lower values of $n$. This confirms the initial motivation to come across eqs.~\eqref{condn_bose}, and one is thus forced to deal with all these conditions in order to classify Weyl-invariant models\footnote{As in the first step of this analysis, the conditions \eqref{condn_bose} actually select only the cases where Lagrangian and non-Lagrangian field equations are not directly equivalent, while an exhaustive classification of the symmetries of the actions should also identify all vanishing $o\,(D)$ components of the $\cF$ tensors.}. On the other hand, whenever one fixes the number of index families $N$, not all shifts \eqref{shiftn_bose} compatible with the Lorentz structure of the field are expected to play a role, simply because the constraints \eqref{constrredb} imply that all combinations of $N+1$ traces of $\cF$ vanish.

We can now show that the same set of conditions emerges even when one considers the variation of the constraints \eqref{constrredb} under these transformations. Let us start with the simplest example: under the shift \eqref{shift1_bose} the symmetrized double traces of $\cF$ vary as
\be \label{constr1_bose}
T_{(\,ij}\,T_{kl\,)}\, \d\, \cF \, = \, T_{(\,ij\,|} \left[\, (\,D-2\,)\,\O_{\,|\,kl\,)}\, + \, S^{\,m}{}_{|\,k}\, \O_{\,l\,)\,m} \,\right] \, + \, \h^{mn}\, T_{(\,ij}\,T_{kl\,)} \, \O_{\,mn} \, ,
\ee
so that they vanish provided eq.~\eqref{cond1_bose} holds and that the parameters satisfy
\be \label{constrO}
T_{(\,ij}\,T_{kl\,)} \, \O_{\,mn} \, = \, 0 \, .
\ee
These last conditions will not play any role when we shall look for the shift symmetries of the equation of motion for $\vf$, but they will be crucial in the identification of the pathological models for which, let us reiterate, \emph{all} field equations must possess a symmetry. Otherwise, the quantities left undetermined by eq.~\eqref{Efield_bose} will be simply forced to vanish by eq.~\eqref{Emult_bose}.

For generic shifts of the form \eqref{shiftn_bose} it is more convenient to proceed in steps. The traces of the $\cF$ variation indeed read
\begin{align}
T_{ab} \, \d \, \cF \, & = \, n\ \h^{i_1j_1}\!\ldots
\h^{i_{n-1}j_{n-1}} \Big[\,D\ \O^{\,(n)}{}_{ab\,,\,i_1j_1,\, \ldots \, ,\,
i_{n-1}j_{n-1}} \nn  \\
&+\,\sum_{r\,=\,1}^{n-1}\,\O^{\,(n)}{}_{i_r\,(\,a\,,\,
b\,)\,j_r,\,\ldots\,,\, i_{s\neq r}j_{r \neq n},\, \ldots}\,+\,
S^{\,k}{}_{(\,a}\, \O^{\,(n)}{}_{b\,)\,k\,,\,i_1j_1,\, \ldots \, ,\,
i_{n-1}j_{n-1}} \,\Big] \nn \\
&+\, \h^{i_1j_1}\ldots\,\h^{i_nj_n}\, T_{ab}\,
\O^{\,(n)}{}_{i_1j_1,\, \ldots \, ,\, i_nj_n} \, ,
\end{align}
and we can simplify them using the relations \eqref{condn_bose}. In fact, the aim is to check whether the constraints \eqref{constrredb} imply \emph{additional} conditions, so that we can restrict the attention only to the cases where the conditions \eqref{condn_bose} are satisfied. When eqs.~\eqref{condn_bose} hold the traces of $\d\,\cF$ simply reduce to
\be
T_{ab} \, \d \cF = \, \h^{i_1j_1}\!\ldots
\h^{i_{n-1}j_{n-1}} \Big[\, 2\,n\ \O^{\,(n)}{}_{ab\,,\,i_1j_1,\, \ldots \, ,\,
i_{n-1}j_{n-1}} + \, \h^{i_nj_n}\, T_{ab}\,
\O^{\,(n)}{}_{i_1j_1,\, \ldots \, ,\, i_nj_n} \,\Big] \, .
\ee
Computing a further trace and using again the relations \eqref{condn_bose} leads eventually to
\begin{align}
& T_{ab}\,T_{cd}\, \d\, \cF \, = \, 2\,n\,(\,n-1\,)\ \h^{i_1j_1}\!\ldots
\h^{i_{n-2}j_{n-2}} \times \nn \\
& \times \Big[\, 2\,\O^{\,(n)}{}_{ab\,,\,cd\,,\,i_1j_1,\, \ldots \, ,\,
i_{n-2}j_{n-2}} - \, \O^{\,(n)}{}_{a\,(\,c\,,\,d\,)\,b\,,\,i_1j_1,\, \ldots \, ,\,
i_{n-2}j_{n-2}} \,\Big] \nn \\[5pt]
& + \, n\ \h^{i_1j_1}\!\ldots \h^{i_{n-1}j_{n-1}} \Big[\, D\, T_{ab}\, \O^{\,(n)}{}_{cd\,,\,i_1j_1,\, \ldots \, ,\, i_{n-1}j_{n-1}} + \, 2\, T_{cd}\, \O^{\,(n)}{}_{ab\,,\,i_1j_1,\, \ldots \, ,\, i_{n-1}j_{n-1}} \nn \\
& - \, T_{(\,a\,|\,(\,c}\, \O^{\,(n)}{}_{d\,)\,|\,b\,)\,,\,i_1j_1,\, \ldots \, ,\, i_{n-1}j_{n-1}} + \, \sum_{r\,=\,1}^{n-1}\, T_{ab}\, \O^{\,(n)}{}_{i_r\,(\,c\,,\,
d\,)\,j_r,\,\ldots\,,\, i_{s\neq r}j_{r \neq n},\, \ldots} \nn \\
& + \, T_{ab}\, S^{\,k}{}_{(\,c}\, \O^{\,(n)}{}_{d\,)\,k\,,\,i_1j_1,\, \ldots \, ,\,
i_{n-1}j_{n-1}} \,\Big] +\, \h^{i_1j_1} \!\ldots\h^{i_nj_n}\, T_{ab}\,T_{cd}\,
\O^{\,(n)}{}_{i_1j_1,\, \ldots \, ,\, i_nj_n} \, .
\end{align}
After symmetrizing in the four family indices $(\,a,b,c,d\,)$ the first group of terms vanishes while the others combine to give
\begin{align}
T_{(\,ab}\,T_{cd\,)}\, \d\, \cF \, & = \, \frac{n}{2}\ \h^{i_1j_1}\!\ldots
\h^{i_{n-1}j_{n-1}}\, T_{(\,ab\,|} \Big[\,(\,D-2\,)\, \O^{\,(n)}{}_{|\,cd\,)\,,\,i_1j_1,\, \ldots \, ,\,
i_{n-1}j_{n-1}} \nn  \\
&+\,\sum_{r\,=\,1}^{n-1}\,\O^{\,(n)}{}_{i_r\,|\,c\,,\,
d\,)\,j_r,\,\ldots\,,\, i_{s\neq r}j_{r \neq n},\, \ldots}\,+\,
S^{\,k}{}_{|\,c}\, \O^{\,(n)}{}_{d\,)\,k\,,\,i_1j_1,\, \ldots \, ,\,
i_{n-1}j_{n-1}} \,\Big] \nn \\
&+\, \h^{i_1j_1}\ldots\,\h^{i_nj_n}\, T_{(\,ab}\,T_{cd\,)}\,
\O^{\,(n)}{}_{i_1j_1,\, \ldots \, ,\, i_nj_n} \, .
\end{align}
In conclusion, when eqs.~\eqref{condn_bose} hold even the constraints \eqref{constrredb} are preserved by a generic shift of the form \eqref{shiftn_bose} provided the parameters also satisfy the conditions
\be \label{constrn_bose}
T_{(\,ab}\,T_{cd\,)}\, \O^{\,(n)}{}_{i_1j_1,\, \ldots \,,\,i_nj_n} \, = \, 0 \, .
\ee

We can now move on to prove that, as expected, all shifts satisfying the conditions \eqref{condn_bose} also leave invariant the field equation \eqref{Efield_bose}. Let us start again from the simplest example, in order to clarify the strategy that we are going to conform to. The shift \eqref{shift1_bose} varies the Einstein-like tensors \eqref{Ebose} as
\begin{align}
\d \, \cE \, & = \, \sum_{p\,=\,1}^N \, \frac{(-1)^{\,p}}{p\,!\,(\,p+1\,)\,!}\ \h^{i_1j_1}
\ldots\, \h^{i_pj_p}\, [\, Y_{\{2^p\}}\, T_{i_1j_1} \ldots\, T_{i_pj_p} \comma \h^{\,kl} \,] \, \O_{\,kl} \nn \\
& + \, \sum_{p\,=\,0}^N \, \frac{(-1)^{\,p}}{p\,!\,(\,p+1\,)\,!}\
\h^{i_1j_1} \ldots\, \h^{i_pj_p}\, \h^{\,kl}
\left(\, Y_{\{2^p\}}\, T_{i_1j_1} \ldots\, T_{i_pj_p} \,\right) \O_{\,kl} \, . \label{shiftE1_bose}
\end{align}
Taking into account the $\{2^{\,p}\}$ Young projection in the family
indices, the commutator appearing in eq.~\eqref{shiftE1_bose} reduces to
\begin{align}
& [\, Y_{\{2^p\}}\, T_{i_1j_1} \ldots\, T_{i_pj_p} \comma \h^{\,kl} \,]\, \O_{\,kl} \nn \\
& = \, Y_{\{2^p\}}\, \sum_{n\,=\,1}^p \, \prod_{r\,\neq\,n}^p \, T_{i_rj_r} \left\{\, (\,D+p-1\,)\, \O_{\,i_nj_n} + \, S^{\,k}{}_{(\,i_n}\, \O_{\,j_n\,)\,k}  \,\right\} \, , \label{commE}
\end{align}
as can be seen resorting to the techniques repeatedly used to manipulate the traces of the Bianchi identities in Sections \ref{sec:labalag-bose} and \ref{sec:minimal-bose} and to the computational rules collected in Appendix \ref{app:identities}. Once the two-column projection of the second line of
eq.~\eqref{shiftE1_bose} is combined with the first, one is
left with
\begin{align} \label{shift_einstein}
\d \, \cE & = \sum_{p\,=\,1}^N \, \frac{(-1)^{\,p}}{p\,!\,(\,p+1\,)\,!} \ \h^{i_1j_1} \!\ldots
\h^{i_pj_p} \, Y_{\{2^p\}} \!\sum_{n\,=\,1}^p \, \prod_{r\,\neq\,n}^p T_{i_rj_r} \left\{\, (\,D-2\,)\,
\O_{\,i_nj_n} \!+\, S^{\,k}{}_{(\,i_n} \O_{\,j_n\,)\,k} \,\right\} \nn \\
& + \sum_{p\,=\,2}^N \, \frac{(-1)^{\,p}}{(\,p\,!\,)^{\,2}}\
\h^{i_1j_1} \!\ldots \h^{i_pj_p}\, Y_{\{4,2^{p-2}\}}\,
\sum_{n\,=\,1}^p \, \Big(\, Y_{\{2^{p-1}\}}\, \prod_{r\,\neq\,n}\,
T_{i_rj_r} \,\Big)\, \O_{\,i_nj_n} \, .
\end{align}
The arguments used to identify the ansatz for the constrained Lagrangians in Section \ref{sec:labalag-bose} now guarantee that the
$\{4,2^{\,p-2}\}$ projection present in eq.~\eqref{shift_einstein}
can be recast in a form where the family indices carried by a pair
of $\h$ tensors are fully symmetrized. Hence, the terms in
the second line can be compensated by a shift of the $\cB_{ijkl}$,
in sharp contrast with the whole first line, that vanishes if one
tries to symmetrize three or more indices. Therefore, these
terms can not be canceled redefining the $\cB_{ijkl}$ and must vanish identically, as they indeed do if the conditions \eqref{cond1_bose} hold. As in the previous analyses, alternative solutions exist
if the contributions appearing in the first line actually embody at least an
additional $\h$ tensor, and this leads again to consider the other shifts of the form \eqref{shiftn_bose}.

To begin with, one can notice that the conditions \eqref{condn_bose} suffice to eliminate all $S^{\,i}{}_j$ operators that would appear in the variation of the Einstein like-tensors. As a result, when eqs.~\eqref{condn_bose} hold all $\O^{\,(n)}$ parameters that carry a Young projection with more than two columns induce a variation of $\cE$ that can be canceled by suitable $\cB_{\,ijkl}$ shifts. On the other hand, for two-column projected Weyl-parameters the invariance of the field equations under shifts satisfying eqs.~\eqref{condn_bose} is less evident, and it rests on the precise form of these relations rather than on the mere possibility of eliminating all $S^{\,i}{}_j$ operators. To treat this problem in full generality it is convenient to introduce the notation
\be
\cO\,(\,\l\,)^{\,kl}{}_{\,ij} \, = \, \frac{\l}{2} \ \d_{\,i}{}^{(\,k}\,\d^{\,l\,)}{}_j \, + \, S^{\,(\,k}{}_{(\,i}\,\d^{\,l\,)}{}_{j\,)} \, .
\ee
Restricting the attention to two-column projected parameters the conditions \eqref{condn_bose} then take the form
\be \label{condn_cond}
\cO\,(\,D-n-1\,)^{\,ab}{}_{\,ij}\,\O^{\,(n)}{}_{ab\,,\,k_1l_1,\, \ldots\, ,\,k_{n-1}l_{n-1}} \, = \, 0 \, ,
\ee
simply because any symmetrization over four family indices vanishes. Moreover, starting from eq.~\eqref{shift_einstein} one can prove via a recursive argument that a shift of the form \eqref{shiftn_bose} induces the variation
\begin{align}
\d\, \cE \, & = \, \sum_{p\,=\,n}^N \, n!\, k_{\,p}\ \h^{i_1j_1} \!\ldots\, \h^{i_pj_p}\, Y_{\{2^p\}}\!\! \sum_{r_1<\,\ldots\,<\,r_n}^p\, \prod_{s\,\neq\,r_q}^p \, T_{i_sj_s}\, \cO\,(\,D-n-1\,)^{\,k_1l_1}{}_{i_{r_1}j_{r_1}} \times \nn \\
& \times\, \cO\,(\,D-n-2\,)^{\,k_2l_2}{}_{i_{r_2}j_{r_2}}
\ldots\, \cO\,(\,D-2\,n\,)^{\,k_nl_n}{}_{i_{r_n}j_{r_n}}\, \O^{\,(n)}{}_{k_1l_1,\, \ldots\, ,\,k_nl_n} \, , \label{shiftEgen} \end{align}
up to terms that carry Young projections with more than two columns. Indeed, we have already seen that these can be compensated by suitable $\cB_{\,ijkl}$ shifts and thus do not have to vanish identically. All crucial contributions are thus displayed in eq.~\eqref{shiftEgen}, whose recursive proof can be obtained via the redefinitions
\be
\O^{\,(n)}{}_{k_1l_1,\, \ldots\, ,\,k_nl_n} \, \to \, \h^{\,ab}\, \O^{\,(n+1)}{}_{ab\,,\,k_1l_1,\, \ldots\, ,\,k_nl_n} \, ,
\ee
and noticing that they imply
\be \label{Oredef}
\cO\,(\,\l\,)^{\,ab}{}_{\,ij}\,\O^{\,(n)}{}_{ab\,,\,k_1l_1,\, \ldots\, ,\,k_{n-1}l_{n-1}} \to \, \h^{\,ab}\, \cO\,(\,\l-2\,)^{\,cd}{}_{\,ij}\,\O^{\,(n+1)}{}_{ab\,,\,cd\,,\,k_1l_1,\, \ldots\, ,\,k_{n-1}l_{n-1}} \, .
\ee
Therefore, one can move the $\h^{\,ab}$ tensor to the left of the $\cO$ operators using \eqref{Oredef}, and then repeat the steps that led to eq.~\eqref{shift_einstein} to conclude the proof.

The expression appearing in eq.~\eqref{shiftEgen} is not normal ordered, but it is rather convenient since it contains the same operator defining the conditions \eqref{condn_cond} that we identified analyzing the Bianchi identities. $\cO\,(\,D-n-1\,)$ does not appear immediately on the left of the Weyl parameter as in eq.~\eqref{condn_cond}, but one can notice that eqs.~\eqref{condn_cond} also imply
\be \label{comm_shift}
S^{\,m}{}_{(\,i}\,\O^{\,(n)}{}_{j\,)\,m\,,\,kl\,,\, \ldots} \, - \, S^{\,m}{}_{(\,k}\,\O^{\,(n)}{}_{l\,)\,m\,,\,ij\,,\, \ldots} \, = \, 0
\ee
via proper reshuffling of their indices. As a consequence, when eqs.~\eqref{condn_cond} hold the $\cO$ operators commute since
\be
[\, \cO\,(\,\l_1\,)^{\,mn}{}_{ij} \comma \cO\,(\,\l_2\,)^{\,pq}{}_{kl} \,] \, \O_{\,mn\,,\,pq\,,\, \ldots} = \, S^{\,m}{}_{(\,i}\,\O_{\,j\,)\,m\,,\,kl\,,\, \ldots} \, - \, S^{\,m}{}_{(\,k}\,\O_{\,l\,)\,m\,,\,ij\,,\, \ldots}  \, .
\ee
This suffices to show that the equation of motion \eqref{Efield_bose} is left invariant by shift transformations of $\cF$ of the form \eqref{shiftn_bose} provided the parameters satisfy the conditions \eqref{condn_bose}.

In eqs.~\eqref{condn_bose} and \eqref{constrn_bose} we have thus identified the set of conditions that select the ``physical'' shift symmetries of the field equations \eqref{Efield_bose} and \eqref{Emult_bose}, and we can now verify that they indeed follow from suitable Weyl-like shifts of the gauge fields. One can begin by considering a Weyl-like transformation of the form
\be \label{shiftvf1}
\d \, \vf \, = \, \h^{\,ij}\, \Th_{\,ij} \, ,
\ee
that shift the Fronsdal-Labastida tensor \eqref{laba-b} as
\be \label{shiftF}
\d \, \cF \, = \, \12 \ \pr^{\,i}\pr^{\,j} \left[\, (\,D-2\,)\,\Th_{\,ij} \, + \, S^{\,k}{}_{(\,i}\, \Th_{\,j\,)\,k} \,\right]\, + \, \h^{\,ij}\, \cF\left(\,\Th_{\,ij}\,\right) \, .
\ee
The $\cF\,(\,\O_{\,ij}\,)$ are simply Fronsdal-Labastida tensors for the $\O_{\,ij}$ parameters, while the conditions that emerge commuting $\h^{\,ij}$ with the derivatives are again those appearing in eqs.~\eqref{bianchi1_bose}, \eqref{constr1_bose} and \eqref{shift_einstein}. As a consequence, if the equations of motion admit a shift symmetry subject to the conditions \eqref{cond1_bose} and \eqref{constrO}, it is \emph{always} possible to realize it in the form
\be \label{shiftF_ind}
\d \, \cF \, = \, \h^{\,ij}\, \cF(\,\Th_{ij}\,)
\ee
starting from the shift \eqref{shiftvf1} of the gauge field. Indeed, the $S^{\,i}{}_j$ operators commute with the Fronsdal-Labastida ones, so that
\be
(\,D-2\,)\,\cF\left(\,\Th_{\,ij}\,\right) \, + \, S^{\,k}{}_{(\,i\,|}\,\cF\left(\,\Th_{\,|\,j\,)\,k}\,\right) \, = \, 0
\ee
if the $\Th_{\,ij}$ satisfy the conditions \eqref{cond1_bose}. Furthermore, an admissible shift of the field should clearly satisfy
\be
T_{(\,ij}\,T_{kl\,)} \, \d \, \vf \, = \, 0 \, .
\ee
When eq.~\eqref{cond1_bose} holds this implies that
\be \label{constrTH_bose}
T_{(\,ij}\,T_{kl\,)} \, \Th_{\,mn} \, = \, 0 \, ,
\ee
and, as a consequence of eq.~\eqref{compute_constrF}, this leads to
\be
T_{(\,ij}\,T_{kl\,)} \, \cF \left(\,\Th_{\,mn}\,\right) \, = \, 0 \, .
\ee
The conditions \eqref{constrTH_bose} then imply that the $\Th_{\,ij}$ satisfy the same properties as gauge fields, so that the $\cF\,(\,\Th_{\,ij}\,)$ are also subject to the Bianchi identities. In conclusion, the induced shifts \eqref{shiftF_ind} fulfill all conditions we selected when working directly in terms of $\cF$ shifts, and they thus leave the field equations \eqref{Efield_bose} and \eqref{Emult_bose} invariant.

Even in this context the terms appearing in eq.~\eqref{shiftF} could well cancel each other, and one is thus led to consider in a recursive fashion also shifts of the form
\be \label{shiftvfn}
\d \, \vf \, = \, \h^{i_1j_1} \!\ldots\, \h^{i_nj_n}\, \Th^{\,(n)}{}_{\,i_1j_1,\, \ldots \,,\,i_nj_n}
\ee
in order to classify the Weyl-like transformations that leave invariant the traceless part of $\cF$. These affect the Fronsdal-Labastida tensor as
\begin{align}
\d \, \cF \, & = \, \frac{n}{2}\ \h^{i_1j_1}\ldots\,
\h^{i_{n-1}j_{n-1}}\, \pr^{\,l}\Big[\,(\,D-2\,)\, \Th^{\,(n)}{}_{kl\,,\,i_1j_1,\, \ldots \, ,\,
i_{n-1}j_{n-1}} \nn  \\
&+\,\sum_{r\,=\,1}^{n-1}\,\Th^{\,(n)}{}_{i_r\,(\,k\,,\,
l\,)\,j_r,\,\ldots\,,\, i_{s\neq r}j_{s \neq r},\, \ldots}\,+\,
S^{\,m}{}_{(\,k}\, \O^{\,(n)}{}_{l\,)\,m\,,\,i_1j_1,\, \ldots \, ,\,
i_{n-1}j_{n-1}} \,\Big] \nn \\
&+\, \h^{i_1j_1}\ldots\,\h^{i_nj_n}\, \cF\left(\,
\Th^{\,(n)}{}_{i_1j_1,\, \ldots \, ,\, i_nj_n} \,\right) \, ,
\end{align}
so that, whenever the conditions \eqref{condn_bose} hold, the shifts \eqref{shiftn_bose} leaving invariant the field equations can be realized in the form
\be \label{shiftFgen}
\d \, \cF \, = \, \h^{i_1j_1} \!\ldots\, \h^{i_nj_n}\, \cF\left(\,\Th^{\,(n)}{}_{\,i_1j_1,\, \ldots \,,\,i_nj_n}\,\right) \, .
\ee
Notice that in the bosonic framework the analogy between shifts of $\cF$ that preserve the Bianchi identities and shifts of $\vf$ that act on $\cF$ as in eq.~\eqref{shiftFgen} is easily seen. Indeed, barring an irrelevant overall gradient, the terms of $\cF$ that do not contain $\Box\,\vf$ build the same combination of differential operators that enter the Bianchi identities. On the other hand, in the fermionic case this analogy is not so direct and these different viewpoints will be of help in the identification of the relevant Weyl-like transformations.


\subsubsection{Fermi fields}\label{sec:weyl-fermi}


In analogy with the bosonic case, all unconstrained Lagrangians that we presented in eqs.~\eqref{lagmult_fermi}, \eqref{lagfield_fermi} and \eqref{lagunc_fermi} give rise to field equations that can be cast in the form
\begin{align}
E_{\,\bar{\psi}} \, & : \ \sum_{p\,,\,q\,=\,0}^N \, \frac{(-1)^{\,p\,+\,\frac{q\,(q+1)}{2}}}{p\,!\,q\,!\,(\,p+q+1\,)\,!} \ \h^{\,p} \, \g^{\,q}\, (\,\g^{\,[\,q\,]}\,\cS^{\,[\,p\,]}\,)\,  - \, \12 \ \h^{ij}\,\g^{\,k}\, \cY_{\,ijk} \, = \, 0 \, , \label{Efield_fermi} \\[5pt]
E_{\,\bar{\cY}} \, & : \ T_{(\,ij}\,\g_{\,k\,)}\,\psi \, = \, 0 \quad \Longrightarrow \quad T_{(\,ij}\,\g_{\,k\,)}\,\cS \, = \, 0 \, , \label{Emult_fermi}
\end{align}
possibly after a partial gauge fixing that eliminates the compensators $\Psi_{\,i}$. Moreover, we have seen that these conditions can also describe the field equations following from the constrained Lagrangians. In that case the second condition holds by definition even off-shell, while the field equation for $\bar{\psi}$ is obtained eliminating the $\cY_{\,ijk}$ tensors from eq.~\eqref{Efield_fermi}. As for bosons, the equivalence of the Lagrangian field equations with the non-Lagrangian Labastida ones
\be \label{nl_fermi}
\cS \, = \, 0
\ee
must thus be verified even at the constrained level. Rather than adding complications, the unconstrained formulation again simplifies the analysis, providing the more manageable form of the field equations displayed in eq.~\eqref{Efield_fermi}.

In this respect Bose and Fermi fields display a similar behavior, since in general eq.~\eqref{Efield_fermi} \emph{does not} reduce directly to the conditions
\be
\cS \, = \, 0 \, , \qquad\qquad \, \cY_{\,ijk} \, = \, 0 \, .
\ee
The existence of non-trivial solutions of the homogeneous equation \eqref{Efield_fermi} is again equivalent to the existence of additional symmetries, and even in this setting we already know an example of this kind. Indeed, the shift symmetry \eqref{lambdalsym} of the Lagrange multipliers induces a symmetry of the field equations under\footnote{In order to lead to a symmetry of the equations of motion, the $M_{\,ijk,\,lm}$ and $N_{\,ijk;\,lm}$ parameters must be $\{4,1\}$ projected and must be related between each other as in eq.~\eqref{MNrel}.}
\be \label{shiftY}
\delta\, \cY_{\,ijk} \, = \, \eta^{\,lm}\, M_{\,ijk,\,lm} \, +\, \g^{\,lm}\, N_{\,ijk;\,lm}
\ee
even when the $\cY_{\,ijk}$ are composite objects. Therefore, in general the fields equations do not set to zero all $\g$-trace components of the $\cY_{\,ijk}$ tensors. On the other hand, the undetermined components do not really enter the linear combinations that are present in the equations of motion or, equivalently, can be gauged away exploiting the symmetry \eqref{shiftY}. As a consequence they are immaterial insofar as the reduction to the Labastida form \eqref{nl_fermi} is concerned. More explicit examples of this statement can be found in \cite{mixed_fermi}, where the field equations of some simple models were reduced to the Labastida form dealing with all their $\g$-traces.

As in the bosonic case, one can then move on to consider more interesting shift symmetries that involve also the $\cS$ portion of eq.~\eqref{Efield_fermi}. In these cases some $o\,(D)$ components of the kinetic tensors are not forced to vanish on-shell by eq.~\eqref{Efield_fermi}, and thus Lagrangian and non-Lagrangian field equations are \emph{not} directly equivalent. However, even in the fermionic framework one has to take into account that the $\cS$ tensor identifies the combination of differential operators acting on the gauge field $\psi$ defined in eq.~\eqref{laba-f}. As a result, in some cases Lagrangian and non-Lagrangian field equations could share additional symmetries and, most importantly, one has to select only the symmetries of the formal expression \eqref{Efield_fermi} that can be realized in terms of the fields. This information can be again codified in the request for $\d\,\cS$ transformations that also preserve the Bianchi identities, that take the form
\be \label{bianchiredf}
\mathscr{B}_i\, : \ \pr_{\,i}\, \cS \, - \, \12 \dsl \, \g_{\,i}\, \cS \, - \, \12 \ \pr^{\,j}\,T_{ij}\, \cS \, - \, \frac{1}{6} \ \pr^{\,j}\,\g_{\,ij}\,\cS \, = \, 0 \, ,
\ee
when eqs.~\eqref{Emult_fermi} hold. Furthermore, the relevant $\cS$ shifts must also preserve the constraints
\be \label{constrredf}
T_{(\,ij}\,\g_{\,k\,)}\, \cS \, = \, 0 \, .
\ee

In principle one should then analyze these conditions following an iterative procedure similar to that developed for Bose fields. On the other hand, in the present framework this would involve a number of technical difficulties. For this reason, we shall present here a detailed analysis of this kind only for the first level of the iteration. For the successive levels we are going to discuss only the variations of the $\cS$ tensors induced by the Weyl-like transformation of the fields. These identify a set of conditions that should naturally lead to symmetries of the actions, and for some partial checks of this conjecture we refer the reader to \cite{mixed_fermi}. There the two-family case is discussed in full detail, and for $N$-family fields some additional checks are presented in Appendix D. In spite of the sharp differences at the technical level, let us stress that no substantial differences with respect to the bosonic setup are expected to be met in the fermionic one. For instance, in Section \ref{sec:class-fermi} we shall show that the conditions we are going to treat admit solutions very similar to those emerging for Bose fields. In particular, let us anticipate that we shall present a rather rich set of Weyl-invariant models, that all ``live'' in $D < \frac{N-2}{2}$ space-time dimensions and do not propagate any local degrees of freedom.

The discussion of Section \ref{sec:weyl-bose} thus suggests to begin by considering Weyl-like transformations
\be \label{shiftpsi1}
\d \, \psi \, = \, \g^{\,i}\, \Th_{\,i} \, ,
\ee
that give rise to a variation of the Fang-Fronsdal-Labastida tensor \eqref{laba-f} of the form
\be \label{shiftS1}
\d \, \cS \, = \, - \, i \, \pr^{\,k} \left[\, (\,D-2\,)\, \Theta_{\,k} \, + \, 2 \, S^{\,l}{}_{\,k}\, \Theta_{\,l} \,\right] \, - \, \g^{\,k}\, \cS\,(\,\Theta_{\,k}\,) \, .
\ee
As in the bosonic case, the formal expression \eqref{Efield_fermi} in principle can admit shift symmetries of the type\footnote{As in the bosonic case, we are discarding here the cases in which $\cS$ vanishes identically. This analysis is thus aimed to select the cases where Lagrangian and non-Lagrangian field equations are not directly equivalent.}
\be \label{shift1_fermi}
\d \, \cS \, = \, \g^{\,i}\, \O_{\,i} \, ,
\ee
so that one is led to consider the conditions
\be \label{cond1_fermi}
(\,D-2\,)\, \Theta_{\,i} \, + \, 2 \, S^{\,j}{}_{\,i}\, \Theta_{\,j} \, = \, 0
\ee
in order to select fields that are good candidates to display an extra symmetry of the action. In these cases one can indeed identify the $\O_{\,i}$ of eq.~\eqref{shift1_fermi} with $- \,\cS\,(\,\Th_{\,i}\,)$ and, since the $S^{\,i}{}_j$ operators commute with $\cS$, the $\O_{\,i}$ also satisfy eq.~\eqref{cond1_fermi}. In order to fully characterize the ansatz one must also take into account that the Weyl-like gauge transformations must preserve the triple $\g$-trace constraints, so that
\be \label{shiftconstr1_fermi}
T_{(\,ij}\,\g_{\,k\,)}\, \d \, \psi \, = \, T_{(\,ij\,|} \Big[\, (\,D-2\,)\, \Th_{\,|\,k\,)} \, + \, 2 \, S^{\,l}{}_{\,|\,k\,)}\, \Th_{\,l} \,\Big] \, - \, \g^{\,l}\, T_{(\,ij}\, \g_{\,k\,)}\, \Th_{\,l}
\ee
must vanish. When eq.~\eqref{cond1_fermi} holds this only implies the additional conditions
\be \label{constr1_fermi}
T_{(\,ij}\,\g_{\,k\,)}\, \Th_{\,l} \, = \, 0 \, ,
\ee
that in analogy with the bosonic case induce similar conditions on the $\O_{\,i}$ via eq.~\eqref{comp_constrS}.

We can now verify that Weyl-like transformations of the form \eqref{shiftpsi1} satisfying the conditions \eqref{cond1_fermi} and \eqref{constr1_fermi} leave invariant the equations of motion. In the $E_{\,\bar{\cY}}$ case eq.~\eqref{shiftconstr1_fermi} already suffices to conclude that this is true, while in the $E_{\,\bar{\psi}}$ case more work is needed. On the other hand, we are dealing with $\d\,\cS$ transformations that can be expressed as $\cS\,(\,\g^{\,i}\,\Theta_{\,i})$ on account of eq.~\eqref{shiftS1}, and that preserve the constraints \eqref{constrredf} on account of eq.~\eqref{shiftconstr1_fermi}. As a result, they must preserve the Bianchi identities as well, and a direct check indeed gives
\begin{align}
\d \, \mathscr{B}_i \, = \, & - \, \12 \dsl \, \Big[\, (\,D-2\,)\, \O_{\,i} \, + \, 2 \, S^{\,j}{}_{\,i}\, \O_{\,j} \,\Big] \nn \\
& - \, \frac{1}{6} \ \pr^{\,j}\,\g_{\,[\,i\,|} \Big[\, (\,D-2\,)\, \O_{\,|\,j\,]} \, + \, 2 \, S^{\,k}{}_{\,|\,j\,]}\, \O_{\,k} \,\Big]\,  + \, \g^{\,j}\, \mathscr{B}_i\, (\,\O_{\,j}\,) \, . \label{shiftB1}
\end{align}
As usual, the $\mathscr{B}_i\,(\,\O_{\,j}\,)$ denote the Bianchi identities for the parameters $\O_{\,j}$, that clearly vanish when $\O_{\,i} = - \, \cS\,(\,\Th_{\,i})$ and the constraints \eqref{constr1_fermi} are satisfied. Since the Bianchi identities provide the frame for the structure of the Rarita-Schwinger-like tensors $\cE$ of eq.~\eqref{Efermi}, the existence of a symmetry of the field equation \eqref{Efield_fermi} is thus really expected. In order to finally prove this statement one can verify that the transformation \eqref{shiftpsi1} shifts $\cE$ as
\begin{align}
& \d \, \cE \, = \sum_{p\,,\,q\,=\,0}^N\!  k_{\,p \comma q} \ \h^{i_1j_1} \!\!\ldots \h^{i_pj_p}\, \g^{\,k_1 \ldots\, k_q}\, [\,Y_{\{2^p,1^q\}}\, T_{i_1j_1} \!\ldots T_{i_pj_p}\, \g_{\,k_1 \ldots\, k_q} \comma \g^{\,l} \,]_{\,(-1)^{\,q+1}} \, \O_{\,l} \nn \\
& + \!\sum_{p\,,\,q\,=\,0}^N (-1)^{\,q} \, k_{\,p \comma q} \, \h^{i_1j_1} \!\!\ldots
\h^{i_pj_p}\, \g^{\,k_1 \ldots\, k_q}\, \g^{\,l}\, Y_{\{3,2^{p-1},1^q\}} \left(\,
Y_{\{2^p,1^q\}}\, T_{i_1j_1} \!\ldots T_{i_pj_p}\, \g_{\,k_1 \ldots\, k_q} \,\right)
\O_{\,l} \nn \\
& + \!\sum_{p\,,\,q\,=\,0}^N\! \frac{(-1)^{\,q}\,q}{p+1}\, k_{\,p \comma q} \ \h^{i_1j_1} \!\!\ldots \h^{i_{p+1}j_{p+1}} \g^{\,k_1 \ldots\, k_{q-1}} Y_{\{2^{p+1},1^{q-1}\}} \sum_{n\,=\,1}^{p+1}\, \prod_{r\,\neq\,n}^p \, T_{i_rj_r} \g_{\,k_1 \ldots\, k_{q-1}\,(\,i_n}\, \O_{\,j_n\,)} \nn \\
& + \!\sum_{p\,,\,q\,=\,0}^N\! \frac{(-1)^{\,q}}{q+1}\, k_{\,p \comma q} \ \h^{i_1j_1} \!\!\ldots \h^{i_pj_p}\, \g^{\,k_1 \ldots\, k_{q+1}}\, Y_{\{2^p,1^{q+1}\}}\, T_{i_1j_1} \!\ldots T_{i_pj_p}\, \g_{\,[\,k_1 \ldots\, k_q} \, \O_{\,k_{q+1}\,]} \, . \label{varEgen}
\end{align}
When one considers the variation of the full equation of motion \eqref{Efield_fermi}, the second line in eq.~\eqref{varEgen} can be compensated by
$\O$-dependent shifts of the $\cY_{\,ijk}$ tensors, on account of its three-column
projection. On the contrary, the other three lines should vanish directly, as they
indeed do provided they combine to rebuild the analogue of the conditions \eqref{cond1_fermi}. To proceed one can prove that, taking into account the two-column projection in the family indices, the (anti-)commutators reduce to
\begin{align}
& [\, Y_{\{2^p,1^q\}} \, T_{i_1j_1} \ldots\, T_{i_pj_p} \, \g_{\,k_1 \ldots\, k_q} \comma \g^{\,l} \,]_{\,(-1)^{q+1}} \, \O_{\,l} \, = \, Y_{\{2^p,1^q\}} \, \sum_{n\,=\,1}^p
 \, \prod_{r\,\neq\,n}^p \, T_{i_rj_r} \, \g_{\,k_1 \ldots\, k_q\,(\,i_n}\, \O_{\,j_n\,)} \nn \\[2pt]
& + \,Y_{\{2^p,1^q\}} \, T_{i_1j_1} \ldots\, T_{i_pj_p}\, \g_{\,[\,k_1 \ldots\, k_{q-1}\,|} \left\{\, (\,D+p+q-1\,)\, \O_{\,|\,k_q\,]}\, + \, 2\, S^{\,l}{}_{\,|\,k_q\,]}\, \O_{\,l} \,\right\} \, . \label{commgproj}
\end{align}
Substituting this identity in eq.~\eqref{varEgen} and combining the terms containing the
same invariant tensors finally gives
\begin{align}
\d \, \cE \, & = \, \sum_{p\,,\,q\,=\,0}^N \, k_{\,p\,,\,q} \ \h^{i_1j_1} \ldots\, \h^{i_pj_p}\, \g^{\,k_1 \ldots\, k_q}\, \times \nn \\
& \times \, Y_{\{2^p,1^q\}}\, T_{i_1j_1} \ldots\, T_{i_pj_p}\, \g_{\,[\,k_1 \ldots\, k_{q-1}\,|} \left\{\, (\,D-2\,)\, \O_{\,|\,k_q\,]}\, + \, 2\, S^{\,l}{}_{\,|\,k_q\,]}\, \O_{\,l} \,\right\} \, + \, \ldots \, , \label{varEred}
\end{align}
where we have omitted the third line of eq.~\eqref{varEgen}, since we have already stressed
that proper shifts of the $\cY_{\,ijk}$ tensors can cancel it. This concludes the proof that all Weyl-like transformations \eqref{shiftpsi1} satisfying the conditions \eqref{cond1_fermi} and \eqref{constr1_fermi} leave invariant the equations of motion \eqref{Efield_fermi} and \eqref{Emult_fermi}.

On the other hand, in analogy with the bosonic framework, the contributions appearing in eq.~\eqref{shiftS1} could in principle cancel each other, rather than vanish independently. As anticipated, one should then recursively consider all Weyl-like transformations
\be\label{varPSI}
\d \, \psi \, = \, \h^{i_1j_1} \!\ldots \h^{i_nj_n}\, \g^{\,k_1 \ldots\, k_m}\,
 \Theta^{\,(\,n \comma m\,)}{}_{\,i_1j_1, \,\ldots\, ,\,i_nj_n;\,k_1 \ldots\, k_m} \, ,
\ee
together with the conditions that lead to corresponding $\cS$ shifts of the form
\be\label{shiftn_fermi}
\d \, \cS \, = \, \h^{i_1j_1} \!\ldots \h^{i_nj_n}\,\g^{\,k_1 \ldots\, k_m}\,
\cS\left(\, \Theta^{\,(\,n \comma m\,)}{}_{\,i_1j_1, \,\ldots\, ,\,i_nj_n;\,k_1 \ldots\, k_m} \,\right) \, .
\ee
However, the explicit expression
\begin{align}
& \d \, \cS \, = \, - \, i\, m \ \h^{i_1j_1} \ldots\, \h^{i_nj_n}\, \g^{\,k_1 \ldots\, k_{m-1}}\, \pr^{\,l}\, \bigg\{\,
(\,D-m-1\,)\, \Theta^{\,(\,n \comma m\,)}{}_{\,i_1j_1, \,\ldots\, ,\,i_nj_n;\,l\,k_1 \ldots\, k_{m-1}} \nn \\
& + \, \sum_{r\,=\,1}^n \, \Theta^{\,(\,n \comma m\,)}{}_{\,\ldots\, ,\,i_{s\neq r}j_{s\neq r},\, \ldots\, ,\,l\,(\,i_r;\,j_r\,)\,k_1 \ldots\, k_{m-1}}
+ \, 2\, S^{\,q}{}_l \, \Theta^{\,(\,n \comma m\,)}{}_{\, i_1j_1,\,\ldots\,,\,i_nj_n;\, q\,k_1 \ldots\, k_{m-1}} \,\bigg\} \nn \\
& - \, i \, \frac{n}{m+1} \ \h^{i_1j_1} \ldots\, \h^{i_{n-1}j_{n-1}} \, \g^{\,k_1 \ldots\, k_{m+1}}\, \pr^{\,l}\,
\Theta^{\,(\,n \comma m\,)}{}_{\,i_1j_1,\,\ldots\,,\,i_{n-1}j_{n-1},\,l\,[\,k_1;\,k_2 \ldots\, k_{m+1}]} \nn \\
& + \, (-1)^{\,m} \, \h^{i_1j_1} \ldots\, \h^{i_nj_n}\, \g^{\,k_1 \ldots\, k_m}\, \cS \left(\,
 \Theta^{\,(\,n \comma m\,)}{}_{\,i_1j_1, \,\ldots\, ,\,i_nj_n;\,k_1 \ldots\, k_m} \,\right) \label{varS}
\end{align}
shows that parameters carrying the same number of family indices but having different symmetries can enter eq.~\eqref{varS} via terms that are saturated with the same combination of invariant tensors. As a consequence, one can obtain less restrictive conditions allowing Weyl-like transformations of the form
\be \label{shiftpsigen}
\d \, \psi \, = \, \sum_{n\,=\,0}^{\left[\frac{m}{2}\right]} \, \h^{i_1j_1} \!\ldots \h^{i_{n}j_{n}}\,
\g^{\,k_1 \ldots\, k_{m-2n}}\, \Theta^{\,(\,n\,,\,m-2n\,)}{}_{\,i_1j_1, \,\ldots\, ,\,i_nj_n;\,k_1 \ldots\, k_{m-2n}} \, ,
\ee
that affect $\cS$ according to
\be \label{shiftSgen}
\d \, \cS \, = \, \sum_{n\,=\,0}^{\left[\frac{m}{2}\right]} \, \h^{i_1j_1} \!\ldots \h^{i_{n}j_{n}}\,
\g^{\,k_1 \ldots\, k_{m-2n}}\, \cS \left(\, \Theta^{\,(\,n\,,\,m-2n\,)}{}_{\,i_1j_1, \,\ldots\, ,\,i_nj_n;\,k_1 \ldots\, k_{m-2n}} \,\right)\, ,
\ee
provided the chains of $n$-dependent conditions
\begin{align}
& \frac{n+1}{(\,m-2\,n-1\,)\,(\,m-2\,n\,)} \ \Theta^{\,(\,n+1 \comma m-2(n+1)\,)}{}_{\,i_1j_1,\,\ldots\,,\,i_nj_n,\,l\,[\,k_1\,;\,k_2 \ldots\, k_{m-2n-1}\,]} \nn \\[4pt]
& + \, (\,D-m+2\,n-1\,)\ \Theta^{\,(\,n \comma m-2n\,)}{}_{\,i_1j_1,\,\ldots\,,\,i_nj_n; \,l\,k_1 \ldots\, k_{m-2n-1}} \nn \\
& + \, \sum_{r\,=\,1}^n \, \Theta^{\,(\,n \comma m-2n\,)}{}_{\,\ldots\, ,\,i_{s\neq r}j_{s\neq r},\, \ldots\, ,\,l\,(\,i_r;\,j_r\,)\,k_1 \ldots\, k_{m-2n-1}} \nn \\
& + \, 2\, S^{\,q}{}_l \, \Theta^{\,(\,n \comma m-2n\,)}{}_{\, i_1j_1,\,\ldots\,,\,i_nj_n;\, q\,k_1 \ldots\, k_{m-2n-1}} = \, 0 \label{chain}
\end{align}
hold for the allowed values of $m$. It is thus natural to conjecture that the conditions \eqref{chain} select models whose actions display Weyl-like symmetries when these are supplemented by the conditions
\be
T_{(\,ab}\,\g_{\,c\,)}\, \Theta^{\,(\,n \comma m\,)}{}_{\,i_1j_1, \,\ldots\, ,\,i_nj_n;\,k_1 \ldots\, k_m} \, = \, 0 \, .
\ee
Nevertheless, the explicit check that these transformations actually leave invariant the equations of motion \eqref{Efield_fermi} and \eqref{Emult_fermi} is a non-trivial technical task, that will not be treated here. In Section \ref{sec:class-fermi} we shall only show that for fully antisymmetric fermions one of the $m$-dependent relations \eqref{chain} admits solutions in the space-time dimensions where the action vanishes. This is a good check because in these cases the $\cS$ tensor does not vanish, so that the Lagrangian can be a total derivative only if it coincides with a vanishing $o\,(D)$ component of the kinetic tensor. This implies the existence of a shift symmetry of the field equations, that should be recovered by this analysis. Moreover, a detailed treatment of all transformations of the form \eqref{shiftpsigen} with up to four family indices can be found in \cite{mixed_fermi}. They suffice to classify all Weyl-like symmetries emerging in the two-family case, and this selected the bound on the number of family indices. Further steps toward a proof of this conjecture are also presented in the Appendix D of \cite{mixed_fermi}, but the final answer is still missing.


\subsection{Pathological cases}\label{sec:classification}


We can now take a closer look at the conditions identified in Section \ref{sec:weyl} that select the models with additional Weyl-like symmetries. For both Bose and Fermi fields one is facing eigenvalue problems for the $S^{\,i}{}_j$ operators, but whose solutions are to be subjected to
further conditions arising from the constraints \eqref{constrredb} or \eqref{constrredf}. For the reader's convenience, let us thus recall here the definition of the $S^{\,i}{}_j$ operators already introduced in Section \ref{sec:lag-eq}:
\be \label{Sij}
S^{\,i}{}_{j} \, \vf \, \equiv \, \vf_{\ldots \,,\, (\, \m^i_1 \ldots\, \m^i_{s_i} | \,,\, \ldots \,,\, |\, \m^i_{s_i+1} \,) \, \m^j_1
\ldots\, \m^j_{s_j-1} \,,\, \ldots}  \, .
\ee
As shown in Appendix \ref{app:MIX}, these operators generate a $gl(N)$ algebra if $N$ index families are present, and introducing a proper dictionary this enables to treat the eigenvalue problems resorting to rather standard techniques. In particular, we shall see that a substantial progress in the classification of the solutions can be attained only exploiting the properties of the $gl(2)$ subalgebras generated by the operators $S^{\,i}{}_j$, $S^{\,j}{}_i$, $S^{\,i}{}_i$ and $S^{\,j}{}_j$ with \emph{fixed} family indices $i$ and $j$. As a consequence, in this section no summations over the indices are left implicit. In dealing with these $gl(2)$ subalgebras it is convenient to turn to the notation
\be
L^{(ij)}_+ \, = \, S^{\,i}{}_j \ , \quad L^{(ij)}_- \, = \, S^{\,j}{}_i \ , \quad
L^{(ij)}_3 \, = \, \frac{1}{2} \left(\, S^{\,i}{}_i \, - \, S^{\,j}{}_j\,
\right) \, , \qquad i\,<\, j \, ,\label{ang_mom}
\ee
that readily recovers the angular momentum algebra
\be
[\, L^{(ij)}_3 \, , \, L^{(ij)}_\pm \, ]\, = \, \pm \, L^{(ij)}_\pm \ , \qquad [\,L^{(ij)}_+ \,
, \, L^{(ij)}_- \, ]\, = \, 2 \, L^{(ij)}_3 \ . \label{ang_momalg}
\ee
Moreover, looking at eqs.~\eqref{Sij} or \eqref{Sdef}, one can recognize that $S^{\,i}{}_i$ and $S^{\,j}{}_j$ act diagonally on any given tensor or spinor-tensor of rank $(s_1,\ldots,s_N)$. For instance,
\be
S^{\,i}{}_{\,i} \, \vf \, = \, s_i \ \vf\, , \qquad\qquad S^{\,j}{}_{j} \, \psi \, = \, s_j \ \psi \, .
\ee
Hence, also $L^{(ij)}_3$ acts diagonally on arbitrary tensors or spinor-tensors, with eigenvalues
\be \label{m_fixed}
m_{\,ij} \, = \, \frac{s_i \, - \, s_j}{2} \ .
\ee
Notice that the spinor index does not play any role in these relations. Actually, all techniques that we are going to present only deal with the vector indices carried by the fields, so that in the following we shall often explicitly refer only to $\vf$. However, in Section \ref{sec:class-fermi} all results will be transferred verbatim to the fermionic framework.

In order to proceed, it is convenient to split the problems into combinations of
simpler ones, expanding in bases on which the ``total angular momentum'' $\mathbf{L}^2_{\,ij}$
acts diagonally. For instance, a gauge field $\vf$ of rank $(s_1,\ldots,s_N)$ can be decomposed
according to
\be \label{decomposition}
\vf \, = \, \sum_{n\,=\,0}^{s_j} \, \vf^{\,\{s_i+s_j-n,\,n\}} \, ,
\ee
where the various components are characterized by the \emph{same} $L^{(ij)}_3$ eigenvalue, that is fixed by the rank of $\vf$ to be that displayed in eq.~\eqref{m_fixed}. On the other hand, the terms $\vf^{\,\{s_i+s_j-n,\,n\}}$ can be related to null eigenvectors of $L^{(ij)}_{+}$ via the standard descent relations
\be
\vf^{\,\{s_i+s_j-n,\,n\}} \, = \, (L^{(ij)}_{-})^{\,s_j-n}\, \widehat{\vf}^{\,\{s_i+s_j-n,\,n\}} \, , \qquad  L^{(ij)}_{+}\, \widehat{\vf}^{\,\{s_i+s_j-n,\,n\}} \, = \, 0 \, ,
\ee
so that they are all eigenvectors of $\mathbf{L}^2_{\,ij}$, but in general with \emph{different}
eigenvalues $\ell_{\,ij}$. The $\widehat \vf$ that lie at the tips of the chains are \emph{irreducible} when considering the permutation group acting \emph{only} on the $i$-th or $j$-th index families since they satisfy eq.~\eqref{condirr} with fixed $i$ and $j$. Indeed, they
are annihilated by $L^{(ij)}_+$, so that a vanishing result obtains if one tries to extend the
symmetrization beyond the $i$-th family. However, the lower members of the
chains, that are built acting on the $\widehat \vf$
with powers of $L^{(ij)}_-$, are still eigenvectors of the corresponding Young projectors acting on the couple of index families under scrutiny. In fact, they commute with $L^{(ij)}_-$ simply because the permutation group acts irreducibly within individual $Y^{(ij)}$ eigenspaces. The terms appearing in the decomposition \eqref{decomposition} can thus be directly identified as
\be
\vf^{\,\{s_i+s_j-n,\,n\}} \, \equiv \, Y^{(ij)}_{\{s_i+s_j-n,\,n\}}\, \vf \, ,
\ee
where $Y^{(ij)}_{\{s_i+s_j-n,\,n\}}$ projects $\vf$ on the $\{s_i+s_j-n,\,n\}$ irreducible representation of the permutation group acting only on the $i$-th and $j$-th index families.
All in all, the operator
\be
L^{(ij)}_- L^{(ij)}_+ \, = \, \mathbf{L}_{\,ij}^2 - (L^{(ij)}_3)^2 - L^{(ij)}_3 \label{totang}
\ee
is diagonal when acting on this basis, whose members $\vf^{\,\{s_i+s_j-n,\,n\}}$ are characterized by
the \emph{fixed} value \eqref{m_fixed} for $m_{\,ij}$ and by a \emph{range} of values of $\ell_{\,ij}$,
\be
\ell_{\,ij}\,(\,n\,) \, = \, \frac{s_i\,+\,s_j}{2} \, - \, n \, , \qquad\qquad \frac{s_i\,-\,s_j}{2} \, \le \, \ell_{\,ij} \, \le \, \frac{s_i\,+\,s_j}{2} \ ,
\ee
so that
\be \label{numcomp}
L^{(ij)}_- L^{(ij)}_+ \, \vf^{\,\{s_i+s_j-n,\,n\}} \, = \, (\,n-s_i-1\,)\,(\,n-s_j\,)\, \vf^{\,\{s_i+s_j-n,\,n\}} \, .
\ee

Let us conclude by remarking that the decomposition \eqref{decomposition} corresponds to a true decomposition in irreducible representations of the permutation group acting on the whole set of space-time indices only in the two-family case. This reason, together with the other technical simplifications that emerge in the two-family case, allowed the exhaustive treatment of this class of fields, in both the bosonic \cite{mixed_bose} and the fermionic cases \cite{mixed_fermi}. On the other hand, to get a clue on the space of solutions here we shall follow an alternative path already outlined in \cite{mixed_fermi} for Fermi fields. When combined with the techniques described in Section \ref{sec:irreducible}, the information provided by the various $su(2)$ subalgebras \eqref{ang_momalg} indeed suffices to solve in the general $N$-family case various eigenvalue problems. Rather than performing an exhaustive analysis, in the following we shall thus present some general classes of solutions, in order to show the richness of the set of Weyl-invariant models in arbitrary dimensions.


\subsubsection{Bose fields}\label{sec:class-bose}


In most of this review we worked with reducible fields. The presence of more than one family of symmetrized indices is indeed the only feature that shapes the structure of constraints, field equations and Lagrangians. On the other hand, one can directly look for \emph{irreducible} solutions of the eigenvalue problems of Section \ref{sec:weyl-bose}, since one can clearly decompose a reducible field in irreducible components and analyze each of them independently. Furthermore, this approach is suggested by a simple observation. In Section \ref{sec:lag-eq} we have seen that in any space-time dimension irreducible two-column Bose fields with Weyl-like symmetries exist, but we also know that fully symmetric bosons admit Weyl-invariant actions only for $s=2$ and in $D=2$. Since all multi-symmetric fields carry a single-row irreducible component, it is evident that the presence of Weyl-like symmetries is intimately related to the detailed irreducible structure of the fields.

When dealing with irreducible fields, one can also take advantage of the irreducibility conditions of eq.~\eqref{condirr}. In fact, they relate the parameters that enter the shifts of Section \ref{sec:weyl-bose}, and one can use them to simplify the corresponding eigenvalue problems. In order to display this mechanism, let us consider the simplest class of Weyl-like transformations, those of the form \eqref{shift1_bose}. As we have seen, they leave invariant the equation of motion \eqref{Efield_bose} provided the conditions \eqref{cond1_bose} hold. For a field $\vf$ of rank $(s_1,\ldots,s_N)$ these give rise to two groups of equations of the form
\begin{alignat}{2}
& (\,D+2\,s_1-6\,)\,\O_{\,ii} \, + \, 2\, \sum_{k\,=\,1}^{i-1} \, S^{\,k}{}_i\,\O_{\,ik} \, + \, 2 \sum_{k\,=\,i+1}^N  S^{\,k}{}_i\,\O_{\,ik} \, = \, 0 \, , & & \quad 1\,\leq\,i\,\leq\,N \, , \nn  \\[1pt]
& (\,D+s_i+s_j-4\,)\,\O_{\,ij} \, + \, \sum_{k\,=\,1}^{i-1} \, S^{\,k}{}_i\,\O_{\,jk} \, + \, \sum_{k\,\neq\,i}^{j-1} \, S^{\,k}{}_j\,\O_{\,ik} \, + \, S^{\,i}{}_j\, \O_{\,ii} \nn & & \\[-3pt]
& + \sum_{k\,=\,i+1}^{j-1}  S^{\,k}{}_i\,\O_{\,jk}\, + \, S^{\,j}{}_i\, \O_{\,jj} \, + \sum_{k\,=\,j+1}^N  S^{\,k}{}_{(\,i}\,\O_{\,j\,)\,k} \, = \, 0 \, , & & \quad  i\,<\,j \, . \label{sistemone}
\end{alignat}
The $S^{\,i}{}_j$ operators modify the number of space-time indices contained in the various families, and this allows the simultaneous presence of more than one parameter in each condition, even if they have a different rank. On the other hand, in order to preserve the irreducibility conditions \eqref{condirr} that project $\vf$ on its $\{s_1,\ldots,s_N\}$ component, the parameters must also satisfy the relations
\be \label{condirrO}
S^{\,k}{}_l\, \O_{\,ij} \, + \, \d^{\,k}{}_{(\,i}\,\O_{\,j\,)\,l} \, = \, 0 \, , \qquad k \, < \, l \, ,
\ee
that actually leave only one independent parameter for each symmetry type. This can be seen resorting to the procedure outlined in Section \ref{sec:irreducible} in the discussion of the structure of the gauge transformations of irreducible fields. At any rate, let us anticipate that these conditions suffice to ``diagonalize'' eqs.~\eqref{sistemone}, reducing the identification of their solutions to simple algebraic equations involving the space-time dimension and the rank of the fields.

Indeed, one can directly use the relations \eqref{condirrO} to simplify some of the terms in eqs.~\eqref{sistemone} since for proper index choices they read
\begin{alignat}{5}
& S^{\,k}{}_i \, \O_{\,jk} & & = \, - \ \O_{\,ij} \, , & & \qquad k\, < \, i \, , \nn \\
& S^{\,i}{}_j \, \O_{\,ii} & & = \, - \, 2\ \O_{\,ij} \, , & & \qquad i\, < \, j \, . \label{condirrexpl}
\end{alignat}
As a result the conditions \eqref{sistemone} can be cast in the form
\begin{alignat}{3}
& \left[\,D+2\,s_1-2\,(\,i+2\,)\,\right]\,\O_{\,ii} \, + \, 2 \sum_{k\,=\,i+1}^N  S^{\,k}{}_i\,\O_{\,ik} \, = \, 0 \, , & & \quad 1\,\leq\,i\,\leq\,N \, , \nn \\[5pt]
& \left[\,D+s_i+s_j-(\,i+j+3\,)\,\right]\,\O_{\,ij} \, + \, \sum_{k\,=\,i+1}^{j-1}  S^{\,k}{}_i\,\O_{\,jk}\, + \, S^{\,j}{}_i\, \O_{\,jj} \nn \\[-3pt]
& + \, \sum_{k\,=\,j+1}^N  S^{\,k}{}_{(\,i}\,\O_{\,j\,)\,k} \, = \, 0 \, , & & \quad i\,<\,j \, . \label{sistemone2}
\end{alignat}
If all parameters are related by eq.~\eqref{condirrO}, even the remaining terms in the first group of equations can be related to $\O_{\,ii}$, while in the second group of equations all terms can be related to $\O_{\,ij}$. To this end one can act with the $S^{\,k}{}_i$ or $S^{\,k}{}_j$ operators appearing in eq.~\eqref{sistemone2} on some of the consequences of eq.~\eqref{condirrO}, obtaining
\begin{alignat}{5}
& S^{\,k}{}_i \, S^{\,i}{}_k \, \O_{\,ij} & & \, = \, - \ S^{\,k}{}_i\, \O_{\,jk} \, , & & \qquad k\, > \, i \, , \nn \\
& S^{\,k}{}_i \, S^{\,i}{}_k \, \O_{\,ii} & & \, = \, - \, 2\, S^{\,k}{}_i\, \O_{\,ik} \, , & & \qquad k\, > \, i \, , \label{condirrexpl2}
\end{alignat}
and similar conditions with $i$ exchanged with $j$. The operators on the left-hand sides are of the form $L^{(ik)}_- L^{(ik)}_+$ and, as we have already seen, standard ``angular momentum'' techniques imply that they act diagonally on eigenvectors of ${\mathbf L}^2_{\,ik}$. Furthermore, the relations \eqref{condirrO} imply
\be \label{irreducibility}
S^{\,k}{}_{\,l}\, \O_{\,ij} \, = \, 0 \, , \qquad k\,<\,l\,, \quad k\,\neq\,i\,,\,j\, ,
\ee
so that the parameters appearing on the right-hand sides of eqs.~\eqref{condirrexpl2} are actually eigenvectors of ${\mathbf L}^2_{\,ik}$ or can be simply related to highest-weight eigenvectors. Acting with $L^{(ik)}_-$ one then remains inside the same $\ell_{\,ik}$ multiplet. As a result, even the left-hand sides are eigenvectors of ${\mathbf L}^2_{\,ik}$ with the same value of $\ell_{\,ik}$. In conclusion, the $L^{(ik)}_- L^{(ik)}_+$ operators appearing in eqs.~\eqref{condirrexpl2} indeed act diagonally, and their eigenvalues can be easily computed exploiting these considerations. For instance, for the first group of conditions \eqref{condirrexpl2},
\be
\ell_{\,ik}\left(\, \O_{\,ij} \,\right) \, = \, \ell_{\,ik}\left(\, \O_{\,jk} \,\right) \, = \, m_{\,ik}\left(\, \O_{\,jk} \,\right) \, = \, \frac{s_i-s_k+1}{2}\, ,
\ee
while
\be
m_{\,ik}\left(\, \O_{\,ij} \,\right) \, = \, \frac{s_i-s_k-1}{2} \, .
\ee
One can eventually use this information to cast eqs.~\eqref{condirrexpl2} in the form
\begin{alignat}{5}
& (\,s_i-s_k+1\,) \, \O_{\,ij} & & \, = \, - \ S^{\,k}{}_i\, \O_{\,jk} \, , & & \qquad i\, < \, k \, , \ k\, \neq \, j \, , \nn \\
& (\,s_i-s_k+2\,) \, \O_{\,ij} & & \, = \, - \ S^{\,j}{}_i\, \O_{\,jj} \, , & & \qquad i\, < \, j \, , \nn \\
& (\,s_i-s_k+1\,) \, \O_{\,ii} & & \, = \, - \, S^{\,k}{}_i\, \O_{\,ik} \, , & & \qquad i\, < \, k \, . \label{condirrexpl3}
\end{alignat}
When dealing with irreducible fields, the conditions \eqref{sistemone} are thus equivalent to
\begin{alignat}{3}
& \left[\,D-2\,(\,N+2\,)+2\left(\,\sum_{k\,=\,i+1}^{N} s_k\, - \, (\,N-i-1\,)\,s_i\,\right)\,\right]\,\O_{\,ii} \, = \, 0 \, , & & \qquad 1\,\leq\,i\,\leq\,N \, , \nn \\[8pt]
& \left[\,D-2\,(\,N+2\,)\, + \sum_{k\,=\,i+1}^{N} s_k \, -\, (\,N-i-1\,)\,s_i \right. \nn \\[-1pt]
& \left. + \sum_{k\,=\,j+1}^{N} s_k \, -\, (\,N-j-1\,)\,s_j \,\right]\,\O_{\,ij} \, = \, 0 \, ,  & & \qquad i\,<\,j \, , \label{sistemone3}
\end{alignat}
so that they reduce to a system of algebraic equations for the coefficients that multiply the parameters. In conclusion, one can identify Weyl-invariant models simply checking whether this homogeneous system admits solutions for proper values of the space-time dimension $D$ and of the $s_i$ that describe the rank of the field $\vf$. Before dealing with this problem, let us recall that this analysis does not suffice to classify all Weyl-invariant models for two reasons. First of all, the manipulations that we performed to reach eqs.~\eqref{sistemone3} are only valid when all $\O_{\,ij}$ are available, and then only when $s_i \geq 2$ for $1 \leq i \leq N$. The other degenerate cases should be treated separately, and one should also discuss the other eigenvalues problems of the form \eqref{condn_bose}. In the following we shall briefly discuss some simple solutions of these further conditions, but the scrutiny of eqs.~\eqref{sistemone3} already suffices to display a rich set of Weyl-invariant models in arbitrary space-time dimensions.

In the study of eqs.~\eqref{sistemone3}, one cannot forget that the irreducibility conditions \eqref{condirrO} played a crucial role in reaching this rewriting of eqs.~\eqref{sistemone}. Therefore one has to look for solutions that are compatible with them. For instance, the irreducibility of $\vf$ implies that if a parameter $\O_{\,ij}$ vanishes, all other parameters $\O_{\,kl}$ with $k \geq i$ and $l \geq j$ must vanish as well. Indeed, in principle $\O_{\,ij}$ carries all irreducible components that are admitted by the $\O_{\,kl}$ with $k \geq i$ and $l \geq j$, and eqs.~\eqref{condirrO} relate all contributions associated to the same Young tableau. Thus, a Weyl-like transformation with a non-trivial $\O_{\,NN}$ parameter must contain all $\O_{\,ij}$. We can then begin by looking for Weyl-like symmetries of this kind, since the condition for $\O_{\,NN}$ in \eqref{sistemone3} takes the simple form
\be
D \, + \, 2\, s_{N} \, - \, 2\,(\,N+2\,)\, = \, 0 \, ,
\ee
and admits solutions for $D \leq 2\,N$. In these cases the other equations reduce to
\begin{align}
& \left(\,\sum_{k\,=\,i+1}^{N-1} s_k\, - \, (\,N-i-1\,)\,s_i\,\right) \,\O_{\,ii} \, = \, 0 \, , \qquad\qquad\quad \, 1\,\leq\,i\,\leq\,N-2\, , \nn \\[8pt]
& \left(\,\sum_{k\,=\,i+1}^{N} \! s_k \, -\, (\,N-i-1\,)\,s_i \, + \sum_{k\,=\,j+1}^{N} \! s_k \, -\, (\,N-j-1\,)\,s_j \,\right)\,\O_{\,ij} \, = \, 0 \, , \label{sistemone_red}
\end{align}
where the latter hold for all $i < j < N$. These conditions iteratively fix the $s_i$ with $i<N$ to be identical, so that a non-trivial solution of eqs.~\eqref{sistemone} exists in $D = 2\,(\,N-s_N+2\,)$ for all $\{s,\ldots,s,s_N\}$-projected fields with $2 \leq s_N \leq N+1$ . This means that these fields admit a shift transformation of $\cF$ that leaves invariant the equation of motion \eqref{Efield_bose} in the corresponding space-time dimension. The irreducibility conditions \eqref{condirrO} also suffice to fix the detailed structure of this shift, since they can be inverted to relate all parameters to $\O_{\,NN}$ via
\begin{alignat}{5}
& \O_{\,ij} & & = \, \frac{1}{(\,s_i-s_j+1\,)\,(\,s_j-s_N+2\,)}\ S^{\,N}{}_i\, S^{\,N}{}_j\, \O_{\,NN} \, , & & \qquad i \,<\, j \,<\, N \, , \nn \\
& \O_{\,iN} & & = \, - \, \frac{1}{s_i-s_N+2}\ S^{\,N}{}_i \, \O_{\,NN} \, , & & \qquad i \,<\,N \, .
\end{alignat}
The Lorentz structure of $\O_{\,NN}$ can be easily extracted from eqs.~\eqref{condirrO}, since they imply
\be
S^{\,k}{}_l\, \O_{\,NN} \, = \, 0 \, , \qquad 1 \, \leq \,k\, <\, l\, \leq\, N\, ,
\ee
so that $\O_{\,NN}$ is a $\{s,\ldots,s,s_N-2\}$-projected tensor. In the two-family case these results agree with those obtained in the complete classification of the Weyl-like symmetries emerging in that setup \cite{mixed_bose}. The analysis of eq.~\eqref{Efield_bose} indeed selected the candidate Weyl-invariant models gathered in Table~\ref{table1_bose}. The first column collects the models where eq.~\eqref{Efield_bose} is left invariant by the shifts of the form \eqref{shift1_bose} that we are considering, and contains $\{s,2\}$ bosons in $D=4$ and $\{s,3\}$ bosons in $D=2$. The second column then collects the models where eq.~\eqref{Efield_bose} is left invariant by shifts of the form \eqref{shiftn_bose} involving parameters with four family indices.

\begin{table}[htb]
\begin{center}
\begin{tabular}{||c||c|c||c|c||}
\cline{2-5}
\multicolumn{1}{c|| }{} & \multicolumn{2}{|c|| }{$\cF^{\, \pe}$} & \multicolumn{2}{|c|| }{$\cF^{\, \pe \pe}$}  \\
\hline
\hline
$D$  &  $s_1$  &  $s_2$   &  $s_1$  &  $s_2$ \\
\hline
\hline
$2$  &  $2$  &  $0$ &   &  \\
\hline
$2$ &  $2$  &  $1$ &   &     \\
\hline
$2$  &  $3$  &  $1$ &   & \\
\hline
$2$  &  $s$  &  $3$ & & \\
\hline
$3$  &  $2$  &  $1$ &  $2$  &  $2$ \\
\hline
$4$  &  $s$  &  $2$ &   & \\
\hline
\end{tabular}
\end{center}\caption{Two-family irreducible bosons with Weyl-like symmetries of eq.~\eqref{Efield_bose}.} \label{table1_bose}
\end{table}

On the other hand, in order to identify a Weyl-like symmetry of the action one must also check whether the corresponding transformations leave invariant the field equations \eqref{Emult_bose}. As we have seen in eq.~\eqref{constr1_bose}, this means that the $\O_{\,ij}$ parameters must also satisfy the constraints
\be \label{constrO2}
T_{(\,ij}\,T_{kl\,)}\, \O_{\,mn} \, = \, 0 \, .
\ee
In general these conditions would simply reduce the number of independent components of the parameters entering the shift \eqref{shift1_bose}, but in some cases they could even trivialize it. Rather than identifying a true Weyl-like symmetry of the action, the corresponding transformation thus only signals an accidental symmetry of the formal expression \eqref{Efield_bose}. In order to clarify this statement let us consider a $\{4,3\}$ field in $D=2$, that provides the simplest example to this effect. Displaying explicitly the space-time indices, the candidate Weyl-like transformation identified in the previous analysis reads
\begin{align}
\d \, \vf_{\,\m_1 \ldots\, \m_4\,,\,\n_1\n_2\n_3} \, & = \, \frac{1}{3} \ \h_{\,(\,\m_1\m_2\,} \O_{\,\m_3\m_4\,)\,(\,\n_1\n_2\,,\,\n_3\,)} \, - \, \frac{1}{3} \ \h_{\,(\,n_1\,|\,(\,\m_1\,} \O_{\,\m_2\m_3\m_4\,)\,|\,\n_2\,,\,\n_3\,)} \nn \\
& +\,  \h_{\,(\,\n_1\n_2\,|\,} \O_{\,\m_1 \ldots\, \m_4\,,\,|\,\n_3\,)} \, , \label{weylexpl}
\end{align}
with a single independent $\{4,1\}$-projected Weyl parameter. On the other hand, in $D=2$ an irreducible $\{4,3\}$ field actually admits only two independent components. Denoting the two space-time directions as $0$ and $1$, one can then choose them to be $\vf_{\,0000\,,\,111}$ and $\vf_{\,1111\,,\,000}$, and their Weyl-like shifts \eqref{weylexpl} explicitly read
\begin{align}
& \d \, \vf_{\,0000\,,\,111} \, = \, 3 \, \left(\, \O_{\,0000\,,\,1} \, - \, 2\ \O_{\,0011\,,\,1} \,\right) \, , \nn \\
& \d \, \vf_{\,1111\,,\,000} \, = \, - \, 3 \, \left(\, \O_{\,1111\,,\,0} \, - \, 2\ \O_{\,1100\,,\,0} \,\right) \, . \label{shiftexpl}
\end{align}
Furthermore, the conditions \eqref{constrO2} simply reduce to the double trace constraint
\be
\h^{\,\l\r}\, \h^{\,\d\s}\, \O_{\,\l\r\d\s\,;\,\n} \, = \, 0 \, ,
\ee
that in components reads
\be
-\, 2\ \O_{\,1100\,,\,0} \, + \, \O_{\,1111\,,\,0} \, = \, 0 \, , \qquad\qquad
\O_{\,0000\,,\,1} \, -\, 2\ \O_{\,0011\,,\,1} \, = \, 0 \, .
\ee
These are exactly the combinations that appear in eq.~\eqref{shiftexpl} and, as a result, a non-trivial shift of the form \eqref{weylexpl} cannot preserve the constraints in $D=2$. A similar behavior is common to all $\{s,3\}$ models in two space-time dimensions, with the exception of the $\{3,3\}$ case, where the fields are actually unconstrained. With the same strategy one can also analyze the $\{s,2\}$ models in $D=4$, but in these cases the constraints \eqref{constrO2} do not suffice to trivialize the $\vf$ or $\cF$ shifts. Thus, for these fields the Weyl-like transformations we have identified leave the actions invariant. In the two-family case all pathological models can be studied with this direct approach and the reader can find in Table~\ref{table2_bose} the full set of two-family Weyl-invariant models that were identified in \cite{mixed_bose}.

\begin{table}[htb]
\begin{center}
\begin{tabular}{||c||c|c||c|c||}
\cline{2-5}
\multicolumn{1}{c|| }{} & \multicolumn{2}{|c|| }{$\cF^{\, \pe}$} & \multicolumn{2}{|c|| }{$\cF^{\, \pe \pe}$}  \\
\hline
\hline
$D$  &  $s_1$  &  $s_2$   &  $s_1$  &  $s_2$ \\
\hline
\hline
$2$  &  $2$  &  $0$ &   &  \\
\hline
$2$ &  $2$  &  $1$ &   &     \\
\hline
$2$  &  $3$  &  $1$ &   & \\
\hline
$2$  &  $3$  &  $3$ & & \\
\hline
$3$  &  $2$  &  $1$ &  $2$  &  $2$ \\
\hline
$4$  &  $s$  &  $2$ &   & \\
\hline
\end{tabular}\end{center}
\caption{Two-family irreducible bosons with Weyl-invariant actions.} \label{table2_bose}
\end{table}

In conclusion, not all solutions of eqs.~\eqref{sistemone} that we have presented effectively identify models that admit Weyl-like symmetries. On the other hand, one can at least select a subclass which does, and which extends the example of two-column Bose fields of Section \ref{sec:lag-eq}. In fact, a general result of representation theory (see, for instance, the first reference in \cite{group}, \textsection $10$-$6$) implies that, for $O\,(n)$ groups, if the total number of boxes in the first two columns of a tableau exceeds $n$, the corresponding traceless tensor vanishes. Thus, in $D=2$ the $\{4,1\}$-projected parameter of eq.~\eqref{weylexpl} does not carry a traceless component. As a result, the constraint \eqref{constrO2} can and indeed do annihilate all its components. On the other hand, the $\{s\}$-projected parameters of the Weyl transformations of $\{s,2\}$ fields in $D=4$ carry also a traceless component, that as such cannot be affected by the constraints \eqref{constrO2}. This pattern extends to the general $N$-family case, where the $\{s,\ldots,s,s_N-2\}$-projected Weyl parameters emerging for $\{s,\ldots,s,s_N\}$ fields in $D = 2\,(\,N-s_N+2\,)$ in general do not admit a traceless component. The only exception is provided by the class of $\{s,\ldots,s,2\}$-projected fields in $D=2\,N$. Therefore, their Weyl-like parameters contain some components that cannot be affected by the constraints \eqref{constrO2}, and eqs.~\eqref{Emult_bose} can be preserved by non-trivial shifts. We can thus conclude that the presence of Weyl-like symmetries is not confined to unconstrained two-column bosons, and the paradigmatic examples of $\{s,\ldots,s,2\}$-projected fields in $D=2\,N$ display all features of generic higher-spin mixed-symmetry fields.

In this fashion we have shown that the conditions \eqref{cond1_bose} admit a rather rich set of solutions in arbitrary dimensions, and we can now show that all shifts of the form \eqref{shiftn_bose} actually play a role. In fact, all conditions \eqref{condn_bose} admit at least one solution. Rather than proposing a detailed treatment similar to that presented in the previous pages, we shall identify simple solutions for all of them confining the attention to two-column Bose fields. For instance, irreducible bosons of the form $\{2^{\,N}\}$ must display Weyl-like symmetries in $N \leq D \leq 2\,N$ since their actions vanish in these space-time dimensions. Indeed, let us recall that, since $\cF$ does not vanish identically, the formal expression \eqref{Ebose} for the Einstein-like tensor can vanish only if it coincides with a vanishing $o\,(D)$ component of the kinetic tensor. Hence, the other components can be arbitrarily shifted and this is the origin of the Weyl-like symmetries that we have to recover from the conditions \eqref{shift1_bose} or \eqref{shiftn_bose}. In the previous analysis we have identified for this class of fields only Weyl-like symmetries in $D = 2\,N$, and we can now show that in the other space-time dimensions the relevant Weyl-like transformations are of the form \eqref{shiftn_bose}.

Let us begin by noticing that in an algebraic language the Weyl-like transformations are to satisfy
\be
S^{\,k}{}_1\, \O_{\,1k} \, = \, - \ \O_{\,11} \, , \qquad
\textrm{for}\ k\ {\rm fixed\ and} \,k\, \neq\,1 \, ,
\ee
ore more generally
\be
S^{\,k}{}_1\, \O_{\,1k\,,\,22\,,\, \ldots \,,\,pp} \, = \, - \
\O_{\,11\,,\,22\,,\, \ldots \,,\,pp} \, , \qquad
\textrm{for}\ k
\ {\rm fixed\ and} \,k\, > \,p \, ,
\ee
for the higher shifts of eq.~\eqref{shiftn_bose}. These are but simple consequences of the vanishing of all symmetrizations over three space-time indices. As a result, the
$\frac{N(N+1)}{2}$ conditions of eq.~\eqref{sistemone}
effectively reduce to the single equation
\be
(\,D-2\,)\, \O_{\,11} \, + \, 2
\, \sum_{k\,=\,2}^N S^{\,k}{}_{1}\, \O_{\,1k} \, = \, 0  \, ,
\ee
and finally to
\be
(\,D-2\,N\,)\ \O_{\,11} \, = \, 0 \, ,
\ee
that recovers the non-trivial solution for $D = 2N$ that we already identified. In a
similar fashion, one can recognize that all equations of the form
\eqref{condn_bose} are equivalent to the conditions
\be \label{cond_2col}
(\,D-2\,N+p-1\,)\ \O_{\,11\,,\,22\,,\, \ldots \,,\,pp} \, = \, 0 \,
,
\ee
so that the $p$-th trace of a $\{2^{\,N}\}$-projected field
$\vf_{\,\m^1_1\m^1_2;\, \ldots\, ;\,\m^N_{\,1}\m^N_{\,2}}$ is not
determined by the Lagrangian field equation in $D =
2\,N\,-\,p\,+\,1$. We have thus recovered the expected Weyl-like symmetries, but in order to do that we had to consider all available conditions of the form \eqref{condn_bose}. Let us stress that the presence of shift symmetries with more ``naked'' invariant tensors could be expected, since lowering the number of space-time dimensions one has to lower the rank of the Weyl-like parameter in order to allow a traceless component for it. Nevertheless, the fact that these solutions are identified by the conditions \eqref{condn_bose} is not at all obvious, since in principle they could emerge as particular solutions of the conditions \eqref{cond1_bose}.

To summarize, even if a full classification of Weyl-invariant models in arbitrary dimensions is deferred to a future work, we have identified two paradigmatic classes of examples and we would like to briefly comment on their properties. First of all, $\{s^{\,N-1},2\}$ fields in $D = 2\,N$ and $\{2^{\,p},1^q\}$ fields in $D \leq 2\,p+q$ do \emph{not} propagate any local degrees of freedom. This can be seen resorting to the standard theorem that we applied to select the $\{s^{\,N-1},2\}$ cases out of the $\{s^{\,N-1},s_N\}$ class. The same is true even for the other models collected in Tables \ref{table1_bose} and \ref{table2_bose}, that complete the full classification when only two index families are present. It is thus reasonable to expect that we have identified a key feature of the pathological models, and that the Lagrangian field equations \eqref{Efield_bose} are \emph{directly} equivalent to the Labastida ones \eqref{nl_bose} in all space-time dimensions $D > \frac{N-2}{2}$. In fact, above this bound $N$-family fields propagate the degrees of freedom of a proper $N$-row representation of the Lorentz group. Moreover, looking at eq.~\eqref{Teta} or at the form taken by the conditions \eqref{condn_bose}, one can recognize that for large $D$ the $S^{\,i}{}_j$ can be discarded, so that the field equations reduce rather directly to the Labastida form. In addition, performing the counting of the propagating degrees of freedom with the techniques of \cite{siegel_count,laba_lag} one should obtain the correct answer of eq.~\eqref{dim2} independently of the value of $D$. The modifications related to the emergence of new Weyl-like gauge transformations are thus expected to occur only outside of the range of validity of eq.~\eqref{dim2}. This expectation is well sustained by the explicit counting of the degrees of freedom propagated by two-family fields that was performed in \cite{mixed_fermi}, extending to a whole class of mixed-symmetry fields the partial results of \cite{laba_lag}.

On the other hand, aside from this common signature the pathological models can display different properties. In eq.~\eqref{cond_2col} we have indeed seen that two-column fields have vanishing actions in all cases where Weyl-like symmetries exist, but in general this is not necessary by any means. Indeed, one can verify that the action of a $\{3,2\}$ field does not vanish in $D=4$. This result can be also recovered noticing that if an action vanishes in a given space-time dimension it must also vanish in all lower ones. No additional subtleties are involved in this example since $\{3,2\}$ fields are actually unconstrained. Therefore, in the mixed-symmetry case the existence of Weyl-like symmetries does not imply in general that the Lagrangian is a total derivative.


\subsubsection{Fermi fields}\label{sec:class-fermi}


Even when dealing with Fermi fields it is convenient to study the eigenvalue problems of Section \ref{sec:weyl-bose} directly in the irreducible case. As for Bose fields, the irreducibility conditions \eqref{condirr} greatly simplify the search for Weyl-invariant actions. On the other hand, we shall again present only some paradigmatic classes of solutions, deferring to a future work a more detailed analysis. In analogy with the bosonic case, we shall thus discuss in some detail the conditions \eqref{cond1_fermi}, that identify Weyl-like transformations of the form \eqref{shift1_fermi}. For the more involved conditions \eqref{chain}, that identify Weyl-like transformations of the form \eqref{shiftSgen}, we shall only present simple solutions associated to fully antisymmetric fields, just to show that they all play a role, rather than looking for new Weyl-invariant models.

Let us thus begin by considering Weyl-like transformations of the form
\be\label{Weyl1G}
\d \, \psi \, = \, \g^{\,i} \, \Theta_{\,i} \, ,
\ee
for which the irreducibility of $\psi$ translates into the conditions
\be \label{irredO}
\d^{\,i}{}_k \, \Theta_{\, j} \, +\, S^{\,i}{}_j \, \Theta_{\, k} \, = \, 0 \, , \qquad i\, <\, j
\ee
that again leave only one independent $\Theta_{\,i}$ parameter for each symmetry type. These are to be combined with the conditions allowing \eqref{Weyl1G} to define a Weyl symmetry, that were given in eq.~\eqref{cond1_fermi} and that one can rewrite in the convenient form
\be \label{shiftcond1gen2}
(\,D+2\, s_i -4\,)\, \Theta_{\,i} + \, 2\, \sum_{j\,=\,1}^{i-1}\, S^{\,j}{}_{\,i}\, \Theta_{\,j} \,+\, 2 \sum_{j\,=\,i+1}^N S^{\,j}{}_{\,i}\, \Theta_{\,j} \, = \, 0 \, , \qquad 1 \, \leq\, N \, .
\ee
As in the preceding section $s_i$ denotes the number of Lorentz labels in the $i$-th family for the spinor-tensor $\psi$. Eqs.~\eqref{irredO} then reduce \eqref{shiftcond1gen2} to
\be \label{shiftcond1gen3}
\left[\,D+2\, s_i -2\,(\,i+1\,)\,\right]\, \Theta_{\,i}\, - \, 2 \sum_{j\,=\,i+1}^N S^{\,j}{}_{\,i}\, S^{\,i}{}_{\,j}\, \Theta_{\,i} \, = \, 0 \, , \qquad 1 \, \leq\, N \, .
\ee

The sum is absent in the last of these equations, that is simply
\be
\left[\,D+2\, s_N -2\,(\,N+1\,)\,\right]\, \Theta_{\,N} \, = \, 0 \, ,
\ee
and admits non-trivial solutions provided the condition
\be\label{DsN}
D+2\, s_N \, = \, 2\,(\,N+1\,)
\ee
holds. On the other hand, the irreducibility conditions \eqref{irredO} again imply that if $\Th_{\,N}$ does not vanish then all $\Theta_{\,i}$ cannot vanish as well. As a consequence, one has to discuss also the other conditions with $i<N$. To this end one can notice that, in analogy with the bosonic case, the irreducibility conditions \eqref{irredO} ensure that the operators $S^{\,j}{}_{\,i}\, S^{\,i}{}_{\,j}$ act \emph{diagonally} on $\Theta_{\,i}$. In fact, $S^{\,j}{}_{\,i}$ and $S^{\,i}{}_{\,j}$ are the $L_-^{(ij)}$ and $L_+^{(ij)}$ operators for the $su(2)$ subalgebra of $gl(N)$ connecting the two rows $i$ and $j$, and eq.~\eqref{irredO} implies that
\be
L_+^{(ij)}\, \Theta_{\,j} = 0 \ , \qquad \Theta_{\,j} \, = \, - \, L_+^{(ij)}\, \Theta_{\,i} \, .
\ee
As a result, $\Theta_{\,i}$ and $\Theta_{\,j}$ have the same ``total angular momentum'' quantum number
\be
\ell_{\,ij}\,(\, \Theta_{\,i} \,) \, = \, \ell_{\,ij}\,(\, \Theta_{\,j} \,) \, = \, \frac{s_i - s_j+1}{2} \, ,
\ee
that is also the ``magnetic'' quantum number of $\Theta_{\,j}$, that lies at the tip of the $su(2)$ chain and lacks one space-time index of the $j$-th row when compared to the gauge field $\psi$. At the same time, the ``magnetic'' quantum number of $\Theta_{\,i}$, that lacks an index in the $i$-th row compared to $\psi$, is
\be
m_{\,ij}\,(\, \Theta_{\,i} \,) \, = \, \frac{s_i - s_j-1}{2} \, ,
\ee
so that finally
\be
S^{\,j}{}_{\,i}\, S^{\,i}{}_{\,j}\, \Theta_{\,i} \, = \, (\,s_i - s_j+1\,)\, \Theta_{\,i} \, .
\ee
In conclusion, when the condition \eqref{DsN} holds the remaining equations \eqref{shiftcond1gen3} with $i<N$ reduce to the chain of conditions
\be \label{conds1gamma}
\left[\, (\,N-i-1\,) \, s_i - \sum_{j\,=\,i+1}^{N-1} s_j \,\right] \, \Theta_{\,i} \, = \, 0\, , \qquad 1 \,\leq\, i \,\leq\, N-2 \, ,
\ee
while the $(N-1)$-th condition is also identically satisfied on account of eq.~\eqref{DsN}. Since under these conditions all $\Th_{\,i}$ must be non-vanishing, a solution obtains only if \emph{all} the coefficients in \eqref{conds1gamma} vanish, which is the case for $\{s,\ldots,s,s_N\}$ irreducible fields with $s$ arbitrary and $s_{N}\leq N$ in the space-time dimensions determined by eq.~\eqref{DsN}. As for Bose fields, the irreducibility conditions also suffice to identify the detailed form of the Weyl-like transformations because they relate the $\Th_{\,i}$ as
\be \label{irrinv}
\Theta_{\,i} \, = \, - \, \frac{1}{s_i-s_j+1}\ S^{\,j}{}_{\,i}\, \Theta_{\,j} \, , \qquad i \,<\, j \, .
\ee

These results agree with those obtained in the exhaustive classification of the shift symmetries of the equation of motion \eqref{Efield_fermi} performed for two-family fields in \cite{mixed_fermi}. The outcome of that analysis is presented in Table~\ref{table1_fermi}, where the reader can recognize the presence of $\{s,2\}$ fields in $D=2$ and of $\{s,1\}$ fields in $D=4$. They appear in the first column, that collects the fields admitting shift symmetries of the form \eqref{Weyl1G}.  The second column then collects the models that admit shift symmetries of the form
\be
\d \, \psi \, = \, \h^{\,ij}\, \Th_{\,ij} \, + \, \g^{\,ij}\, \widetilde{\Th}_{\,ij} \, ,
\ee
while the third column collects finally the models that admit shift symmetries of the form
\be
\d \, \psi \, = \, \h^{\,ij}\, \g^{\,k}\, \Th_{\,ij\,;\,k} \, .
\ee

\begin{table}[htb]
\begin{center}
\begin{tabular}{||c||c|c||c|c||c|c||}
\cline{2-7}
\multicolumn{1}{c|| }{} & \multicolumn{2}{|c|| }{$\ssl$} & \multicolumn{2}{|c|| }{$\cS^{\, \pe}$} &
\multicolumn{2}{|c||}{$\ssl^{\, \pe}$} \\
\hline
\hline
$D$  & $s_1$ & $s_2$ & $s_1$ & $s_2$ & $s_1$ & $s_2$ \\
\hline
\hline
$2$  &  $1$  &  $0$ & 2 & 1 & $s$ & $3$ \\
\hline
$2$  &  $1$  &  $1$ &   & & & \\
\hline
$2$  &  $s$  &  $2$ &   & & & \\
\hline
$3$  &   &   & 1 & 1 & & \\
\hline
$4$  &  $s$  &  $1$ &   & & & \\
\hline
\end{tabular}
\end{center}
\caption{Two-family irreducible fermions with Weyl-like symmetries of eq.~\eqref{Efield_fermi}.}
\label{table1_fermi}
\end{table}

As in the bosonic case, this first step does not suffice to identify Weyl-like symmetries of the actions: the solutions we identified are also to preserve the Labastida constraints \eqref{constrpsi}. In eq.~\eqref{shiftconstr1_fermi} we have seen that, whenever eqs.~\eqref{shiftcond1gen2} hold, for the Weyl-like shifts that we are considering
this is guaranteed by the additional conditions
\be \label{triplered}
T_{(\,ij}\, \g_{\,k\,)}\, \Theta_{\,l} \, = \, 0 \, .
\ee
On the other hand, the conditions \eqref{irrinv} imply that all $\Theta_{\, l}$
parameters have the same $gl(D)$ structure as the irreducible $\Theta_{N}$, that in this
sense represents the most convenient choice for the single independent quantity left over
in eq.~\eqref{shiftcond1gen2}. Therefore, the transformations of the $\{s,\ldots,s,1 \}$
fields in $D = 2\,N$ rest on a $\{s,\ldots,s \}$ conformal Weyl parameter, that in this
space-time dimension admits a $\g$-traceless component that as such cannot be annihilated by
the triple $\g$-trace constraints \eqref{triplered}. This is consistent with the fact
that for two-family fields one can check explicitly that $\{s,1\}$ fields admit Weyl-like
symmetries of the type \eqref{Weyl1G} compatible with the Labastida constraints. On the
contrary, for higher values of $s_N$ the standard theorem already recalled in Section \ref{sec:class-bose} states that a $\g$-traceless component does not exist for $\{s,\ldots,s,s_N-1\}$-projected parameters in $D =
2\,(\,N-s_N+1\,)$. As a consequence, the previous considerations protecting this type of shift are no
more available and indeed in the $\{s,2\}$ case in $D=2$ one can verify that the shift transformations are trivialized by the constraints \eqref{triplered}. As first recognized in \cite{mixed_fermi}, at two-families a cancelation of this kind also concerns the $\{s,3\}$ case in $D=2$, so that the true Weyl-invariant models are only those displayed in Table~\ref{table2_fermi}. While we do not have a complete argument to this effect, the explicit analysis of the two-family case thus suggests that the triple $\g$-trace conditions could eliminate at least some of the solutions \eqref{DsN} in low enough space-time dimensions. Nevertheless, as in the bosonic case we have at least identified a rather general class of solutions that extends the simple example of fully antisymmetric Fermi fields, that are actually unconstrained fields and thus do not display all features of generic mixed-symmetry fermions.

\begin{table}[htb]
\begin{center}
\begin{tabular}{||c||c|c||c|c||}
\cline{2-5}
\multicolumn{1}{c|| }{} & \multicolumn{2}{|c|| }{$\ssl$} & \multicolumn{2}{|c|| }{$\cS^{\, \pe}$} \\
\hline \hline
$D$  & $s_1$ & $s_2$ & $s_1$ & $s_2$ \\
\hline \hline
$2$  &  $ 1 $  &  $ 0 $ &  &   \\
\hline
$2$  &  $2$  &  $2$ &  &  \\
\hline
$3$  &  &  & 1 & 1 \\
\hline
$4$ &  $s$  &  $1$ &   & \\
\hline
\end{tabular}
\end{center}
\caption{Two-family irreducible fermions with Weyl-invariant actions.}
\label{table2_fermi}
\end{table}

Let us support these statements by displaying the simplest case were the constraints \eqref{triplered} obstruct the promotion of the symmetries of eq.~\eqref{Efield_fermi} to symmetries of the action. It is provided by a $\{3,2\}$ field, whose $E_{\,\bar{\psi}}$ field equation is left invariant by the shift transformation
\be
\d \, \psi_{\,\m_1\m_2\m_3\,,\,\n_1\n_2} \, = \, - \, \12\ \g_{\,(\,\m_1} \Th_{\,\m_2\m_3\,)\,(\,\n_1\,,\,\n_2\,)} \, + \, \g_{\,(\,\n_1\,|} \Th_{\,\m_1\m_2\m_3\,,\,|\,\n_2\,)} \, ,
\ee
that involves a single $\{3,1\}$-projected parameter. Even in this example the irreducibility conditions leave only two independents components in $D=2$. Let us denote the two space-time directions as $0$ and $1$, and let us choose them to be $\psi_{\,000\,,\,11}$ and $\psi_{\,111\,,\,00}$. Their Weyl-like transformations then read
\begin{align}
& \d \, \psi_{\,000\,,\,11} \, = \, 2\ \g_{\,1}\, \Th_{\,000\,,\,1} \, - \, 3\ \g_{\,0}\, \Th_{\,001\,,\,1} \, , \nn \\
& \d \, \psi_{\,111\,,\,00} \, = \, 2\ \g_{\,0}\, \Th_{\,111\,,\,0} \, + \, 3\ \g_{\,1}\, \Th_{\,001\,,\,1} \, , \label{shiftexplf}
\end{align}
where we have chosen to identify the three independent components of the parameter with $\Th_{\,000\,,\,1}$, $\Th_{\,111\,,\,0}$ and $\Th_{\,001\,,\,1}$. Furthermore, in this context the conditions \eqref{triplered} reduce to the constraint
\be
\h^{\,\l\r}\,\g^{\,\s}\, \Th_{\,\l\r\s\,,\,\n} \, = \, 0 \, ,
\ee
that implies the relations
\begin{align}
& \g_{\,1}\, \left(\, \Th_{\,111\,,\,0} \, + \, \frac{1}{3}\, \Th_{\,000\,,\,1} \,\right) \, + \, \g_{\,0}\, \Th_{\,001\,,\,1} \, = \, 0 \, , \nn \\
& \g_{\,0}\, \left(\, \Th_{\,000\,,\,1} \, + \, \frac{1}{3}\, \Th_{\,111\,,\,0} \,\right) \, - \, \g_{\,1}\, \Th_{\,001\,,\,1} \, = \, 0 \, . \label{relex}
\end{align}
Resorting to a standard representations of the two-dimensional $\g$-matrices of the form
\be
\g_{\,0} \, = \, \left(\begin{array}{cc} 0 & 1 \\ -1 & 0 \end{array}\right) \, , \qquad\qquad \g_{\,1} \, = \, \left(\begin{array}{cc} 0 & 1 \\ 1 & 0 \end{array}\right) \, ,
\ee
one can then recognize that the conditions \eqref{relex} imply
\be
\Th_{\,000\,,\,1} =  \left(\!\!\begin{array}{c} f_{1}(t,x) \\ f_{2}(t,x) \end{array}\!\!\right) \, \Rightarrow \ \Th_{\,111\,,\,0} = \, - \left(\!\!\begin{array}{c} f_{1}(t,x) \\ f_{2}(t,x) \end{array}\!\!\right) , \
\Th_{\,001\,,\,1} = \, \frac{2}{3} \left(\!\!\begin{array}{c} -\,f_{1}(t,x) \\ f_{2}(t,x) \end{array}\!\!\right) .
\ee
Finally, substituting this result in eqs.~\eqref{shiftexplf} one can notice that it eliminates the combinations of the parameter components that they contain.

We can now close this section by showing that the conditions \eqref{chain} admit a solution. Indeed, in Section~\ref{sec:lag-eq} we pointed out that the actions for fully antisymmetric Fermi fields vanish manifestly for $N \leq D \leq 2\,N$, and we noticed that the formal expressions for their field equations in terms of $\cS$ must possess Weyl-like symmetries. On the other hand, with the previous analysis we have identified a Weyl-like symmetry for these fields only in $D=2\,N$, and we can now recover the missing symmetries solving the conditions \eqref{chain} for spinor-tensor forms. First of all, in this case the $\h^{ij}$ tensors do not enter Lagrangians and field equations, so that the
chains of conditions \eqref{chain} reduce to the far simpler independent equations
\be \label{eqantisymm}
(\,D-q-1\,)\ \Theta^{\,(\,q\,)}{}_{\,k_1 \ldots\, k_{q-1}l} + 2 \, S^{\,m}{}_l \, \Theta^{\,(\,q\,)}{}_{\,k_1 \ldots\, k_{q-1}m} = \, 0 \, ,
\ee
whose solutions identify Weyl-like shifts of the form
\be \d \, \psi \, = \, \g^{\,k_1 \ldots\, k_q}\, \Theta^{\,(\,q\,)}{}_{\,k_1 \ldots\,
k_q} \, . \ee
Let us also stress that no other conditions are to be imposed on these parameters, because all spinor-tensor forms are \emph{unconstrained}.
Furthermore, for this type of fields $S^{\,m}{}_l\, \psi = 0$ for all $l\neq m$,
since it is clearly impossible to symmetrize their space-time indices. As a result for all $l \neq m$
\be
S^{\,m}{}_l \, \d\, \psi \, = \, \g^{\,k_1 \ldots\, k_q} \left(\, S^{\,m}{}_l\, \Theta^{\,(\,q\,)}{}_{\,k_1 \ldots\, k_q} + \, (-1)^{\,q+1}\, \d^{\,m}{}_{[\,k_1} \Theta^{\,(\,q\,)}{}_{\,k_2 \ldots\, k_q\,]\,l} \,\right) = \, 0 \, ,
\ee
so that eq.~\eqref{eqantisymm} becomes
\be \label{eqantisymmbis}
(\,D-2\,N\!+q-1\,)\ \Theta^{\,(\,q\,)}{}_{\,k_1 \ldots\, k_{q-1}l} \, = \, 0 \, .
\ee
Fully antisymmetric fields can admit a single independent $q$-th $\g$-trace,
that as we now see is left undetermined in $D = 2\,N\!-q+1$,
since it can be shifted by non-vanishing $\Theta^{\,(\,q\,)}$ parameters.
Therefore, as expected, we are recovering the existence of
Weyl-like symmetries for these fields for $N+1 \leq D \leq 2\,N$. Notice, finally,
that the Lagrangians \eqref{antilagf} are total derivatives also for $D = N$, but in these cases the
$\cS$ spinor-tensors vanish as well, as one can see computing explicitly their only
available component. This explains why these special space-time dimensions do not emerge
from the previous discussion.

The final considerations of the previous section can be repeated almost verbatim also in the fermionic case. Even in this framework we cannot present a full classification of the Weyl-invariant models, but we have identified two paradigmatic classes of examples: $\{s^{\,N-1},1\}$ fields in $D=2\,N$ and fully antisymmetric  $\{1^{N}\}$ fields in $N \leq D \leq 2\,N$. In both cases and in all additional examples contained in Tables \ref{table1_fermi} and \ref{table2_fermi} no local degrees of freedom are involved. Furthermore, looking at eq.~\eqref{anticomgamma} and at the structure of the conditions \eqref{chain}, one can realize that for large $D$ the $S^{\,i}{}_j$ operators cannot obstruct the reduction of the Lagrangian field equations to the Labastida form \eqref{nl_fermi}. As a result, it is natural to expect that the lack of propagating degrees of freedom be a key signature of the pathological models, and that for $D > \frac{N-2}{2}$ the two forms of the field equations are directly equivalent. Even if no degrees of freedom are involved, the two classes of examples that we have selected can present rather different behaviors. In fact, the Lagrangians of fully antisymmetric fermions are total derivatives when they admit Weyl-like symmetries but, for instance, this is not the case for $\{3,1\}$ fermions in $D=4$.


\section{Conclusions}\label{chap:conclusions}


This review presents three alternative local and covariant Lagrangian formulations
for fields transforming in arbitrary representations of the Lorentz group in Minkowski
backgrounds of generic dimension. The various options detailed here can be
regarded as natural extensions of the metric formulation of linearized gravity,
since they all reduce to it in the spin-$2$ case. In the metric formalism the graviton
polarizations are described in arbitrary space-time dimensions via a symmetric tensor
$h_{\,\m\n}$ that varies under linearized diffeomorphisms as
\be \d\, h_{\,\m\n} \,=\, \pr_{\,(\,\m}\, \x_{\,\n\,)} \ . \ee
However, moving to generic
representations of the Lorentz group the natural extension of this setup, involving
multi-symmetric tensors $\vf_{\,\m_1 \ldots\, \m_{s_1},\, \n_1 \ldots\, \n_{s_2},
\,\ldots}$ subjected to the gauge transformations
\be
\d \, \vf_{\,\m_1 \ldots\, \m_{s_1},\, \n_1 \ldots\, \n_{s_2}, \,\ldots} =\, \pr_{\,(\,\m_1\,} \L^{(1)}{}_{\!\m_2 \ldots\, \m_s\,)\,,\,\n_1 \ldots\, \n_{s_2},\,\ldots} + \, \pr_{\,(\,\n_1\,|\,} \L^{(2)}{}_{\!\m_1 \ldots\, \m_s\,,\,|\,\n_2 \ldots\, \n_{s_2}\,)\,,\,\ldots} + \ldots
\ee
or their fermionic analogues, does not lead to a \emph{local} description of the dynamics. This is the very origin of the three complementary approaches reviewed here, since a local metric-like Lagrangian formulation can be achieved either constraining fields and gauge parameters or adding auxiliary fields.

The constrained formulation reviewed in Section \ref{chap:constrained} for Bose fields took its final form in the late eighties, with the identification of the Lagrangians of eq.~\eqref{lagconstr_bose} by Labastida \cite{laba_lag}. In the compact notation adopted in this review, it involves multi-symmetric tensors $\vf$ transforming under gauge transformations as
\be
\d\, \vf \, = \, \pr^{\,i}\, \L_{\,i} \, ,
\ee
but fields and gauge parameters are subjected to the constraints
\be \label{finconstrb}
T_{(\,ij}\, T_{kl\,)}\, \vf \, = \, 0 \, , \qquad\qquad T_{(\,ij}\, \L_{\,k\,)}\, = \, 0 \, .
\ee
In the same years Labastida also proposed field equations forcing multi-symmetric spinor-tensors $\psi$ to describe free massless modes \cite{laba_fermi,laba_lag}. To this end, he identified the gauge transformations
\be
\d \, \psi \, = \, \pr^{\,i}\, \e_{\,i}
\ee
and the constraints
\be \label{finconstrf}
T_{(\,ij}\, \g_{\,k\,)}\, \psi \, = \, 0 \, , \qquad\qquad \, \g_{\,(\,i}\, \e_{\,j\,)}\, = \, 0 \, ,
\ee
but did not relate his field equations to an action principle. The constrained
formulation of the dynamics was then completed twenty years later in \cite{mixed_fermi},
with the identification of the Lagrangians \eqref{lagconstr_fermi}. The choice of working
with multi-symmetric (spinor-)tensors, that was adopted also in this review, is merely
conventional. Indeed, in the third reference of \cite{nonlocal_mixed} and in
\cite{mixed_bose,mixed_fermi} it was pointed out how one can easily adapt the formalism
to describe fields carrying also antisymmetric sets of indices.

The constrained formulation involves the minimal field content leading to a \emph{local}
covariant description, but does not link directly to the \emph{non-local}
geometrical description \cite{nonlocal,nonlocal_mixed,fs1,fms1,dario_massive} involving
the curvatures of de Wit and Freedman \cite{dewit_fr} or their mixed-symmetry
generalizations \cite{nonlocal_mixed}. In order to fill this gap, in
\cite{mixed_bose,mixed_fermi} we also completed the alternative program initiated by
Francia and Sagnotti in \cite{fs2}. Its goal is the construction of local Lagrangians
that do not rest upon the constraints \eqref{finconstrb} or \eqref{finconstrf}, via the
introduction of a minimal number of auxiliary fields. As reviewed in Section
\ref{sec:minimal}, the treatment of fully-symmetric fields developed in
\cite{fs2,fs3,fms1} can be extended to arbitrary mixed-symmetry Bose fields introducing a
set of ``composite'' compensators $\a_{\,ijk}(\,\Phi\,)$ and a set of Lagrange
multipliers $\b_{\,ijkl}$. For Fermi fields the same result can be achieved introducing a
set of ``composite'' compensators $\x_{\,ij}(\,\Psi\,)$ and a set of Lagrange multipliers
$\l_{\,ijk}$. The $\a_{\,ijk}$ and the $\x_{\,ij}$ cannot be regarded as
independent fields on account of the linear dependence of the Labastida constraints on
the gauge parameters, and this leads to the introduction of the ``fundamental''
compensators $\Phi_{\,i}$ and $\Psi_{\,i}$. At any rate, the labeling of the auxiliary
fields with the family indices used throughout this review makes it manifest that
their number depends only on the number of index families carried by the fields and not
on their detailed Lorentz structure. Moreover, the ``composite'' compensators and the
Lagrange multipliers share the index structure of the constraints on gauge parameters and
fields, so that this field content is expected to be the minimal one leading to an
unconstrained covariant formulation of the dynamics.

However, the minimal formulation of Section \ref{sec:minimal} contains
higher-derivative terms involving the compensators. These can be gauged away via a
partial gauge fixing that does not spoil the constrained gauge symmetry, and therefore
are not expected to create any problem. Nevertheless, their presence stimulated the
development of the third approach described in this review, and first presented in
\cite{mixed_bose,mixed_fermi} further elaborating the treatment of fully symmetric bosons
presented in \cite{dario_massive}. In fact, in Section \ref{sec:ordinary} it is shown how
it is possible to obtain a truly conventional description of the free theory, without
resorting to higher-derivative terms but still involving a number of auxiliary fields
only depending on the number of index families. Clearly, in principle one can deal even
with alternative setups, adding more auxiliary fields with suitable Stueckelberg-like
gauge symmetries. Various examples are provided by \cite{brst,brst_mixed,buch_tripl}.
Even if at the free level it is not possible to select a ``best'' solution among
all these equivalent formulations, let us stress again that the unconstrained frameworks
reviewed here have a minimal field content among those that turn into the
geometrical non-local formulation after the auxiliary fields are eliminated. On the
contrary, the field content of the constrained formulation is too limited to allow a
direct link with higher-spin geometry.

Aside from describing the construction of field equations and Lagrangians in these three
setups, this review also illustrates some peculiar properties of their structure. The
most striking one is the emergence, in particular low space-time dimensions, of Weyl-like
symmetries of the form
\be
\d \, \vf \, = \, \h^{\,ij}\, \O_{\,ij} \, , \qquad\qquad \d \, \psi \, = \, \g^{\,i}\, \O_{\,i}\, .
\ee
Qualitatively they are allowed by the rather complicated form taken by the Lagrangians,
that contain a number of \mbox{($\g$-)}traces of the gauge fields that grows
proportionally to the number of index families that they carry. A full classification of
these pathological cases, that somehow generalize the example of two-dimensional gravity,
is still missing. Yet, the various classes of examples presented in Section
\ref{chap:weyl} and the complete classification for two-family fields presented in
\cite{mixed_bose,mixed_fermi} vindicate the expectation that Weyl-like symmetries
are confined to space-time dimensions were the corresponding fields do not propagate any
degrees of freedom. At any rate, the analogy with two-dimensional gravity suggests that
these models could present nonetheless a rich structure, that it would be worth
investigating in the future.

Looking for a better characterization and understanding of Weyl-invariant models is only
one of the directions along which the work presented in this review could be further
developed. First of all, it would be interesting to adapt the results reviewed here to
backgrounds of constant curvature. This could provide a way to make contact with the
Vasiliev construction and with its possible mixed-symmetry extensions. As repeatedly
stressed in this review, all string spectra contain mixed-symmetry excitations, and this
line of development could then be crucial in order to understand the interplay between
the Vasiliev setup, its possible generalizations involving mixed-symmetry fields
and String Theory. Furthermore, the completion of the free theory for both Bose and Fermi
fields naturally calls for a study of the properties of supermultiplets involving even
mixed-symmetry fields. This requires the crossing of the $\cN = 8$ border either via the
introduction of higher-spin supermultiplets or even, possibly, via the introduction
of higher-spin supercharges.

The real challenge is eventually the study of the interactions of systems of
mixed-symmetry fields. Progress in this direction is expected to be slower and more
problematic, but the definite hope is that the metric-formulation of the dynamics
presented in this review can provide useful insights to this end.


\section*{Acknowledgments}


This review draws its origin from the author's Ph.D. Thesis, that was defended at Scuola Normale Superiore di Pisa under the supervision of A.~Sagnotti. I wish to express my gratitude to him for the frequent discussions, the collaboration and the suggestions emerged from his careful reading of this manuscript. I am also most grateful to D.~Francia and J.~Mourad for discussions and collaboration, and I would like to thank E.~Sezgin and P.~Sundell for the reading of the manuscript and for their suggestions. Finally, my gratitude goes to the APC-Paris VII, the \'Ecole Polyt\'echnique, the Chalmers University of Technology, and the ``Galileo
Galilei Institute'' of Florence for their kind hospitality while the research project reviewed here was in progress. The present work was supported in part by Scuola Normale Superiore, by INFN, by the MIUR-PRIN contract 2007-5ATT78, by the EU contracts MRTN-CT-2004-503369 and MRTN-CT-2004-512194, by the NATO grant PST.CLG.978785, and by the ERC Advanced Investigator Grant no. 226455 ``Supersymmetry, Quantum Gravity and Gauge Fields'' (SUPERFIELDS).


\newpage

\begin{appendix}


\section{Notation and conventions}\label{app:MIX}


The gauge fields treated in this review are multi-symmetric tensors $\vf_{\,\mu_1 \ldots\, \mu_{s_1},\,\nu_1 \ldots\,
\nu_{s_2}\,,\,\ldots\,}$ or spinor tensors $\psi^{\,\a}{}_{\,\mu_1 \ldots\, \mu_{s_1},\,\nu_1 \ldots\,
\nu_{s_2}\,,\,\ldots\,}$, whose different groups of indices are fully \emph{symmetric} under interchanges of pairs belonging to the same set, but have no prescribed symmetry relating different sets. They are thus reducible $gl(D)$ tensors or spinor-tensors. As a
result, they are perhaps less familiar than Young projected
\mbox{(spinor-)}tensors, but are most convenient for the present discussion
and play a natural role in String Theory. For instance, for bosonic strings or for the five
superstring models these reducible tensors accompany
products of bosonic oscillators of the type $\alpha_{\,-1}^{\,\mu_1
\phantom{\mu_{s_1}}} \hspace{-10pt} \ldots\,
\alpha_{\,-1}^{\,\mu_{s_1}} \, \alpha_{\,-2}^{\,\nu_1
\phantom{\mu_{s_1}}} \hspace{-10pt} \ldots\,
\alpha_{\,-2}^{\,\nu_{s_2}} \ldots\,$, that are only symmetric under
interchanges of pairs of identical oscillators. Let us stress that this is only a conventional choice, and one can work as conveniently in an antisymmetric basis, dealing with multi-forms \cite{multiforms} rather than with multi-symmetric tensors. Both conventions can be used to describe arbitrary representations of the Lorentz group, and in general massive superstring spectra contain spinor-tensors carrying both symmetric and
antisymmetric index sets. At any rate, in this review we only deal with the symmetric convention, while a description of how the various formulations presented here adapt to multi-forms can be found in \cite{mixed_bose,mixed_fermi}. On the other hand, in Sections \ref{sec:irreducible} and \ref{sec:minimal-irreducible} the modifications that one has to take into account when dealing with irreducible \mbox{(spinor-)}tensors are described in some detail.

Turning to the details of the notation, the reader should be aware that the ``mostly plus'' convention for the space-time signature is used throughout this review, as in \cite{mixed_bose,mixed_fermi,review_SIGRAV}. We also resort to the compact notation introduced in these papers, that extends the index-free notation developed for fully symmetric fields in \cite{nonlocal,fs1,st,fs2,fs3,fms1,dario_massive}. It eliminates
all space-time indices from tensor relations, so that the gauge fields $\vf_{\,\mu_1 \ldots\, \mu_{s_1},\,\nu_1 \ldots\,
\nu_{s_2}\,,\,\ldots\,}$ and $\psi^{\,\a}{}_{\,\mu_1 \ldots\, \mu_{s_1},\,\nu_1 \ldots\,
\nu_{s_2}\,,\,\ldots\,}$ are simply denoted by $\vf$ and $\psi$. When needed, in the reducible case their rank is denoted in general by the list of integer numbers $(s_1,s_2,\ldots\,)$, both in the bosonic and in the fermionic case, the only exception being the example of fully symmetric fields. For them we resort to the standard convention, and the single label denoting the spin takes as usual both integer and half-integer values. Aside from these basic fields, Lagrangians and field equations involve gradients, divergences and traces of $\vf$ or $\psi$, as
well as $\g$-traces of $\psi$ in the fermionic case. Furthermore, the Lagrangians also contain Minkowski metric tensors related to one (or two) of the previous index sets and $\g$-matrices. As a result, \emph{``family
indices"} are often needed in order to specify the sets to which some
tensor indices belong. As in \cite{mixed_bose,mixed_fermi,review_SIGRAV}, these family indices
are here denoted by small-case Latin letters, and the Einstein
convention for summing over pairs of them is used throughout. It
actually proves helpful to be slightly more precise: \emph{upper}
family indices are thus reserved for operators, like a gradient,
which \emph{add} space-time indices, while \emph{lower} family
indices are used for operators, like a divergence, which
\emph{remove} them. For instance gradients, divergences and traces
of a field $\vf$ are denoted concisely by $\pr^{\, i} \, \vf$,
$\pr_{\, i}\, \vf$ and $T_{ij} \, \vf$. This shorthand notation
suffices to identify the detailed meaning of these symbols, so that
\begin{eqnarray}
\pr^{\, i} \, \vf & \equiv & \pr_{\,(\,\m^i_1|} \, \vf_{\,\ldots \,,\, | \, \m^i_2 \, \ldots \, \m^i_{s_i+1} ) \,,\, \ldots} \ , \nonumber \\
\pr_{\, i}\, \vf & \equiv & \pr^{\,\l} \, \vf_{\,\ldots \,,\, \l \, \m^i_1 \, \ldots \, \m^i_{s_i-1} \,,\, \ldots} \ , \nonumber \\
T_{ij} \, \vf & \equiv & \vf^{\phantom{\,\ldots \,,\,} \l}{}_{\hspace{-19pt} \ldots \,,\,\phantom{\l} \ \m^i_1 \, \ldots \, \m^i_{s_i-1} \,,\, \ldots \,,\, \l \,
\m^j_1 \ldots \, \m^j_{s_j-1} \,,\, \ldots} \ . \label{operations}
\end{eqnarray}
The couple of parentheses that appears in the first line of eq.~\eqref{operations} denotes a symmetrization of the indices it encloses. Let us stress that in this review the symmetrizations are \emph{not} of unit strength, but involve nonetheless the minimum possible number of terms. Thus, for instance,
$T_{(\,ij}\, T_{kl\,)}$ here stands for $T_{ij}\,T_{kl}+T_{ik}\,T_{jl}+T_{il}\,T_{jk}$. Moreover, we also use
square brackets to denote antisymmetrizations.

An explicit description of the remaining operations similar to that in eq.~\eqref{operations} can be found in the main body of this review, where they are introduced for the first time. On the other hand, \emph{all computational rules follow from the algebra of these operators}, and we shall thus momentarily resort to an oscillator realization that it is convenient to derive it. To this end, in the present symmetric convention one can introduce commuting vectors $u^{\,i\,\m}$ to cast generic gauge fields in the form
\begin{alignat}{2}
& \vf & & \equiv \, \frac{1}{s_{1}\,! \,\ldots\, s_{N}\,!} \ u^{\,1\,\m_1} \ldots\, u^{\,1\,\m_{s_1}} \, u^{\,2\,\n_1} \ldots\, u^{\,2\,\n_{s_2}} \ldots \ \vf_{\,\m_1 \ldots\, \m_{s_1},\,\n_1 \ldots\, \n_{s_2}\,,\, \ldots} \, , \nn \\
& \psi & & \equiv \, \frac{1}{s_{1}\,! \,\ldots\, s_{N}\,!} \ u^{\,1\,\m_1} \ldots\, u^{\,1\,\m_{s_1}} \, u^{\,2\,\n_1} \ldots\, u^{\,2\,\n_{s_2}} \ldots \ \psi_{\,\m_1 \ldots\, \m_{s_1},\,\n_1 \ldots\, \n_{s_2}\,,\, \ldots} \, .
\end{alignat}
Divergences, gradients, $\g$-traces and ``$\g$-matrices'' can thus be described via
\begin{alignat}{2}
& \pr_{\,i} \, \equiv \, \pr^{\,\m} \, \frac{\pr}{\pr \, u^{\,i\,\m}} \, , \qquad\qquad & & \pr^{\,i} \, \equiv \, \pr_{\,\m} \, u^{\,i\,\m} \, , \nn \\
& \g_{\,i} \, \equiv \, \g^{\,\m} \, \frac{\pr}{\pr \, u^{\,i\,\m}} \, , \qquad\qquad & & \g^{\,i} \, \equiv \, \g_{\,\m} \, u^{\,i\,\m} \, , \label{op_comm}
\end{alignat}
while traces and ``metric tensors'' can be introduced either via the commutation relations of the family $\g$-matrices
\be
\left\{\, \g_{\,i} \comma \g_{\,j} \,\right\} \, = \, 2\ T_{ij} \, , \qquad\qquad \left\{\, \g^{\,i} \comma \g^{\,j} \,\right\} \, = \, 4\ \h^{\,ij} \, ,
\ee
or directly as
\be \label{T_eta}
T_{ij}\, \equiv \, \h^{\,\m\n}\, \frac{\pr}{\pr \, u^{\,i\,\m}}\, \frac{\pr}{\pr \, u^{\,j\,\n}}\, , \qquad\qquad \h^{\,ij} \, \equiv \, \12 \ \h_{\,\m\n} \, u^{\,i\,\m}\, u^{\,j\,\n}\, .
\ee
Interestingly, the algebra of the $T_{ij}$ and $\h^{\,ij}$ operators or, equivalently, that of the $\g_{\,i}$ and $\g^{\,i}$ operators close only introducing the new operators
\be \label{Sdef}
S^{\,i}{}_j \, \equiv \, \h_{\,\m}{}^{\n}\, u^{\,i\,\m}\, \frac{\pr}{\pr \, u^{\,j\,\n}} \, .
\ee
For $i = j$ these act as number operators and count the indices in the $i$-th family,
\be \label{diagonalS}
S^{\,i}{}_{\,i} \, \vf \, = \, s_i \ \vf\, , \qquad\qquad S^{\,i}{}_{\,i} \, \psi \, = \, s_i \ \psi \, ,
\ee
while for $i \neq j$ their net effect is to displace indices from the $i$-th to the $j$-th index families. Using the natural convention
\be
\left[\, \frac{\pr}{\pr \, u^{\,i\,\m}} \comma u^{\,j\,\n} \,\right] \, = \, \h_{\,\m}{}^{\,\n} \, \d_{\,i}{}^{\,j} \, ,
\ee
one can finally compute the algebra of the various operators used in this review. Focusing for a while on those that do not involve $\g$-matrices it reads
\begin{align}
&[ \, \pr_{\,i} \comma \pr^{\,j} \, ] \ = \ \Box \, \d_{\,i}{}^{\,j} \, , \\[0pt]
&[\, T_{ij} \comma \pr^{\,k} \,] \ = \ \pr_{\,(\,i}\,\d_{\,j\,)}{}^k \, , \\[0pt]
&[\, \pr_{\,k} \comma \h^{\,ij} \,] \ = \ \12 \ \d_{\,k}{}^{\,(\,i} \, \pr^{\,i\,)} \, , \\[0pt]
&[\, T_{ij} \comma \h^{\,kl} \,] \ = \ \frac{D}{2} \ \d_{\,i}{}^{\,(\,k\,}\,\d_{\,j}{}^{\,l\,)} \, + \,
\12 \left( \, \d_{\,i}{}^{\,(\,k}S^{\,l\,)}{}_j \, + \, \d_{\,j}{}^{\,(\,k}S^{\,l\,)}{}_i \, \right) ,  \label{Teta} \\[0pt]
&[\, S^{\,i}{}_j \comma \h^{\,kl} \,] \ = \ \d_{\,j}{}^{\,(\,k} \, \h^{\,l\,)\,i} \, , \\[0pt]
&[\, T_{ij} \comma S^{\,k}{}_l \,] \ = \ T_{l\,(\,i}\, \d_{\,j\,)}{}^{\,k} \, ,  \\[0pt]
&[\, S^{\,i}{}_j \comma \pr^{\,k} \,] \ = \ \pr^{\,i} \, \d_{\,j}{}^{\,k}  \, ,  \\[0pt]
&[\, \pr_{\,k} \comma S^{\,i}{}_j \,] \ = \ \d_{\,k}{}^{\,i} \, \pr_{\,j}   \, , \\[0pt]
&[\, S^{\,i}{}_j \comma S^{\,k}{}_l \,] \ = \ \d_{\,j}{}^{\,k}S^{\,i}{}_l \, - \, \d_{\,l}{}^{\,i}S^{\,k}{}_j \, . \label{SS}
\end{align}
Notice that the commutators of the $S^{\,i}{}_{j}$ operators, the key novelty of the mixed-symmetry case, build the $gl(N)$ Lie algebra if $N$ index families are
present. The other relevant commutation relations are
\begin{align}
&[\, \g_{\,i} \comma \pr^{\,j} \,] \ = \ \, \dsl \ \d_{\,i}{}^{\,j} \, , \\[0pt]
&[\, \pr_{\,i} \comma \g^{\,j} \,] \ = \ \, \dsl \ \d_{\,i}{}^{\,j} \, , \\[0pt]
&[\, \g_{\,i} \comma \h^{\,jk} \,] \ = \ \12 \ \d_{\,i}{}^{\,(\,j}\, \g^{\,k\,)} \, ,  \\[0pt]
&[\, T_{ij} \comma \g^{\,k} \,] \ = \ \g_{\,(\,i}\, \d_{\,j\,)}{}^{\,k} \, ,  \\[0pt]
&[\, S^{\,i}{}_{j} \comma \g^{\,k} \,] \ = \ \g^{\,i} \, \d_{\,j}{}^{\,k} \, ,  \\[0pt]
&[\, \g_{\,k} \comma S^{\,i}{}_j \,] \ = \ \d_{\,k}{}^{\,i} \,\g_{\,j} \, ,
\end{align}
while the basic anticommutators are
\begin{align}
&\{\, \g_{\,i} \comma \g^{\,j} \,\} \, = \, D \, \d_{\,i}{}^{\,j} \, + \, 2 \, S^{\,j}{}_{\,i} \, ,  \label{anticomgamma}  \\[0pt]
&\{ \,\g_{\,i} \, , \dsl \, \} \ = \ 2 \, \pr_{\,i} \, , \\[0pt]
&\{ \dsl \,,\, \g^{\,i} \, \} \ = \ 2 \, \pr^{\,i} \, .
\end{align}
Altogether these relations give rise to the more involved computational rules of Appendix \ref{app:identities}, that describes the (anti)commutators involving different copies of these objects that are needed to derive the results presented in this review. In this respect, let us stress that in these cases we always use an antisymmetric basis for ``family'' $\g$-matrices, that can be obtained joining the $T_{ij}$ operators with the fully antisymmetric combinations
\be \label{anti_down}
\g_{\,i_1 \ldots \, i_n} \, \equiv \, \frac{1}{n!} \ \g_{\,[\,i_1} \, \g_{\phantom{[}i_2} \ldots \, \g_{\,i_n \, ]} \, ,
\ee
and joining the $\h^{\,ij}$ operators with the corresponding objects with raised indices,
\be \label{anti_up}
\g^{\,i_1 \ldots \, i_n} \, \equiv \, \frac{1}{n!} \ \g^{\,[\,i_1} \, \g^{\phantom{[}i_2} \ldots \, \g^{\,i_n \, ]} \, .
\ee
Furthermore, different groups of \emph{symmetrized} indices borne by a given quantity are separated by colons, while a semicolon signals the beginning of a group of \emph{antisymmetrized} indices.

In order to further simplify the combinatorics, as in
\cite{mixed_bose,mixed_fermi} it also proves convenient to introduce the scalar
product
\be \label{scalar} \bra \vf \comma \phi \ket \, \equiv \, \frac{1}{s_1! \ldots s_n!} \ \vf_{\,\m^1_1 \ldots \, \m^1_{s_1} \, , \, \ldots \, , \,
\m^n_1 \ldots \, \m^n_{s_n}} \, \phi^{\ \m^1_1 \ldots \, \m^1_{s_1} \, , \, \ldots \, , \, \, \m^n_1 \ldots \, \m^n_{s_n}} \, \equiv \,
\frac{1}{s_1! \, \ldots s_n!} \ \vf \, \phi \, ,
\ee
that can be adopted even in the fermionic case in the form
\be \bra \bar{\psi} \comma \c \ket \, \equiv \, \frac{1}{s_1! \ldots s_n!} \ \bar{\psi}_{\,\m^1_1 \ldots \, \m^1_{s_1} \, , \, \ldots \, , \,
\m^n_1 \ldots \, \m^n_{s_n}} \, \c^{\,\m^1_1 \ldots \, \m^1_{s_1} \, , \, \ldots \, , \, \, \m^n_1 \ldots \, \m^n_{s_n}} \, \equiv \,
\frac{1}{s_1! \, \ldots s_n!} \ \bar{\psi} \, \c \, .
\ee
Inside the brackets it is then possible to integrate by parts and to
turn $\h$'s into traces without introducing any $s_i$-dependent
combinatoric factors, since for instance
\be \label{deriv}
\bra \vf \comma \pr^{\,i} \, \phi \ket \, \equiv \,  \frac{1}{s_1! \ldots s_n!} \ \vf \ \pr^{\,i} \, \phi \, = \, - \, \frac{s_i}{s_1! \ldots s_n!} \
(\,\pr_{\,i}\, \vf\,) \, \phi \, \equiv \, - \, \bra \pr_{\,i}\, \vf \comma \phi \ket \, , \ee
and, if $i \neq j$,
\be
\bra \vf \comma \h^{\,ij} \, \phi \ket \, \equiv \, \frac{1}{s_1! \ldots s_n!} \ \vf \ \h^{\,ij} \, \phi \, = \, \12 \, \frac{s_i \, s_j}{s_1! \ldots
s_n!} \ (\,T_{ij} \, \vf\,) \, \phi \, \equiv \, \12 \, \bra T_{ij} \, \vf
\comma \phi \ket \, , \label{byparts}
\ee
where the reader should notice the somewhat unusual factor
$\frac{1}{2}$, originating from our choice of normalization
in eq.~\eqref{T_eta}. In a similar fashion, for
$\g$-matrices
\be
\bra \bar{\psi} \comma \g^{\,i} \, \c \ket \, \equiv \, \frac{1}{s_1! \ldots s_n!} \ \bar{\psi} \ \g^{\,i} \, \c \, = \, \frac{s_i}{s_1! \ldots s_n!} \ (\,\bar{\psi}\,\g_{\,i}\,) \, \c \, \equiv \, \bra \bar{\psi}\, \g_{\,i} \comma \c \ket \, .
\ee
However, in the main body of this review we are ignoring, for
simplicity, the overall factors $\prod_{i=1}^N s_i!$, that should
accompany the Lagrangians of multi-symmetric tensors or spinor-tensors to grant
them the conventional normalization.

Finally, in this review we make extensive use of a number of standard
tools related to the symmetric group. These include, in particular,
the Young projectors $Y$, that are used repeatedly to separate
irreducible components in family-index space. In the text the
various components are often specified by ordered lists of the
lengths of the rows for the associated diagrams enclosed between braces. For
instance, the $\{3,2\}$ component corresponds to
\be
\begin{picture}(0,25)(20,0)
\multiframe(0,10.5)(10.5,0){1}(10,10){}
\multiframe(10.5,10.5)(10.5,0){1}(10,10){}
\multiframe(21,10.5)(10.5,0){1}(10,10){}
\multiframe(0,0)(10,0){1}(10,10){}
\multiframe(10.5,0)(10,0){1}(10,10){}
\end{picture}
\ee
In general the Young projectors $Y$ can be built combining the
contributions of different Young tableaux, that can be identified
associating integer labels to the tensor indices to be projected and
allowing within the given graph all their arrangements such that
these integers grow from left to right and from top to bottom. In
some cases, however, this simple procedure can fail to
produce an orthogonal decomposition, which can still be attained by
a further Graham-Schmidt orthogonalization. This difficulty is not
present if, for any pair of tableaux, there is at least a couple of
indices belonging to a row of the first that lie in the same column
of the second, and vice versa. Let us stress that this
problem is never to be faced in the constructions presented in this review, as a result of
the particular symmetry properties of the involved objects.

Moreover, in this review we adopt the conventions
of \cite{mixed_bose,mixed_fermi} and Young tableaux are then defined in the
\emph{symmetric} basis. Thus, the projector corresponding to a
tableau $\t$ containing $n$ boxes takes the form
\be
Y_\tau \, = \, \frac{\lambda(\tau)}{n\,!} \ S\, A \, , \label{young_proj_gen}
\ee
where $S$ and $A$ are the corresponding products of ``row symmetrizers''
and ``column antisymmetrizers''. As pointed out in \cite{mixed_bose,mixed_fermi} the
\emph{antisymmetric} basis, where the roles of $S$ and $A$ are
interchanged, is more convenient when dealing with multi-forms rather than with multi-symmetric fields and, as a consequence, it is never used in this review. At any rate, the function $\lambda(\tau)$ appearing in eq.~\eqref{young_proj_gen} gives the dimension of the
associated representation of the symmetric group, that can be
computed, for instance, counting the standard ways of filling the
boxes of the corresponding diagram with the numbers $1,2,\ldots,n$
in increasing order from left to right and from top to bottom. We
often denote the ratio that appears in eq.~\eqref{young_proj_gen} as
\be \label{hooklenght}
\frac{\lambda(\tau)}{n\,!} \, = \, \frac{1}{h(\tau)} \, ,
\ee
where $h(\t)$ is the ``hook length'' of the corresponding tableau.
Similar techniques are used in Section
\ref{sec:irreducible} to identify irreducible Lorentz
tensors.

The hook length $h$ also enters the formulae for the dimensions
of $gl(D)$ and $o(D)$ (spinorial) representations, that might prove useful to the reader. Thus, the number of components of an irreducible $gl(D)$ tensor whose vector indices are $\{s_1, \ldots ,s_N\}$ projected is
\be
\dim_{\,gl(D)}\, [\{s_1,\ldots,s_N\}] \, = \, \frac{1}{h} \, \prod_{i\,=\,1}^N \, \frac{(\,D+s_i-i\,)\,!}{(\,D-i\,)\,!} \, .
\ee
For spinor-tensors this value must be multiplied by the number of their spinorial components, that is $2^{\,[\frac{D}{2}]}$ in the Dirac case. The number of independent components of the corresponding $\{s_1, \ldots ,s_N\}$-projected $o(D)$ tensor is given by
\be \label{dim2}
\dim_{\,T}\, [\{s_1,\ldots,s_N\}] \, = \, \frac{1}{h} \, \prod_{i\,=\,1}^N \, \frac{(\,D+s_i-N-i-1\,)\,!}{(\,D-2\,i\,)\,!} \, \prod_{j\,=\,i}^N \, (\,D+s_i+s_j-i-j\,) \, ,
\ee
while for the $o(D)$ spinor-tensor with
the same Young projection in its vector indices it reads
\be \label{dim1}
\dim_{\,S}\, [\{s_1,\ldots,s_N\}] \, = \, \frac{2^{\,[\frac{D}{2}]}}{h} \, \prod_{i\,=\,1}^N \, \frac{(\,D+s_i-N-i\,)\,!}{(\,D-2\,i\,)\,!} \, \prod_{j\,=\,i+1}^N \, (\,D+s_i+s_j-i-j+1\,) \, .
\ee

We refrain from adding further details, since these and other related standard facts are discussed extensively in the literature, and in particular in \cite{group}.


\section{Useful identities}\label{app:identities}


Using repeatedly the (anti)commutation rules of Appendix
\ref{app:MIX} yields the useful identities
\begin{align}
& [\,\pr_{\,l} \, , \, \h^{i_1j_1} \ldots\, \h^{i_pj_p} \,] \, = \, \12 \, \sum_{n\,=\,1}^p \, \prod_{r\,\neq\,n}^p \, \h^{i_rj_r} \, \d_{\,l}{}^{\,(\,i_n}\, \pr^{\,j_n\,)} \, , \\[2pt]
& [\,T_{i_1j_1} \ldots\, T_{i_pj_p} \, , \, \pr^{\,l}\,] \, = \, \sum_{n\,=\,1}^p \, \prod_{r\,\neq\,n}^p \, T_{i_rj_r} \, \pr_{\,(\,i_n}\, \d_{\,j_n\,)}{}^{\,l} \, ,
\end{align}
that are widely used in Section \ref{sec:labalag-bose}
and in Section \ref{sec:labalag-fermi} in the construction of the constrained Lagrangians. To built the Lagrangians for Fermi fields we also abide to the identities
\begin{align}
& [\,\pr_{\,l} \, , \, \g^{\,k_1 \ldots\, k_q} \,] \, = \, (-1)^{\,q+1} \ \d_{\,l}{}^{\,[\,k_1}\, \g^{\,k_2 \ldots\, k_q\,]} \dsl \, + \, \d_{\,l}{}^{\,[\,k_1\,} \pr^{\,k_2}\, \g^{\,k_3 \ldots\, k_q\,]} \, , \\[5pt]
& [\,\g_{\,k_1 \ldots\, k_q} \, , \, \pr^{\,l}\,] \, = \, (-1)^{\,q+1} \dsl \ \g_{\,[\,k_1 \ldots\, k_{q-1} \,} \d_{\,k_q\,]}{}^{\,l} \, + \, \g_{\,[\,k_1 \ldots\, k_{q-2}\,} \pr_{\,k_{q-1}\,} \d_{\,k_q\,]}{}^{\,l} \, , \\[5pt]
& [\,\g_{\,k_1 \ldots\, k_{q}} \, , \dsl \ ]_{\,(-1)^{\,q+1}} = \, 2 \ \g_{\,[\,k_1 \ldots\, k_{q-1}\,} \pr_{\,k_q\,]} \, , \\[5pt]
& [\, \dsl \, , \, \g^{\,k_1 \ldots\, k_q} \,]_{\,(-1)^{\,q+1}} = \, 2 \ \pr^{\,[\,k_1}\, \g^{\,k_2 \ldots\, k_q\,]} \, ,
\end{align}
where the symbols $[\, \comma \,]_{\,(-1)^{\,q+1}}$ can
denote both commutators and anticommutators, depending on the sign
of $(-1)^{\,q+1}$.

For the reduction of the equation of motion to the Labastida form discussed in Section \ref{sec:weyl}, Fermi fields require the knowledge of a set of (anti)commutation relations that can be conveniently presented introducing the new operators
\be \label{newgamma}
\g^{\,k_1 \ldots\, k_q}{}_{\,l} \, = \, \g^{\,k_1 \ldots\, k_q}\, \g_{\,l} \, - \, \g^{\,[\,k_1 \ldots\, k_{q-1}\,} S^{\,k_q\,]}{}_{\,l} \, .
\ee
They appear in fact in the identities
\begin{align}
& [\,\g_{\,l} \, , \, \g^{\,k_1 \ldots\, k_q} \,]_{\,(-1)^{\,q+1}} = \, (\,D-q+1\,)\ \d_{\,l}{}^{\,[\,k_1}\, \g^{\,k_2 \ldots\, k_q\,]} \, + \, 2\, (-1)^{\,q+1}\, \g^{\,[\,k_1 \ldots\, k_{q-1}\,} S^{\,k_q\,]}{}_{\,l} \, , \\[5pt]
& [\,\g_{\,k_1 \ldots\, k_{q}} \, , \, \g^{\,l} \,]_{\,(-1)^{\,q+1}} = \, (\,D-q+1\,)\ \g_{\,[\,k_1 \ldots\, k_{q-1}\,} \d_{\,k_q\,]}{}^{\,l} \, + \, 2 \, (-1)^{q+1}\, S^{\,l}{}_{\,[\,k_1\,} \g_{\,k_2 \ldots\, k_q\,]}  \, , \\[5pt]
& [\,T_{\,lm} \, , \, \g^{\,k_1 \ldots\, k_q} \,] \, = \, (-1)^{\,q+1} \, \d_{\,(\,l\,|}{}^{\,[\,k_1}\, \g^{\,k_2 \ldots\, k_q\,]}{}_{\,|\,m\,)} \, , \displaybreak \\[5pt]
& [\,\g_{\,lm} \, , \, \g^{\,k_1 \ldots\, k_q} \,] \, = \, - \, (\,D-q+1\,)\,(\,D-q+2\,)\ \d_{\,l}{}^{\,[\,k_1}\, \d_{\,m}{}^{\,k_2}\, \g^{\,k_3 \ldots\, k_q\,]} \nn \\[2pt]
& \phantom{[\,\g_{\,lm} \, , \, \g^{\,k_1 \ldots\, k_q} \,] \,} + \, (-1)^{\,q} \, (\,D-q+2\,)\ \d_{\,[\,l\,|}{}^{\,[\,k_1}\, \g^{\,k_2 \ldots\, k_q\,]}{}_{\,|\,m\,]} + \, 2 \ \g^{\,[\,k_1 \ldots\, k_{q-1}\,|}{}_{\,[\,l}\, S^{\,|\,k_q\,]}{}_{\,m\,]} \nn \\[2pt]
& \phantom{[\,\g_{\,lm} \, , \, \g^{\,k_1 \ldots\, k_q} \,] \,} + \, (-1)^{\,q+1} \, (\,D-q+1\,)\ \d_{\,[\,l\,|}{}^{\,[\,k_1}\, \g^{\,k_2 \ldots\, k_{q-1}\,} S^{\,k_q\,]}{}_{\,|\,m\,]} \, , \\[4pt]
& [\,S^{\,l}{}_{m} \, , \, \g^{\,k_1 \ldots\, k_q} \,] \, = \, (-1)^{\,q+1} \, \d_{\,m}{}^{\,[\,k_1}\, \g^{\,k_2 \ldots\, k_q\,]\,l} \, , \\[5pt]
& [\,\g_{\,k_1 \ldots\, k_q} \, , \, S^{\,l}{}_{m} \,] \, = \, (-1)^{\,q+1} \, \g_{\,m\,[\,k_1 \ldots\, k_{q-1}\,} \d_{\,k_q\,]}{}^{\,l} \, .
\end{align}
These are used together with the corresponding commutators involving products of metric tensors:
\begin{align}
& [\,\g_{\,l} \, , \, \h^{i_1j_1} \ldots\, \h^{i_pj_p} \,] \, = \, \12 \, \sum_{n\,=\,1}^p \, \prod_{r\,\neq\,n}^p \, \h^{i_rj_r} \, \d_{\,l}{}^{\,(\,i_n}\, \g^{\,j_n\,)} \, , \\[2pt]
& [\,T_{\,lm} \, , \, \h^{i_1j_1} \ldots\, \h^{i_pj_p} \,] \, = \, \12 \, \sum_{n\,=\,1}^p \, \Big\{ \ D \, \prod_{r\,\neq\,n}^p \h^{i_rj_r}\, \d_{\,l}{}^{\,(\,i_n}\,  \d_{\,m}{}^{\,j_n\,)} \nn \\
& \phantom{[\,T_{\,lm} \, ,} + \, \sum_{m \, < \, n} \, \prod_{r\,\neq\,m,n}^p \!\!\h^{i_rj_r} \d_{\,(\,l\,|}{}^{(\,i_m\,} \h^{\,j_m\,)\,(\,i_n\,} \d^{\,j_n\,)}{}_{\,|\,m\,)} + \prod_{r\,\neq\,n}^p \h^{i_rj_r} \, \d_{\,(\,l\,|}{}^{(\,i_n\,} S^{\,j_n\,)}{}_{|\,m\,)} \, \Big\} \, , \\[2pt]
& [\,\g_{\,lm} \, , \, \h^{i_1j_1} \ldots\, \h^{i_pj_p} \,] \, = \, \frac{1}{4} \, \sum_{n\,=\,1}^p \, \Big\{\, \sum_{m \, < \, n} \, \prod_{r\,\neq\,m,n}^p \!\!\h^{i_rj_r} \, \d_{\,[\,l\,|}{}^{(\,i_m\,} \g^{\,j_m\,)\,(\,i_n\,} \d^{\,j_n\,)}{}_{\,|\,m\,]} \nn \\
& \phantom{[\,\g_{\,lm} \, , \, \h^{i_1j_1} \ldots\, \h^{i_pj_p} \,] \,} + \, 2\, \prod_{r\,\neq\,n}^p \h^{i_rj_r} \, \d_{\,[\,l\,|}{}^{(\,i_n\,} \g^{\,j_n\,)}{}_{|\,m\,]} \, \Big\} \, .
\end{align}
Finally, when dealing with Fermi fields one has to make a wide use of the following composition rules
for family $\g$-matrices:
\begin{align}
& \g_{\,k_1 \ldots\, k_q} \, \g_{\,l} \, = \, \g_{\,k_1 \ldots\, k_q \, l} \, + \, \g_{\,[\,k_1 \ldots\, k_{q-1}} \, T_{k_q\,]\,l} \, , \label{compgamma1} \\[3pt]
& \g_{\,k_1 \ldots\, k_q} \, \g_{\,lm} \, = \, \g_{\,k_1 \ldots\, k_q\, lm} \, + \left(\, \g_{\,[\,k_1 \ldots\, k_{q-1}\,|\,m} \, T_{\,|\,k_q\,]\,l} \, - \, \g_{\,[\,k_1 \ldots\, k_{q-1}\,|\,l} \, T_{\,|\,k_q\,]\,m} \,\right) \nonumber \\
& \phantom{\g_{\,k_1 \ldots\, k_q} \, \g_{\,lm} \,} + \, \g_{\,[\,k_1 \ldots\, k_{q-2}} \, T_{k_{q-1}\,|\,m}\, T_{\,|\,k_q\,]\,l} \, . \label{compgamma2}
\end{align}
Similar rules also apply when one raises indices, but the
normalization of $\h^{\,ij}$ introduced in eq.~\eqref{T_eta}
requires that this operation be supplemented by the substitution
\be
T_{ij} \, \to \, 2 \, \h^{ij} \, .
\ee
In order to derive the equations of motion for Fermi fields, or in order to impose
the condition \eqref{selfadjferm} of self-adjointness of the
constrained Lagrangians, it is also convenient to keep in mind the
further, standard relation
\be
\g_{\,0} \, (\g_{\,k_1 \ldots\, k_q})^{\,\dagger}\, \g_{\,0} \, = \, (-1)^{\frac{(q-1)(q-2)}{2}} \, \g_{\,k_1 \ldots\, k_q} \, .
\ee

The computational rules collected in these first two appendices suffice to compute generic \mbox{($\g$-)}traces
of the kinetic tensors $\cF$ and $\cS$ and of the Bianchi
identities. In particular, starting from the Fronsdal-Labastida
tensor
\be
\cF \, = \, \Box \, \vf \, - \, \pr^{\,i}\pr_{\,i} \, \vf \, + \, \12 \ \pr^{\,i}\pr^{\,j} \, T_{ij} \,
\vf\, ,
\ee
for Bose fields one can obtain
\begin{align}
& \prod_{r\,=\,1}^p \, T_{i_rj_r} \, \cF \, = \, (\,p+1\,) \ \Box  \prod_{r\,=\,1}^p T_{i_rj_r} \, \vf \, - \, 2 \sum_{n\,=\,1}^p \, \pr_{\,i_n}\pr_{\,j_n} \prod_{r\,\neq\,n}^p T_{i_rj_r} \, \vf \nn \\
& + \, \sum_{n\,=\,1}^p \, \sum_{m\,<\,n} \left(\, \pr_{\,i_n}\pr_{\,(\,i_m}\, T_{j_m\,)\,j_n} + \, \pr_{\,j_n}\pr_{\,(\,i_m}\, T_{j_m\,)\,i_n} \,\right) \!\prod_{r\,\neq\,m\,,\,n}^p\! T_{i_rj_r} \, \vf \nn \\
& - \, \pr^{\,k} \bigg[\ \pr_{\,k} \prod_{r\,=\,1}^p T_{i_rj_r} \, \vf \, - \sum_{n\,=\,1}^p \, \pr_{\,(\,i_n}\,T_{\,j_n\,)\,k} \prod_{r\,\neq\,n}^p T_{i_rj_r} \, \vf \ \bigg] + \, \12 \ \pr^{\,k}\pr^{\,l} \, T_{kl} \, \prod_{r\,=\,1}^p T_{i_rj_r} \, \vf \, . \label{fgentrace}
\end{align}
This result is at the root of the derivation, in
eq.~\eqref{selfadjbose}, of the constrained Lagrangians via the
identification of a constrained self-adjoint Einstein-like operator. As a consequence, it is also crucial in the derivation of the equations of motion that follow from the various bosonic Lagrangians presented in this review. Furthermore, starting from the Bianchi identities
\be
\mathscr{B}_i\, : \ \pr_{\,i}\, \cF \, - \, \12 \ \pr^{\,j}\, T_{ij} \, \cF \, = \, - \, \frac{1}{12} \ \pr^{\,j}\pr^{\,k}\pr^{\,l} \, T_{(\,ij}\,T_{kl\,)} \, \vf \, ,
\ee %
in the constrained theory one can obtain the result
\begin{align}
T_{i_1j_1} \ldots\, T_{i_pj_p}\, \mathscr{B}_k\, : \ \ & \pr_{\,k}\, T_{i_1j_1} \ldots\, T_{i_pj_p} \, \cF \, - \, \12 \, \sum_{n\,=\,1}^p \, \pr_{\,(\,i_n}\,T_{j_n\,)\,k}\, \prod_{r\,\neq\,n}^p \, T_{i_rj_r}\, \cF \nn \\
& - \, \12 \ \pr^{\,l}\, T_{i_1j_1} \ldots\, T_{i_pj_p}\, T_{kl}\, \cF \, = \, 0 \, ,
\end{align}
that was already displayed in eq.~\eqref{trbianchinp-b}, and that plays a crucial role in the derivation of the constrained Lagrangians via the request of gauge invariance.

In a similar fashion, starting from the Fang-Fronsdal-Labastida tensor
\be \cS \, = \, i \left\{\, \dsl \, \psi \, - \, \pr^{\,i} \,
\g_{\,i}\, \psi \,\right\}\, , \ee
for Fermi fields one can obtain
\begin{align}
& \prod_{r\,=\,1}^p \, T_{i_rj_r} \, \g_{\,k_1 \ldots\, k_q} \, \cS \, = \, (-1)^{\,q} \, i \, (\,q+1\,) \dsl \ \g_{k_1 \ldots\, k_q} \prod_{r\,=\,1}^p \, T_{i_rj_r} \, \psi \nn \\
& + \,  i\,(\,q+1\,) \, \g_{\,[\,k_1 \ldots\, k_{q-1}\,} \pr_{\,k_q\,]} \prod_{r\,=\,1}^p \, T_{i_rj_r} \, \psi \, - \,i\, \sum_{n\,=\,1}^p \, \g_{\,[\,k_1 \ldots\, k_{q-1}\,}  T_{k_q\,]\,(\,i_n} \, \pr_{\,j_n\,)} \prod_{r\,\ne\,n}^p \, T_{i_rj_r} \, \psi \nn \\
& - \, i \, \sum_{n\,=\,1}^p \, \g_{\,k_1 \ldots\, k_q \, (\,i_n} \, \pr_{\,j_n\,)} \prod_{r\,\ne\,n}^p \, T_{i_rj_r} \, \psi \, - \, i\ \pr^{\,l} \, \g_{\,[\,k_1 \ldots\, k_{q-1}\,} T_{k_q\,]\,l} \prod_{r\,=\,1}^p \, T_{i_rj_r} \, \psi \nn \\
&  - \, i\ \pr^{\,l} \, \g_{k_1 \ldots\, k_q l} \prod_{r\,=\,1}^p \, T_{i_rj_r} \, \psi \, . \label{sgentrace}
\end{align}
In analogy with the bosonic case, this result is at the root of the derivation, in
eq.~\eqref{selfadjferm}, of the constrained Lagrangians via the
identification of a constrained self-adjoint Rarita-Schwinger-like operator. As a consequence, it is also crucial in the derivation of the equations of motion that follow from the various fermionic Lagrangians presented in this review. The $\g$-traces of the Bianchi identities
\be \label{bianchif_app}
\mathscr{B}_i\, : \ \pr_{\,i}\, \cS \, - \, \12 \dsl \, \g_{\,i}\, \cS \, - \, \12 \ \pr^{\,j}\,T_{ij}\, \cS \, - \, \frac{1}{6} \ \pr^{\,j}\,\g_{\,ij}\,\cS \, = \, \frac{i}{6} \ \pr^{\,j}\pr^{\,k}\, T_{(\,ij}\,\g_{\,k\,)}\, \psi \, ,
\ee
can be also easily computed applying the rules displayed in this appendix. However, even at the constrained level the result takes the far more involved form
\begin{align}
& (-1)^{\,q} \prod_{r\,=\,1}^{p} T_{i_r j_r} \, \g_{\,k_1 \ldots\, k_q} \, \mathscr{B}_l \, : \ \frac{1}{6} \, \sum_{n\,=\,1}^{p} \, \pr_{\,(\,i_n\,}\g_{\,j_n)\,k_1 \ldots\, k_q\,l}\, \prod_{r\,\ne\,n}^{p} T_{i_r j_r} \, \cS \nn \\
& + \, \pr_{\,[\,k_1\,}\g_{\,k_2 \ldots\, k_q\,l\ ]} \prod_{r\,=\,1}^{p} T_{i_r j_r} \, \cS \, + \, \frac{q-1}{6} \ \pr_{\,[\,k_1\,}\g_{\,k_2 \ldots\, k_q\,]\,l} \prod_{r\,=\,1}^{p} T_{i_r j_r} \, \cS \nn \\
& - \, \frac{(-1)^{\,q}}{2} \sum_{n\,=\,1}^{p} \g_{\,k_1 \ldots\, k_q}\,\pr_{\,(\,i_n}\,T_{j_n\,)\,l}\! \prod_{r\,\ne\,n}^{p} T_{i_r j_r} \, \cS \, + \, \frac{1}{6} \, \sum_{n\,=\,1}^{p} \pr_{\,(\,i_n\,}\g_{\,j_n\,)\,[\,k_1 \ldots\, k_{q-1}\,} T_{\,k_q\,]\,l}\! \prod_{r\,\ne\,n}^{p} T_{i_r j_r} \, \cS \nn \\
& - \, \frac{1}{6} \, \sum_{n\,=\,1}^{p} \, \g_{\,l\,[\,k_1 \ldots\, k_{q-1}\,} T_{\,k_q\,]\,(\,i_n}\,\pr_{\,j_n\,)} \prod_{r\,\ne\,n}^{p} T_{i_r j_r} \, \cS \, + \, \frac{q+1}{6} \ \pr_{\,[\,k_1\,} \g_{\,k_2 \ldots\, k_{q-1}\,} T_{\,k_q\,]\,l} \, \prod_{r\,=\,1}^{p} T_{i_r j_r} \, \cS \nn \\
& + \frac{(-1)^{\,q}}{6} \, \sum_{n\,=\,1}^{p} \, \g_{\,[\,k_1 \ldots\, k_{q-2}\,} T_{\,k_{q-1}\,|\,l}\,T_{\,|\,k_q\,]\,(\,i_n}\,\pr_{\,j_n\,)} \prod_{r\,\ne\,n}^{p} T_{i_r j_r} \, \cS \nn \\
& - \, \dsl \left\{\, \frac{q+3}{6} \ \g_{\,k_1 \ldots\, k_q \, l} + \, \frac{q-1}{6} \ \g_{\,[\,k_1 \ldots\, k_{q-1}\,} T_{\,k_q\,]\,l} \,\right\}\! \prod_{r\,=\,1}^{p} T_{i_r j_r}\, \cS  \nn \\
& - \frac{(-1)^{\,q}}{6} \ \pr^{\,m} \, \g_{\,k_1 \ldots k_q \,lm} \prod_{r\,=\,1}^{p} T_{i_r j_r}\, \cS - \, \frac{(-1)^{\,q}}{2} \, \pr^{\,m} \, \g_{\,k_1 \ldots k_q} \, T_{lm} \prod_{r\,=\,1}^{p} T_{i_r j_r} \, \cS \nn \\
& - \frac{1}{6} \, \pr^{\,m} \!\left(\, \g_{\,l\,[\,k_1 \ldots\, k_{q-1}\,} T_{\,k_q\,]\,m}  - \, \g_{\,m\,[\,k_1 \ldots\, k_{q-1}\,} T_{\,k_q\,]\,l} \,\right) \prod_{r\,=\,1}^{p} T_{i_r j_r} \, \cS \nn \\
& + \, \frac{(-1)^{\,q}}{6} \ \pr^{\,m} \, \g_{\,[\,k_1 \ldots\, k_{q-2}\,} T_{\,k_{q-1}\,|\,l}\,T_{\,|\,k_q\,]\,m} \prod_{r\,=\,1}^{p} T_{i_r j_r} \, \cS \, = \, 0 \, . \label{gtrace_bianchi}
\end{align}
The projections of this expression that allow to build the Lagrangians via the request of gauge invariance will be computed in the next appendix.


\section{Proofs of some results of Section \ref{chap:unconstrained}}\label{app:proofs}


In this last appendix we would like to provide the proofs of some results that we used in Section \ref{chap:unconstrained} to fix the structure of the unconstrained Lagrangians. In particular we shall deal with
\begin{itemize}
\item the structure of the compensator terms of the bosonic Lagrangians, that is fixed by eq.~\eqref{varunc-b2};
\item the structure of the compensator terms of the fermionic Lagrangians, that is fixed by eq.~\eqref{idconstrf};
\item the two-column projections of the multiple $\g$-traces of the Bianchi identities for Fermi fields, that lead to eq.~\eqref{finalbianchif} and to eq.~\eqref{trbianchiuncf}.
\end{itemize}

\subsubsection*{Proof of eq.~\eqref{varunc-b2}}

For Bose fields the structure of the compensator terms entering the unconstrained Lagrangians is fixed by the identity
\begin{align}
& \bra\, Y_{\{3,2^{\,p-1}\}}\, T_{i_1j_1} \ldots\, T_{i_pj_p}\, \L_{\,k} \,\comma\, \pr_{\,k}\, \cA^{\,[\,p\,]}{}_{\,i_1j_1,\,\ldots\,,\,i_pj_p} \,\ket \nn \\
& = \, \frac{p}{p+2}\ \bra\, T_{i_2j_2} \ldots\, T_{i_pj_p}\, T_{(\,i_1j_1}\, \L_{\,k\,)} \,\comma\, \pr_{\,k}\, \cA^{\,[\,p\,]}{}_{\,i_1j_1,\,\ldots\,,\,i_pj_p} \,\ket \, , \label{startingb}
\end{align}
that allows to rewrite the terms in the gauge variation \eqref{varunc-b} in a way that manifestly contains symmetrized traces of the gauge parameters. The scalar product plays a crucial role here, since makes it possible to extract the $\{3,2^{\,p-1}\}$ component enforcing the projection associated to a \emph{single}
Young tableau. In fact, while the full $\{3,2^{p-1}\}$ projection results from a sum of different tableaux, one can choose them in such a way that only one of them contributes to the scalar product. In particular, one needs only the tableau
\be \label{youngid2b}
\textrm{
\begin{picture}(30,50)(0,0)
\multiframe(0,35)(15,0){1}(15,15){{\footnotesize $i_1$}}
\multiframe(15.5,35)(15,0){1}(15,15){{\footnotesize $j_1$}}
\multiframe(31,35)(15,0){1}(15,15){{\footnotesize $k$}}
\multiframe(0,15)(15,0){1}(15,19.5){\vspace{7pt}$\vdots$}
\multiframe(15.5,15)(15,0){1}(15,19.5){\vspace{7pt}$\vdots$}
\multiframe(0,0)(15,0){1}(15,14.8){{\footnotesize $i_p$}}
\multiframe(15.5,0)(15,0){1}(15,14.8){{\footnotesize $j_p$}}
\end{picture}
}
\ee
which corresponds to the choice of standard labeling $i_n=2\,n-1$,
$j_n=2\,n$ for $1 \leq n \leq p$ and $k=2\,p+1$.
In fact, any symmetrization involving three indices of the set
$(\,i_m,j_n\,)$ in the right entry of the scalar product would induce a symmetrization beyond a given line of the tableau
\be \label{youngid2b2}
\textrm{
\begin{picture}(30,50)(0,0)
\multiframe(0,35)(15,0){1}(15,15){{\footnotesize $i_1$}}
\multiframe(15.5,35)(15,0){1}(15,15){{\footnotesize $j_1$}}
\multiframe(0,15)(15,0){1}(15,19.5){\vspace{7pt}$\vdots$}
\multiframe(15.5,15)(15,0){1}(15,19.5){\vspace{7pt}$\vdots$}
\multiframe(0,0)(15,0){1}(15,14.8){{\footnotesize $i_p$}}
\multiframe(15.5,0)(15,0){1}(15,14.8){{\footnotesize $j_p$}}
\end{picture}
}
\ee
that identifies the structure of $\cA^{\,[\,p\,]}{}_{i_1j_1,\,\ldots\,,\,i_pj_p}$. In order to proceed, it is then convenient to apply the projection associated to the tableau \eqref{youngid2b} directly on the right entry of the scalar product. In fact, the operations needed to build it differ from those
implied by the tableau \eqref{youngid2b2} only in the symmetrization
of the three indices $(\,i_1,j_1,k\,)$. More precisely, denoting the
tableau \eqref{youngid2b} by $\t_1$ and the tableau
\eqref{youngid2b2} by $\t_2$ and using the notation of
eq.~\eqref{young_proj_gen} one can recognize that acting on products
of traces
\be
Y_{\t_2} \, = \, 2 \, \frac{\l(\t_2)}{(\,2\,p\,)\,!} \, \widetilde{S}_{\,\t_2} \, A_{\,\t_2} \, .
\ee
Notice that the product $\widetilde{S}_{\,\t_2}$ of ``row symmetrizers'' does not
include the operator $S_{(i_1,j_1)}$ because this symmetrization is
already induced by the others. When applied to the expression $\partial_{\,k}\, T_{i_1j_1} \ldots\,
T_{i_pj_p}$, the product of the two Young projectors
$Y_{\t_1}$ and $Y_{\t_2}$ then gives
\be \label{youngtableaux12}
Y_{\t_1} \, Y_{\t_2} = \, \frac{\l(\t_1)}{(\,2\,p+1\,)\,!} \,
S_{\,(i_1,\,j_1,\,k)} \, \widetilde{S}_{\,\t_2} \, A_{\,\t_2} \, Y_{\,\t_2} =
\, \frac{\l(\t_1)}{2\,(\,2\,p+1\,)\,\l(\t_2)} \,
S_{\,(i_1,\,j_1,\,k)} \, (\,Y_{\t_2}\,)^{\,2} .
\ee
The ratio that appears in eq.~\eqref{youngtableaux12} is
\be
\frac{\l(\t_1)}{(\,2\,p+1\,)\, \l(\t_2)} \, = \, \frac{p}{p+2} \, ,
\ee
so that
\be
Y_{\t_1} \, \pr_{\,k} \, \cA^{\,[\,p\,]}{}_{i_1j_1\,,\,\ldots\,
,\,i_pj_p} \, = \, \frac{p}{p+2} \ \pr_{\,(\,k} \,
\cA^{\,[\,p\,]}{}_{i_1j_1\,),\,i_2j_2\,,\, \ldots\, ,\,i_pj_p} \, .
\ee

\subsubsection*{Proof of eq.~\eqref{idconstrf}}

For Fermi fields the structure of the compensator terms entering the unconstrained Lagrangians is fixed by the identity
\begin{align}
& \bra Y_{\{3,2^{p-1},1^q\}} \, T_{i_1j_1} \ldots T_{i_pj_p} \, \bar{\e}_{\,l} \, \g_{\,k_1 \ldots\, k_q} \comma \pr_{\,l} \, (\,\g^{\,[\,q\,]}\,\cW^{\,[\,p\,]}\,)_{\, i_1 j_1,\,\ldots\,,\,i_p j_p \, ;\, k_1 \ldots\, k_q} \ket \nn \\[5pt]
& = \, \bra \, (\,p+1\,) \ T_{i_2j_2} \ldots T_{i_pj_p}\, T_{(\,i_1j_1} \, \bar{\e}_{\,l\,)} \, \g_{\,k_1 \ldots\, k_q} \, - \, T_{i_2j_2} \ldots T_{i_pj_p} \, T_{\,[\,k_1\,|\,(\,l} \, \bar{\e}_{\,i_1} \, \g_{\,j_1\,)}\, \g_{\,|\,k_2 \ldots\, k_q\,]} \nn \comma \\
& \frac{p}{3\, (\,p+1\,)(\,p+q+2\,)} \ \pr_{\,(\,l\,|} \,
(\,\g^{\,[\,q\,]}\,\cW^{\,[\,p\,]}\,)_{\,|\, i_1
j_1\,),\,\ldots\,,\,i_p j_p \, ;\, k_1 \ldots\, k_q} \ket\, . \label{startingf}
\end{align}%
that appears in eq.~\eqref{idconstrf}. As in eq.~\eqref{startingb}, the presence of the scalar product makes it possible to extract the $\{3,2^{p-1}\}$ component acting on the left entry with the projector associated to the \emph{single} Young tableau
\be \label{proof_tableau1}
\textrm{
\begin{picture}(30,100)(0,0)
\multiframe(0,85)(15,0){1}(15,15){{\footnotesize $i_1$}}
\multiframe(15.5,85)(15,0){1}(15,15){{\footnotesize $j_1$}}
\multiframe(31,85)(15,0){1}(15,15){{\footnotesize $l$}}
\multiframe(0,65)(15,0){1}(15,19.5){\vspace{7pt}$\vdots$}
\multiframe(15.5,65)(15,0){1}(15,19.5){\vspace{7pt}$\vdots$}
\multiframe(0,50)(15,0){1}(15,14.8){{\footnotesize $i_p$}}
\multiframe(15.5,50)(15,0){1}(15,14.8){{\footnotesize $j_p$}}
\multiframe(0,35)(15,0){1}(15,14.8){{\footnotesize $k_1$}}
\multiframe(0,15)(15,0){1}(15,19.5){\vspace{7pt}$\vdots$}
\multiframe(0,0)(15,0){1}(15,14.8){{\footnotesize $k_q$}}
\end{picture}
}
\ee
In fact, one can decompose $Y_{\{3,2^{p-1}\}}$ into a sum of Young
tableaux where only the projector associated to
\eqref{proof_tableau1} is not annihilated when contracted against the
right entry of the scalar product, where
$(\g^{\,[\,q\,]}\,\cW^{\,[\,p\,]})$ carries the projection
\be \label{proof_tableau2}
\textrm{
\begin{picture}(30,100)(0,0)
\multiframe(0,85)(15,0){1}(15,15){{\footnotesize $i_1$}}
\multiframe(15.5,85)(15,0){1}(15,15){{\footnotesize $j_1$}}
\multiframe(0,65)(15,0){1}(15,19.5){\vspace{7pt}$\vdots$}
\multiframe(15.5,65)(15,0){1}(15,19.5){\vspace{7pt}$\vdots$}
\multiframe(0,50)(15,0){1}(15,14.8){{\footnotesize $i_p$}}
\multiframe(15.5,50)(15,0){1}(15,14.8){{\footnotesize $j_p$}}
\multiframe(0,35)(15,0){1}(15,14.8){{\footnotesize $k_1$}}
\multiframe(0,15)(15,0){1}(15,19.5){\vspace{7pt}$\vdots$}
\multiframe(0,0)(15,0){1}(15,14.8){{\footnotesize $k_q$}}
\end{picture}
}\ee
since all other tableaux would result in symmetrizations beyond a given a line.

In analogy with the previous discussion, the operations needed to build the projection
associated to the tableau \eqref{proof_tableau1} differ from those
implied by the tableau \eqref{proof_tableau2} only in the symmetrization
of the three indices $(i_1,j_1,l)$. More precisely, denoting the
tableau \eqref{proof_tableau1} by $\t_1$ and the tableau
\eqref{proof_tableau2} by $\t_2$ and using the notation of
eq.~\eqref{young_proj_gen}, when acting on the left entry of eq.~\eqref{startingf}
\be
Y_{\t_2} \, = \, \frac{2}{h(\t_2)} \, {\widetilde S}_{\,\t_2} \, A_{\,\t_2} \, .
\ee
The product ${\widetilde S}_{\,\t_2}$ of ``row symmetrizers'' does not
include the operator $S_{(i_1,j_1)}$ simply because this symmetrization is
already induced by the others. Acting with $Y_{\t_1}$ gives instead
\be
Y_{\t_1} \, Y_{\t_2} \, = \, \frac{1}{h(\tau_1)} \,
S_{\,(i_1,\,j_1,\,l)} \, {\widetilde S}_{\,\t_2} \, A_{\,\t_2} \, Y_{\,\t_2} \, =
\, \frac{h(\tau_2)}{2\,h(\tau_1)} \,
S_{\,(i_1,\,j_1,\,l)} \, (\,Y_{\t_2}\,)^{\,2} \ ,
\ee
so that, taking into account the ratio between the hook lengths, one ends up with
\begin{align}
& \bra Y_{\{3,2^{p-1},1^q\}} \, T_{i_1j_1} \ldots T_{i_pj_p} \, \bar{\e}_{\,l} \, \g_{\,k_1 \ldots\, k_q} \comma \pr_{\,l} \, (\,\g^{\,[\,q\,]}\,\cW^{\,[\,p\,]}\,)_{\, i_1 j_1,\,\ldots\,,\,i_p j_p \, ;\, k_1 \,\ldots\, k_q} \ket \\
& = \frac{p\,(\,p+q+1\,)}{(p+1)(p+q+2)} \, \bra Y_{\t_2} \, T_{i_1j_1} \!\ldots T_{i_pj_p} \, \bar{\e}_{\,l} \, \g_{\,k_1 \ldots\, k_q} \comma \pr_{\,(\,l\,|} \, (\,\g^{\,[\,q\,]}\cW^{\,[\,p\,]}\,)_{\,|\, i_1 j_1\,)\,,\,\ldots\,,\,i_p j_p \, ;\, k_1 \,\ldots\, k_q} \ket \, . \nn
\end{align}
In the bosonic case these steps sufficed to recover the Labastida
constraints on the gauge parameters, due to the presence of only
identical $T$ tensors. However, here one needs to refine the arguments,
because the family indices $(i_1,j_1)$ that are symmetrized with $l$
can be carried both by a trace $T$ and by an antisymmetrized $\g$-trace,
as can be easily seen absorbing the symmetrizations induced by
$\t_2$ in the right entry of the scalar product, so that
\begin{align}
& \bra Y_{\{3,2^{p-1},1^q\}} \, T_{i_1j_1} \ldots T_{i_pj_p} \, \bar{\e}_{\,l} \, \g_{\,k_1 \ldots\, k_q} \comma \pr_{\,l} \, (\,\g^{\,[\,q\,]}\,\cW^{\,[\,p\,]}\,)_{\, i_1 j_1,\,\ldots\,,\,i_p j_p \, ;\, k_1 \,\ldots\, k_q} \ket \nn \\
& = \, \bra \frac{2^{\,p}\,(\,q+1\,)\,!}{(\,p+q+1\,)\,!} \ T_{[\,i_1\,|\,j_1} \ldots\, T_{|\,i_p\,|\,j_p}\, \bar{\e}_{\,l}\, \g_{\,|\,k_1 \ldots\, k_q\,]} \comma \nn \\
& \phantom{=\, \bra} \frac{p\,(\,p+q+1\,)}{(\,p+1\,)(\,p+q+2\,)} \ \pr_{\,(\,l\,|} \, (\,\g^{\,[\,q\,]}\,\cW^{\,[\,p\,]}\,)_{\,|\, i_1 j_1\,)\,,\,\ldots\,,\,i_p j_p \, ;\, k_1 \,\ldots\, k_q} \, \ket \, . \label{scalar_antisymm}
\end{align}

In order to proceed, it is convenient to separate the terms in the
antisymmetrization that differ in the position of $(\,i_1,j_1\,)$,
\begin{align}
& T_{[\,i_1\,|\,j_1} \ldots T_{|\,i_p\,|\,j_p}\, \bar{\e}_{\,l}\, \g_{\,|\,k_1 \ldots\, k_q\,]} \, = \, T_{i_1j_1}\,T_{[\,i_2\,|\,j_2} \ldots\, T_{|\,i_p\,|\,j_p} \, \bar{\e}_{\,l}\, \g_{\,|\,k_1 \ldots\, k_q\,]} \nn \\
& - \, \sum_{n\,=\,2}^p\, (-1)^{\,n\,p}\ T_{i_1j_n}\,T_{j_1\,[\,i_2}\, T_{i_3\,|\,j_{n+1}} \ldots T_{|\,i_p\,|\,j_{n-1}} \, \bar{\e}_{\,l}\, \g_{\,|\,k_1 \ldots\, k_q\,]} \nn \\
& - \, T_{[\,i_2\,|\,j_2} \ldots\, T_{|\,i_p\,|\,j_p}\, T_{|\,k_1\,|\,j_1} \, \bar{\e}_{\,l}\, \g_{\,i_1\,|\,k_1 \ldots\, k_q\,]} \, . \label{expanti}
\end{align}
Forcing again the symmetrization in $(\,i_1,j_1,l\,)$, in the scalar
product the second group of terms in eq.~\eqref{expanti} becomes proportional
to the first term, because
\begin{align}
& \sum_{n\,=\,2}^p\, (-1)^{\,n\,p}\ T_{[\,i_2\,|\,j_n} \ldots\, T_{|\,i_{p-1}\,|\,j_{n-2}} \ldots T_{|\,i_p\,|\,(\,i_1} \, T_{j_1\,|\,j_{n-1}} \bar{\e}_{\,|\,l\,)}\, \g_{\,|\,k_1 \ldots\, k_q\,]} \nn \\
= \, & \sum_{n\,=\,2}^p\, (-1)^{\,n\,p}\ T_{[\,i_2\,|\,j_n} \ldots\, T_{|\,i_{p-1}\,|\,j_{n-2}} \ldots T_{|\,i_p\,|\,(\,i_1} \, T_{j_1l} \, \bar{\e}_{\,j_{n-1}\,)}\, \g_{\,|\,k_1 \ldots\, k_q\,]} \nn \\
- \, & \sum_{n\,=\,2}^p\, (-1)^{\,n\,p}\ T_{(\,i_1j_1}\,T_{l\,)\,[\,i_2}\,T_{i_3\,|\,j_{n+1}} \ldots\, T_{|\,i_p\,|\,j_{n-1}}\, \bar{\e}_{\,j_n}\, \g_{\,|\,k_1 \ldots\, k_q\,]} \nn \\
- \, & (\,p-1\,) \, T_{[\,i_2\,|\,j_n} \ldots\,
T_{|\,i_p\,|\,j_{n-1}}\, T_{(\,i_1j_1\,} \bar{\e}_{\,l\,)}\,
\g_{\,|\,k_1 \ldots\, k_q\,]}\, ,
\end{align}
and all antisymmetrizations over four indices (including those
induced on a couple of traces by a symmetrization over three
indices) are annihilated by the right entry of the scalar product
\eqref{scalar_antisymm}, so that only the last term survives.

In conclusion, the left entry of eq.~\eqref{scalar_antisymm} gives
rise to two terms when one forces the symmetrization in
$(\,i_1,j_1,l\,)$,
\begin{align}
T_{[\,i_1\,|\,j_1} \ldots\, T_{|\,i_p\,|\,j_p}\, \bar{\e}_{\,l}\, \g_{\,|\,k_1 \ldots\, k_q\,]} \, & \longrightarrow \,  \frac{1}{6}\, \Big\{\, (\,p+1\,)\, T_{[\,i_2\,|\,j_n} \ldots\, T_{|\,i_p\,|\,j_{n-1}}\, T_{(\,i_1j_1\,} \bar{\e}_{\,l\,)}\, \g_{\,|\,k_1 \ldots\, k_q\,]} \nn \\
& - \, T_{[\,i_2\,|\,j_n} \ldots\, T_{|\,i_p\,|\,j_{n-1}}\, T_{|\,k_1\,|\,(\,i_1} \bar{\e}_{\,j_1}\, \g_{\,l\,)\,|\,k_1 \ldots\, k_q\,]} \,\Big\} \, ,
\end{align}
and imposing the remaining symmetrizations one can recover in each
of them the projection associated to the Young tableau
\be \label{proof_tableau3}
\textrm{
\begin{picture}(30,100)(0,0)
\multiframe(0,85)(15,0){1}(15,15){{\footnotesize $i_2$}}
\multiframe(15.5,85)(15,0){1}(15,15){{\footnotesize $j_2$}}
\multiframe(0,65)(15,0){1}(15,19.5){\vspace{7pt}$\vdots$}
\multiframe(15.5,65)(15,0){1}(15,19.5){\vspace{7pt}$\vdots$}
\multiframe(0,50)(15,0){1}(15,14.8){{\footnotesize $i_p$}}
\multiframe(15.5,50)(15,0){1}(15,14.8){{\footnotesize $j_p$}}
\multiframe(0,35)(15,0){1}(15,14.7){{\footnotesize $k_1$}}
\multiframe(0,15)(15,0){1}(15,19.5){\vspace{7pt}$\vdots$}
\multiframe(0,0)(15,0){1}(15,14.8){{\footnotesize $k_q$}}
\end{picture}
}\ee
However, this projection is automatically enforced by the right
entry of the scalar product, so that
\begin{align}
& \bra Y_{\{3,2^{p-1},1^q\}} \, T_{i_1j_1} \ldots T_{i_pj_p} \, \bar{\e}_{\,l} \, \g_{\,k_1 \ldots\, k_q} \comma \pr_{\,l} \, (\,\g^{\,[\,q\,]}\,\cW^{\,[\,p\,]}\,)_{\, i_1 j_1,\,\ldots\,,\,i_p j_p \, ;\, k_1 \,\ldots\, k_q} \ket \nn \\
& = \, \frac{p\,(\,p+q+1\,)}{(\,p+1\,)(\,p+q+2\,)} \ \bra\, \frac{1}{3\,(\,p+q+1\,)} \, Y_{\t_3} \, \Big\{\, (\,p+1\,)\, T_{[\,i_1\,|\,j_1} \ldots\, T_{|\,i_p\,|\,j_p}\, \bar{\e}_{\,l}\, \g_{\,|\,k_1 \ldots\, k_q\,]} \nn \\
& - \, T_{[\,i_2\,|\,j_n} \ldots\, T_{|\,i_p\,|\,j_{n-1}}\, T_{|\,k_1\,|\,(\,i_1} \bar{\e}_{\,j_1}\, \g_{\,l\,)\,|\,k_1 \ldots\, k_q\,]} \,\Big\} \comma \pr_{\,(\,l\,|} \, (\,\g^{\,[\,q\,]}\,\cW^{\,[\,p\,]}\,)_{\,|\, i_1 j_1\,)\,,\,\ldots\,,\,i_p j_p \, ;\, k_1 \,\ldots\, k_q} \ket \nn \\
& = \, \bra \, (\,p+1\,) \ T_{i_2j_2} \ldots T_{i_pj_p}\, T_{(\,i_1j_1} \, \bar{\e}_{\,l\,)} \, \g_{\,k_1 \ldots\, k_q} \, - \, T_{i_2j_2} \ldots T_{i_pj_p} \, T_{\,[\,k_1\,|\,(\,i_1} \, \bar{\e}_{\,j_1} \, \g_{\,l\,)\,|\,k_2 \ldots\, k_q\,]} \comma \nn \\
& \frac{p}{3\, (\,p+1\,)(\,p+q+2\,)} \ \pr_{\,(\,l\,|} \,
(\,\g^{\,[\,q\,]}\,\cW^{\,[\,p\,]}\,)_{\,|\, i_1
j_1\,),\,\ldots\,,\,i_p j_p \, ;\, k_1 \,\ldots\, k_q} \ket\, , \label{laststep}
\end{align}
where $\t_3$ denotes the tableau of eq.~\eqref{proof_tableau3} and
the last two lines lead to the result in
eqs.~\eqref{idconstrf} or \eqref{startingf}, using eq.~\eqref{compgamma1}.

\subsubsection*{Two-column projected consequences of the Bianchi identities for Fermi fields}

In Sections \ref{sec:labalag-fermi} and \ref{sec:minimal-fermi} we presented some two-column projected consequences of the Bianchi identities \eqref{bianchif_app}, and now we would like to show how they can be extracted from their multiple $\g$-traces \eqref{gtrace_bianchi}. As anticipated in Section \ref{sec:minimal-fermi}, in the fermionic case eqs.~\eqref{gtrace_bianchi} actually admit the couple of two-column projections that can be obtained from the action of an already two-column projected combination of $\g$-traces. More precisely, acting with $T_{i_1j_1} \ldots\, T_{i_pj_p}\, \g_{\,k_1 \ldots\, k_q}$ on the $\mathscr{B}_l$ of eq.~\eqref{bianchif_app} one can obtain both two-column projections appearing in
\be
\{2^{\,p},1^q\} \, \otimes \, \{1\} \, = \, \{2^{\,p},1^{q+1}\} \, \oplus \, \{2^{\,p+1},1^{q-1}\} \, \oplus \, \{3,2^{\,p-1},1^q\} \, .
\ee
On the other hand, it is convenient to deal with Young projections containing the maximum number of antisymmetrizations compatible with the tensorial structure of the expressions under scrutiny. In this case one is led to consider the $\{2^{\,p},1^{q+1}\}$ projection of \eqref{gtrace_bianchi}, that can be computed acting with the
projector associated to the \emph{single} Young tableau
\be \label{proof_tableau4}
\textrm{
\begin{picture}(30,115)(0,0)
\multiframe(0,100)(15,0){1}(15,15){{\footnotesize $i_1$}}
\multiframe(15.5,100)(15,0){1}(15,15){{\footnotesize $j_1$}}
\multiframe(0,80)(15,0){1}(15,19.5){\vspace{7pt}$\vdots$}
\multiframe(15.5,80)(15,0){1}(15,19.5){\vspace{7pt}$\vdots$}
\multiframe(0,65)(15,0){1}(15,14.8){{\footnotesize $i_p$}}
\multiframe(15.5,65)(15,0){1}(15,14.8){{\footnotesize $j_p$}}
\multiframe(0,50)(15,0){1}(15,14.8){{\footnotesize $k_1$}}
\multiframe(0,30)(15,0){1}(15,19.5){\vspace{7pt}$\vdots$}
\multiframe(0,15)(15,0){1}(15,14.8){{\footnotesize $k_q$}}
\multiframe(0,0)(15,0){1}(15,14.8){{\footnotesize $l$}}
\end{picture}
}\ee
on account of the symmetry properties of the combination $T_{i_1j_1} \ldots \, T_{i_pj_p} \, \g_{\,k_1 \ldots\, k_q} \, \mathscr{B}_l$. One can thus combine the terms of \eqref{gtrace_bianchi} resorting to techniques similar to those adopted to deal with the traces of the bosonic Bianchi identities in Section \ref{sec:minimal-bose}. First of all, the structure of the tableau \eqref{proof_tableau4} implies that all symmetrizations in the generic set of three indices $(\,i_m,j_m,k_n\,)$ vanish. One can use this fact to immediately relate various terms as in eq.~\eqref{manipulation}. However, for Fermi fields this kind of manipulations does not suffice to make all contributions with the same number of traces and an antisymmetrized $\g$-trace with a given number of indices proportional. Still, one can also consider that, in order to build the projection associated to \eqref{proof_tableau4}, an antisymmetrization over $l$ and all the $k_n$ indices is required. The $\{2^{\,p},1^{q+1}\}$ projections of expressions that become proportional after this operation remain proportional with the same overall factor even after the full projection is enforced. For instance, denoting the column antisymmetrizer with $A_{\{2^p,1^q\}}$ as in eq.~\eqref{young_proj_gen}, one can easily recognize that
\begin{align}
& A_{\{2^p,1^q\}}\, \pr_{\,l} \, T_{i_1j_1} \ldots\, T_{i_pj_p}\, \g_{\,k_1 \ldots\, k_q} \, \cS \, = \, p\,!\,(\,q-1\,)\,! \ \pr_{\,[\,l} \, T_{i_1\,|\,j_1} \ldots\, T_{|\,i_p\,|\,j_p}\, \g_{\,|\,k_1 \ldots\, k_q\,]} \, \cS \, , \nn \\
& A_{\{2^p,1^q\}}\, \pr_{\,[\,k_1\,}\g_{\,k_2 \ldots\, k_q\,]\,l} \prod_{r\,=\,1}^{p} T_{i_r j_r} \, \cS \, = \, (-1)^{\,q}\, p\,!\,q\,!\ \pr_{\,[\,l} \, T_{i_1\,|\,j_1} \ldots\, T_{|\,i_p\,|\,j_p}\, \g_{\,|\,k_1 \ldots\, k_q\,]} \, \cS \, ,
\end{align}
and this implies
\be
Y_{\{2^p,1^q\}}\, \pr_{\,[\,k_1\,}\g_{\,k_2 \ldots\, k_q\,]\,l} \prod_{r\,=\,1}^{p} T_{i_r j_r} \, \cS \, = \, (-1)^{\,q}\, q\, Y_{\{2^p,1^q\}}\, \pr_{\,l} \, T_{i_1j_1} \ldots\, T_{i_pj_p}\, \g_{\,k_1 \ldots\, k_q} \, \cS \, .
\ee

Using these techniques one can finally obtain
\begin{align}
& Y_{\{2^p,1^{q+1}\}}\, T_{i_1j_1} \ldots\, T_{i_pj_p} \, \g_{\,k_1 \ldots\, k_q}\, \mathscr{B}_l \, : \nn \\[4pt]
& \left[\, (\,q+1\,) \,+\, \frac{q\,(\,q-1\,)}{6} \,+\, \frac{p}{2} \,+\, \frac{p\,q}{6} \,\right] \, Y_{\{\,2^p,\,1^{q+1}\}} \, \pr_{\,l} \, T_{i_1j_1} \ldots\, T_{i_pj_p}\, \g_{\,k_1 \ldots\, k_q} \, \cS \nn \\
& + \, \frac{1}{6} \ Y_{\{\,2^p,\,1^{q+1}\}} \, \sum_{n\,=\,1}^{p}\, \pr_{\,(\,i_n\,|}\, \prod_{r\,\neq\,n}^p \, T_{i_rj_r} \, \g_{\,|\,j_n\,)\,l\,k_1 \ldots\, k_q}\, \cS \nn \\[1pt]
& - \, \frac{q+3}{6} \ Y_{\{\,2^p,\,1^{q+1}\}}\, \Big\{\! \dsl \ T_{i_1j_1} \ldots\, T_{i_pj_p}\, \g_{\,l\,k_1 \ldots\, k_q} \, \cS  + \pr^{\,m}\, T_{i_1j_1} \ldots\, T_{i_pj_p}\, T_{lm}\, \g_{\,k_1 \ldots\, k_q} \, \cS \,\Big\} \nn \\[4pt]
& - \, \frac{1}{6} \ Y_{\{\,2^p,\,1^{q+1}\}} \, \pr^{\,m}\, T_{i_1j_1} \ldots\, T_{i_pj_p}\, \g_{\,k_1 \ldots\, k_q\, lm} \, \cS \, = \, 0 \, . \label{firstprojbianchi}
\end{align}
As in eq.~\eqref{expansion}, in the two divergence terms only the two-column
projected combination of traces and antisymmetrized $\g$-traces can
contribute to the full two-column projection involving also the
divergence. In a similar fashion, a two-column projection in all
indices is enforced in the first gradient term. In fact, the
manifest symmetries for its lower family indices are
\be \label{manifestsym}
\{2^{\,p},1^{q+1}\} \, \otimes \, \{1\} \, = \, \{2^{\,p},1^{q+2}\} \, \oplus \, \{2^{\,p+1},1^q\} \, \oplus \, \{3,2^{\,p-1},1^{q+1}\} \, ,
\ee
but, on the other hand, this term originates from the product of $(p+1)$ traces and a
$q$-fold $\g$-trace. As a result, the irreducible components with less than four columns that it
admits come from the decomposition
\be
\{2^{\,p+1}\} \, \otimes \, \{1^q\} \, = \, \{2^{\,p+1},1^q\} \, \oplus \, \{3,2^{\,p},1^{q-1}\} \, \oplus \, \{3^2,2^{\,p-1},1^{q-2}\} \, \oplus \, \ldots \ ,
\ee
so that the corresponding term is actually $\{2^{p+1},1^q\}$
projected. This is no longer true for the last term of
eq.~\eqref{firstprojbianchi}, where the components displayed in
eq.~\eqref{manifestsym} and those carried by the product of $p$
traces and a $(q+2)$-fold $\g$-trace,
\be
\{2^{\,p}\} \, \otimes \, \{1^{q+2}\} \, = \, \{2^{\,p},1^{q+2}\} \, \oplus \, \{3,2^{\,p-1},1^{q+1}\} \, \oplus \, \{3^2,2^{\,p-2},1^q\} \, \oplus \, \ldots \ ,
\ee
share two admissible components. However, in the constrained setting
one can elude this problem and simply select the $\{2^p,1^{q+2}\}$
component, because it is the only non-vanishing one. Therefore
eq.~\eqref{firstprojbianchi} directly leads to
eq.~\eqref{finalbianchif}, that suffices to annihilate the gauge variation of the constrained Lagrangians.

On the contrary, in the unconstrained setting the situation is more subtle, since one is left with
\begin{align}
& \frac{(\,q+3\,)\,(\,p+q+2\,)}{6} \ Y_{\{\,2^p,\,1^{q+1}\}} \, \pr_{\,l} \, (\,\g^{\,[\,q\,]}\,\cS^{\,[\,p\,]}\,)_{\, i_1 j_1,\,\ldots\,,\,i_p j_p \, ;\, k_1 \ldots\, k_q} \nn \\
+ \ & \frac{1}{6} \ Y_{\{\,2^p,\,1^{q+1}\}} \, \sum_{n\,=\,1}^{p}\, \pr_{\,(\,i_n\,|}\, (\,\g^{\,[\,q+2\,]}\,\cS^{\,[\,p-1\,]}\,)_{\,\ldots\,,\,i_{r\,\ne\,n} j_{r\,\ne\,n} \,,\,\ldots\, ;\,|\,j_n\,)\,l \, k_1 \ldots\, k_q} \nn \\[2pt]
- \ & \frac{q+3}{6} \,\Big\{\! \dsl \, (\,\g^{\,[\,q+1\,]}\,\cS^{\,[\,p\,]}\,)_{\, i_1 j_1,\,\ldots\,,\,i_p j_p \, ;\, l\,k_1 \ldots\, k_q} - \, \pr^{\,m}\, (\,\g^{\,[\,q\,]}\,\cS^{\,[\,p+1\,]}\,)_{\, i_1 j_1,\,\ldots\,,\,i_p j_p\,,\,lm \, ;\, k_1 \ldots\, k_q} \Big\} \nn \\[5pt]
- \ & \frac{1}{6} \ Y_{\{\,2^p,\,1^{q+1}\}} \, \pr^{\,m}\, T_{i_1j_1} \ldots\, T_{i_pj_p}\, \g_{\,k_1 \ldots\, k_q\, lm} \, \cS \nn \\[5pt]
= \ & \frac{i}{6} \ Y_{\{\,2^p,\,1^{q+1}\}} \, T_{i_1j_1} \ldots\, T_{i_pj_p}\, \g_{\,k_1 \ldots\, k_q} \, \pr^{\,m}\pr^{\,n} \, T_{(\,lm}\,\g_{\,n\,)}\, \psi \, . \label{projunc}
\end{align}
In order to reproduce the expression that appears in the gauge variation \eqref{unc_finalvar}, one should extract the $\{3,2^{\,p-1},1^{q+1}\}$ component from the last gradient term. Then one should express it as a combination of $\g$-traces of $T_{(\,ij}\,\g_{\,k\,)}\,\psi$ in order to fix the structure of the gauge transformations of the Lagrange multipliers $\l_{\,ijk}$. Rather than following this path, it is more convenient to exploit the possibility of producing different two-column projected consequences of the Bianchi identities. In fact, one can consider
\be
Y_{\{2^p,1^{q+1}\}}\, \sum_{n\,=\,1}^p \, \prod_{r\,\ne\,n}^p \, T_{i_rj_r} \, \g_{\,k_1
\ldots\, k_q\,l\,(\,i_n} \mathscr{B}_{j_n\,)} \, ,
\ee
that displays the same manifest symmetries as $Y_{\{2^p,1^{q+1}\}}\, T_{i_1j_1} \ldots\, T_{i_pj_p} \, \g_{\,k_1 \ldots\, k_q}\, \mathscr{B}_l$. Using the techniques described at the beginning of this appendix it can be cast in the form
\begin{align}
& p\,(\,q+1\,) \, \bigg\{\, \frac{(\,q+3\,)\,(\,p+q+2\,)}{6} \ Y_{\{\,2^p,\,1^{q+1}\}} \, \pr_{\,l} \, (\,\g^{\,[\,q\,]}\,\cS^{\,[\,p\,]}\,)_{\, i_1 j_1,\,\ldots\,,\,i_p j_p \, ;\, k_1 \ldots\, k_q} \nn \\
& + \, \frac{1}{6} \ Y_{\{\,2^p,\,1^{q+1}\}} \, \sum_{n\,=\,1}^{p}\, \pr_{\,(\,i_n\,|}\, (\,\g^{\,[\,q+2\,]}\,\cS^{\,[\,p-1\,]}\,)_{\,\ldots\,,\,i_{r\,\ne\,n} j_{r\,\ne\,n} \,,\,\ldots\, ;\,|\,j_n\,)\,l \, k_1 \ldots\, k_q} \nn \\[2pt]
& - \, \frac{q+3}{6}\, \Big[\! \dsl \, (\,\g^{\,[\,q+1\,]}\,\cS^{\,[\,p\,]}\,)_{\, i_1 j_1,\,\ldots\,,\,i_p j_p \, ;\, lk_1 \ldots\, k_q} + \, \pr^{\,m} (\,\g^{\,[\,q\,]}\,\cS^{\,[\,p+1\,]}\,)_{\, i_1 j_1,\,\ldots\,,\,i_p j_p\,,\,lm \, ;\, k_1 \ldots\, k_q} \,\Big] \bigg\} \nn \\[5pt]
& - \, \frac{p\,(\,q+3\,)}{6}\ Y_{\{2^p,1^{q+1}\}}\, \pr^{\,m}\, T_{i_1j_1} \ldots\, T_{i_pj_p} \, \g_{\,k_1 \ldots\, k_q\, lm}\, \cS \nn \\
& + \, \frac{(-1)^{\,q}}{3}\ Y_{\{2^p,1^{q+1}\}}\, \pr^{\, m}\, \sum_{n\,=\,1}^p \, \prod_{r\,\neq\,n}^p \, T_{i_rj_r} \, T_{m\,(\,i_n}\, \g_{\,j_n\,)\,k_1 \ldots\, k_q\, l} \, \cS \nn \\
& = \, \frac{i}{6} \ Y_{\{2^p,1^{q+1}\}} \, \sum_{n\,=\,1}^p \, \prod_{r\,\ne\,n}^p \, T_{i_rj_r} \, \g_{\,k_1 \ldots\, k_q\,l\,(\,i_n\,|}\, \pr^{\,m}\pr^{\,n} \left(\, T_{\,|\,j_n\,)\,(\,m}\,\g_{\,n\,)} \, + \, T_{\,mn}\,\g_{\,|\,j_n\,)} \,\right) \, . \label{projunc2}
\end{align}
The terms that already came in the form $(\,\g^{\,[\,m\,]}\,\cS^{\,[\,n\,]}\,)$ in eq.~\eqref{projunc} enter exactly the same combination also here, but one can rearrange these two expressions in order to project the gradient terms that do not appear in the desired form.
Indeed, in the combination
\be
Y_{\{2^p,1^{q+1}\}} \bigg\{\, (\,p+2\,)\,(\,q+3\,)\ T_{i_1j_1} \!\ldots T_{i_pj_p} \, \g_{\,k_1 \ldots\, k_q} \mathscr{B}_l \, - \, \sum_{n\,=\,1}^p \, \prod_{r\,\ne\,n}^p \, T_{i_rj_r} \, \g_{\,k_1
\ldots\, k_q\,l\,(\,i_n} \mathscr{B}_{j_n\,)} \,\bigg\}
\ee
the terms with $p$ traces and a $(q+2)$-fold $\g$-trace,
\begin{align}
& - \, \frac{1}{3} \ Y_{\{2^p,1^{q+1}\}} \, \pr^{\,m} \bigg\{\, (\,q+3\,)\ T_{i_1j_1} \ldots\, T_{i_pj_p}\, \g_{\,k_1 \ldots\, k_q\, lm}\, \cS \nn \\
& + \, (-1)^{\,q+1}\, \sum_{n\,=\,1}^p\, \prod_{r\,\neq\,n}^p\, T_{i_rj_r}\, T_{m\,(\,i_n}\, \g_{\,j_n\,)\,k_1 \ldots\, k_q\, l}\, \cS \,\bigg\} \, ,
\end{align}
are actually $\{2^{\,p},1^{q+2}\}$ projected. In order to verify this statement, notice that one can impose the $\{2^{\,p},1^{q+1}\}$ projection again via the single Young tableau \eqref{proof_tableau4}. Then, using the notation of eq.~\eqref{young_proj_gen} and applying the column antisymmetrizer associated to \eqref{proof_tableau4} to the expression within parentheses gives
\begin{align}
& A_{\{2^p,1^{q+1}\}}\, \pr^{\,m} \bigg\{\, (\,q+3\,)\ T_{i_1j_1} \!\ldots T_{i_pj_p}\, \g_{\,k_1 \ldots\, k_q\, lm}\, \cS + \, \ldots \,\bigg\} \nn \\
& = \, p\,!\,(\,q+1\,)\,!\ \pr^{\,m} \bigg\{\, (\,q+3\,)\ T_{[\,i_1\,|\,j_1} \ldots\, T_{|\,i_p\,|\,j_p}\, \g_{\,|\,k_1 \ldots\, k_q\, l\,]\,m}\, \cS \nn \\
& + \, (-1)^{\,q+1} \, (\,q+2\,)\, \sum_{n\,=\,1}^p \, T_{[\,i_1\,|\,j_n} \ldots\, T_{|\,i_{p-1}\,|\,j_{n-2}}\, T_{m\,j_{n-1}}\, \g_{\,|\,i_p\,k_1 \ldots\, k_q\,l\,]}\, \cS \nn \\
& + \, (-1)^{\,q+1}\, \sum_{n\,=\,1}^p\, T_{[\,i_1\,|\,j_n} \ldots\, T_{|\,i_{p-1}\,|\,j_n-2}\,T_{m\,|\,i_p\,|}\, \g_{\,j_{n-1}\,|\,k_1 \ldots\, k_q\, l\,]}\, \cS \, \bigg\} \, . \label{anti0}
\end{align}
When acting with the row symmetrizer $S$, the last two terms
become proportional, since they only differ because of an
antisymmetrization over all the $i_n$ indices and a single $j_n$ index,
that is annihilated by $S$. In conclusion, the right-hand side of
eq.~\eqref{anti0} is equivalent to
\be \label{anti1}
(\,q+3\,)\,p\,!\,(\,q+1\,)\,!\ \pr^{\,m}\, T_{[\,i_1\,|\,j_1} \ldots\, T_{|\,i_p\,|\,j_p}\, \g_{\,|\,k_1 \ldots\, k_q\, lm\,]}\, \cS \, ,
\ee
and the remaining symmetrizations induced by $S$ build an expression
proportional to that associated to a Young tableau with the index
$m$ attached at the end of the longer column of
\eqref{proof_tableau4}. Due to the symmetry properties of the
underlying combination, this suffices to build the Young projector
$Y_{\{2^p,1^{q+2}\}}$. In order to fix the proportionality factor
and to reach the result displayed in eq.~\eqref{trbianchiuncf}, one
can then compare eq.~\eqref{anti1} with an expression obtained
acting with the column antisymmetrizer of the $\{2^p,1^{q+2}\}$
projector,
\be \label{anti2}
A_{\{2^p,1^{q+2}\}}\, T_{i_1j_1} \!\ldots T_{i_pj_p}\, \g_{\,k_1 \ldots\, k_q\, lm}\, \cS = \, p\,!\,(\,q+2\,)\,!\ T_{[\,i_1\,|\,j_1} \!\ldots T_{|\,i_p\,|\,j_p}\, \g_{\,|\,k_1 \ldots\, k_q\, lm\,]}\, \cS \, .
\ee
Taking into account the hook lengths of the corresponding
expressions,
\be
h\,(\,\{2^p,1^{q+1}\}\,) \, = \, \frac{p\,!\,(\,p+q+2\,)\,!}{q+2} \, , \qquad h\,(\,\{2^p,1^{q+2}\}\,) \, = \, \frac{p\,!\,(\,p+q+3\,)\,!}{q+3} \, ,
\ee
finally gives
\begin{align}
& Y_{\{2^p,1^{q+1}\}} \, \pr^{\,m} \bigg\{\, (\,q+3\,)\ T_{i_1j_1} \!\ldots T_{i_pj_p}\, \g_{\,k_1 \ldots\, k_q\, lm}\, \cS - \sum_{n\,=\,1}^p \prod_{r\,\neq\,n}^p T_{i_rj_r}\, T_{m\,(\,i_n\,} \g_{\,j_n\,)\,l\,k_1 \ldots\, k_q}\, \cS \,\bigg\} \nn \\
& = \, (\,p+q+3\,)\ \pr^{\,m} \, Y_{\{2^p,1^{q+2}\}}\, T_{i_1j_1} \!\ldots T_{i_pj_p}\, \g_{\,k_1 \ldots\, k_q\, lm}\, \cS \, ,
\end{align}
that when added to the other contributions contained in
eqs.~\eqref{projunc} and \eqref{projunc2} proves the result
in eq.~\eqref{trbianchiuncf}.

\end{appendix}



\begin{thebibliography}{200}

\bibitem{veneziano}
G.~Veneziano,
  ``Construction of a crossing - symmetric, Regge behaved amplitude for linearly rising trajectories,''
  Nuovo Cim.\  A {\bf 57} (1968) 190.

\bibitem{yoneda}
T.~Yoneya,
  ``Quantum gravity and the zero slope limit of the generalized Virasoro model,''
  Lett.\ Nuovo Cim.\  {\bf 8} (1973) 951.

\bibitem{scherk_schwarz}
J.~Scherk and J.~H.~Schwarz,
  ``Dual Models For Nonhadrons,''
  Nucl.\ Phys.\  B {\bf 81} (1974) 118;

\bibitem{uv}
M.~H.~Goroff and A.~Sagnotti,
  ``Quantum Gravity At Two Loops,''
  Phys.\ Lett.\  B {\bf 160} (1985) 81,
  ``The Ultraviolet Behavior Of Einstein Gravity,''
  Nucl.\ Phys.\  B {\bf 266} (1986) 709.

\bibitem{review}
A.~Sagnotti and A.~Sevrin,
  ``Strings, gravity and particle physics,''
  Riv.\ Nuovo Cim.\  {\bf 31} (2008) 423
  [arXiv:hep-ex/0209011].

\bibitem{books}
M.~B.~Green, J.~H.~Schwarz and E.~Witten,
  ``Superstring Theory,'' 2 vols. (Cambridge Univ. Press, Cambridge, UK, 1987);\\
J.~Polchinski,
  ``String theory,'' 2 vols. (Cambridge Univ. Press, Cambridge, UK, 1998);\\
B.~Zwiebach,
  ``A first course in string theory'' (Cambridge Univ. Press, Cambridge, UK, 2004);\\
K.~Becker, M.~Becker and J.~H.~Schwarz,
   ``String theory and M-theory: A modern introduction'' (Cambridge Univ. Press, Cambridge, UK, 2007);\\
E.~Kiritsis,
   ``String theory in a nutshell'' (Princeton Univ. Press, Princeton, NJ, USA, 2007).

\bibitem{free_sft}
M.~Kato and K.~Ogawa,
  ``Covariant Quantization Of String Based On Brs Invariance,''
  Nucl.\ Phys.\ B {\bf 212} (1983) 443;\\
W.~Siegel,
  ``Classical Superstring Mechanics,''
  Nucl.\ Phys.\ B {\bf 263} (1986) 93;\\
W.~Siegel and B.~Zwiebach, ``Gauge String Fields,''
  Nucl. Phys. B {\bf 263} (1986) 105;\\
T.~Banks and M.~E.~Peskin,
  ``Gauge Invariance Of String Fields,''
  Nucl.\ Phys.\ B {\bf 264} (1986) 513;\\
N.~Ohta,
  ``Covariant Quantization Of Superstrings Based On Brs Invariance,''
  Phys.\ Rev.\ D {\bf 33} (1986) 1681,
  ``Covariant Second Quantization Of Superstrings,''
  Phys.\ Rev.\ Lett.\  {\bf 56} (1986) 440
  [Erratum-ibid.\  {\bf 56} (1986) 1316];\\
A.~Neveu, H.~Nicolai and P.~C.~West,
  ``Gauge Covariant Local Formulation Of Free Strings And Superstrings,''
  Nucl.\ Phys.\ B {\bf 264} (1986) 573;\\
A.~Neveu and P.~C.~West,
  ``Gauge Covariant Local Formulation Of Bosonic Strings,''
  Nucl.\ Phys.\  B {\bf 268} (1986) 125.

\bibitem{open_sft}
E.~Witten,
  ``Noncommutative Geometry And String Field Theory,''
  Nucl.\ Phys.\ B {\bf 268} (1986) 253.

\bibitem{closed_sft}
B.~Zwiebach,
  ``Closed string field theory: Quantum action and the B-V master equation,''
  Nucl.\ Phys.\  B {\bf 390} (1993) 33
  [arXiv:hep-th/9206084].

\bibitem{sugra8}
E.~Cremmer, B.~Julia and J.~Scherk,
  ``Supergravity theory in 11 dimensions,''
  Phys.\ Lett.\  B {\bf 76} (1978) 409;\\
E.~Cremmer and B.~Julia,
  ``The SO(8) Supergravity,''
  Nucl.\ Phys.\  B {\bf 159} (1979) 141.

\bibitem{loop_sugra}
Z.~Bern, J.~J.~Carrasco, L.~J.~Dixon, H.~Johansson and R.~Roiban,
  ``The Ultraviolet Behavior of N=8 Supergravity at Four Loops,''
  arXiv:0905.2326 [hep-th].

\bibitem{reviewHS}
X.~Bekaert, I.~L.~Buchbinder, A.~Pashnev and M.~Tsulaia,
  ``On higher spin theory: Strings, BRST, dimensional reductions,''
  Class.\ Quant.\ Grav.\  {\bf 21} (2004) S1457
  [arXiv:hep-th/0312252];\\
D.~Sorokin,
  ``Introduction to the classical theory of higher spins,''
  AIP Conf.\ Proc.\  {\bf 767} (2005) 172
  [arXiv:hep-th/0405069];\\
N.~Bouatta, G.~Compere and A.~Sagnotti,
  ``An introduction to free higher-spin fields,''
  arXiv:hep-th/0409068;\\
A.~Sagnotti, E.~Sezgin and P.~Sundell,
  ``On higher spins with a strong Sp(2,R) condition,''
  arXiv:hep-th/0501156;\\
A.~Fotopoulos and M.~Tsulaia,
  ``Gauge Invariant Lagrangians for Free and Interacting Higher Spin Fields. A Review of the BRST formulation,''
  arXiv:0805.1346 [hep-th].

\bibitem{reviews_vasiliev}
M.~A.~Vasiliev,
  ``Higher-spin gauge theories in four, three and two dimensions,''
  Int.\ J.\ Mod.\ Phys.\ D {\bf 5} (1996) 763
  [arXiv:hep-th/9611024],
  ``Higher spin gauge theories in various dimensions,''
  Fortsch.\ Phys.\  {\bf 52} (2004) 702
  [arXiv:hep-th/0401177];\\
X.~Bekaert, S.~Cnockaert, C.~Iazeolla and M.~A.~Vasiliev,
  ``Nonlinear higher spin theories in various dimensions,''
  arXiv:hep-th/0503128.

\bibitem{majorana}
E.~Majorana,
  ``Teoria relativistica di particelle con momento intrinseco arbitrario,''
  Nuovo Cim.\  {\bf 9} (1932) 335
  [English translation in G.F. Bassani (ed.), ``Ettore Majorana Scientific Papers'' (SIF, Bologna, IT and Springer, Berlin, DE, 2006)].

\bibitem{dirac}
P.~A.~M.~Dirac,
  ``Relativistic wave equations,''
  Proc.\ Roy.\ Soc.\ Lond.\  {\bf 155A} (1936) 447.

\bibitem{wigner}
E.~P.~Wigner,
  ``On Unitary Representations Of The Inhomogeneous Lorentz Group,''
  Annals Math.\  {\bf 40} (1939) 149
  [Nucl.\ Phys.\ Proc.\ Suppl.\  {\bf 6} (1989) 9].

\bibitem{bargmann_wigner}
V.~Bargmann and E.~P.~Wigner,
  ``Group Theoretical Discussion Of Relativistic Wave Equations,''
  Proc.\ Nat.\ Acad.\ Sci.\  {\bf 34} (1948) 211.

\bibitem{fierz}
M.~Fierz,
  ``Uber die relativistiche theorie kraftefreier teilchen mit beliebigem spin,''
  Helv.\ Phys.\ Acta {\bf 12} (1939) 3.

\bibitem{fierz_pauli}
M.~Fierz and W.~Pauli,
  ``On relativistic wave equations for particles of arbitrary spin in an electromagnetic field,''
  Proc.\ Roy.\ Soc.\ Lond.\  A {\bf 173} (1939) 211.

\bibitem{rarita_schwinger}
W.~Rarita and J.~Schwinger,
  ``On a theory of particles with half integral spin,''
  Phys.\ Rev.\  {\bf 60} (1941) 61.

\bibitem{moldauer}
E.~S.~Fradkin,
  ``On the theory of particles with higher spins. (In Russian),''
  JETP {\bf 20} (1950) 27;\\
P.~A.~Moldauer and K.~M.~Case,
  ``Properties of Half-Integral Spin Dirac-Fierz-Pauli Particles,''
  Phys.\ Rev.\  {\bf 102} (1956) 279;\\
C.~Fronsdal,
  ``On the theory of higher spin fields,''
  Nuovo Cim. Suppl.\  {\bf 9} (1958) 416.

\bibitem{chang}
S.~J.~Chang,
  ``Lagrange Formulation for Systems with Higher Spin,''
  Phys.\ Rev.\  {\bf 161} (1967) 1308.

\bibitem{singh_bose}
L.~P.~S.~Singh and C.~R.~Hagen,
  ``Lagrangian formulation for arbitrary spin. 1. The boson case,''
  Phys.\ Rev.\  D {\bf 9} (1974) 898.

\bibitem{singh_fermi}
L.~P.~S.~Singh and C.~R.~Hagen,
  ``Lagrangian formulation for arbitrary spin. 2. The fermion case,''
  Phys.\ Rev.\  D {\bf 9} (1974) 910.

\bibitem{fronsdal}
C.~Fronsdal,
  ``Massless Fields With Integer Spin,''
  Phys.\ Rev.\ D {\bf 18} (1978) 3624.

\bibitem{fang_fronsdal}
J.~Fang and C.~Fronsdal,
  ``Massless Fields With Half Integral Spin,''
  Phys.\ Rev.\ D {\bf 18} (1978) 3630.

\bibitem{curt_gauge}
T.~Curtright,
  ``Massless Field Supermultiplets With Arbitrary Spin,''
  Phys.\ Lett.\  B {\bf 85} (1979) 219.

\bibitem{dewit_fr}
B.~de Wit and D.~Z.~Freedman,
  ``Systematics Of Higher Spin Gauge Fields,''
  Phys.\ Rev.\  D {\bf 21} (1980) 358.

\bibitem{vasiliev_frame}
M.~A.~Vasiliev,
  ``'Gauge' Form Of Description Of Massless Fields With Arbitrary Spin. (In Russian),''
  Yad.\ Fiz.\  {\bf 32} (1980) 855
  [Sov.\ J.\ Nucl.\ Phys.\  {\bf 32} (1980) 439],
  ``Free Massless Fermionic Fields Of Arbitrary Spin In D-Dimensional De Sitter Space,''
  Nucl.\ Phys.\  B {\bf 301} (1988) 26;\\
V.~E.~Lopatin and M.~A.~Vasiliev,
  ``Free Massless Bosonic Fields Of Arbitrary Spin In d-Dimensional De Sitter Space,''
  Mod.\ Phys.\ Lett.\  A {\bf 3} (1988) 257.

\bibitem{mcdowell}
S.~W.~MacDowell and F.~Mansouri,
  ``Unified Geometric Theory Of Gravity And Supergravity,''
  Phys.\ Rev.\ Lett.\  {\bf 38} (1977) 739
  [Erratum-ibid.\  {\bf 38} (1977) 1376].

\bibitem{kaluza_klein}
T.~Kaluza,
  ``Zum Unitätsproblem in der Physik,''
  Sitzungsber.\ Preuss.\ Akad.\ Wiss.\ Berlin (Math.\ Phys.) (1921) 966;\\
O.~Klein,
  ``Quantentheorie und f\"unfdimensionale Relativit\"atstheorie,''
  Z.\ Phys.\  {\bf 37} (1926) 895
  [Surveys High Energ.\ Phys.\  {\bf 5} (1986) 241];\\
A.~Einstein and P.~Bergmann,
  ``On A Generalization Of Kaluza's Theory Of Electricity,''
  Annals Math.\  {\bf 39} (1938) 683.

\bibitem{curt_mixed}
T.~L.~Curtright and P.~G.~O.~Freund,
  ``Massive Dual Fields,''
  Nucl.\ Phys.\  B {\bf 172} (1980) 413;\\
R.~Delbourgo and P.~D.~Jarvis,
  ``Exotic Gauge Potential Representations And Ghost Counting Via Orthosymplectic Brs Supersymmetry,''
  J.\ Phys.\ A  {\bf 16} (1983) L275;\\
T.~Curtright,
  ``Generalized Gauge Fields,''
  Phys.\ Lett.\  B {\bf 165} (1985) 304.

\bibitem{mixed1}
C.~S.~Aulakh, I.~G.~Koh and S.~Ouvry,
  ``Higher Spin Fields With Mixed Symmetry,''
  Phys.\ Lett.\  B {\bf 173} (1986) 284;\\
J.~M.~F.~Labastida and T.~R.~Morris,
  ``Massless Mixed Symmetry Bosonic Free Fields,''
  Phys.\ Lett.\  B {\bf 180} (1986) 101;\\
K.~S.~Chung, C.~W.~Han, J.~K.~Kim and I.~G.~Koh,
  ``Brs Structure And Gauge Invariant Actions For Higher Spin Fields,''
  Phys.\ Rev.\  D {\bf 37}, 1079 (1988).

\bibitem{mixed2}
S.~Ouvry and J.~Stern,
  ``Gauge Fields Of Any Spin And Symmetry,''
  Phys.\ Lett.\  B {\bf 177} (1986) 335.

\bibitem{triplets}
A.~K.~H.~Bengtsson,
  ``A Unified Action For Higher Spin Gauge Bosons From Covariant String Theory,''
  Phys.\ Lett.\ B {\bf 182} (1986) 321;\\
M. Henneaux and C. Teitelboim,
  in ``Quantum Mechanics of Fundamental Systems, 2'',
  eds. C. Teitelboim and J. Zanelli (Plenum Press, New York, 1988), p. 113.

\bibitem{siegel}
W.~Siegel and B.~Zwiebach,
  ``Gauge String Fields from the Light Cone,''
  Nucl.\ Phys.\  B {\bf 282} (1987) 125;\\
W.~Siegel,
  ``Gauging Ramond String Fields Via OSP(1,1/2),''
  Nucl.\ Phys.\  B {\bf 284} (1987) 632.

\bibitem{laba_bose}
J.~M.~F.~Labastida,
  ``Massless Bosonic Free Fields,''
  Phys.\ Rev.\ Lett.\  {\bf 58} (1987) 531.

\bibitem{laba_fermi}
J.~M.~F.~Labastida,
  ``Massless Fermionic Free Fields,''
  Phys.\ Lett.\  B {\bf 186} (1987) 365.

\bibitem{laba_lag}
J.~M.~F.~Labastida,
  ``Massless Particles In Arbitrary Representations Of The Lorentz Group,''
  Nucl.\ Phys.\  B {\bf 322} (1989) 185.

\bibitem{mixed_bose}
A.~Campoleoni, D.~Francia, J.~Mourad and A.~Sagnotti,
  ``Unconstrained Higher Spins of Mixed Symmetry. I. Bose Fields,''
  Nucl.\ Phys.\  B {\bf 815} (2009) 289
  [arXiv:0810.4350 [hep-th]].

\bibitem{mixed_fermi}
A.~Campoleoni, D.~Francia, J.~Mourad and A.~Sagnotti,
  ``Unconstrained Higher Spins of Mixed Symmetry. II. Fermi Fields,''
  Nucl.\ Phys.\  B {\bf 828} (2010) 405
  [arXiv:0904.4447 [hep-th]].

\bibitem{fs1}
D.~Francia and A.~Sagnotti,
  ``On the geometry of higher-spin gauge fields,''
   Class.\ Quant.\ Grav.\  {\bf 20} (2003) S473 [arXiv:hep-th/0212185].

\bibitem{st}
A.~Sagnotti and M.~Tsulaia,
  ``On higher spins and the tensionless limit of string theory,''
  Nucl.\ Phys.\ B {\bf 682} (2004) 83 [arXiv:hep-th/0311257].

\bibitem{orientifolds}
A.~Sagnotti, ``Open Strings And Their Symmetry Groups,'' in Cargese '87, ``Non-Perturbative Quantum Field
Theory,'' eds. G. Mack et al (Pergamon Press, 1988), p. 521,
arXiv:hep-th/0208020;\\
G.~Pradisi and A.~Sagnotti,
``Open String Orbifolds,''
Phys.\ Lett.\ B {\bf 216} (1989) 59;\\
P.~Horava,
``Strings On World Sheet Orbifolds,''
Nucl.\ Phys.\ B {\bf 327} (1989) 461,
``Background Duality Of Open String Models,''
Phys.\ Lett.\ B {\bf 231} (1989) 251;\\
M.~Bianchi and A.~Sagnotti,
``On The Systematics Of Open String Theories,''
Phys.\ Lett.\ B {\bf 247} (1990) 517,
``Twist Symmetry And Open String Wilson Lines,''
Nucl.\ Phys.\ B {\bf 361} (1991) 519;\\
M.~Bianchi, G.~Pradisi and A.~Sagnotti,
``Toroidal compactification and symmetry breaking in open string theories,''
Nucl.\ Phys.\ B {\bf 376} (1992) 365;\\
A.~Sagnotti,
``A Note on the Green-Schwarz mechanism in open string theories,''
Phys.\ Lett.\  B {\bf 294}, 196 (1992)
[arXiv:hep-th/9210127];
\\
For reviews see: E.~Dudas,
``Theory and phenomenology of type I strings and M-theory,''
Class.\ Quant.\ Grav.\  {\bf 17}, (2000) R41 [arXiv:hep-ph/0006190];\\
C.~Angelantonj and A.~Sagnotti,
``Open strings,''
Phys.\ Rept.\  {\bf 371} (2002) 1 [Erratum-ibid.\  {\bf 376} (2003)
339] [arXiv:hep-th/0204089];\\
R.~Blumenhagen, B.~Kors, D.~Lust and S.~Stieberger,
``Four-dimensional String Compactifications with D-Branes, Orientifolds and Fluxes,''
Phys.\ Rept.\  {\bf 445} (2007) 1
[arXiv:hep-th/0610327].

\bibitem{0b}\
L.~J.~Dixon and J.~A.~Harvey,
 ``String Theories In Ten-Dimensions Without Space-Time Supersymmetry,''
 Nucl.\ Phys.\  B {\bf 274} (1986) 93;\\
N.~Seiberg and E.~Witten,
 ``Spin Structures In String Theory,''
 Nucl.\ Phys.\  B {\bf 276}, 272 (1986).

\bibitem{gso}
F.~Gliozzi, J.~Scherk and D.~I.~Olive,
  ``Supersymmetry, Supergravity Theories And The Dual Spinor Model,''
  Nucl.\ Phys.\  B {\bf 122} (1977) 253.

\bibitem{brst}
A.~Pashnev and M.~Tsulaia,
  ``Dimensional reduction and BRST approach to the description of a Regge  trajectory,''
  Mod.\ Phys.\ Lett.\ A {\bf 12} (1997) 861 [arXiv:hep-th/9703010],
  ``Description of the higher massless irreducible integer spins in the BRST approach,''
  Mod.\ Phys.\ Lett.\ A {\bf 13} (1998) 1853 [arXiv:hep-th/9803207];\\
C.~Burdik, A.~Pashnev and M.~Tsulaia,
  ``The Lagrangian description of representations of the Poincare group,''
  Nucl.\ Phys.\ Proc.\ Suppl.\  {\bf 102} (2001) 285
  [arXiv:hep-th/0103143];\\
I.~L.~Buchbinder, A.~Pashnev and M.~Tsulaia,
  ``Lagrangian formulation of the massless higher integer spin fields in the AdS background,''
  Phys.\ Lett.\ B {\bf 523} (2001) 338 [arXiv:hep-th/0109067],
  ``Massless higher spin fields in the AdS background and BRST constructions for nonlinear algebras,''
  arXiv:hep-th/0206026;\\
I.~L.~Buchbinder, V.~A.~Krykhtin and A.~Pashnev,
  ``BRST approach to Lagrangian construction for fermionic massless higher spin fields,''
  Nucl.\ Phys.\ B {\bf 711} (2005) 367 [arXiv:hep-th/0410215];\\
I.~L.~Buchbinder, V.~A.~Krykhtin and P.~M.~Lavrov,
  ``Gauge invariant Lagrangian formulation of higher spin massive bosonic field theory in AdS space,''
  Nucl.\ Phys.\  B {\bf 762} (2007) 344
  [arXiv:hep-th/0608005].
  

\bibitem{nonlocal}
D.~Francia and A.~Sagnotti,
  ``Free geometric equations for higher spins,''
  Phys.\ Lett.\ B {\bf 543} (2002) 303 [arXiv:hep-th/0207002].

\bibitem{damour_deser}
T.~Damour and S.~Deser,
  ``'Geometry' Of Spin 3 Gauge Theories,''
  Annales Poincare Phys.\ Theor.\  {\bf 47} (1987) 277.

\bibitem{tesi_dario}
D.~Francia, Ph.D. Thesis, Univ. Roma III (2006).

\bibitem{fms1}
D.~Francia, J.~Mourad and A.~Sagnotti,
  ``Current exchanges and unconstrained higher spins,''
  Nucl.\ Phys.\  B {\bf 773} (2007) 203
  [arXiv:hep-th/0701163].

\bibitem{nonlocal_mixed}
X.~Bekaert and N.~Boulanger,
  ``Tensor gauge fields in arbitrary representations of GL(D,R): Duality and Poincare lemma,''
  Commun.\ Math.\ Phys.\  {\bf 245} (2004) 27
  [arXiv:hep-th/0208058],
  ``On geometric equations and duality for free higher spins,''
  Phys.\ Lett.\ B {\bf 561} (2003) 183 [arXiv:hep-th/0301243],
  ``Tensor gauge fields in arbitrary representations of GL(D,R). II: Quadratic actions,''
  Commun.\ Math.\ Phys.\  {\bf 271} (2007) 723
  [arXiv:hep-th/0606198];\\
P.~de Medeiros and C.~Hull,
  ``Geometric second order field equations for general tensor gauge fields,''
  JHEP {\bf 0305} (2003) 019 [arXiv:hep-th/0303036].

\bibitem{dario_massive}
D.~Francia,
  ``Geometric Lagrangians for massive higher-spin fields,''
  Nucl.\ Phys.\  B {\bf 796} (2008) 77
  [arXiv:0710.5378 [hep-th]],
  ``Geometric massive higher spins and current exchanges,''
  Fortsch.\ Phys.\  {\bf 56} (2008) 800
  [arXiv:0804.2857 [hep-th]].

\bibitem{brst_mixed}
R.~R.~Metsaev,
  ``Massless mixed symmetry bosonic free fields in d-dimensional anti-de Sitter space-time,''
  Phys.\ Lett.\  B {\bf 354} (1995) 78;\\
C.~Burdik, A.~Pashnev and M.~Tsulaia,
  ``On the mixed symmetry irreducible representations of the Poincare group in the BRST approach,''
  Mod.\ Phys.\ Lett.\  A {\bf 16} (2001) 731
  [arXiv:hep-th/0101201];\\
G.~Bonelli,
  ``On the covariant quantization of tensionless bosonic strings in AdS spacetime,''
  JHEP {\bf 0311} (2003) 028
  [arXiv:hep-th/0309222];\\
I.~L.~Buchbinder, V.~A.~Krykhtin and H.~Takata,
  ``Gauge invariant Lagrangian construction for massive higher spin fermionic fields,''
  Phys.\ Lett.\  B {\bf 641} (2006) 386
  [arXiv:hep-th/0603212],
  ``Gauge invariant Lagrangian construction for massive bosonic mixed symmetry higher spin fields,''
  Phys.\ Lett.\  B {\bf 656} (2007) 253
  [arXiv:0707.2181 [hep-th]];\\
P.~Y.~Moshin and A.~A.~Reshetnyak,
  ``BRST approach to Lagrangian formulation for mixed-symmetry fermionic higher-spin fields,''
  JHEP {\bf 0710} (2007) 040
  [arXiv:0707.0386 [hep-th]];\\
I.~L.~Buchbinder, V.~A.~Krykhtin and L.~L.~Ryskina,
  ``BRST approach to Lagrangian formulation of bosonic totally antisymmeric tensor fields in curved space,''
  Mod.\ Phys.\ Lett.\  A {\bf 24} (2009) 401
  [arXiv:0810.3467 [hep-th]],
  ``Lagrangian formulation of massive fermionic totally antisymmetric tensor field theory in $AdS_d$ space,''
  arXiv:0902.1471 [hep-th];\\
K.~B.~Alkalaev, M.~Grigoriev and I.~Y.~Tipunin,
  ``Massless Poincare modules and gauge invariant equations,''
  arXiv:0811.3999 [hep-th];\\
K.~B.~Alkalaev and M.~Grigoriev,
  ``Unified BRST description of AdS gauge fields,''
  arXiv:0910.2690 [hep-th].

\bibitem{schwinger}
J.~Schwinger, ``Particles, sources, and fields'' (Addison-Wesley,
Reading, MA, USA, 1970).

\bibitem{fs2}
D.~Francia and A.~Sagnotti,
  ``Minimal local Lagrangians for higher-spin geometry,''
  Phys.\ Lett.\ B {\bf 624} (2005) 93 [arXiv:hep-th/0507144].

\bibitem{fs3}
D.~Francia and A.~Sagnotti,
  ``Higher-spin geometry and string theory,''
  J.\ Phys.\ Conf.\ Ser.\  {\bf 33} (2006) 57 [arXiv:hep-th/0601199].

\bibitem{fms2}
D.~Francia, J.~Mourad and A.~Sagnotti,
  ``(A)dS exchanges and partially-massless higher spins,''
  Nucl.\ Phys.\  B {\bf 804} (2008) 383
  [arXiv:0803.3832 [hep-th]].

\bibitem{buch_tripl}
I.~L.~Buchbinder, A.~V.~Galajinsky and V.~A.~Krykhtin,
  ``Quartet unconstrained formulation for massless higher spin fields,''
  Nucl.\ Phys.\  B {\bf 779} (2007) 155
  [arXiv:hep-th/0702161].

\bibitem{mixed_frame}
L.~Brink, R.~R.~Metsaev and M.~A.~Vasiliev,
  ``How massless are massless fields in AdS(d),''
  Nucl.\ Phys.\  B {\bf 586} (2000) 183
  [arXiv:hep-th/0005136];\\
Yu.~M.~Zinoviev,
  ``On massive mixed symmetry tensor fields in Minkowski space and (A)dS,''
  arXiv:hep-th/0211233,
  ``First order formalism for mixed symmetry tensor fields,''
  arXiv:hep-th/0304067,
  ``First order formalism for massive mixed symmetry tensor fields in Minkowski and (A)dS spaces,''
  arXiv:hep-th/0306292;\\
K.~B.~Alkalaev, O.~V.~Shaynkman and M.~A.~Vasiliev,
  ``On the frame-like formulation of mixed-symmetry massless fields in (A)dS(d),''
  Nucl.\ Phys.\  B {\bf 692} (2004) 363
  [arXiv:hep-th/0311164],
  ``Lagrangian formulation for free mixed-symmetry bosonic gauge fields in (A)dS(d),''
  JHEP {\bf 0508} (2005) 069
  [arXiv:hep-th/0501108],
  ``Frame-like formulation for free mixed-symmetry bosonic massless higher-spin fields in AdS(d),''
  arXiv:hep-th/0601225;\\
K.~B.~Alkalaev,
  ``Two-column higher spin massless fields in AdS(d),''
  Theor.\ Math.\ Phys.\  {\bf 140} (2004) 1253
  [Teor.\ Mat.\ Fiz.\  {\bf 140} (2004) 424]
  [arXiv:hep-th/0311212];\\
Yu.~M.~Zinoviev,
  ``Frame-like gauge invariant formulation for massive high spin particles,''
  Nucl.\ Phys.\  B {\bf 808} (2009) 185
  [arXiv:0808.1778 [hep-th]],
  ``Toward frame-like gauge invariant formulation for massive mixed symmetry bosonic fields,''
  Nucl.\ Phys.\  B {\bf 812} (2009) 46
  [arXiv:0809.3287 [hep-th]],
  ``Note on antisymmetric spin-tensors,''
  JHEP {\bf 0904} (2009) 035
  [arXiv:0903.0262 [hep-th]],
  ``Frame-like gauge invariant formulation for mixed symmetry fermionic fields,''
  arXiv:0904.0549 [hep-th],
  ``Towards frame-like gauge invariant formulation for massive mixed symmetry bosonic fields. II. General Young tableau with two rows,''
  arXiv:0907.2140 [hep-th].


\bibitem{mixed_unfolded}
E.~D.~Skvortsov,
  ``Mixed-Symmetry Massless Fields in Minkowski space Unfolded,''
  JHEP {\bf 0807} (2008) 004
  [arXiv:0801.2268 [hep-th]],
  ``Frame-like Actions for Massless Mixed-Symmetry Fields in Minkowski space,''
  Nucl.\ Phys.\  B {\bf 808} (2009) 569
  [arXiv:0807.0903 [hep-th]],
  ``Gauge fields in (anti)-de Sitter space and Connections of its symmetry algebra,''
  arXiv:0904.2919 [hep-th];\\
N.~Boulanger, C.~Iazeolla and P.~Sundell,
  ``Unfolding Mixed-Symmetry Fields in AdS and the BMV Conjecture: I. General Formalism,''
  JHEP {\bf 0907} (2009) 013
  [arXiv:0812.3615 [hep-th]],
  ``Unfolding Mixed-Symmetry Fields in AdS and the BMV Conjecture: II. Oscillator Realization,''
  JHEP {\bf 0907} (2009) 014
  [arXiv:0812.4438 [hep-th]];\\
M.~A.~Vasiliev,
  ``Bosonic conformal higher--spin fields of any symmetry,''
  arXiv:0909.5226 [hep-th].

\bibitem{vasiliev}
M.~A.~Vasiliev,
  ``Consistent equation for interacting gauge fields of all spins in (3+1)-dimensions,''
  Phys.\ Lett.\ B {\bf 243} (1990) 378,
  ``Properties of equations of motion of interacting gauge fields of all spins in (3+1)-dimensions,''
  Class.\ Quant.\ Grav.\  {\bf 8} (1991) 1387,
  ``More On Equations Of Motion For Interacting Massless Fields Of All Spins In (3+1)-Dimensions,''
  Phys.\ Lett.\ B {\bf 285} (1992) 225.

\bibitem{vasiliev_final}
M.~A.~Vasiliev,
  ``Nonlinear equations for symmetric massless higher spin fields in (A)dS(d),''
  Phys.\ Lett.\  B {\bf 567} (2003) 139
  [arXiv:hep-th/0304049].

\bibitem{demedeiros}
P.~de Medeiros,
  ``Massive gauge-invariant field theories on spaces of constant curvature,''
  Class.\ Quant.\ Grav.\  {\bf 21} (2004) 2571
  [arXiv:hep-th/0311254].

\bibitem{worldline}
R.~Marnelius,
  ``Manifestly Conformal Covariant Description Of Spinning And Charged Particles,''
  Phys.\ Rev.\  D {\bf 20} (1979) 2091;\\
V.~D.~Gershun and V.~I.~Tkach,
  ``Classical And Quantum Dynamics Of Particles With Arbitrary Spin,''
  JETP Lett.\ {\bf 29} (1979) 288 [Pisma Zh.\ Eksp.\ Teor.\ Fiz.\ {\bf 29} (1979) 320];\\
P.~S.~Howe, S.~Penati, M.~Pernici and P.~K.~Townsend,
  ``Wave Equations For Arbitrary Spin From Quantization Of The Extended Supersymmetric Spinning Particle,''
  Phys.\ Lett.\  B {\bf 215} (1988) 555,
 ``A Particle Mechanics Description Of Antisymmetric Tensor Fields,''
  Class.\ Quant.\ Grav.\  {\bf 6} (1989) 1125;\\
W.~Siegel,
  ``Conformal Invariance Of Extended Spinning Particle Mechanics,''
  Int.\ J.\ Mod.\ Phys.\  A {\bf 3} (1988) 2713;\\
R.~Marnelius and U.~Martensson,
  ``Derivation Of Manifestly Covariant Quantum Models For Spinning Relativistic Particles,''
  Nucl.\ Phys.\  B {\bf 335} (1990) 395;\\
I.~A.~Bandos and J.~Lukierski,
  ``Tensorial central charges and new superparticle models with fundamental spinor coordinates,''
  Mod.\ Phys.\ Lett.\  A {\bf 14}, 1257 (1999)
  [arXiv:hep-th/9811022];\\
I.~A.~Bandos, J.~Lukierski and D.~P.~Sorokin,
  ``Superparticle models with tensorial central charges,''
  Phys.\ Rev.\  D {\bf 61}, 045002 (2000)
  [arXiv:hep-th/9904109];\\
I.~Bandos, X.~Bekaert, J.~A.~de Azcarraga, D.~Sorokin and M.~Tsulaia,
  ``Dynamics of higher spin fields and tensorial space,''
  JHEP {\bf 0505}, 031 (2005)
  [arXiv:hep-th/0501113];\\
F.~Bastianelli, O.~Corradini and E.~Latini,
  ``Higher spin fields from a worldline perspective,''
  JHEP {\bf 0702} (2007) 072
  [arXiv:hep-th/0701055],
  ``Spinning particles and higher spin fields on (A)dS backgrounds,''
  arXiv:0810.0188 [hep-th];\\
F.~Bastianelli, O.~Corradini and A.~Waldron,
  ``Detours and Paths: BRST Complexes and Worldline Formalism,''
  JHEP {\bf 0905} (2009) 017
  [arXiv:0902.0530 [hep-th]];\\
R.~Marnelius,
  ``Lagrangian higher spin field theories from the O(N) extended supersymmetric particle,''
  arXiv:0906.2084 [hep-th];\\
D.~Cherney, E.~Latini and A.~Waldron,
  ``Generalized Einstein Operator Generating Functions,''
  arXiv:0909.4578 [hep-th].


\bibitem{mixed_lightcone}
R.~R.~Metsaev,
  ``Cubic interaction vertices of totally symmetric and mixed symmetry massless representations of the Poincare group in D = 6 space-time,''
  Phys.\ Lett.\  B {\bf 309} (1993) 39,
  ``Cubic interaction vertices for massive and massless higher spin fields,''
  Nucl.\ Phys.\  B {\bf 759} (2006) 147
  [arXiv:hep-th/0512342],
  ``Cubic interaction vertices for fermionic and bosonic arbitrary spin fields,''
  arXiv:0712.3526 [hep-th].

\bibitem{majorana_upgrade}
P.~A.~Horvathy, M.~S.~Plyushchay and M.~Valenzuela,
  ``Bosonized supersymmetry from the Majorana-Dirac-Staunton theory and massive higher-spin fields,''
  Phys.\ Rev.\  D {\bf 77} (2008) 025017
  [arXiv:0710.1394 [hep-th]];\\
X.~Bekaert, M.~R.~de Traubenberg and M.~Valenzuela,
  ``An infinite supermultiplet of massive higher-spin fields,''
  JHEP {\bf 0905} (2009) 118
  [arXiv:0904.2533 [hep-th]].

\bibitem{velo_zwanziger}
G.~Velo and D.~Zwanziger,
  ``Propagation And Quantization Of Rarita-Schwinger Waves In An External Electromagnetic Potential,''
  Phys.\ Rev.\  {\bf 186} (1969) 1337,
  ``Noncausality and other defects of interaction lagrangians for particles with spin one and higher,''
  Phys.\ Rev.\  {\bf 188} (1969) 2218,
  ``Fallacy Of Perturbative Methods For Higher Spin Equations,''
  Lett.\ Nuovo Cim.\  {\bf 15} (1976) 39.

\bibitem{nappi}
P.~C.~Argyres and C.~R.~Nappi,
  ``Massive Spin-2 Bosonic String States In An Electromagnetic Background,''
  Phys.\ Lett.\  B {\bf 224} (1989) 89.

\bibitem{porrati_raman}
M.~Porrati and R.~Rahman,
  ``Causal Propagation of a Charged Spin 3/2 Field in an External Electromagnetic Background,''
  Phys.\ Rev.\  D {\bf 80} (2009) 025009
  [arXiv:0906.1432 [hep-th]].

\bibitem{aragone_deser1}
C.~Aragone,
  ``Inconsistency of interaction between the gravitational field and the tensor meson field,''
  Nuovo Cim.\  A {\bf 64} (1969) 841;\\
C.~Aragone and S.~Deser,
  ``Constraints on gravitationally coupled tensor fields,''
  Nuovo Cim.\  A {\bf 3} (1971) 709.

\bibitem{weinberg}
S.~Weinberg,
  ``Photons And Gravitons In S Matrix Theory: Derivation Of Charge Conservation And Equality Of Gravitational And Inertial Mass,''
  Phys.\ Rev.\  {\bf 135} (1964) B1049.

\bibitem{coleman_mandula}
S.~R.~Coleman and J.~Mandula,
  ``All Possible Symmetries Of The S Matrix,''
  Phys.\ Rev.\  {\bf 159} (1967) 1251.

\bibitem{sugra}
D.~Z.~Freedman, P.~van Nieuwenhuizen and S.~Ferrara,
  ``Progress Toward A Theory Of Supergravity,''
  Phys.\ Rev.\  D {\bf 13} (1976) 3214;\\
S.~Deser and B.~Zumino,
  ``Consistent Supergravity,''
  Phys.\ Lett.\  B {\bf 62} (1976) 335.

\bibitem{aragone_deser2}
C.~Aragone and S.~Deser,
  ``Consistency Problems Of Hypergravity,''
  Phys.\ Lett.\  B {\bf 86} (1979) 161;\\
F.~A.~Berends, J.~W.~van Holten, B.~de Wit and P.~van Nieuwenhuizen,
  ``On Spin 5/2 Gauge Fields,''
  J.\ Phys.\ A  {\bf 13} (1980) 1643.

\bibitem{aragone_deser3}
C.~Aragone and S.~Deser,
  ``Consistency Problems Of Spin-2 Gravity Coupling,''
  Nuovo Cim.\  B {\bf 57} (1980) 33;\\
C.~Aragone and H.~La Roche,
  ``Massless Second Order Tetradic Spin 3 Fields And Higher Helicity Bosons,''
  Nuovo Cim.\  A {\bf 72} (1982) 149;\\
S.~Deser and Z.~Yang,
  ``Inconsistency Of Spin 4 - Spin-2 Gauge Field Couplings,''
  Class.\ Quant.\ Grav.\  {\bf 7} (1990) 1491.

\bibitem{haag}
R.~Haag, J.~T.~Lopuszanski and M.~Sohnius,
  ``All Possible Generators Of Supersymmetries Of The S Matrix,''
  Nucl.\ Phys.\  B {\bf 88}, 257 (1975).

\bibitem{nogo_new}
S.~Weinberg and E.~Witten,
  ``Limits On Massless Particles,''
  Phys.\ Lett.\  B {\bf 96} (1980) 59;\\
P.~Benincasa and F.~Cachazo,
  ``Consistency Conditions on the S-Matrix of Massless Particles,''
  arXiv:0705.4305 [hep-th];\\
M.~Porrati and R.~Rahman,
  ``Intrinsic Cutoff and Acausality for Massive Spin 2 Fields Coupled to Electromagnetism,''
  Nucl.\ Phys.\  B {\bf 801} (2008) 174
  [arXiv:0801.2581 [hep-th]],
  ``A Model Independent Ultraviolet Cutoff for Theories with Charged Massive Higher Spin Fields,''
  Nucl.\ Phys.\  B {\bf 814} (2009) 370
  [arXiv:0812.4254 [hep-th]];\\
M.~Porrati,
  ``Universal Limits on Massless High-Spin Particles,''
  Phys.\ Rev.\  D {\bf 78} (2008) 065016
  [arXiv:0804.4672 [hep-th]].

\bibitem{bbb}
A.~K.~H.~Bengtsson, I.~Bengtsson and L.~Brink,
  ``Cubic Interaction Terms For Arbitrary Spin,''
  Nucl.\ Phys.\  B {\bf 227} (1983) 31,
  ``Cubic Interaction Terms For Arbitrarily Extended Supermultiplets,''
  Nucl.\ Phys.\  B {\bf 227} (1983) 41.

\bibitem{vertices_light}
A.~K.~H.~Bengtsson, I.~Bengtsson and N.~Linden,
  ``Interacting Higher Spin Gauge Fields On The Light Front,''
  Class.\ Quant.\ Grav.\  {\bf 4} (1987) 1333.

\bibitem{bbvd1}
F.~A.~Berends, G.~J.~H.~Burgers and H.~Van Dam,
  ``On Spin Three Selfinteractions,''
  Z.\ Phys.\  C {\bf 24} (1984) 247.

\bibitem{bbvd2}
F.~A.~Berends, G.~J.~H.~Burgers and H.~van Dam,
  ``On The Theoretical Problems In Constructing Interactions Involving Higher Spin Massless Particles,''
  Nucl.\ Phys.\  B {\bf 260} (1985) 295.

\bibitem{bbvd3}
F.~A.~Berends, G.~J.~H.~Burgers and H.~van Dam,
  ``Explicit Construction Of Conserved Currents For Massless Fields Of Arbitrary Spin,''
  Nucl.\ Phys.\  B {\bf 271} (1986) 429.

\bibitem{vertices_brst}
I.~G.~Koh and S.~Ouvry,
  ``Interacting Gauge Fields Of Any Spin And Symmetry,''
  Phys.\ Lett.\  B {\bf 179} (1986) 115
  [Erratum-ibid.\  {\bf 183B} (1987) 434];\\
A.~K.~H.~Bengtsson,
  ``BRST Approach To Interacting Higher Spin Gauge Fields,''
  Class.\ Quant.\ Grav.\  {\bf 5} (1988) 437.

\bibitem{nogo_spin3}
A.~K.~H.~Bengtsson,
  ``On Gauge Invariance For Spin 3 Fields,''
  Phys.\ Rev.\  D {\bf 32} (1985) 2031;\\
A.~K.~H.~Bengtsson and I.~Bengtsson,
  ``Massless Higher Spin Fields Revisited,''
  Class.\ Quant.\ Grav.\  {\bf 3} (1986) 927.

\bibitem{interactions_recent}
G.~Bonelli,
  ``On the tensionless limit of bosonic strings, infinite symmetries and higher spins,''
  Nucl.\ Phys.\  B {\bf 669} (2003) 159
  [arXiv:hep-th/0305155];\\
X.~Bekaert, N.~Boulanger and S.~Cnockaert,
  ``Spin three gauge theory revisited,''
  JHEP {\bf 0601} (2006) 052
  [arXiv:hep-th/0508048];\\
N.~Boulanger, S.~Leclercq and S.~Cnockaert,
  ``Parity violating vertices for spin-3 gauge fields,''
  Phys.\ Rev.\  D {\bf 73} (2006) 065019
  [arXiv:hep-th/0509118];\\
I.~L.~Buchbinder, A.~Fotopoulos, A.~C.~Petkou and M.~Tsulaia,
  ``Constructing the cubic interaction vertex of higher spin gauge fields,''
  Phys.\ Rev.\  D {\bf 74} (2006) 105018
  [arXiv:hep-th/0609082];\\
Yu.~M.~Zinoviev,
  ``On massive spin 2 interactions,''
  Nucl.\ Phys.\  B {\bf 770} (2007) 83
  [arXiv:hep-th/0609170],
  ``On spin 2 electromagnetic interactions,''
  Mod.\ Phys.\ Lett.\  A {\bf 24} (2009) 17
  [arXiv:0806.4030 [hep-th]],
  ``On massive spin 2 electromagnetic interactions,''
  arXiv:0901.3462 [hep-th];\\
A.~Fotopoulos and M.~Tsulaia,
  ``Interacting Higher Spins and the High Energy Limit of the Bosonic String,''
  Phys.\ Rev.\  D {\bf 76} (2007) 025014
  [arXiv:0705.2939 [hep-th]];\\
A.~Fotopoulos, N.~Irges, A.~C.~Petkou and M.~Tsulaia,
  ``Higher-Spin Gauge Fields Interacting with Scalars: The Lagrangian Cubic Vertex,''
  JHEP {\bf 0710} (2007) 021
  [arXiv:0708.1399 [hep-th]];\\
X.~Bekaert, E.~Joung and J.~Mourad,
  ``On higher spin interactions with matter,''
  JHEP {\bf 0905} (2009) 126
  [arXiv:0903.3338 [hep-th]].

\bibitem{fv}
E.~S.~Fradkin and M.~A.~Vasiliev,
  ``On the Gravitational Interaction of Massless Higher Spin Fields,''
  Phys.\ Lett.\  B {\bf 189} (1987) 89,
  ``Cubic Interaction in Extended Theories of Massless Higher Spin Fields,''
  Nucl.\ Phys.\  B {\bf 291} (1987) 141.

\bibitem{seeds}
N.~Boulanger and S.~Leclercq,
  ``Consistent couplings between spin-2 and spin-3 massless fields,''
  JHEP {\bf 0611} (2006) 034
  [arXiv:hep-th/0609221];\\
Yu.~M.~Zinoviev,
  ``On spin 3 interacting with gravity,''
  Class.\ Quant.\ Grav.\  {\bf 26} (2009) 035022
  [arXiv:0805.2226 [hep-th]];\\
N.~Boulanger, S.~Leclercq and P.~Sundell,
  ``On The Uniqueness of Minimal Coupling in Higher-Spin Gauge Theory,''
  JHEP {\bf 0808} (2008) 056
  [arXiv:0805.2764 [hep-th]].

\bibitem{HSalgebra}
E.~S.~Fradkin and M.~A.~Vasiliev,
  ``Candidate To The Role Of Higher Spin Symmetry,''
  Annals Phys.\  {\bf 177} (1987) 63;\\
M.~A.~Vasiliev,
  ``Free Massless Fields Of Arbitrary Spin In The De Sitter Space And Initial Data For A Higher Spin Superalgebra,''
  Fortsch.\ Phys.\  {\bf 35} (1987) 741
  [Yad.\ Fiz.\  {\bf 45} (1987) 1784],
  ``Consistent Equations For Interacting Massless Fields Of All Spins In The First Order In Curvatures,''
  Annals Phys.\  {\bf 190} (1989) 59.

\bibitem{ss}
E.~Sezgin and P.~Sundell,
  ``On curvature expansion of higher spin gauge theory,''
  Class.\ Quant.\ Grav.\  {\bf 18} (2001) 3241
  [arXiv:hep-th/0012168],
  ``Analysis of higher spin field equations in four dimensions,''
  JHEP {\bf 0207} (2002) 055
  [arXiv:hep-th/0205132].

\bibitem{solutions}
E.~Sezgin and P.~Sundell,
  ``An exact solution of 4D higher-spin gauge theory,''
  Nucl.\ Phys.\  B {\bf 762} (2007) 1
  [arXiv:hep-th/0508158];\\
C.~Iazeolla, E.~Sezgin and P.~Sundell,
  ``Real Forms of Complex Higher Spin Field Equations and New Exact Solutions,''
  Nucl.\ Phys.\  B {\bf 791} (2008) 231
  [arXiv:0706.2983 [hep-th]];\\
V.~E.~Didenko and M.~A.~Vasiliev,
  ``Schwarzschild black hole in 4d higher-spin gauge theory,''
  arXiv:0906.3898 [hep-th].

\bibitem{mixed_nogo}
X.~Bekaert, N.~Boulanger and M.~Henneaux,
  ``Consistent deformations of dual formulations of linearized gravity: A no-go result,''
  Phys.\ Rev.\  D {\bf 67} (2003) 044010
  [arXiv:hep-th/0210278];\\
N.~Boulanger and S.~Cnockaert,
  ``Consistent deformations of (p,p)-type gauge field theories,''
  JHEP {\bf 0403} (2004) 031
  [arXiv:hep-th/0402180];\\
X.~Bekaert, N.~Boulanger and S.~Cnockaert,
  ``No self-interaction for two-column massless fields,''
  J.\ Math.\ Phys.\  {\bf 46} (2005) 012303
  [arXiv:hep-th/0407102].

\bibitem{scattering}
D.~Amati, M.~Ciafaloni and G.~Veneziano,
  ``Superstring Collisions at Planckian Energies,''
  Phys.\ Lett.\  B {\bf 197} (1987) 81,
  ``Classical and Quantum Gravity Effects from Planckian Energy Superstring Collisions,''
  Int.\ J.\ Mod.\ Phys.\  A {\bf 3} (1988) 1615,
  ``Can Space-Time Be Probed Below The String Size?,''
  Phys.\ Lett.\  B {\bf 216} (1989) 41;\\
D.~J.~Gross and P.~F.~Mende,
  ``The High-Energy Behavior of String Scattering Amplitudes,''
  Phys.\ Lett.\  B {\bf 197} (1987) 129,
  ``String Theory Beyond the Planck Scale,''
  Nucl.\ Phys.\  B {\bf 303} (1988) 407;\\
D.~J.~Gross,
  ``High-Energy Symmetries Of String Theory,''
  Phys.\ Rev.\ Lett.\  {\bf 60} (1988) 1229;\\
D.~J.~Gross and J.~L.~Manes,
  ``The High-Energy Behavior Of Open String Scattering,''
  Nucl.\ Phys.\  B {\bf 326} (1989) 73;\\
N.~Moeller and P.~C.~West,
  ``Arbitrary four string scattering at high energy and fixed angle,''
  Nucl.\ Phys.\  B {\bf 729} (2005) 1
  [arXiv:hep-th/0507152].

\bibitem{maldacena}
J.~M.~Maldacena,
  ``The large N limit of superconformal field theories and supergravity,''
  Adv.\ Theor.\ Math.\ Phys.\  {\bf 2} (1998) 231
  [Int.\ J.\ Theor.\ Phys.\  {\bf 38} (1999) 1113]
  [arXiv:hep-th/9711200];\\
S.~S.~Gubser, I.~R.~Klebanov and A.~M.~Polyakov,
  ``Gauge theory correlators from non-critical string theory,''
  Phys.\ Lett.\  B {\bf 428} (1998) 105
  [arXiv:hep-th/9802109];\\
E.~Witten,
  ``Anti-de Sitter space and holography,''
  Adv.\ Theor.\ Math.\ Phys.\  {\bf 2} (1998) 253
  [arXiv:hep-th/9802150].

\bibitem{sundborg}
B.~Sundborg,
  ``Stringy gravity, interacting tensionless strings and massless higher spins,''
  Nucl.\ Phys.\ Proc.\ Suppl.\  {\bf 102} (2001) 113
  [arXiv:hep-th/0103247];\\
E.~Sezgin and P.~Sundell,
  ``Doubletons and 5D higher spin gauge theory,''
  JHEP {\bf 0109} (2001) 036
  [arXiv:hep-th/0105001],
  ``Massless higher spins and holography,''
  Nucl.\ Phys.\  B {\bf 644} (2002) 303
  [Erratum-ibid.\  B {\bf 660} (2003) 403]
  [arXiv:hep-th/0205131].

\bibitem{anselmi}
D.~Anselmi,
  ``Theory of higher spin tensor currents and central charges,''
  Nucl.\ Phys.\  B {\bf 541} (1999) 323
  [arXiv:hep-th/9808004].

\bibitem{bouffe}
A.~Mikhailov,
  ``Notes on higher spin symmetries,''
  arXiv:hep-th/0201019;\\
I.~R.~Klebanov and A.~M.~Polyakov,
  ``AdS dual of the critical O(N) vector model,''
  Phys.\ Lett.\  B {\bf 550} (2002) 213
  [arXiv:hep-th/0210114];\\
M.~Bianchi, J.~F.~Morales and H.~Samtleben,
  ``On stringy AdS(5) x S**5 and higher spin holography,''
  JHEP {\bf 0307} (2003) 062
  [arXiv:hep-th/0305052];\\
N.~Beisert, M.~Bianchi, J.~F.~Morales and H.~Samtleben,
  ``On the spectrum of AdS/CFT beyond supergravity,''
  JHEP {\bf 0402} (2004) 001
  [arXiv:hep-th/0310292],
  ``Higher spin symmetry and N = 4 SYM,''
  JHEP {\bf 0407} (2004) 058
  [arXiv:hep-th/0405057];\\
G.~Bonelli,
  ``On the boundary gauge dual of closed tensionless free strings in AdS,''
  JHEP {\bf 0411} (2004) 059
  [arXiv:hep-th/0407144];\\
M.~Bianchi,
  ``Higher spins and stringy AdS(5) x S(5),''
  Fortsch.\ Phys.\  {\bf 53} (2005) 665
  [arXiv:hep-th/0409304];\\
A.~C.~Petkou,
  ``Holography, duality and higher-spin theories,''
  arXiv:hep-th/0410116.

\bibitem{review_SIGRAV}
A.~Campoleoni,
  ``Lagrangian formulations for Bose and Fermi higher-spin fields of mixed symmetry,''
  arXiv:0905.1472 [hep-th].

\bibitem{reduction}
T.~R.~Govindarajan, S.~D.~Rindani and M.~Sivakumar,
  ``Dimensional Reduction And Theories With Massive Gauge Fields,''
  Phys.\ Rev.\  D {\bf 32} (1985) 454;\\
S.~D.~Rindani and M.~Sivakumar,
  ``Gauge - Invariant Description Of Massive Higher - Spin Particles By Dimensional Reduction,''
  Phys.\ Rev.\  D {\bf 32} (1985) 3238;\\
C.~Aragone, S.~Deser and Z.~Yang,
  ``Massive Higher Spin From Dimensional Reduction Of Gauge Fields,''
  Annals Phys.\  {\bf 179} (1987) 76;\\
S.~D.~Rindani, D.~Sahdev and M.~Sivakumar,
  ``Dimensional reduction of symmetric higher spin actions. 1. Bosons,''
  Mod.\ Phys.\ Lett.\  A {\bf 4} (1989) 265,
  ``Dimensional Reduction Of Symmetric Higher Spin Actions. 2: Fermions,''
  Mod.\ Phys.\ Lett.\  A {\bf 4} (1989) 275.

\bibitem{siegel_count}
W.~Siegel,
  ``Hidden Ghosts,''
  Phys.\ Lett.\  B {\bf 93} (1980) 170.

\bibitem{group}
M.~Hamermesh,
  ``Group theory and its applications to physical problems''
  (Dover Publications, New York, NY, USA, 1969);\\
I.~V.~Schensted,
   ``A course on the application of group theory to Quantum Mechanics''
   (NEO Press, Peaks Island, ME, USA, 1976);\\
B.E.~Sagan.
   ``The symmetric group,'' second edition
   (Springer-Verlag, New York, NY, USA, 2001).\\
See also: X.~Bekaert and N.~Boulanger,
  ``The unitary representations of the Poincare group in any spacetime dimension,''
  arXiv:hep-th/0611263.

\bibitem{plethysm}
R.~C.~King,
  ``Branching rules for classical Lie groups using tensor and spinor methods,''
  J. Phys. A: Math. Gen. {\bf 8} (1975) 429.

\bibitem{sorokvas}
   D.~P.~Sorokin and M.~A.~Vasiliev,
  ``Reducible higher-spin multiplets in flat and AdS spaces and their geometric frame-like formulation,''
  Nucl.\ Phys.\  B {\bf 809} (2009) 110
  [arXiv:0807.0206 [hep-th]].

\bibitem{multiforms}
M.~Dubois-Violette and M.~Henneaux,
  ``Tensor fields of mixed Young symmetry type and N-complexes,''
  Commun.\ Math.\ Phys.\  {\bf 226} (2002) 393
  [arXiv:math/0110088];\\
P.~de Medeiros and C.~Hull,
  ``Exotic tensor gauge theory and duality,''
  Commun.\ Math.\ Phys.\  {\bf 235} (2003) 255
  [arXiv:hep-th/0208155].


\end{thebibliography}
\end{document}